\documentclass[10pt,leqno]{amsart}
\usepackage{amsmath,amsthm,amsfonts}
\usepackage{latexsym}
\usepackage{amssymb}
\usepackage[dvips]{epsfig}

\renewcommand{\thesection}{\arabic{section}}

\newtheorem{theorem}{Theorem}[section]
\newtheorem{lemma}[theorem]{Lemma}
\newtheorem{prop}[theorem]{Proposition}
\newtheorem{defi}[theorem]{Definition}

\newtheorem{corollary}[theorem]{Corollary}
\theoremstyle{remark}
\newtheorem{remark}[theorem]{Remark}

\renewcommand{\theequation}{\thesection .\arabic{equation}}
\let\subs\subsection
\renewcommand\subsection{\setcounter{equation}{0}
\gdef\theequation{\thesubsection \arabic{equation}}\subs}
\let\sect\section
\renewcommand\section{\setcounter{equation}{0}
\gdef\theequation{\thesection .\arabic{equation}}\sect}

\newcommand{\cA}{{\mathcal{A}}}
\newcommand{\cB}{{\mathcal{B}}}
\newcommand{\cD}{{\mathcal{D}}}
\newcommand{\cE}{{\mathcal{E}}}
\newcommand{\cG}{{\mathcal{G}}}

\newcommand{\cJ}{{\mathcal{J}}}
\newcommand{\cN}{{\mathcal{N}}}

\newcommand{\cW}{{\mathcal{W}}}

\newcommand{\cT}{{\mathcal{T}}}
\newcommand{\cS}{{\mathcal{S}}}
\newcommand{\cF}{{\mathcal{F}}}
\newcommand{\cK}{{\mathcal{K}}}
\newcommand{\cL}{{\mathcal{L}}}
\newcommand{\cP}{{\mathcal{P}}}

\newcommand{\IC}{{\mathbb{C}}}

\newcommand{\IR}{{\mathbb{R}}}
\newcommand{\TT}{{\mathbb{T}}}

\newcommand{\tor}{\TT}
\newcommand{\ZZ}{{\mathbb{Z}}}
\newcommand{\IZ}{{\mathbb{Z}}}
\newcommand{\Laplace}{\triangle}

\newcommand{\ue}{{\underline{e}}}
\newcommand{\ulm}{{\underline{m}}}
\newcommand{\ur}{{\underline{r}}}
\newcommand{\uu}{{\underline{u}}}
\newcommand{\uv}{{\underline{v}}}

\newcommand{\uw}{{\underline{w}}}
\newcommand{\uz}{{\underline{z}}}
\newcommand{\up}{\uu^+}
\newcommand{\um}{\uu^-}
\newcommand{\vp}{\uv^+}
\newcommand{\vm}{\uv^-}
\newcommand{\bp}{b^{(+,+)}}
\newcommand{\bmp}{b^{(-,+)}}
\newcommand{\bpm}{b^{(+,-)}}
\newcommand{\bmm}{b^{(-,-)}}

\newcommand{\be}{\begin{eqnarray}}
\newcommand{\ee}{\end{eqnarray}}

\newcommand{\BMO}{{\rm{BMO}}}
\newcommand{\diam}{\mathop{\rm{diam}}}
\newcommand{\degg}{\mathop{\rm{deg}}}
\newcommand{\dist}{\mathop{\rm{dist}}}
\newcommand{\disc}{\mathop{\rm{disc}}}

\newcommand{\mes}{\mathop{\rm{mes}\, }}
\newcommand{\compl}{\mathop{\rm{compl}}}
\renewcommand{\mod}{{\rm{mod}\, }}
\newcommand{\rsp}{\mathop{\rm{sp}}}

\newcommand{\tr}{\mathop{\rm{tr}}}
\newcommand{\Res}{\mathop{\rm{Res}}}
\newcommand{\Ree}{\mathop{\rm{Re}}}
\newcommand{\car}{\mathop{\rm{Car}}\nolimits}
\newcommand{\imm}{\mathop{\rm{Im}}}
\newcommand{\spec}{\mathop{\rm{spec}}}
\newcommand{\const}{\mathop{\rm{const}}}

\newcommand{\tilw}{\tilde{w}}

\newcommand{\capo}{\cA_{\rho_0}}

\newcommand{\vep}{{\varepsilon}}

\newcommand{\vz}{\varphi(z)}
\newcommand{\bs}{{\backslash}}

\newcommand{\doppelint}{{\displaystyle{\int\!\!\!\!\int}}}

\newcommand{\strich}{-\!\!\!\!-\!\!\!\!-}
\newcommand{\niint}{{{\strich\!\!\!\!\!\!\!\!\!\doppelint}}}
\newcommand{\strichint}{\mathop{\niint}}
\newcommand{\nn}{\nonumber}
\newcommand{\rank}{\mathop{\rm rank}}

\newcommand{\notint}{{-\!\!\!\!\!\!\int}}
\newcommand{\nint}{\mathop{\notint}}
\newcommand{\la}{\langle}
\newcommand{\ra}{\rangle}
\def\beeq{\begin{equation}}
\def\eneq{\end{equation}}
\def\eps{\varepsilon}
\def\les{\lesssim}

\def\cI{{\mathfrak{I}}}
\def\cZ{{\mathcal Z}}
\def\bm{\begin{matrix}}
\def\endm{\end{matrix}}
\def\cS{{\mathcal S}}
\def\mape{\bigl(e(x),\omega, E+i\eta\bigr)}
\def\mapen{\bigl(e(x), \omega\bigr)- E-i\eta}
\def\Mat{\left[\begin{matrix} 1&0\\0&0\end{matrix}\right]}

\begin{document}
\title[Fine properties of the IDS and separation of eigenvalues]{Fine
properties of the integrated density of states and a quantitative
separation property of the Dirichlet eigenvalues}

\author{Michael Goldstein and Wilhelm Schlag}

\address{Dept.\ of Mathematics, University of Toronto, Toronto, Ontario, Canada M5S 1A1}

\email{gold@math.toronto.edu}

\address{The University of Chicago, Department of Mathematics, 5734 South University Avenue, Chicago, IL 60637, U.S.A.}

\email{schlag@math.uchicago.edu}

\thanks{The first author was partially supported by an NSERC grant. The second author was
partially supported by the NSF, DMS-0300081, and a Sloan fellowship.
The authors wish to thank Jossi Avron, Yakov Sinai, and Thomas Spencer for helpful
discussions.
Part of this work was
done at the ESI in Vienna, the IAS in Princeton, at Caltech, and at the University of Toronto.
The authors are grateful to these institutions for their hospitality.}

\date{}

\begin{abstract}
We consider one-dimensional difference Schr\"odinger equations
\[ [H(x,
\omega)\varphi](n) \equiv -\varphi(n-1)- \varphi(n+1) + V(x +
n\omega)\varphi(n) = E\varphi(n),
\] $n \in \ZZ$, $x, \omega \in [0, 1]$
with real analytic function $V(x)$.  Suppose $V(x)$ is a small
perturbation of a trigonometric polynomial $V_0(x)$ of degree $k_0$,
and assume positive Lyapunov exponents and Diophantine $\omega$. We
prove that the  integrated density of states $\cN$ is H\"older
$\frac{1}{2k_0}-\kappa$ continuous for any $\kappa>0$. Moreover, we
show that $\cN$ is absolutely continuous for a.e.~$\omega$. Our
approach is via finite volume bounds. I.e., we study the eigenvalues
of the problem $H(x, \omega)\varphi = E\varphi$ on a finite interval
$[1, N]$ with Dirichlet boundary conditions.  Then the averaged
number of these Dirichlet eigenvalues which fall into an interval
$(E-\eta, E+\eta)$ with $\eta \asymp N^{-1+\delta}$, $0<\delta \ll
1$ does not exceed $N \eta^{\frac{1}{2k_0}-\kappa}$, $\kappa>0$.
Moreover, for $\omega \notin \Omega(\vep)$, $\mes \Omega(\vep) <
\vep$ and $E \notin \cE_\omega(\vep)$, $\mes \cE_\omega(\vep) <
\vep$, this averaged number does not exceed $\exp\left((\log
\vep^{-1})^A\right)\eta N$, for any $\eta > N^{-1+b}$, $b > 0$. For
the integrated density of states $\cN(\cdot)$ of the problem
$H(x,\omega)\varphi = E\varphi$ this implies that $\cN(E + \eta) -
\cN(E - \eta) \le \exp\left((\log \vep^{-1})^A\right)\eta$ for any
$E \notin \cE_\omega(\vep)$. To investigate the distribution of the
Dirichlet eigenvalues of $H(x, \omega) \varphi = E\varphi$ on a
finite interval $[1, N]$ we study the distribution of the zeros of
the characteristic determinants $f_N(\cdot, \omega, E)$ with
complexified phase $x$, and frozen $\omega, E$.  We prove
equidistribution of these zeros in some annulus $\cA_\rho= \left\{z
\in \IC: 1 - \rho < |z| < 1 + \rho\right\}$ and show also that no
more than $2k_0$ of them fall into any disk of radius
$\exp\left(-(\log N)^A\right)$, $A \gg 1$. In addition, we obtain
the lower bound $e^{-N^\delta}$  (with $\delta>0$ arbitrary) for the
separation of the eigenvalues of the Dirichlet eigenvalues over the
interval $[0,N]$. This necessarily requires the removal of a small
set of energies.
\end{abstract}

\maketitle

\section{Introduction and statement of the main
results}\label{sec:intro}

During the last few years several methods based on averages of
subharmonic functions have been developed for quasi-periodic
Schr\"odinger equations on $\ZZ^1$.  These methods have also been
applied to Schr\"odinger equations on $\ZZ^1$ with a potential given
by values of a real function along the trajectories of the skew
shift on the two--dimensional torus $\TT^2$, see \cite{BG},
\cite{GS}, \cite{BGS}.  These equations as well as more general
Schr\"odinger equations with potentials defined by  some dynamical
system are recognized as relevant object in the theory of quantum
disordered systems starting from the famous work by
Anderson~\cite{And} and Harper~\cite{Har} (see for instance the
monographs \cite{Bou2}, \cite{FP}, \cite{CFKS}, \cite{CL} for more
history). It was realized by Sinai that the phenomenon discovered in
Anderson's work suggested a  mathematical program related to
fundamental problems in disordered and dynamical systems, KAM
theory, and analysis. The papers by Dinaburg and Sinai \cite{DS},
Goldsheid, Molchanov, Pastur \cite{GMP}, Fr\"ohlich, Spencer
\cite{FS1}, \cite{FS2},  Sinai \cite{Si1}, Fr\"ohlich, Spencer,
Wittwer \cite{FSW} established a series of fundamental results in
the rigorous theory of this phenomenon which are now considered
classical. Several other important contributions to this area can be
found in the references to this paper. The methods of \cite{BG},
\cite{GS}, \cite{BGS} allow one to deduce some
 information about the eigenvalues and eigenfunctions of these difference
 equations.  In particular, in the quasi-periodic case we know due
to these methods that positive Lyapunov exponents lead to exponential
decay of the corresponding eigenfunctions which is called {\it Anderson
localization}.  However, even in the simplest case
of a one-dimensional shift we have no satisfactory description of the mechanism
which forces the Lyapunov exponent to be positive, let alone why this mechanism fails
in certain regimes when the Lyapunov exponent deteriorates.  Moreover,
we do not have satisfactory answers to this question in the most
studied case of the equation discovered by Harper \cite{Har}, called also
almost Mathieu equation, which is
\begin{equation}
\label{eq:1.1}
-\varphi(n+1) - \varphi(n-1) + \lambda\cos (2\pi(x +n\omega)) \varphi(n) =
E\varphi(n),\quad n \in \ZZ\ .
\end{equation}
These questions appear to be relevant to the further development of the
theory of Schr\"odinger equations with a potential defined by some
dynamical system.  Among the goals of such a development we would like to mention
the positivity of the Lyapunov exponent for any disorder in the case
of the skew shift, as well as positivity of the Lyapunov exponent for
potentials given by the so-called standard map, see \cite{Si3}.

We believe  that the central  object underlying
the mechanism responsible for the
positivity of the Lyapunov exponent consists of the so-called {\it integrated
density of states\/}. It is defined as follows:  Consider the one-dimensional
difference Schr\"odinger equation
\begin{equation}
\label{eq:1.2}
-\varphi(n+1)-\varphi(n-1) + \lambda v(n) \varphi(n) = E\varphi(n),\quad
n \in \ZZ^1\ ,
\end{equation}
here $\varphi(n)$, $n \in \ZZ$, is an unknown function, $v(n)$ is a given
real function called the potential, $\lambda$ is a real parameter, $E$
is a spectral parameter.  Assume that $v(n)$ is given by a
measure preserving ergodic transformation $T: X \to X$ of a measure
space $(X, \mu)$, i.e. $v(n) = v(n, x) = V(T^n x)$, $n \in \ZZ$, $x
\in X$, where $V(x)$ is a real function on $X$.  Let $E_{\Lambda, j}(x,
\lambda)$, $j = 1, 2, \dots, |\Lambda|$ be the eigenvalues of equation
(\ref{eq:1.2}) with $v(n) = v(n, x)$ on a finite interval $\Lambda = [a, b]
\in \ZZ$ with zero boundary conditions $\varphi(a-1) = \varphi(b+1) =
0$, $|\Lambda| = b-a +1$.  The distributions
\begin{equation}
\label{eq:1.3}
\cN_\Lambda(E,\lambda) = |\Lambda|^{-1} \sum_{E_{\Lambda, j}(x, \lambda) < E}\ 1
\end{equation}
converge to some distribution $\cN(d\cdot, \lambda)$ on $\IR$ when
$a \to - \infty$, $b \to +\infty$.  This limiting distribution does not
depend on $x \in X$ for a.~a. $x \in X$, and it is called the integrated
density of states (IDS), see \cite{CFKS} for details.  Apart from its
significance for the description of the quantum
system, the integrated density of states is related in a simple manner
to the Lyapunov exponent of \eqref{eq:1.2} and also
a key object in the spectral problem for (\ref{eq:1.2}).  The Lyapunov exponent
of (\ref{eq:1.2}) is defined as follows.  Given the initial data $\varphi(0),
\varphi(1)$, the solution of the difference equation (\ref{eq:1.2}) for $n > 0$ with
these data can be expressed in the form
\begin{equation}
\nn
\begin{bmatrix}
\varphi(n+1)\\ \varphi(n)\end{bmatrix} = M_n \begin{bmatrix}
\varphi(1)\\ \varphi(0)\end{bmatrix}\ ,
\end{equation}
where $M_n$ is the so-called monodromy matrix
\begin{equation}
\label{eq:Mnprod}
M_n = \prod^1_{k=n} A_k\ ,\quad A_k = \begin{bmatrix}
\lambda v(k) -E & -1\\
1 & 0\end{bmatrix}\ .
\end{equation}
For the case $v(n) = V\left(T^n x\right)$, $n \in \ZZ$, $x \in X$ with
an ergodic measure preserving automorphism $T$, the following limit
\begin{equation}
\label{eq:lap}
L(E) = \lim_{n\to + \infty} {1\over n} \log \|M_n(x, E)\| = \lim_{n\to
\infty} \int_X {1\over n} \log\|M_n(\xi, n)\| d\xi
\end{equation}
exists for a.~a. $x \in X$ by the F\"urstenberg-Kesten theorem~\cite{FurKes} and it is called
the {\em Lyapunov exponent}.
The
relation between the Lyapunov exponent and the IDS is given by the Thouless
formula
\begin{equation}\nn
L(E) = \int \log |E - E_1| \cN(dE_1)
\end{equation}

In basic terms, the idea behind the approach introduced in
\cite{BG} and \cite{GS} is as follows.  Regularity properties of
the IDS $\cN(E)$ can be studied by analyzing  the function $u_n(x,
E) = n^{-1} \log \|M_n(x,E)\|$ in the $x$-variable for fixed $E$.
This is due to the fact that the variable $E$ enters here as a
spectral parameter of the linear problem (\ref{eq:1.2}) on a
finite intervals $[a, b]$, while the entries of the monodromy
$M_n$ are as follows:
\begin{equation}
\label{eq:Mn_def}
M_n = \begin{bmatrix}
f_{[1, n]} (E) & - f_{[2, n]} (E)\\
f_{[1, n-1]} (E) & - f_{[2, n-1]}(E)\end{bmatrix}
\end{equation}
Here $f_{[a, b]}(E)$ stands for the characteristic polynomial of
the problem (\ref{eq:1.2}) on the interval $[a, b]$ with zero boundary conditions
$\psi(a-1)=0$, $\psi(b+1) =0$, i.e.,
\begin{equation}
\label{eq:f_def}
f^{(D)}_{[a,b]}(E)\equiv f_{[a,b]}(E) = \begin{vmatrix}
v(a) -E & -1 & 0 & \cdots & 0\\
-1 & v(a+1)-E & -1 & \cdots & 0\\
\\
\\
0 & \cdots & 0 & -1 & v(b) -E\end{vmatrix}
\end{equation}
The main information on $u_n(x)$ used here consists of certain
estimates on the measure of the deviations of $u_n(x)$ from its average
in $X$.  These are the so-called large deviation theorems introduced in
\cite{BG} and \cite{GS}.

\begin{figure}[ht]
\centerline{\hbox{\vbox{ \epsfxsize= 13.0 truecm \epsfysize= 13.0
truecm \epsfbox{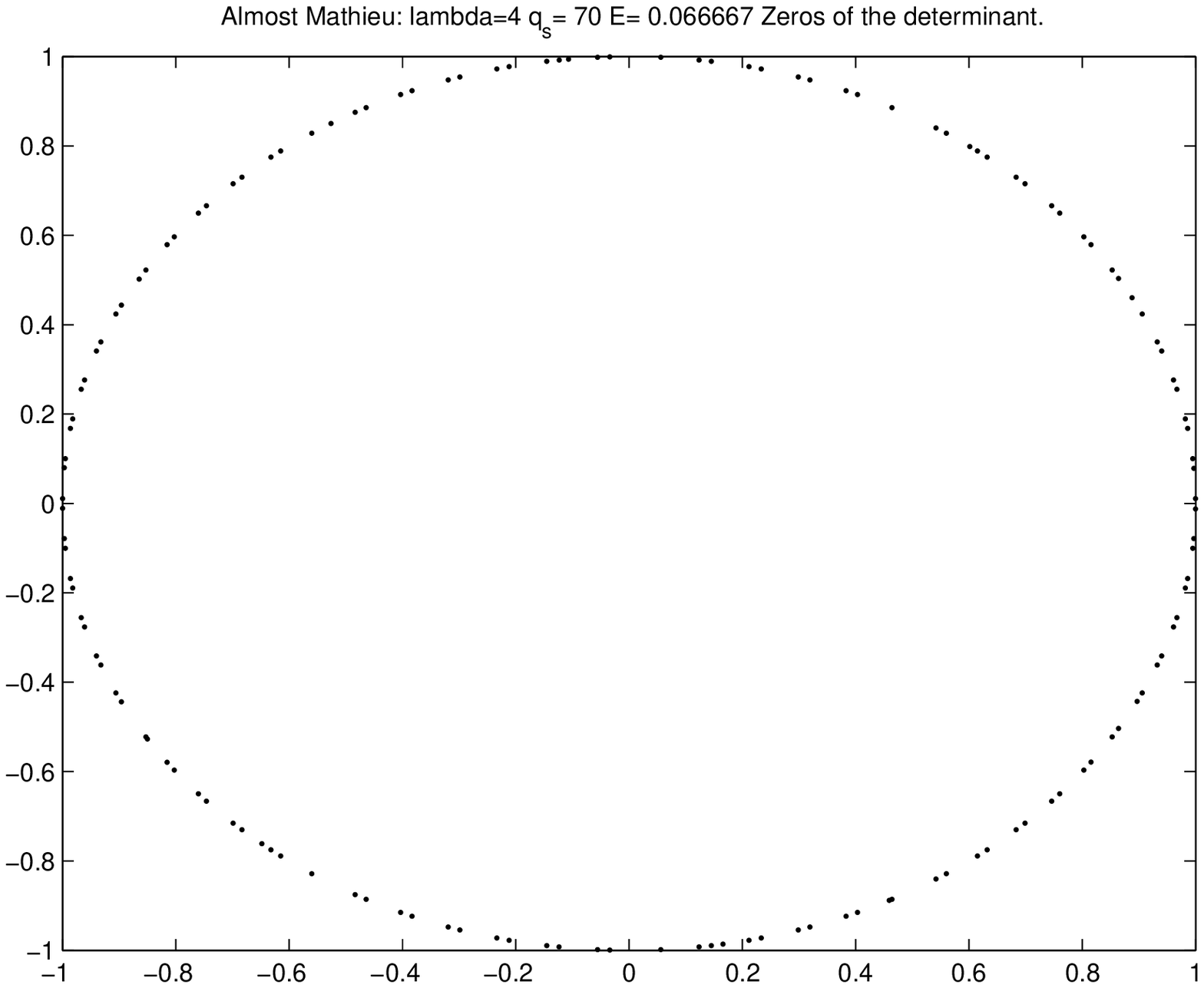}}}} 
\label{fig:1}
\end{figure}

In this paper we develop these large deviation theorems further and thus achieve
a higher level of resolution of these methods.
Moreover, such theorems are required not just for the norm of the matrix $M_n$
but rather for its entries, i.e., for the characteristic determinants.
These large deviation theorems are intimately related to the distribution of the
zeros of the function $f_{[1,n]}$ in the phase variable. In fact, these zeros turn
out to be relatively uniformly distributed along certain circles.
Figures~\ref{fig:1} and~2 are examples of
the zeros of the determinants
$f_{[1, n]}(z, E)$ for the almost Mathieu problem (\ref{eq:1.1}) in the complex
$z$--plane, where $z = e(x) \equiv \exp(2\pi ix)$ with complexified
phase $x$. In these pictures we chose $\omega=\sqrt{2}$ and $n=q_s=70$ the denominator of a convergent
of~$\omega$, as well as $\lambda=4$ and two different values of~$E$.

We now state our main results. Throughout this paper, we will be considering the measure
and complexities of sets $\cS\subset\IC$. More precisely,
\[ \mes(\cS)\le \alpha,\quad \compl(\cS)\le A\]
means that there is $\tilde\cS$ with $\cS\subset\tilde \cS\subset\IC$ with the
property that \[\cS=\bigcup_{j=1}^A\cD(z_j,r_j), \qquad \sum_{j=1}^A \mes(\cD(z_j,r_j))\le \alpha. \]
Here $\cD$ is a disk in the complex plane. Another piece of notation is $\tor_{c,a}$.
This refers to the set of all $\omega\in\tor$ which satisfy the Diophantine condition~\eqref{eq:diophant}.
Many of the constants appearing in this paper depend on the parameters $a,c$.

\begin{theorem}
\label{thm:1} Let $V_0\bigl(e(x)\bigr) = \sum\limits^{k_0}_{-k_0}
v(k) e(kx)$ be a trigonometric polynomial, $v(-k) =
\overline{v(k)}$, $-k_0 \le k \le k_0$. Let $L(E, \omega_0)$ be the
Lyapunov exponent defined as in (\ref{eq:lap}) for $V = V_0$ and
some $\omega_0 \in \tor_{c,a}$. Assume that it exceeds $\gamma_0$
for all $E \in (E',E'')$.
\begin{enumerate}
\item[{\rm{(1)}}]
Given $\rho_0 > 0$ there exists $\tau_0 = \tau_0
(\lambda,V_0,\omega_0,\gamma_0,\rho_0)$ with the following property:
 for any 1-periodic, analytic function $V(e(x+iy))$, $- \rho_0 <
y < \rho_0 $ assuming real values when $y=0$ and  deviating from
$V_0(e(x))$ by at most $\tau_0$,
 any $\omega
\in \tor_{c,a} \cap (\omega_0 - \tau_0,\omega_0 + \tau_0) $, and any
$E \in (E',E'')$, with $\eta = N^{-1+\delta}$, $\delta \ll 1$, $N
\gg 1$, one has
\end{enumerate}
\begin{equation}
\int_{\tor} \#\left(\rsp\bigl(H_N(x, \omega) \bigr) \cap \bigl(E -
\eta, E + \eta\bigr)\right) dx   \le \eta^{\frac{1}{2k_0}-\eps}
\cdot N \label{eq:1.8}
\end{equation}
\begin{enumerate}
\item[]
with some constant $1 \ll B$ and arbitrary $\eps>0$.

\item[{\rm{(2)}}] The IDS  $\cN(\cdot)$ satisfies, for any small
$\eps>0$,
$$
\cN(E+ \eta) - \cN(E - \eta) \le \eta^{\frac{1}{2k_0}-\eps}\ ,
$$
for all $E\in (E',E'')$ and all small $\eta>0$.
\end{enumerate}
\end{theorem}

For the case of the almost Mathieu equation~\eqref{eq:1.1} (which
corresponds to $k_0=1$) and large $\lambda$, Bourgain~\cite{Bou} had
previously obtained a H\"older-$(\frac12-\eps)$ result for the IDS,
which is known to be optimal in those regimes, see~\cite{Si1}. See
also~\cite{Bou2}. Next, we show that the IDS is  Lipschitz at most
energies.

\begin{theorem}
\label{thm:2} Let $V(x)$ be real analytic.  Assume $L(\omega_0, E)
\ge \gamma_0 > 0$ for some $\omega_0 \in \tor_{c,a}$ and all $E \in
(E', E'')$ and fix $b>0$ small.  There exist $N_0 = N_0(\lambda , V,
\gamma_0,b,c,a)$ , $\tau_0 = \tau_0(\lambda, V, \gamma_0,b,c,a) > 0$
so that:

For any $\vep > 0$, there exists $\Omega(\vep) \subset \tor$, $\mes
\Omega(\vep) < \vep$ such that for any $\omega \in (\omega_0 -
\tau_0, \omega_0 + \tau_0) \cap \left(\tor_{c,a} \setminus
\Omega(\vep)\right)$, there exists $\cE_\omega(\vep) \subset \IR$,
$\mes \cE_\omega(\vep) < \vep$ such that for any $N > N_0$ and any
$E \in (E', E'') \setminus \cE_\omega(\vep)$ and any $\eta >
1/N(\log N)^{1+b}$, one has
\begin{equation}
\label{eq:1.9} \int_\tor \#\left(\rsp\left(H_N(x, \omega)\right)\cap
\left(E - \eta, E+\eta\right)\right)\,dx \le \exp\left((\log
\vep^{-1})^A\right)\eta N\ .
\end{equation}
In particular, the IDS satisfies
\begin{equation}\nn
 \cN(E + \eta) - \cN(E - \eta) \le \exp\left((\log
\vep^{-1})^A\right)\eta
\end{equation}
for any $E \in (E', E'') \setminus \cE_\omega(\vep)$, $\eta > 0$.
\end{theorem}

The proof of Theorem~\ref{thm:2} establishes the estimate
\eqref{eq:1.9} for any $E \in \IR \setminus \cE_{\omega}(\vep)$,
with very detailed description of $\cE_{\omega}(\vep)$ as a union of
intervals of different scales. That allows one to combine the
estimate \eqref{eq:1.9} with (\ref{eq:1.8}) of Theorem~\ref{thm:1}
to prove the following

\begin{theorem}
\label{thm:3} As before, assume that the Lyapunov exponent is
positive on $(E',E'')$. Then for almost all $\omega \in (\omega_0 -
\tau_0, \omega_0 + \tau_0)$ the IDS $\cN(E)$ is absolutely
continuous on $(E', E'')$. In particular, if $L(\omega_0, E)
> 0$ for any $E$, then $\cN(E)$ is absolutely continuous everywhere.
\end{theorem}

The proof of this theorem proceeds by combining the Lipschitz bound
of Theorem~\ref{thm:2} with the H\"older bound of
Theorem~\ref{thm:1}. Note that this requires detailed information on
the size and complexity of those sets on which the IDS is not
Lipschitz (the {\em exceptional set}) as can be seen from the
example of a Cantor staircase function. Indeed, in that case there
is a uniform H\"older bound with an exponent that equals the
Hausdorff dimension of the Cantor set. However, in our case the
exceptional set has Hausdorff dimension zero, whereas the H\"older
exponent is fixed and positive.

\begin{theorem}
\label{thm:4} Using the notations of Theorem \ref{thm:1} there
exists $k_0(\lambda, V) \le 2 \degg V_0$ with the following
property: for all $E \in \IR$, $s\in\ZZ$
 and $\omega \in
\tor_{c,a}$ and any $x_0 \in \tor$ there exists $s^{-},
s^{+}$ with $|s-s^{\pm}| < \exp\bigl((\log s)^{\delta}\bigr)$ such that the Dirichlet
determinant $f_{[-s^-, s^+]}(\cdot, \omega, E)$ has no more that
$k_0(\lambda, V)$ zeros in $\cD\bigl(e(x_0), r_0\bigr)$,
$r_0 \asymp \exp\left(-(\log s)^A\right)$.
\end{theorem}


%

\begin{figure}[ht]
\centerline{\hbox{\vbox{ \epsfxsize= 13.0 truecm \epsfysize= 13.0
truecm \epsfbox{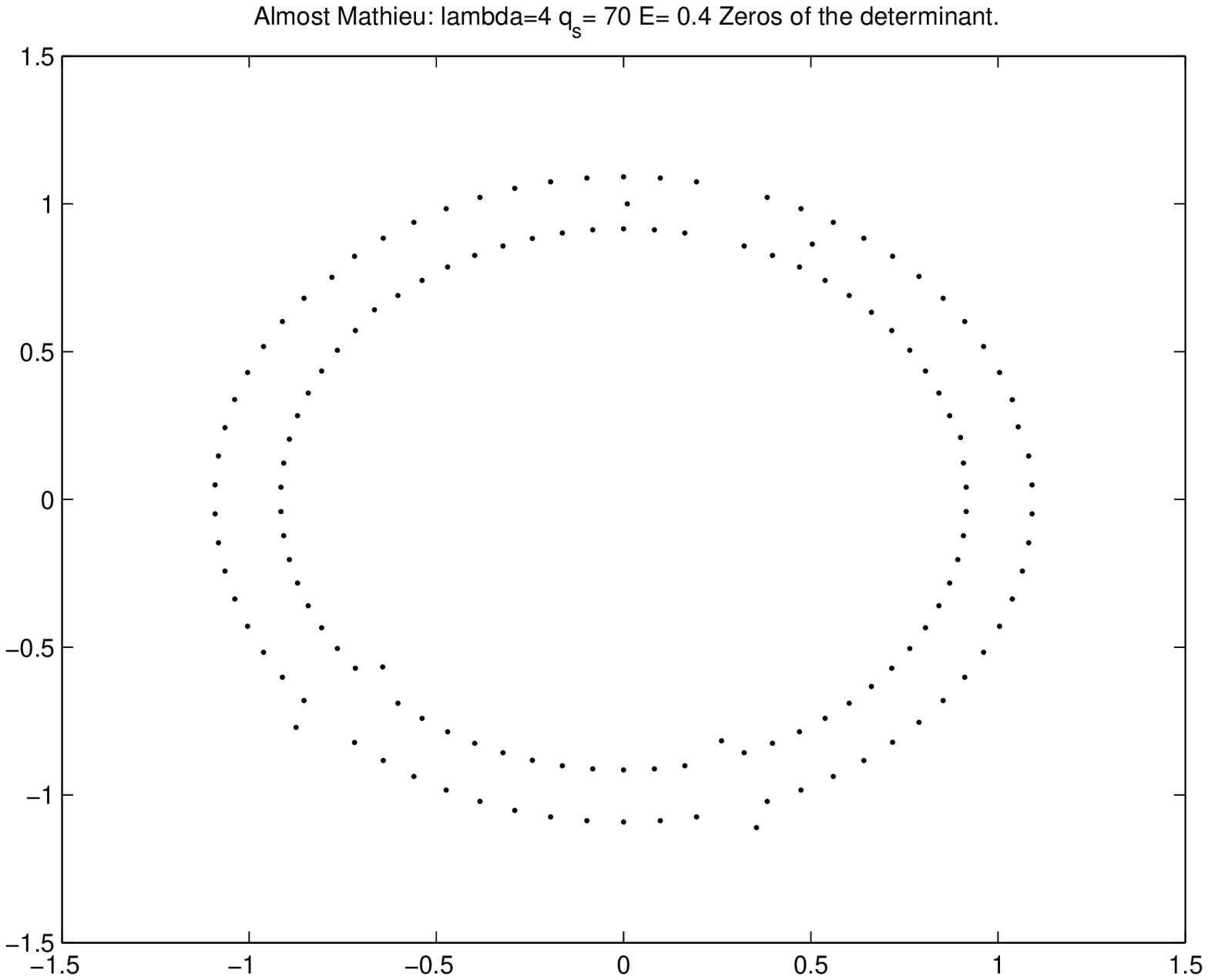}}}} 
\label{fig:2}
\end{figure}

The estimates  on the number of eigenvalues falling into a small
interval stated  in Theorems \ref{thm:1}, \ref{thm:2} are based on an
analysis of the Green function
$G_N(x, \omega, E+ i\eta)(m, n) = \bigl(H_N(x,\omega) - E-i\eta\bigr)(m,
n)$.  To evaluate the trace $\tr\left(\bigl(H_N(x, \omega ) -
E-i\eta\bigr)^{-1}\right)$ we study $G_N(x, \omega, E+ i\eta)(m, m)$ as
a function of complexified phases $x$.  By Cramer's rule
\begin{equation}
\label{eq:cramer}
G_N(x, \omega, E+ i\eta)(m, m) = {f_{[1, m]}\bigl(e(x), \omega, E+
i\eta\bigr) f_{[m+1, N]} \bigl(e(x), \omega, E+ i\eta\bigr)\over f_{[1,
N]} \bigl(e(x), \omega, E+ i\eta\bigr)}\ ,
\end{equation} where $f_{[a, b]}$ are defined as in (1.9) with
$$
v(k) = V\bigl(e(x+ k\omega)\bigr)\ .
$$
We derive bounds on  the meromorphic function on the
right-hand side of (\ref{eq:cramer}) by studying the distribution of the zeros of
the Dirichlet determinants $f_{[a, b]} (\cdot, \omega, E+ i\eta)$.  On
the other hand, in view of (\ref{eq:f_def}), $f_{[a, b]}$ is one of the entries of
the monodromy $M_{[a, b]}$.  Since $M_{[a,b]}$ is a product of matrices
$A_k$, see~\eqref{eq:Mn_def}, one can expect that there is an intrinsic
factorization of $f_{[a, b]}$ related to this matrix product.
This observation specifies the first objective of this
work.  We develop some machinery that allows
for the implementation of such a  factorization.
The first basic component here
was found in~\cite{GS}. It is a property of deterministic
matrix products of unimodular $2\times2$ matrices,
which goes by the name of {\em avalanche principle},
see Section~\ref{sec:impure}. To some extent, this avalanche principle
can be viewed as a form of multi-scale analysis for matrix products.
Indeed, it is typically used to reduce some property of a long matrix product
to properties of shorter products.
In order to apply this mechanism, it is
essential that the Lyapunov exponents are positive. Moreover, the
large deviation theorems are essential to verify that the short blocks
appearing in the avalanche principle satisfy all needed conditions.

The second basic component in our technology are
these large deviation theorems.  In Sections~\ref{sec:dets} and~\ref{sec:impure}
we develop first the large deviation estimates for $\log \big
| f_{[1,  N]}(\cdot, \omega, E)\big |$ instead of $\log \big \|
M_N(\cdot, \omega, E)\big \|$ and then obtain an {\sl avalanche
principle expansion\/} for $\log \big |f_{[1, N]}(\cdot, \omega, E)\big
|$. Each term (up to a finite number of exceptions) which appears in this additive
expansion is of the form  $\pm \log \big \| M_{[s_j, s_{j+1}]}(\cdot,
\omega, E)\big \|$ with $s_{j+1} - s_j \asymp \bigl(\log N\bigr)^A$, $A
> 1$. The Cartan estimate for subharmonic functions is of basic importance in this context.
It states that a logarithmic potential
\[ w(z) =\int \log|z-\zeta|\, \mu(d\zeta) \]
of a positive measure $\mu$ on $\IC$ satisfies
\beeq
\label{eq:cartan} w(z) \gtrsim  -H\|\mu\| \text{\ \ for all\ \ }z\in \IC\setminus \bigcup_{j} \cD(z_j,r_j)
\eneq
where $\sum_j r_j \les e^{-H}$ for any $H\ge1$. Since a general subharmonic function $u$ is the sum
of such a logarithmic potential and a harmonic function by the Riesz representation theorem,
we can apply this bound to such $u$. An example is of course the function
\[ u(z)= \log\| M_{[a,b]}(z,\omega,E)\| \]
which extends from the real axis to a subharmonic function on some strip due to our
assumption of analyticity of $V$.  Another important issue here is the number of disks
needed in~\eqref{eq:cartan}. If $u(z)=\log|f(z)|$ with analytic $f(z)$, then locally it is possible to
control this number by means of the number of zeros. This remark is relevant in view of
the complexity bounds in our theorems.

It is important to mention here that the methods of
Sections~\ref{sec:dets}, \ref{sec:impure} apply to a wide class of
transformations $T$, provided the large deviation estimates are
valid for the monodromy matrices generated by $T$,
see~\eqref{eq:Mnprod}. In particular, we remark that the avalanche
principle expansion from Section~\ref{sec:impure} applies to
so-called skew-shift $T \,: \tor^2 \to \tor^2$, $T(x_1, x_2) = (x_1
+\omega, x_2 + x_1)\;\mod\,\IZ^2$ provided the disorder is large,
see~\cite{BGS}.

One of our first applications of this
avalanche principle expansion consists of a
uniform upper estimate for $\log \big | f_N(\cdot, \omega, E)\big |$ by
\begin{equation}
\nn
\int_\tor \log \big |f_N \bigl(e(x), \omega, E\bigr)\big|\,dx +
O\bigl((\log N)^A\bigr)\ .
\end{equation}
This estimate, which we prove in Section~\ref{sec:upper}, is an important
technical tool in the development of our method.  It implies in particular that the
number of zeros of $f_N(\cdot, \omega, E)$ in a disk of radius $N^{-1}$
is bounded by $(\log N)^A.$ We remark that this estimate is not an optimal one.
Combining this upper bound and the large deviation theorem leads to bounds on
the number of zeros of $f_N(\cdot, \omega, E)$ in a small disk by means of Jensen's formula.
The elementary corollaries of Jensen's formula needed for this purpose are
discussed in Section~\ref{sec:jensen}.

In Section~\ref{sec:weier} we continue our analysis of the function $f_N(\cdot, \omega, E)$ by
developing a local factorization of $f_N(\cdot, \omega, E)$  using the
Weierstrass preparation theorem and the estimates of the preceding sections.  In
particular, the aforementioned bound on the number of zeros guarantees
that the polynomial factor in the Weierstrass preparation theorem
is of degree at most $(\log N)^C$.

Section~\ref{sec:resultant} introduces our method of eliminating ``bad''
frequencies $\omega$. More precisely, we need to eliminate the set $\Omega_\ell\subset\tor$
of those frequencies with the property that the two determinants
\beeq
\label{eq:2det}
 f_{\ell_1}(\cdot,\omega,E),\qquad  f_{\ell_2}(\cdot e(t\omega),\omega,E)
\eneq
with $\ell_1\asymp\ell_2\asymp\ell$ and $t>\exp((\log\ell)^A)$ have close zeros. Indeed,
we show that there is a small set $\Omega_\ell$ such that for all $\omega\in\tor\setminus\Omega_\ell$
there is a small set $\cE_{\ell,\omega}\subset\IR$ so that the aforementioned zeros are separated
by an amount $\exp(-(\log\ell)^C)$, say, provided
$E\in\IR\setminus \cE_{\ell,\omega}$.

The relevance of this issue can be seen as follows. Let $z_0\in\IC$ and $r_0 \le \exp \bigl(-(\log N)^A\bigr)$, $A > 1$.
Then the number of terms
\begin{equation}
\label{eq:MSJ}
\pm \log \big \| M_{[s_j, s_{j+1}]} (\cdot, \omega, E) \big \|
\end{equation}
arising in the avalanche principle expansion of $\log \big |f_N(\cdot, \omega,E)\big |$
which contribute to Jensen's average
\begin{align}
\label{eq:JENS}
&\frac{1}{2\pi}\int_{0}^{2\pi} \log {\big |f_N(z_0+r_0e^{i\theta}, \omega, E)\big |\over \big
|f_N(z_0, \omega, E)\big |}\, d\theta \\
&= \frac{1}{2\pi}\int_{0}^{2\pi}\log \big |f_N(z_0+r_0e^{i\theta}, \omega, E)\big |\,d\theta-
\log\big|f_N(z_0, \omega, E)\big | \nn
\end{align}
should be much smaller than the total number of terms in this expansion.
More precisely, this is the case for most $\omega$ and $E$. To see this, take the difference of the
expansions of both logarithms in~\eqref{eq:JENS}. Then, since
\[ M_{[s_j, s_{j+1}]} (\cdot, \omega, E) = M_{[s_j+t, s_{j+1}+t]} (\cdot e(-t\omega), \omega, E) \text{\ \ for all\ \ }t\in\ZZ,\]
there can only be a large number of such contributing terms~\eqref{eq:MSJ} if there
is a large collection of $t$ for which the pairs in~\eqref{eq:2det} have zeros inside the disk~$\cD(z_0,r_0)$.

This issue of close zeros of determinants (\ref{eq:2det}) is
called a {\sl double resonance\/} and is essential to the theory
of Anderson localization. In \cite{BG}, \cite{BGS}, it was
analyzed on the level of the monodromy matrices rather than the
determinants by means of a geometrical method. This method
exploited the fact that the large deviation sets for the monodromy
matrices considered as subsets of the $(\omega,E)$-plane are
semi-algebraic of a certain degree.

In this paper we develop a more quantitative method of analyzing
the resonance sets which is  based on the classical resultants and discriminants
from the theory of polynomials. The polynomials in question are
those which arise in the factorization of the determinants via the
Weierstrass preparation theorem. This method turns the set of
resonances into a set on which some analytic function (for the
case of double resonances it is the resultant) attains small
values. Cartan's estimate applied to this analytic function leads
to bounds on the measure and complexity of the set $\Omega_\ell$.
For double resonances this program is carried out in
Sections~\ref{sec:resultant} and~\ref{sec:double}.

In Sections~\ref{sec:localize} and~\ref{sec:minimal} we address the question
of the distance between the Dirichlet eigenvalues, which is of basic importance to
all the main results in this paper.
The eigenvalues $E^{(N)}_j(x, \omega)$, $j = 1, 2, \dots, N$ of $H_{[1,
N]}(x, \omega)$ are simple due to the Dirichlet boundary conditions.
We derive a quantitative estimate of the form
\begin{equation}
\label{eq:EjNsep} \Big | E_j^{(N)}(x, \omega) -
E_k^{(N)}(x,\omega)\Big | \ge \tau(N)
\end{equation}
which holds for all $x\in\tor$, all $\omega\in\tor_{c,a}$ outside of
some small bad set and for all $E_j^{(N)}(x,\omega)$ which do not
belong to some small set of bad energies. To prove this we need a
rather accurate description of the normalized Dirichlet
eigenfunctions $\psi_j^{(N)} (x, \omega, n)$, $n \in [1, N]$ of
$H_N(x, \omega)$. That is done in Section~\ref{sec:localize}.  In
Section~\ref{sec:minimal} we prove that (\ref{eq:EjNsep}) is valid
with $\tau(N) = \exp\bigl(-N^\delta\bigr)$, $\delta \ll 1$, provided
$E_j^{(N)}(x, \omega)$ is outside of some set $\cE_{N, \omega}$,
$\mes \cE_{N,\omega} < \exp\bigl(-N^{\delta_1}\bigr)$ for some
$\delta_1\ll\delta$. The main idea here is that if (\ref{eq:EjNsep})
fails, then $\psi_j^{(N)}$ and $\psi_k^{(N)}$ have to be close which
is impossible due to the orthogonality of $\psi_j^{(N)}$ and
$\psi_k^{(N)}$. It is important to note that the smallest distance
of two eigenvalues $E_j^{(N)}$ can be as small as $e^{-cN}$, see
Sinai~\cite{Si1}. Hence, the subexponential bound we are claiming
can only hold after the removal of some exceptional set of energies.
Estimate (\ref{eq:EjNsep}) allows us to show that up to some small
exceptional pieces the graphs of the functions $x\mapsto
E_j^{(N)}(x, \omega)$  have sufficiently steep slopes. More
precisely, we show that
\begin{equation}
\label{eq:bjslope} \big |\partial_x E_j^{(N)} \big | >
e^{-N^{\delta_2}}
\end{equation}
provided $E_j^{(N)}(x, \omega)$ is outside of some set $\cE'_{N,
\omega}$, $\mes \cE'_{N, \omega} < e^{-N^{\delta_3}}$ where
$0<\delta_3<\delta_2<\delta_1$. This result, based on
(\ref{eq:EjNsep}) and Sard-type arguments is proven in
Section~\ref{sec:mobil}. The hardest part here is to control the
complexity of these Sard-sets. The proof of (\ref{eq:bjslope}) also
shows that the zeros of $f_N(\cdot, \omega, E)$ are separated by
$e^{-N^{\delta_3}}$, provided $E \notin \cE'_{N, \omega}$.

To establish Theorems~\ref{thm:1}, \ref{thm:4},  we  need another
technical component consisting of Harnack-type inequalities for the
functions $\log \big \|M_N(\cdot, \omega, E)\big \|$. This then
leads to fine estimates of the Jensen averages of these functions
based on the avalanche principle expansion.  We show that due to the
fact that $M_N(z, \omega, E)$ are $2\times 2$--unimodular and
analytic the Harnack inequality for the subharmonic function  $\big
\|M_N(\cdot, \omega, E)\big \|$ is as good as for the logarithm of
an analytic function, up to terms of second order. The latter refers
to  the radius of a disk where one of the entries of $M_N(\cdot,
\omega, E)$ is free of zeros.  Due to this property there is no
``substantial accumulation'' of negative error terms in the Jensen's
average of the avalanche principle expansion.  All that is done in
Sections~\ref{sec:harnack} and~\ref{sec:Jav}.

We establish Theorem~\ref{thm:4} in Section~\ref{sec:Th3}.  In
Section~\ref{sec:monster} we explain how to relate the estimates on
resolvents at complex energies to the number of eigenvalues falling
into small interval. Expressions of the form $w_m(z) = \log{\big
\|M_{2m}(z, \omega, E)\big \|\over \big
\|M_m\left(ze(m\omega),\omega, E\right) \big \|\, \big \|M_m(z,
\omega, E)\big \|}$ are closely related to resolvents on the one
hand and to the avalanche principle expansion on the other.  We
refer to $w_m$ and $W_m = \exp(-w_m)$ as concatenation terms.

In Section~\ref{sec:16} we prove that the Riesz measure for $w_m(z)$
is ``almost'' non-negative.  Using that fact we establish in
Section~\ref{sec:17} a relation between $w_m(z)$ and $w_{\ulm}(z)$
for consecutive scales $\ulm \ll m$.  Finally, in Sections
\ref{sec:18}, \ref{sec:19} we prove Theorems \ref{thm:1},
\ref{thm:2}, and \ref{thm:3},  respectively, using the estimates of
Section~\ref{sec:17}.

\section{A large deviation theorem for the entries of the monodromy}\label{sec:dets}

\noindent
Suppose $T:\tor^d\to\tor^d$ is a measure preserving transformation and $V:\tor^d\to\IR$ is analytic and real-valued.
The propagator matrix of the family of discrete Schr\"odinger equations,
\begin{equation}
\label{eq:schr}
 -\psi_{n+1}-\psi_{n-1} + v(n,x)\psi(n) = E\psi_n
\end{equation}
where $v(n,x)=V(T^nx)$ for $x\in\tor^d$, is of the form
\[ M_n(x,E) = \prod_{j=n}^1 A(T^j(x),E) \]
with the $2\times 2$ matrix
\[ A(x,E) = \left[ \begin{array}{cc} V(x) - E & -1 \\
                                      1 & 0
                   \end{array}
            \right].
\]
A basic quantity in this context is the {\em Lyapunov exponent} defined as
\[ L(E)=\lim_{n\to\infty} L_n(E)= \inf_{n} L_n(E) \text{\ \ where\ \ }L_n(E) = \frac{1}{n} \int_{\tor^d} \log\|M_n(x,E)\|\,dx.\]
For certain transformations~$T$, as for example shifts and skew shifts (the latter only for large disorders), large deviation theorems of the form
\begin{equation}
\label{eq:LDE}
\mes\Bigl[ x\in\tor^d\:\Big|\: |\log\|M_n(x,E)\|-nL_n(E)| > n^{1-\tau} \Bigr] \le C\exp(-cn^{\sigma})
\end{equation}
are known, where $\sigma,\tau>0$ are positive parameters.
See~\cite{BG}, \cite{GS}, \cite{BS}, \cite{BGS} for the proofs of
such estimates as well as applications to  Anderson localization and
other results for discrete Schr\"odinger equations with
deterministic potentials.

\noindent In this section we show that $M_n(x,E)$ in \eqref{eq:LDE} can be replaced with any of its entries.
Recall that
\be
M_n(x,E) = \left[\begin{array}{cc} f_n(x,E) & -f_{n-1}(Tx,E) \\
                                 f_{n-1}(x,E) & -f_{n-2}(Tx,E) \\
                                     \end{array} \right]
\label{mondet} \ee
where
\be && f_n(x,E) = \det\left[
\begin{array}{ccccccccc}
                         v(1,x)-E & -1 & 0 & 0 & . & . & . & . &  0    \\
                        -1 &  v(2,x)-E & -1 & 0 & 0 & . & . & . &  0 \\
                        0 & -1 &  v(3,x)-E & -1 & 0 & 0 &  . & . & 0 \\
                        . & . & . & . & . & . & . & . & . \\
                        . & . & . & . & . & . & . & . & . \\
                        . & . & . & . & . & . & . & . & . \\
                        . & . & . & . & . & . & . & . & . \\
                        0 & 0 & . & . & . & . & . & -1 &  v(n,x)-E
\end{array} \right].
\label{fn}\ee
It is customary to denote the matrix on the right-hand side as $H_{[1,n]}(x)-E$ so that one as
$f_n(x,E)=\det(H_{[1,n]}(x)-E)$.
We want to emphasize that we do not prove large deviation theorems as in~\eqref{eq:LDE} in this section.
Rather, we assume that they hold for the matrices~$M_n(x,E)$, and then show how to deduce a similar
estimate for the determinants~$f_n(x,E)$. For this reason, we work in a rather general setting,
assuming only minimal properties of~$T$. In later sections we present applications of these results on
the determinants.

\noindent In the study of \eqref{eq:schr} properties of subharmonic functions have played an important role,
mainly since
\[(z_1,z_2,\ldots,z_d)\mapsto \log\|M_n(z_1,\ldots,z_d,E)\|\]
is subharmonic in each variable separately on a neighborhood of~$\tor^d$. Here we have abused notation slightly by considering $V$ and thus $M_n$ as functions of $e(x):= e^{2\pi ix}$ rather than~$x$ itself so that
various subharmonic extensions are defined on the annulus
\begin{equation} \label{eq:arho}
\cA_\rho := \{ z \in \IC\: |\: 1 -\rho < |z| < 1 + \rho \}
\end{equation}
or products thereof. We will continue the practice of passing between $x$ and $e(x)$
without any further notice. Another comment on notation: $C$ denotes a numerical constant whose value can
change from line to line. It will be usually clear from the context what these constants depend on.
Occasionally we also use $a\lesssim b$ (or $a\ll b$) to denote $a\le Cb$ (with $C$ very small)
and $a\gtrsim b$ (or $a\gg b$) for $a\ge Cb$ (with $C$ very large). Finally $a\asymp b$ means $a\lesssim b$ and $a\gtrsim b$ simultaneously.

\noindent In the following lemma we collect some of the basic properties of subharmonic functions.
Some of the details in the following proof and many other facts can be found in the books of Koosis~\cite{koosis} and Levin~\cite{levin}.

\begin{lemma}
\label{lem:riesz}
Let $u:\Omega\to \IR$ be a subharmonic function on a domain $\Omega\subset\IC$.
Suppose that $\partial \Omega$ consists of finitely many piece-wise $C^1$ curves.
There exists a positive measure $\mu$ on~$\Omega$ such that for any $\Omega_1\Subset \Omega$
(i.e., $\Omega_1$ is a compactly contained subregion of~$\Omega$)
\begin{equation}
\label{eq:rieszrep}
u(z) = \int_{\Omega_1} \log|z-\zeta|\,d\mu(\zeta) + h(z)
\end{equation}
where $h$ is harmonic on~$\Omega_1$ and $\mu$ is unique with this property.
Moreover, $\mu$ and $h$ satisfy the bounds
\be
\mu(\Omega_1) &\le& C(\Omega,\Omega_1)\,(\sup_{\Omega} u - \sup_{\Omega_1} u) \label{eq:mubound} \\
\|h-\sup_{\Omega_1}u\|_{L^\infty(\Omega_2)} &\le& C(\Omega,\Omega_1,\Omega_2)\,(\sup_{\Omega} u - \sup_{\Omega_1} u) \label{eq:hsup}
\ee
for any $\Omega_2\Subset\Omega_1$. Finally, small values of the logarithmic potential in~\eqref{eq:rieszrep}
are controlled by means of Cartan's theorem:
For any $0<S<1$ there exist disks $\{D(z_j,r_j)\}_{j=1}^\infty\subset\IC$ with the property that
\be
 \sum_j r_j &\le& 5S \label{eq:smalldisks}\\
 \int_{\Omega_1} \log|z-\zeta|\,d\mu(\zeta) &>& - \mu(\Omega_1)\log\frac{e}{S} \text{\ \ for all\ \ }z\in \IC\setminus \bigcup_{j=1}^\infty D(z_j,r_j).\label{eq:lowerbd}
\ee
\end{lemma}
\begin{proof}
Choose $\Omega_1^*$ such that $\Omega_1\Subset\Omega_1^* \Subset \Omega$ and so that $\partial\Omega_1^*$
consists of finitely many $C^\infty$ curves. Let $G(z,w)$ denote the Green's function with respect
to~$\Omega_1^*$, i.e., $(z,w)\mapsto G(z,w)$ is $C^\infty$ on $\{(z,w)\in \Omega_1^*\times\Omega_1^*\:|\:z\not= w\}$, $G(z,w)=0$ if $z\in\Omega_1^*$ and $w\in\partial\Omega_1^*$, $-\Laplace_z G(z,w)= \delta_w(z)$. It is well-known that $G$ exists, that $G\ge0$, and that $G(z,w)=-\log|z-w|+H(z,w)$, where $H(z,w)$ is $C^\infty$ on~$\Omega_1^*\times\Omega_1^*$ and harmonic in each variable. If $u$ is subharmonic on $\Omega$, then
\begin{equation}
\label{eq:genjensen}
-\int_{\partial \Omega_1^*} u(y) \frac{\partial}{\partial n}G(z,y)\,d\sigma(y) - u(z) = \int_{\Omega_1^*} G(z,\zeta)\,d\mu(\zeta)
\end{equation}
for some unique positive measure $\mu$ on $\Omega$. If $u\in C^2(\Omega)$, then $\mu=\Laplace u$ and~\eqref{eq:genjensen} is Green's theorem, whereas the general case follows by approximation, see~\cite{levin} and~\cite{koosis}.
One obtains from~\eqref{eq:genjensen} that
\be
 u(z) &=& -\int_{\Omega_1} G(z,\zeta)\,d\mu(\zeta) - \int_{\Omega_1^*\setminus\Omega_1} G(z,\zeta)\,d\mu(\zeta)
          - \int_{\partial \Omega_1^*} u(y) \frac{\partial}{\partial n}G(z,y)\,d\sigma(y) \nn \\
      &=& \int_{\Omega_1} \log|z-\zeta|\,d\mu(\zeta) + h(z), \label{eq:rieszOm1}
\ee
where
\begin{equation}
\label{eq:hdef}
 h(z) = -\int_{\Omega_1} H(z,\zeta)\,d\mu(\zeta)-\int_{\Omega_1^*\setminus\Omega_1} G(z,\zeta)\,d\mu(\zeta)
          - \int_{\partial \Omega_1^*} u(y) \frac{\partial}{\partial n}G(z,y)\,d\sigma(y).
\end{equation}
Since $\inf_{z,w\in \Omega_1} G(z,w)>0$ and $\frac{\partial}{\partial n} G(z,w)\le0$,
it follows from~\eqref{eq:genjensen} that
\begin{equation}
\label{eq:massbound}
\mu(\Omega_1)\le C(\Omega_1,\Omega) (\sup_\Omega u - \sup_{\Omega_1} u),
\end{equation}
as claimed.
Similarly,
\begin{equation}
\label{eq:massbound'}
\mu(\Omega_1^*)\le C(\Omega_1^*,\Omega) (\sup_\Omega u - \sup_{\Omega_1^*} u)\le C(\Omega_1^*,\Omega) (\sup_\Omega u - \sup_{\Omega_1} u) .
\end{equation}
Clearly, $h$ is harmonic on~$\Omega_1$ and the first two integrals in~\eqref{eq:hdef} are
 bounded in absolute value by $C(\Omega_2,\Omega_1,\Omega_1^*)(\mu(\Omega_1)+\mu(\Omega_1^*))$ on~$\Omega_2$,
and thus controlled by~\eqref{eq:massbound} and~\eqref{eq:massbound'}. It remains to control the last term in~\eqref{eq:hdef}, i.e.,
\[ h_0(z) := \int_{\partial \Omega_1^*} u(y) \frac{-\partial G(z,y)}{\partial n}\,d\sigma(y).\]
Denote $M:=\sup_\Omega u$. Applying Harnack's inequality to the nonnegative harmonic function $M-h_0\ge$
yields
\[ \sup_{\Omega_1} (M-h_0) \le C(\Omega_1,\Omega_1^*) \inf_{\Omega_1} (M-h_0) \]
which simplifies to
\[ \inf_{\Omega_1} h_0  \ge -(C-1)M + C \sup_{\Omega_1} h_0.\]
It follows from \eqref{eq:genjensen} that $h_0(z)\ge u(z)$ for all $z\in\Omega_1^*$.
Thus the previous line implies that
\[ \inf_{\Omega_1} h_0 - \sup_{\Omega_1} u \ge -(C-1)(M -\sup_{\Omega_1} u),\]
whereas clearly
\[ \sup_{\Omega_1} h_0 - \sup_{\Omega_1} u \le M -\sup_{\Omega_1} u,\]
and \eqref{eq:hsup} follows.
Finally, Cartan's theorem can be found in~\cite{levin}, page~76, as stated.
\end{proof}

\noindent The following lemma relates the supremum of a subharmonic function on $\cA_\rho$ to its supremum
over~$\tor$ by means of Cartan's estimate. We first state what we mean by a separately subharmonic function.

\begin{defi} Suppose $u:\Omega_1\times\ldots\times \Omega_d \to \IR\cup\{-\infty\}$ is continuous.
Then $u$~is said to be separately subharmonic, if for any $1\le j\le d$ and $z_k\in \Omega_k$ for~$k\not=j$
the function
\[ z\mapsto u(z_1,\ldots,z_{k-1},z,z_{k+1},\ldots,z_d)\]
is  subharmonic in~$z\in\Omega_j$.
\end{defi}

\noindent The continuity assumption is convenient but not absolutely necessary.
It does not follow from the subharmonicity alone, as that only requires upper semi-continuity.
One advantage of our assumption is that $\sup_{x\in K} u(x,z_2,\ldots,z_d)$ is again separately subharmonic in
$(z_2,\ldots,z_d)$ for any compact~$K\Subset \Omega_1$.

\begin{lemma}
\label{lem:cartansup}
Let $u(z_1,\dots, z_d)$ be a separately subharmonic function on $\prod_{\rho,d} := \underbrace{\cA_\rho \times \cdots \times\cA_\rho}_d $.
Suppose that $\sup\limits_{\prod_{\rho,d}} u \le M$.
There are constants $C_d$, $C_{\rho,d}$ such that,
if for some $0<\delta<1$ and some~$L$
\begin{equation}\label{eq2.2}
\mes [ x \in \TT^d \:|\: u < - L ] > \delta \ ,
\end{equation}
then
\begin{equation}\nn
\sup\limits_{\tor^d} u \le C_{\rho,d}\, M -
\frac{L}{C_{\rho,d} \log^d \big( \frac{C_d}{\delta}\big) } \ .
\end{equation}
\end{lemma}
\begin{proof}
We use induction in the dimension $d$.  Let
$\cB = \{x \in \TT^d \:|\: u < -L \}$ where $x=(x_1,x_2,\ldots,x_d)$. Then~\eqref{eq2.2} implies that
\begin{equation}\label{eq2.4}
\mes \left[ (x_1 ,\dots, x_d) \in \TT^{d-1}\: \Big| \:
| \cB(x_2 ,\dots, x_d) | > \frac{\delta}{2} \right]
\ge \frac{\delta}{2},
\end{equation}
where $\cB(x_2 ,\dots, x_d)$ denotes the slice of $\cB$ for fixed $(x_2 ,\dots, x_d)$.
Fix any such $(x_2,\dots,x_d)$ so that
\[| \cB(x_2 ,\dots, x_d) | > \frac{\delta}{2}.\]
Then $u_1 (z) := u(z,x_2,\dots, x_d)$ is
subharmonic on $\cA_\rho$ and $\sup\limits_{\cA_\rho} u_1 \le M$.  If
$\sup\limits_\TT u_1 > -m$, then by Lemma~\ref{lem:riesz} with $\Omega=\cA_\rho$
there is the Riesz representation
\begin{equation}\nn
u_1 (z) = \int_{\cA_{\rho / 2}}
\log | z - \zeta | d\mu (\zeta ) + h(z)
\end{equation}
where
\begin{equation}\nn
\mu (\cA_{\rho / 2} ) + \| h \|_{L^\infty (\cA_{\rho /4} )} \le
C_\rho (m+ 2M) \ .
\end{equation}
Thus, by Cartan's estimate
\begin{equation}\nn
\mes \left[ x\in \TT \:\Big|\: \int_{\cA_{\rho / 2}}  \log
|x -\zeta | d\mu (\zeta) < -C_\rho (m+2M) \log \frac{e}{S} \right]
\le 10 S \ ,
\end{equation}
so
\begin{equation}\nn
\mes \left[ x\in \TT \:\Big|\: u_1 (x) <  -C_\rho (m+2M) \log
\left( \frac{20e^2}{\delta} \right)  \right] \le
\frac{\delta}{2} \ .
\end{equation}
By the choice of $(x_2,\ldots,x_d)$, necessarily $-C_\rho (m+2M) \log \left( \frac{20e^2}{\delta}\right) \le -L$. Thus
\begin{equation}\label{eq2.9}
\sup\limits_{x\in \TT} u_1 (x) \le 2M -
\frac{L}{C_\rho \log
\left( \frac{20e^2}{\delta}\right)} =: -L' \ .
\end{equation}
Let
\[u_2 (z_2 ,\dots, z_d) = \sup\limits_{x\in \TT}
u_1 (x,z_2,\dots,z_d).\]
Clearly, $u_2$ is subharmonic on
$\prod_{\rho,d-1} $ and  $\sup\limits_{\prod_{\rho,d-1}} u_2 \le M$.
Thus, by~(\ref{eq2.4}) and~(\ref{eq2.9})
\begin{equation}\nn
\mes \left[ (x_2,\dots,x_d) \in \TT^{d-1}
\:\Big|\: u_2 (x_2,\dots,x_d) \le -L'\right] \ge
\frac{\delta}{2} \ .
\end{equation}
By the inductive assumption therefore
\begin{equation}\nn
\sup\limits_{\TT^{d-1}} u_2 \le C_{\rho,d-1} M -
\frac{L'}{C_{\rho,d-1}\log^{d-1}\left(\frac{2C_{d-1}}{\delta}\right)}
\end{equation}
and hence, by~(\ref{eq2.9})
\begin{equation}\nn
\sup\limits_{\TT^d} u \le C_{\rho,d} M -
\frac{L}{C_{\rho,d}\log^{d}\left(\frac{C_d}{\delta}\right)}
\end{equation}
with $C_d = 20 e^2 \cdot 2^{d-1} $.
\end{proof}

\noindent We now turn to properties of matrices in $SL(2,\IR)$ (we refer to these matrices as
{\em unimodular}  matrices). It follows from
polar decomposition that for any $A \in SL (2,\IR)$ there are unit vectors
$\uu^+_A$, $\uu^-_A$, $\uv^+_A$, $\uv^-_A$
so that $A\uu^+_A = \| A \| \uv^+_A $, $A\uu^-_A = \| A \|^{-1} \uv^-_A$.
Moreover, $\uu^+_A \perp \uu^-_A$ and $\uv^+_A \perp \uv^-_A$. The following lemma
deals with the question of stability of these contracting and expanding directions.
It will play an important role in the passage from the matrix $M_n$ to its entries.

\begin{lemma}
\label{lem:stabil}
For any $A$, $B \in SL (2,\IR)$
\begin{eqnarray}
|B\uu^-_{AB} \wedge \uu^-_A | \le \| A\|^{-2} \|B\| \ ,
&|\uu^-_{BA} \wedge \uu^-_A | \le \| A\|^{-2} \| B\|^2 \label{eq2.13}\\
|\uv^+_{AB} \wedge \uv^+_A | \le \| A\|^{-2} \|B\|^2 \ ,
&|\uv^+_{BA} \wedge B\uv^+_A | \le \| A\|^{-2} \| B\| \ . \label{eq2.14}
\end{eqnarray}
\end{lemma}
\begin{proof}
Let $\uu^-_A = c_1 \uu^+_{BA} + c_2 \uu^-_{BA}$. Then
\begin{equation}\nn
BA \uu^-_A  = c_1 \| BA \| \uv^+_{BA} + c_2 \| BA \|^{-1}
\uv^-_{BA} \ ,
\end{equation}
so that
\begin{eqnarray}\nn
| \uu^-_A \wedge \uu^-_{BA} | = | c_1 | &=& \| BA\|^{-1}
|BA \uu^-_A \wedge \uv^-_{BA} |
\le \|BA\|^{-1} \|B\| \|A\|^{-1} \\
&\le & \| B\| \, \|A\|^{-1} \|B\| \|A\|^{-1}
=  \| A \|^{-2} \| B\|^2 \ . \nonumber
\end{eqnarray}
Here we have used that $\| A \| \le \| B^{-1} \| \,
\| BA \| = \| B \| \, \| BA \| $ (one has $\|B\|=\|B^{-1}\|$ for
any unimodular~$B$).
Similarly, (with different $c_1$, $c_2$)
\begin{eqnarray}
\nn
&& B\uu^{-}_{AB} = c_1 \uu^+_{A}  + c_2 \uu^-_A \text{\ \ which implies \ \ } AB \uu^-_{AB} = c_1 \| A \| \uv^+_A +
c_2 \| A \|^{-1} \uv^-_A.
\end{eqnarray}
Hence
\begin{eqnarray}\nn
| B \uu^-_{AB} \wedge \uu^-_A | = |c_1 |
&=& \| A\|^{-1} |AB \uu^-_{AB} \wedge \uv^-_A |\\
&\le &\| A\|^{-1} \| AB\|^{-1} \le \| A\|^{-2} \| B \|,
\nonumber
\end{eqnarray}
where we used $\|A\|\le \|AB\|\|B^{-1}\|=\|AB\|\|B\|$.
It is easy to see from the definitions that $\uv^+_A = \uu^-_{A^{-1}} $.
The second inequality in~(\ref{eq2.13}) therefore implies the first in~(\ref{eq2.14}).
Finally, the first inequality in~(\ref{eq2.13}) yields
\[ |B^{-1}\uv^+_{BA} \wedge \uv^+_A | \le \| A\|^{-2} \| B\|. \]
Using the fact that unimodular matrices preserve areas allows one to pass to
\[ |\uv^+_{BA} \wedge B\uv^+_A | \le \| A\|^{-2} \| B\|, \]
as desired.
\end{proof}

\noindent As already mentioned above, only rather general properties of the measure
preserving transformation~$T:\tor^d\to\tor^d$ will be assumed in this section.
For the following technical lemma we require that for any disk $D\subset \TT^d$
\begin{equation}\label{eq2.19}
\sup\limits_{x\in \TT^d} \min
\{ n \ge 1 \:|\: T^{2n+1}x \in D \} \le
C (\diam D)^{-A}
\end{equation}
with some constants $A$, $C$. For the case of the shift
$x \mapsto x + \omega \;(\mod \ZZ^d)$
or the skew shift $(x,y) \mapsto (x+ y, y+ \omega)$
one can take $A= d + \vep$ for typical (in measure) $\omega\in\tor^d$.

\begin{lemma}
\label{lem:ergod}
Let $T:\tor^d\to\tor^d$ be as in \eqref{eq2.19}.
Suppose $f \in C^1 (\TT^d)$ satisfies
$|f| + |\nabla f| \le K$ for some~$K\ge1$.
Let $0<\vep < K^{-2A} $ and assume that
\begin{equation}\nn
\sup\limits_{\TT^d} | f (f\circ T) - 1 | \le \vep.
\end{equation}
Then
\[\sup_{\TT^d} | f^2 -1 |
\le C(\vep K^{2A} )^\frac{1}{1+A}.\]
\end{lemma}
\begin{proof}
Let $f(x) (f\circ T) (x) =: 1 + \rho (x)$. Then
$\| \rho \|_\infty \le \vep $ and
\begin{equation}\nn
f \circ T^{2n} (x) = f(x)
\prod\limits_{\ell=0}^{n-1}
\frac{1+\rho (T^{2\ell + 1} x)}
{1+\rho (T^{2\ell} x)}  =: f(x)
\big(P_{2n} (x) \big)^{-1}.
\end{equation}
Moreover,
\begin{equation}\nn
f(T^{2n+1} x) f(x) = P_{2n} (x)
\big( 1+ \rho (T^{2n} x \big)) \ .
\end{equation}
Thus
\begin{eqnarray}\nn
|f^2 (x) - 1 |
&\le & |f(x) |\, |f(T^{2n+1} x ) - f(x) |
+ |P_{2n} (x)
\big( 1+ \rho (T^{2n} x) \big) - 1 |\\
&\le & K^2 \| T^{2n+1} x - x \|_{\tor^d} + C \vep n
\nonumber
\end{eqnarray}
provided $\vep n \ll 1$.  Applying~\eqref{eq2.19} with $D=D(x,\delta)$ and minimizing over~$\delta$
leads to
\begin{equation}\nn
\sup\limits_{\TT^d} |f^2 -1| \le
C (K^2 \delta + \vep \delta^{-A} ) =
C \vep^{\frac{1}{1+A}} K^{\frac{2A}{1+A}} \ ,
\end{equation}
and the lemma follows.
\end{proof}

\noindent Now suppose there is a large deviation theorem for the
monodromy of the form ($1> \sigma > 0 $ fixed)
\begin{equation}\label{eq2.25}
\mes \left[ x \in \TT^d \:\Big|\:
\bigl|  \log  \| M_N  (x , E) \|
- N\,L_N (E) \bigr| > N^{1-\sigma} \right]
\le C \exp (-N^\sigma )
\end{equation}
for for all positive integers $N$. Let $\cB_\cN$ be the set on the
left-hand side of~(\ref{eq2.25}) and denote
\begin{equation}\label{eq2.26}
\cG_N = \TT^d \bs \bigcup\limits_{|\ell|\le N }
T^\ell \cB_N
\end{equation}
so that $|\tor^d \bs \cG_N | = CN \exp (-N^\sigma) $.
We shall also assume that the Lyapunov exponents are positive, i.e., $\inf_{E}\,L(E) > \gamma > 0$.

\begin{lemma}
\label{lem:basic1}
Fix some $E$ and let $f_\ell (x) =
\det (H_{[1,\ell]} (x) - E)$ be as in~\eqref{fn}. Then
\begin{equation}\label{eq2.27}
\mes \left[ x \in \TT^d \:\Big|\:
|f_\ell (x ) |  \le \exp (-\ell^{d+2}) \right]
\le \exp (-\ell)
\end{equation}
for large $\ell$, i.e., $\ell \ge \ell_0 (V,\gamma) $.
\end{lemma}
\begin{proof}
Suppose~(\ref{eq2.27}) fails and set $u(x) = \log | f_\ell (x) |$.  Applying Lemma~\ref{lem:cartansup} to $u$
with $M \sim \ell$ and $\delta = e^{-\ell}$ yields
\begin{equation}\nn
\sup\limits_{\TT^d } u \le C\ell
- \frac{\ell^{d+2}}{C\log^d \left( \frac{C}{\delta} \right)}
\le -C_1 \ell^2
\end{equation}
for some constant $C_1$.  In other words,
$\sup_{x \in \TT^d} | f_\ell | \le \exp
(-C_1 \ell^2) $. Since
\begin{equation}\label{eq2.29}
M_\ell (x) = \left[ \begin{array}{ll}
f_\ell     &-f_{\ell-1} \circ T \\
f_{\ell-1} &-f_{\ell-2} \circ T \end{array} \right]
{\hbox{\quad with\quad }} \det\, M_{\ell} = 1 \ ,
\end{equation}
one has
\begin{equation}\label{eq2.30}
\sup_x | f_{\ell-1}(x) f_{\ell -1 } \circ T(x)-1 | \le \exp (-C_1
\ell^2 / 2)
\end{equation}
for large $\ell$.  Now apply Lemma~\ref{lem:ergod} with $K \sim e^\ell$
and $\vep $ given by the right-hand side of~(\ref{eq2.30})
to conclude $\sup\limits_{\TT^d} | f_{\ell-1} | \le 2 $
for large $\ell$.  Thus
\begin{equation} \nn
f_\ell = V\circ T^\ell \cdot f_{\ell-1} - f_{\ell -2 }
\end{equation}
implies that
\begin{equation}\nn
\sup\limits_{\TT^d} | f_{\ell-2}| \le C(1 + \| V \|_\infty ) \ .
\end{equation}
In particular,
\begin{equation}\nn
\sup\limits_{x \in \TT^d} \| M_\ell (x) \|
\le C(1 + \| V \|_\infty ) \ ,
\end{equation}
which contradicts $\displaystyle \int_{\TT^d} \log
\| M_\ell (x) \| dx > \gamma \cdot \ell $ for
$\ell$ large.
\end{proof}

\begin{lemma}
\label{lem:basic2}
Let $\ell_0$ be as in Lemma~\ref{lem:basic1}.
Fix some $E$ and let $f_\ell (x) = \det (H_\ell (x) - E)$.
Then for any $\ell_0 \le \ell \le N^{\frac{1-\sigma}{2}}$
\begin{equation}\label{eq2.34}
\mes \left[ x \in \TT^d \:\Big|\:
| f_\ell (x)  | \le \exp
(-N^{1-\sigma}) \right]
\le \exp \left( - N^{\frac{1-\sigma}{d}} \ell^{-\frac{2}{d}}
\right)
\end{equation}
provided $N \ge N_0 (V,\gamma) $ is large.
\end{lemma}
\begin{proof}
Assume this fails for some $\ell$ and $N$ as in the statement of the lemma.
As before,  Lemma~\ref{lem:cartansup} applied to  to $u(x) = \log
| f_\ell (x) | $, $M \sim \ell$ and with
$\delta = $ right-hand side of~(\ref{eq2.34}) yields
\begin{equation}\nn
\sup\limits_{\TT^d} u \le C\ell -
\frac{N^{1-\sigma}}{C\log^d \left( \frac{C}{\delta}\right)}
\le -C_1 \ell^2
\end{equation}
for some constant $C_1$.  In other words,
\begin{equation}\nn
\sup\limits_{x \in \TT^d}
|f_\ell (x)| \le \exp (-C_1 \ell^2 ) \ .
\end{equation}
This leads to a contradiction as in Lemma~\ref{lem:basic1}.
\end{proof}

\noindent The next lemma gives us control over three determinants.
In what follows we will show how to use it to obtain a large deviation
estimate for a single determinant.

\begin{lemma}
\label{lem:3small}
Fix some $E$ and let $f_\ell$ be as above. Assume that \eqref{eq:LDE} holds.  Then
\begin{eqnarray}
\label{eq2.37}
\mes\Big[ x \in \TT^d
\:\Big|\: |f_N (x) | + |f_N (T^{j_1} x) | +
| f_N (T^{j_2} x) |
&\le & \exp (N L_N(E) - 100 N^{1-\sigma} ) \Big] \\
&\le & \exp (-N^{\frac{1-\sigma}{2d} \wedge \sigma} )
\nonumber
\end{eqnarray}
for any $\ell_0 \le j_1 \le j_1 + \ell_0 \le j_2 \le
N^{\frac{1-\sigma}{8}}$ and $N\ge N_1(V,\gamma)$. Moreover, to obtain~\eqref{eq2.37} for some~$N$ only
requires~\eqref{eq:LDE} with the same~$N$.
\end{lemma}
\begin{proof}
We will fix $E$ and suppress $E$ from most of the notation. The reader should keep in
mind that basically everything in the proof depends on~$E$.
In view of~(\ref{eq2.29}),
$f_N (x) = M_N (x  ) \ue_1\wedge\ue_2$.
If
\[ | M_N (x ) \ue_1 \wedge \ue_2 | \le \exp
(NL_N(E) - 100 N^{1-\sigma} ),\]
then (with $\uu^+_N = \uu^+_N (x)$ etc.)
\begin{eqnarray}
\label{eq2.38}
&&\| M_N (x ) \|\: |\uu^+_N \cdot \ue_1 |
| \uv^+_N \wedge \ue_2 | -
\| M_N (x  ) \|^{-1}\:
| \uu^-_N \cdot \ue_1 |
| \uv^-_N \wedge \ue_2 | \\
&&\le \exp (NL_N(E) - 100N^{1-\sigma} ) \ . \nonumber
\end{eqnarray}
Assume that $x \in \cG_N$, as in~(\ref{eq2.26}). Then (recalling that
$\uu^+_N \perp \uu^-_N$) one has
\begin{equation}\nn
| \uu^-_N (x) \wedge \ue_1 |\,| \uv^+_N (x) \wedge \ue_2 |
\le \exp (-99 N^{1-\sigma}) \ ,
\end{equation}
so either
\begin{equation}\label{eq2.40}
|\uu^-_N (x) \wedge \ue_1 |
\le \exp (-40 N^{1-\sigma}) {\hbox{ \quad or \quad}}
| \uv^+_N (x) \wedge \ue_2 |
\le \exp (-40 N^{1-\sigma}) \ .
\end{equation}
Suppose~(\ref{eq2.37}) fails.  Then also
\begin{eqnarray}\label{eq2.41}
\mes \Big[ x \in \cG_N \:\Big|\:
|f_N (x) | + |f_N (T^{j_1} x) |
+ |f_N (T^{j_2} x) |
&\le & \exp (NL_N(E) - 100N^{1-\sigma}) \Big]\\
&> & \exp \Big( -N^{\frac{1-\sigma}{2d} \wedge \sigma}\Big)\ .
\nonumber
\end{eqnarray}
Let $x$ be in the set on the left-hand side of~(\ref{eq2.41}).
In view of~(\ref{eq2.40}) either the first inequality in
(\ref{eq2.40}) has to occur for two points among
$x$, $T^{j_1} x$, $T^{j_2} x$ or the second.
In the former case we have
\begin{equation}\label{eq2.42}
| \uu^-_N (x) \wedge \ue_1 | \le
\vep := \exp (-40 N^{1-\sigma})
{\hbox{\quad and \quad }}
| \uu^-_N (T^{j_1} x) \wedge \ue_1 | \le \vep \ ,
{\hbox{\quad say.}}
\end{equation}
We now compare $M_{j_1}(x)\uu^-_N (x)$ and $\uu^-_N (T^{j_1} x)$.
Using  the simple fact that
\begin{equation}
\label{eq:2dreieck}
|w_1\wedge w_2| \le |w_1\wedge w_3| + |w_1|\,\|w_2\pm w_3\| \le |w_1\wedge w_3| + C\,|w_2\wedge w_3|
\end{equation}
for any three unit vectors $w_1,w_2,w_3$ in the plane, one obtains that
\be
&& | \uu^-_N (T^{j_1} x) \wedge M_{j_1} (x ) \uu^-_N (x) | \nn \\
&\le & |\uu^-_N (T^{j_1} x) \wedge M_{j_1}(x ) \uu^-_{N+j_1} (x) | + C\, | M_{j_1}(x )\uu^-_{N+j_1} (x) \wedge
M_{j_1} (x ) \uu^-_N (x ) | \nn \\
&\le & |\uu^-_N (T^{j_1} x) \wedge M_{j_1}(x ) \uu^-_{N+j_1} (x) |
+ C\| M_{j_1}(x ) \|\: |\uu^-_{N+j_1} (x) \wedge \uu^-_N (x ) |. \label{eq:2triangle}
\ee
Since $M_{N+j_1}(x )= M_N(T^{j_1}x ) M_{j_1}(x )$, one can apply
Lemma~\ref{lem:stabil} with $A=M_N(T^{j_1}x )$ and $B=M_{j_1}(x )$ to~\eqref{eq:2triangle}, which yields
\begin{eqnarray}
&& | \uu^-_N (T^{j_1} x) \wedge M_{j_1} (x ) \uu^-_N (x) | \nn \\
&\lesssim & \| M_N (T^{j_1} x ) \|^{-2}
\| M_{j_1} (x ) \| + \| M_{j_1} (x ) \|\:\| M_{j_1} (T^N x ) \|^2\:\| M_N (x ) \|^{-2} \nonumber \\
&\lesssim &\exp (-2N L_N(E) + 2N^{1-\sigma} ) \exp (C j_1 ) \label{eq2.43}\\
&\lesssim &\exp (-2N L_N(E) + 3N^{1-\sigma} ) \le \exp(-N\gamma)\ . \label{eq2.44}
\end{eqnarray}
To pass to~(\ref{eq2.43}) one uses that $x \in \cG_N$.
Combining~(\ref{eq2.42}) and~(\ref{eq2.44}) shows that  for large
$N \ge N_0 (\gamma, \sigma)$,
\begin{eqnarray}\label{eq2.45}
&&|\ue_1 \wedge M_{j_1} (x ) \ue_1 |
\le C(1+ \| M_{j_1} (x ) \| ) \vep + \exp (-N\gamma ) \text{\ \ \ or\ \ \ }
| f_{j_1 -1} (x) | \le C \exp (-20 N^{1-\sigma}) \ .
\end{eqnarray}
Now suppose that the second inequality in~(\ref{eq2.40}) occurs
twice. Then
\begin{equation}\label{eq2.46}
|\uv^+_N (x) \wedge \ue_2 | \le \vep
{\hbox{\quad and \quad }}
|\uv^+_N (T^{j_2} x) \wedge \ue_2 | \le \vep \ ,
\end{equation}
say.
We use~\eqref{eq:2dreieck} and Lemma~\ref{lem:stabil} to compare
$\uv^+_N (T^{j_2} x)$ and $M_{j_2}(T^Nx ) \uv^+_N (x)$. In this case, we invoke
the second inequality in~(\ref{eq2.14}) from Lemma~\ref{lem:stabil} with $B=M_{j_1}(T^Nx)$ and~$A=M_N(x)$.
Hence
\begin{eqnarray}
&&| \uv^+_N (T^{j_2} x) \wedge M_{j_2}(T^Nx ) \uv^+_N (x) | \nn \\
&& \lesssim | \uv^+_{N+j_2}  (x) \wedge M_{j_2}(T^N x ) \uv^+_N (x) |
 + |\uv^+_{N+j_2} (x) \wedge \uv^+_N (T^{j_2} x) |  \nonumber\\
&&\lesssim \| M_N (x) \|^{-2} \| M_{j_2} (T^N x) \|
+ \| M_N (T^{j_2} x)\|^{-2}
\| M_{j_2} (x) \|^2  \nonumber\\
&&\lesssim \exp (-2N L_N(E) + 2N^{1-\sigma} ) \exp (C j_2 ) \ . \nn
\end{eqnarray}
For the last inequality one again uses that
$x \in \cG_N$.  Combining~(\ref{eq2.45}) and~(\ref{eq2.46}) yields
\begin{eqnarray}
\nn
| \ue_2 \wedge M_{j_2} (T^N x ) \ue_2 |
\le C(1 + \| M_{j_2} (T^N x ) \| )
\vep + \exp (-N \gamma )
\end{eqnarray}
or
\begin{eqnarray}\nn
|f_{j_2 - 1} (T^{N+1} x) | \le C \exp
(-20 N^{1-\sigma })
\end{eqnarray}
for large $N$.  The conclusion from the preceding is that
(\ref{eq2.41}) implies that for some choice of $\ell$
from $j_1 - 1$, $j_2 - j_1 - 1$,
\begin{eqnarray}\label{eq2.50}
\mes \big[ x\in\tor^d \:|\: | f_\ell (x ) |
\le C \exp (-20 N^{1-\sigma }) \big] >
\exp (-N^{\tau} )
\end{eqnarray}
where we have set $\tau = \frac{1-\sigma}{2d} \wedge \sigma$ for
simplicity.  Since
$\ell < N^{\frac{1-\sigma}{8}}$ and
$N^{\frac{1-\sigma}{d}}$.
$\left( N^{\frac{1-\sigma}{8}}\right)^{-\frac{2}{d}} =
N^{\frac{3(1-\sigma)}{4d}}> N^\tau $,
(\ref{eq2.50}) contradicts~(\ref{eq2.34}).
\end{proof}

\noindent Lemma~\eqref{lem:3small} implies the following: Let
\be
\Omega_N
&:=& \Bigg\{
x \in \cG_N \Big|\
\min\limits_{0<j_1 < j_2 < j_3 \le N^{\frac{1-\sigma}{8}}}
\ (|f_N (T^{j_1} x)| + | f_N (T^{j_2} x) | + | f_N (T^{j_3} x) | ) > e^{N L_N - 100 N^{1-\sigma}} \nn\\
&& \qquad\qquad\qquad\qquad {\hbox{ with }} j_2 - j_1 \ge \ell_0 , j_3 - j_2 \ge \ell_0
\Bigg\} \ . \label{eq2.51}
\end{eqnarray}
Then $\mes (\TT^d \bs \Omega_N ) \le \exp (-N^{\tau}) $,
where $\tau = \frac{1-\sigma}{2d}  \wedge \sigma $. Recall that $\sigma>0$ is
the exponent appearing in~\eqref{eq2.25} and that $\gamma>0$ is the lower bound on
the Lyapunov exponent. We now show how to pass from~\eqref{eq2.51} to a lower bound on the average
of the determinants~$f_N(x)$.

\begin{lemma}
\label{lem:meanlower}
There exists some constant $\kappa > 0$ (depending on $\sigma$ in \eqref{eq:LDE}) such that
\begin{equation}\label{eq2.52}
\int\limits_{\TT^d}
\frac{1}{N} \log | \det (H_{[1,N]} (x) - E ) | \, dx  > L_N (E) - N^{-\kappa}
\end{equation}
for $N \ge N_0 (\sigma, V , \gamma ) $. Furthermore, \eqref{eq:LDE} for some~$N$ implies~\eqref{eq2.52}
with the same~$N$.
\end{lemma}
\begin{proof}
It follows from Lemma~\ref{lem:basic2}  that the integral in~(\ref{eq2.52}) is
finite.  Let $u(x) = \frac{1}{N} $ $\log | f_N (x)| $,
and set $M = N^{(1-\sigma)/8}$. Then
\begin{eqnarray}
\langle u \rangle
&:=& \int_{\TT^d} \ u(x)\, dx
= \frac1{M} \sum\limits_{k=1}^M \int_{\TT^d} \
u(T^{k\ell_0 } x)\, dx \nonumber \\
&\ge & \int_{\Omega_N}
\Big\{ \frac{M-2}{M} (L_N - 100N^{-\sigma} ) +
\frac{2}{M} \inf\limits_{1\le k \le M } u(T^{k\ell_0 } x)
\Big\}\, dx \label{eq2.53}\\
&&\qquad + \frac{1}{M} \sum\limits_{k=1}^M
\int_{\TT^d \bs \Omega_N }\  u(T^{k\ell_0 } x )\, dx \ .
\nonumber
\end{eqnarray}
By Lemma~\ref{lem:riesz} one has the Riesz representation  (for small $\rho>0$)
\begin{equation}\nn
u(z) = \int_{\cA_\rho} \log | z - \zeta | d\mu (\zeta) + h(z)
\end{equation}
where $\mu (\cA_\rho) + \| h \|_{L^\infty (\cA_{\rho / 2}) }
\le C (2S - \langle u \rangle )$.  Here
\[ S:= \sup_{z\in\cA_{2\rho}} u(z)\]
and we have used that $\sup_{\TT^d} u \ge \langle u\rangle $.
By Cartan's estimate in~Lemma~\ref{lem:riesz} there is a set $\cB = \cB_\vep \subset
\TT^d $ of measure not exceeding $N \cdot \exp (-N^\vep) $ such that for any small $\vep > 0$
\begin{equation}\nn
\inf\limits_{1\le k \le M }
u(T^{k\ell_0} x ) > - C (2S  - \langle u \rangle )
N^\vep
\end{equation}
for all $x \in \TT^d \bs \cB$. Therefore,
\begin{eqnarray}
\int_{\Omega_N} \
\inf\limits_{1\le k \le M } \
u(T^{k\ell_0} x )\, dx
& > & \int_{\Omega_N \bs \cB} \ - C (2S -\langle u \rangle)
N^\vep\, dx +
\int_{\Omega_N \cap \cB}\  \inf\limits_{1\le k\le M} \
u (T^{k\ell_0 } x )\, dx \nonumber \\
& > & - C (2S -\langle u \rangle) N^\vep -
\sum\limits_{k=1}^M \
\int_{\Omega_N \cap \cB} \ | u(T^{k\ell_0 } x ) |\, dx \ .
\nn
\end{eqnarray}
Combining this with~(\ref{eq2.53}) leads to
\begin{eqnarray}\label{eq2.57}
\langle u \rangle &\ge &
\left( 1 - \frac{2}{M} \right)
(L_N - 100 N^{-\sigma})
| \Omega_N | - \frac{CN^\vep}{M}
( 2S - \langle u \rangle ) \\
&& - \frac{2}{M} \sum\limits_{k=1}^M \
\int\limits_{\cB \cup \Omega^c_N} \
| u (T^{k\ell_0} x ) |\, dx \ . \nonumber
\end{eqnarray}
It remains to estimate the integral in~(\ref{eq2.57}).
By Lemma~\ref{lem:basic2},
\begin{equation}\nn
\| u \|_{L^2 (\TT^2)} \le CN^b
\end{equation}
with some constant $b> 0$.  Therefore,
\begin{equation}\nn
\int_{\cB\cup \Omega^c_N } | u (T^{k\ell_0 } x) |\, dx \le | \cB
\cup \Omega^c_N |^\frac12 \| u \|_{L^2 (\TT^d)} \le C N^b \exp
(-N^\vep ) \ ,
\end{equation}
and the lemma follows.
\end{proof}

\noindent We now derive the large deviation theorem (LDT) for the determinants. In addition to~\eqref{eq2.25}
we assume a uniform upper bound on the functions~$\frac{1}{N} \log \| M_N (x , E)\|$. This is a
mild assumption that is satisfied in all cases we are interested in, see for example Section~\ref{sec:upper}
below.

\begin{prop}\label{prop:thm2.1}
Fix a large positive integer~$N$.
In addition to the large deviation theorem for the monodromy matrices~(\ref{eq2.25}) with that choice of~$N$
assume the uniform upper bound
\begin{equation}\label{eq2.60}
\sup\limits_{\TT^d}\frac{1}{N} \log
\| M_N (x , E) \| < L_N (E) + N^{-\kappa}
\end{equation}
with the same~$N$. Then for some small constant $\tau > 0$ (depending on~$\sigma$ in~\eqref{eq2.25}),
\begin{equation}\label{eq2.61}
\mes \left[ x \in \TT^d \:\Big|\: \frac1{N} \log |\det ( H_{[1,N]} (x) - E)|
< L_N (E) - N^{-\tau } \right] \le e^{-N^\tau}.
\end{equation}
\end{prop}
\begin{proof}
For simplicity, we shall set $d=2$.  The general case is similar.
In view of~(\ref{eq2.52}), \eqref{mondet}, and~(\ref{eq2.60}),
$u= \frac{1}{N} \log | f_N |$ satisfies
\begin{equation}\nn
\left\{ \begin{array}{l}
\langle u \rangle > L_N - N^{-\kappa} \\
\\
\sup\limits_{\TT^d} u < L_N + N^{-\kappa}.
\end{array}
\right.
\end{equation}
Let $v(z) = \int_\TT u(z,y)\, dy $.
Then $\sup\limits_{\TT} v \le \sup\limits_{\tor^2} u < L_N + N^{-\kappa} $
and $\langle v \rangle = \langle u \rangle > L_N - N^{-\kappa} $.
This implies that
\begin{equation}
\label{eq:vL1}
\|v- \langle v \rangle\|_1 \le CN^{-\kappa}.
\end{equation}
Since in particular $\sup\limits_{\TT} v \ge \langle v\rangle > \gamma / 2 > 0$
if $N$ is large, and  also $\sup_{\cA_\rho} u \le C$, one concludes from Lemma~\ref{lem:riesz} that
the Riesz measure and the harmonic part of~$v$ on $\cA_{\rho/4}$ have size~$O(1)$.
We are in a position to apply the ``BMO splitting lemma'', Lemma~2.3 in~\cite{BGS}, which shows that
\begin{equation}\label{eq2.63}
\| v \|_{\BMO (\TT)} \le C \| v - \langle v\rangle \|^{1/2}_1
\le CN^{-\kappa/2}\ .
\end{equation}
Lemma~2.3 in \cite{BGS} requires boundedness of the subharmonic function, which does not hold here.
However, all that was used in the proof of that lemma was a bound on the Riesz mass and the harmonic part,
and we provided both.
Let
\begin{equation}\nn
\cG = \Big\{ x \in \TT \:\Big|\: v(x) =
\int_\tor \, u(x,y) \, dy > L_N - N^{-\kappa /4}
\Big\} \ .
\end{equation}
By~(\ref{eq2.63}),
\begin{equation}\label{eq2.65}
\mes (\TT \bs \cG ) \le \exp ( -N^{\kappa/4}) \ .
\end{equation}
Fix some $x \in \cG$.  Then
\begin{equation}\nn
\sup\limits_{y\in\tor} u(x,y) \le
\sup\limits_{\TT^2} u < L_N + N^{-\kappa} <
\int_\TT \ u(x,y) \,dy + 2N^{-\kappa/4} \ .
\end{equation}
Since also $\sup\limits_{y \in \TT} u(x,y) \ge 0$ and
$\sup\limits_{z\in \cA_\rho} u(x,z) \le C$,
the BMO splitting lemma, Lemma~2.3 in~\cite{BGS},  again implies that
\begin{equation}\nn
\| u(x ,\cdot ) \|_{\BMO (\TT)} \le CN^{-\kappa / 8}
\end{equation}
for any $x \in \cG$ and thus, by the John-Nirenberg inequality,
\begin{equation}\label{eq2.68}
\mes \left[ y \in \TT \:\Big|\: u(x,y) < L_N - CN^{-\kappa / 16} \right] \le
\exp \Big( -N^{\kappa / 16} \Big)
\end{equation}
for $N$ large. The theorem follows from~(\ref{eq2.65}) and~(\ref{eq2.68})  via Fubini.
\end{proof}

\begin{defi} Let $H \gg 1$.  For an arbitrary subset $\cB \subset \cD(z_0,
1)\subset \IC$ we say that $\cB \in \car_1(H, K)$ if $\cB\subset
\bigcup\limits^{j_0}_{j=1} \cD(z_j, r_j)$ with $j_0 \le K$, and
\begin{equation}
\sum_j\, r_j < e^{-H}\ .
\end{equation}
If $d$ is a positive integer greater than one and $\cB \subset
\prod\limits_{i=1}^d \cD(z_{i,0}, 1)\subset \IC^d$ then we define
inductively that $\cB\in \car_d(H, K)$ if for any $1 \le j \le d$ there
exists $\cB_j \subset \cD(z_{j,0}, 1)\subset \IC, \cB_j \in \car_1(H,
K)$ so that $\cB_z^{(j)} \in \car_{d-1}(H, K)$ for any $z \in \IC
\setminus \cB_j$,  here $\cB_z^{(j)} = \left\{(z_1, \dots, z_d) \in \cB:
z_j = z\right\}$.
\end{defi}

 \begin{remark} (a) This definition is consistent with the notation of
Theorem~4 in Levin's book~\cite{levin}, p.~79. \\
(b) It is important in the definition of $\car_d(H,K)$ for $d>1$ that we control
both the measure and the complexity of each slice $\cB_z^{(j)}$, $1\le j\le d$.
\end{remark}

The following lemma is a straightforward consequence of this definition.

\begin{lemma}
\label{lem:cart_12}
\begin{enumerate}
\item[{\rm{(1)}}] Let $\cB_j \in \car_d(H, K)$, $\cB_j \subset
\prod\limits^d_{\ell=1} \cD(z_{\ell,0}, 1)$, $j = 1, 2, \dots, T$.
Then $\cB = \bigcup\limits_j\, \cB_j \in \car_d\bigl(H - \log T,
TK\bigr)$.

\item[{\rm{(2)}}] Let $\cB \in \car_d(H, K)$, $\cB \subset
\prod\limits^d_{j=1} \cD\bigl(z_{j,0}, 1\bigr)$.  Then there exists
$\cB' \in \car_{d-1}(H, K)$, $\cB' \subset \prod\limits^d_{j=2}
\cD\bigl(z_{j, 0}, 1\bigr)$, such that $\cB_{(w_2, \dots, w_d)} \in
\car_1(H, K)$, for any $(w_2, \dots, w_d) \in \cB'$.
\end{enumerate}
\end{lemma}

\noindent The following Lemma~\ref{lem:high_cart} is a generalization of the usual Cartan estimate
to several variables.

\begin{lemma}
\label{lem:high_cart}
 Let $\varphi(z_1, \dots, z_d)$ be an analytic function defined
in a polydisk $\cP = \prod\limits^d_{j=1} \cD(z_{j,0}, 1)$, $z_{j,0} \in
\IC$.  Let $M \ge \sup\limits_{\uz\in\cP} \log |\varphi(\uz)|$,  $m \le \log
\bigl |\varphi(\uz_0)\bigr |$, $\uz_0 = (z_{1,0},\dots, z_{d,0})$.  Given $H
\gg 1$ there exists a set $\cB \subset \cP$,  $\cB \in
\car_d\left(H^{1/d}, K\right)$, $K = C_d H(M - m)$,  such that
\beeq
\label{eq:cart_bd}
\log \bigl | \varphi(z)\bigr | > M-C_d H(M-m)
\eneq
for any $z \in \prod^d_{j=1} \cD(z_{j,0}, 1/6)\setminus \cB$.
\end{lemma}
\begin{proof}
The proof goes by induction over $d$. For $d = 1$ the assertion is
Cartan's estimate for analytic functions. Indeed, Theorem~4 on
page~79 in~\cite{levin} applied to $f(z)=e^{-m}\varphi(z)$ yields
that
\[ \log \bigl | \varphi(z)\bigr | > m-C H(M-m)=M -(CH+1)(M-m) \]
holds outside of a collection of disks $\{\cD(a_k,r_k)\}_{k=1}^K$
with $\sum_{k=1}^K r_k\lesssim \exp(-H)$. Increasing the constant $C$ leads to~\eqref{eq:cart_bd}.
Moreover, $K/5$ cannot exceed the number of zeros of the function $\varphi(z)$
in the disk $\cD(z_{1,0},1)$, which is in turn estimated by Jensen's formula, see~\eqref{eq:jensen},  as $\lesssim M-m$. Although this
bound on $K$ is not explicitly stated in Theorem~4 in~\cite{levin}, it can be deduced from the proofs
of Theorems~3 and~4 (see also the discussion of (2) on page~78). Indeed, one can assume that
each of the disks $\cD(a_k,r_k)$ contains a zero of $\varphi$, and it is shown in the proof of Theorem~3
that no point is contained in more than five of these disks. Hence we have proved the $d=1$ case with
a bad set $\cB\in \car_1(H,C(M-m))$, which is slightly better than stated above (the $H$ dependence of $K$ appears
if $d>1$ and we will ignore some slight improvements that are possible to the statement of the lemma due to this issue).

In the general case take $1 \le j \le d$ and consider $\psi(z) = \varphi\left(z_{1,0},
\dots, z_{j-1,0}, z, z_{j+1,0}, \dots, z_{n,0}\right)$.  Due to the $d=1$ case
there exists $\cB^{(j)} \in \car_1\left(H^{{1/d}},
C_1(M-m)\right)$,  such that
$$
\log \bigl |\psi(z)\bigr | > M-C_1 H^{1/d}(M-m)
$$
for any $z \in \cD\left(z_{j,0}, 1/6\right) \setminus \cB^{(j)}$.  Take
arbitrary $z_{j,1} \in \cD\left(z_{j,0}, 1/6\right) \setminus \cB^{(j)}$
and consider the function
\[\chi\left(z_1, z_2, \dots, z_{j-1}, z_{j+1}, \dots, z_d\right) =
\varphi\left(z_1, \dots, z_{j-1}, z_{j1}, z_{j+1}, \dots, z_d\right)\]
 in the
polydisk $\prod\limits_{i\ne j} \cD\bigl(z_{i, 0}, 1\bigr)$.  Then
\begin{align*}
\sup \log
\bigl | \chi(z_1, \dots, z_{j-1}, z_{j+1}, \dots, z_d)\bigr | &\le M, \\
\log \bigl |\chi(z_{1,0}, \dots, z_{j-1,0}, z_{j,1}, z_{j+1,0}, \dots,
z_{d,0}) \bigr | &>  M-CH^{1/d}(M-m).
\end{align*}
Thus $\chi$ satisfies the conditions
of the lemma with the same $M$ and with $m$ replaced with
\[M-CH^{1/d}(M-m) .\] We now apply the inductive assumption for $d-1$ and with
$H$ replaced with $H^{\frac{d-1}{d}}$ to finish the proof.
\end{proof}

\begin{remark}
\label{rem:diff_rad}
The radius $1/6$ in Lemma~\ref{lem:high_cart} was chosen in order
to allow the use of Theorem~4 in~\cite{levin} as stated. However, it is straightforward
to obtain the following stronger statement: Given $\eps>0$, the lower bound~\eqref{eq:cart_bd}
is valid for all $z \in \prod^d_{j=1} \cD(z_{j,0}, 1-\eps)\setminus \cB$. The influence of $\eps$ is
only felt in the constants $C_d$. This can by seen by making some modifications to the proof
of Theorem~4 in~\cite{levin} and to the proof of Lemma~\ref{lem:high_cart}.
\end{remark}

Later we will need the following general assertion which is a combination of the Cartan-type
estimate of the previous lemma and Jensen's formula on the zeros of analytic functions.

\begin{lemma}
\label{lem:cart_zero}
Fix some $\uw_0=(w_{1,0}, w_{2,0}, \dots, w_{d,0})\in{\mathbb C}^d$ and suppose that $f(\uw)$ is an analytic function in
$\cP = \prod\limits^d_{j=1} D(w_{j,0},1)$.  Assume that
$ M \ge \sup_{\uw\in \cP} \log |f(\uw)|$, 
and let
$m \le  \log |f(\uw_1)|$ for some $ \uw_1 = (w_{1,1}, w_{2,1}, \dots, w_{d,1}) \in \prod\limits^d_{j=1} \cD(w_{j,0}, 1/2)$.
Given $H \gg 1$ there exists $\cB'_H \subset \cP' = \prod\limits^d_{j=2} \cD(w_{j,0}, 3/4)$, $\cB'_H \in
\car_{d-1} \left(H^{1/d}, K\right)$, $K = CH(M - m)$ such that for any $\uw' = (w_2, \dots, w_d) \in \cP' \setminus \cB'_H$ the
following holds: if
\[ \log|f(\tilw_1, \uw') | < M-C_dH(M-m) \text{\ \ for some\ \ }\tilde{w}_1\in\cD(w_{1,0},1/2),\]
then there exists $\hat w_1$ with $|\hat w_1 - \tilw_1| \lesssim e^{-H^{\frac1d}}$ such that $f(\hat w_1, \uw') = 0$.
\end{lemma}
\begin{proof} Due to Lemma~\ref{lem:high_cart}, and Remark~\ref{rem:diff_rad},
there exists $\cB_H \subset \cP$, $\cB_H \in
\car_d\bigl(H^{1/d}, K\bigr)$, $K = C_dH(M-m)$ such that for any $\uw \in \prod\limits^d_{j=1} \cD(w_{j, 0}, 3/4)\setminus \cB_H$
one has
\begin{equation}
\label{eq:fuw}
\log \big | f(\uw) \big | > M - C_d H(M - m)\ .
\end{equation}
By Lemma~\ref{lem:cart_12}, part (2), there exists $\cB'_H \subset
\prod\limits^d_{j=2} \cD\bigl(w_{j, 0}, 1\bigr)$, $\cB'_H \in
\car_{d-1} (H^{\frac{1}{d}},K)$ such that $\bigl(\cB_H\bigr)_{\uw'} \in
\car_1(H^{\frac{1}{d}},K)$ for any $\uw'=(w_2, \dots, w_d) \in \cB'_H$. Here $(\cB)_{\uw'}$
stands for the $\uw'$--section of $\cB$. Assume
$$
\log \big |f(\tilde w_1, \uw') \big | < M-C_dH(M-m)
$$
for some $\tilde w_1 \in \cD(w_{1,0},1/2)$, and $\uw'\in
\cP' \setminus \cB'_H$. Since $\bigl(\cB_H\bigr)_{\uw'} \in
\car_1(H^{\frac1d},K)$ there exists $r \lesssim \exp\bigl(-H^{1/d}\bigr)$ such that
\begin{equation}\nn
\left\{z: |z - \tilde w_1| = r\right\} \cap \bigl(\cB_H\bigr)_{\uw'} = \emptyset\ .
\end{equation}
Then in view of \eqref{eq:fuw},
$$
\log \big | f(z,\uw')|  > M - C_d H(M - m)
$$
for any $|z - \tilde w_1| = r$.  It follows from Jensen's formula, see \eqref{eq:jensen},  that
$f(\cdot, \uw')$ has at least one zero in the disk
$\cD(\tilde w_1, r)$, as claimed.
\end{proof}


\section{Large deviation theorems and the avalanche principle expansion for monodromies with impurities}
\label{sec:impure}

\noindent For the remainder of this paper we let $T:\tor\to \tor$ be the one-dimensional shift, i.e.,
$T(x)= x+\omega\;(\mod 1)$. In this section, we need to assume that~$\omega$ satisfies the Diophantine condition
\begin{equation}
\label{eq:diophant}
\|n\omega\|\geq \frac{c}{n(\log n)^{a}} \mbox{\ \ \ for all $n\ge1$}.
\end{equation}
and some $a>1$.
It is well-know that a.e.~$\omega$ satisfies this condition with some $c=c(\omega)>0$.
We denote the class of $\omega$ satisfying~\eqref{eq:diophant} by $\tor_{c,a}$.
For many applications in this paper
one can relax~\eqref{eq:diophant} considerably.
Another standing assumption we make is that the Lyapunov exponents are positive, i.e.,
$\inf_{E\in\IR} L(E)>\gamma>0$.
Recall from~\cite{GS} that under these assumptions there is the large deviation theorem
\begin{equation}
\label{eq:LDEom}
\mes[x\in\tor\:|\: |\log\|M_n(x,\omega,E)\| - nL_n(\omega,E)| > \delta n] \le C\exp(-c\delta n),
\end{equation}
for any $\delta  > (\log n)^B/n$ where $B,c, C$ are constants.

In this section we apply the results from the previous section to
study the following products:  For any  $1\le k_1 < \cdots < k_t \le n$ let
\begin{eqnarray}\label{eq3.1}
&& M^{(k_1,\dots,k_t)}_n (x,\omega, E) :=
M_{n-k_t}(x+k_t \omega,\omega, E)
\left[\begin{array}{rr} -1 &0\\ 0&0 \end{array} \right]
M_{k_t - k_{t-1} -1}(x+k_{t-1} \omega,\omega, E)
\left[\begin{array}{rr} -1 &0\\ 0&0 \end{array} \right] \\
&&\qquad \qquad \cdots M_{k_2 - k_1 -1}(x+k_1 \omega,\omega, E)
\left[\begin{array}{rr} -1 &0\\ 0&0 \end{array} \right]
M_{k_1-1} (x,\omega, E). \nonumber
\end{eqnarray}
We refer to these matrices as {\em monodromies with impurities}. The latter ones being the matrices
$\left[\begin{array}{rr} -1 &0\\ 0&0 \end{array} \right]$
at positions $k_1, k_2,\ldots, k_t$.
Matrices as in~\eqref{eq3.1} arise from taking derivatives of $M_N(x,\omega,E)$ in the parameters.
Furthermore, we consider the averages
\begin{equation}\label{eq3.2}
L^{(k_1,\dots,k_t)}_n  (\omega,E) =
\frac{1}{n} \int_0^1 \log \| M^{(k_1,\dots,k_t)}_n \
(x,\omega, E) \|\, dx \ .
\end{equation}

\noindent
In what follows, we will use the notation $f_n(x)$ for the determinants from the previous section
without further notice, see~\eqref{fn}. Our first result is a large deviation theorem for the
matrices in~\eqref{eq3.1}, followed by a uniform upper bound.

\begin{prop}
\label{prop:impureLDT}
For any positive integer $t$ there exist $0<\sigma<1$ and a constant $C=C(t)$ such that
\begin{equation}\nn
\mes\bigl[x\in\tor\:|\:|\log \| M^{(k_1,\dots, k_t)}_n (x,\omega,E) \| -  n L_n (\omega,E)| > Cn^{1-\sigma} \bigr]
\le C\exp(-n^\sigma)
\end{equation}
for all $n$ and any choice of $1\le k_1<k_2<\ldots< k_t\le n$.
\end{prop}
\begin{proof}
First, observe that
\begin{equation}\nn
\left[ \begin{array}{rr} -1 &0\\ 0 &0 \end{array} \right]
M_\ell (x,E) =
\left[ \begin{array}{cc} -f_\ell (x,E) &f_{\ell-1} (x+\omega,E)\\
0&0 \end{array} \right],
\end{equation}
see \eqref{mondet}. Thus (with $E$ fixed, and omitting $\omega,E$ from the variables for simplicity)
\begin{eqnarray}
&& M^{(k_1,\dots,k_t)}_n (x)  \nonumber \\
&&= \left[ \begin{array}{cc}
    f_{n-k_t}(x+k_t \omega) &-f_{n-k_t-1}(x+k_t \omega + \omega )\\
    f_{n-k_t-1}(x+k_t \omega) &-f_{n-k_t-2}(x+k_t \omega + \omega)
    \end{array}\right] \nonumber \\
&&\quad \left[ \begin{array}{cc}
    -f_{k_t- k_{t-1} -1}(x+k_{t-1}\omega) &f_{k_t- k_{t-1} -2}
    (x+k_t \omega + \omega )\\ 0&0 \end{array}\right] \nonumber \\
&&\cdots \left[ \begin{array}{cc}
   -f_{k_2-k_1-1}(x+k_1\omega) &f_{k_2-k_1-2}(x+k_1\omega +\omega)\\
   0&0 \end{array}\right]
   \left[ \begin{array}{cc}
   -f_{k_1-1} (x) &f_{k_1 -2 } (x+ \omega) \\
   0&0\\ \end{array} \right] \nonumber \\
&&= \left[ \begin{array}{cc}
\pm f_{n-k_t} (x+k_t \omega) f_{k_t -k_{t-1} -1} (x+ k_{t-1}\omega)
\cdots f_{k_2 - k_1- 1} (x+ k_1 \omega) f_{k_1-1} (x) & \star\\
\star & \star \\ \end{array} \right] \ .
\end{eqnarray}
In particular,
\begin{equation}\label{eq3.6}
\quad \| M_n^{(k_1,\dots,k_t)} (x) \| \ge
\Big| f_{n-k_t} (x+k_t \omega)
f_{k_t - k_{t-1}-1 } (x+k_{t-1} \omega)
\dots f_{k_2-k_1 -1} (x+k_1\omega) f_{k_1 -1} (x) \Big| \ .
\end{equation}
The large deviation theorem for the entries, Proposition~\ref{prop:thm2.1},  states that
\begin{equation}\nn
\log | f_m (x) | \ge mL_m - C m^{1-\tau }
\end{equation}
up to a set of $x$ of measure $\le \exp (-m^\tau)$ for some
constant $\tau > 0$.  Taking $\tau > 0$ small as we may,
we now distinguish between $k_{i+1} - k_i - 1\le n^{1-2\tau}$
and $k_{i+1} - k_i - 1 > n^{1-2\tau}$ in
(\ref{eq3.6}).
In the latter case,
\begin{equation}\nn
\log |f_{k_{i+1} - k_i - 1} (x) |
> (k_{i+1} - k_i - 1)
L_{k_{i+1}-k_i - 1} - Cn^{1-\tau}
\end{equation}
up to a set of $x$ of measure
$\le \exp (-n^{(1-2\tau) \tau })$.
If $m \le n^{1-2\tau}$, then one can directly invoke Cartan's estimate, see~Lemma~\ref{lem:riesz},
to conclude that
\begin{eqnarray}\nn
\mes \big[x\in \tor^d \:|\: \log | f_m (x) | < - n^{1-\tau} \big]
&\le & C \exp \left(-\frac{n^{1-\tau}}{Cm}\right) \\
&\le & C \exp \big(- c\,n^\tau \big) \nonumber
\end{eqnarray}
since $z\mapsto \log | f_m (z) | $ has Riesz mass $\le C m \le C n^{1-2\tau } $. In view of the preceding,
one has the following lower bound from~(\ref{eq3.6}):
\begin{eqnarray}\label{eq3.10}
\log \| M_n^{k_1,\dots,k_t} (x) \|
&\ge& (n-k_t) L_{n-k_t} + (k_t - k_{t-1} -1) L_{k_t - k_{t-1} -1} \nonumber \\
&& + \cdots + (k_2 - k_2 -1) L_{k_2 - k_2 - 1}
+ (k_1 - 1) L_{k_1 -1} -C (s+1) n^{1-\tau }
\end{eqnarray}
up to a set of $x$ of measure $\le C(s+1) \exp ( - n^{(1-2\tau) \tau })$.
Invoking the rate of convergence estimate
\begin{equation}
\sup_{E\in I}|L_m(\omega,E)-L_{n}(\omega,E)| \le \frac{C_0}{n}\ ,\quad m
\ge n\ ,
\label{eq:rate}
\end{equation}
for the Lyapunov exponents, see Theorem~5.1 in~\cite{GS},
 one concludes that the right-hand side in~(\ref{eq3.10})
is at least $nL_n - C(s+1) n^{1-\tau}$.
Here $I$ is a bounded interval, and $C_0=C_0(I,\gamma,\omega,V)$.
The lemma now follows with $\sigma = (1-2\tau) \tau $.
\end{proof}

\begin{lemma}
\label{lem:impureupper}
Let $M^{(k_1,\dots,k_t)}(x,\omega, E)$ be as in~(\ref{eq3.1}).
Then
\begin{equation}\label{eq3.32}
\sup\limits_x \| M_n^{(k_1,\dots, k_t)} (x,\omega,E) \|
\le n\, L_n(\omega,E) + C t n^{1-\sigma}.
\end{equation}
\end{lemma}
\begin{proof}
Clearly, by the definition stated in~(\ref{eq3.1}),
\begin{equation}\nn
\sup\limits_x \| M_n^{(k_1,\dots,k_t)} (x,E) \| \le
\sup\limits_x \| M_{n-k_t} (x) \|
\sup\limits_x \| M_{k_t - k_{t-1}} (x) \|
\cdot \dots \cdot
\sup\limits_x \| M_{k_1-1} (x) \| \ .
\end{equation}
Hence, by the uniform upper bound in \cite{BG}  (see Lemma~2.1 in that paper),
\begin{eqnarray}
\nn
\sup\limits_x \log \| M^{(k_1,\dots,k_t)}_n (x,E) \|
&\le& (n-k_t) L_{n-k_t} + C\,n^{1-\sigma}
+ (k_t - k_{t-1} - 1) L_{k_t - k_{t-1} - 1}\\
&&\quad + C\,n^{1-\sigma}  + \cdots + (k_2 - k_1 -1 ) L_{k_2 - k_1 - 1} +
C\,n^{1-\sigma} \nonumber\\
&\le& n\, L_n (E) + Ct\,n^{1-\sigma} \nonumber
\end{eqnarray}
where the final inequality follows from the rate of convergence estimate~\eqref{eq:rate}.
\end{proof}

\noindent In this paper we make repeated use of the avalanche principle from~\cite{GS}.
Recall that this device allows one to compare the norm of a long product of
unimodular matrices to the product of the norms of its factors, provided the factors are
large and adjacent factors do not cancel each other pairwise. For our present purposes we will
need to generalize this statement to non-unimodular matrices, cf.~\eqref{eq3.1}. Since the proof from~\cite{GS}
carries over without any changes, we do not present it. Recall the following facts, that are easily seen
by means of polar decomposition (see also Lemma~\ref{lem:stabil} above for the unimodular case):
For any $2\times 2$ matrix~$K$, we denote the normalized eigenvectors of $\sqrt{K^*K}$ by $\up_K$ and $\um_K$,
respectively. One has $K\up_K=\|K\|\vp_K$ and $K\um_K=|\det K|\|K\|^{-1}\vm_K$ where $\vp_K$ and $\vm_K$
are unit vectors. Both $\up_K\perp \um_K$, and $\vp_K\perp\vm_K$.
Given two unimodular $2\times2$ matrices $K$ and $M$, we let $\bp(K,M)=\vp_K\cdot \up_M$ and similarly
for $\bpm, \bmp$, and $\bmm$. These quantities are only defined up to a sign.

\begin{prop}
\label{prop:AP}
Let $A_1,\ldots,A_n$ be a sequence of  $2\times 2$--matrices whose determinants satisfy
\begin{equation}
\label{eq:detsmall}
\max\limits_{1\le j\le n}|\det A_j|\le 1.
\end{equation}
Suppose that
\be
&&\min_{1\le j\le n}\|A_j\|\ge\mu>n\mbox{\ \ \ and}\label{large}\\
   &&\max_{1\le j<n}[\log\|A_{j+1}\|+\log\|A_j\|-\log\|A_{j+1}A_{j}\|]<\frac12\log\mu\label{diff}.
\ee
Then
\begin{equation}
\Bigl|\log\|A_n\cdot\ldots\cdot A_1\|+\sum_{j=2}^{n-1} \log\|A_j\|-\sum_{j=1}^{n-1}\log\|A_{j+1}A_{j}\|\Bigr|
< C\frac{n}{\mu}
\label{eq:AP}
\end{equation}
with some absolute constant $C$.
\end{prop}

\noindent The case of $\max\limits_{1\le j\le n}|\det A_j|\le \kappa$ with some $\kappa>1$ is reduced to
$\kappa=1$ as in~\eqref{eq:detsmall} by means of rescaling.
A typical application Proposition~\ref{prop:AP} is as follows. The reader should not be mislead by the slightly
cumbersome formulation of Proposition~\ref{prop:APrep} below. It simply captures the differences between a monodromy
matrix with some impurities and a monodromy matrix of the same length without impurities in terms of matrices
``from a previous scale'', i.e., of much shorter length.
This is of course accomplished by applying the previous proposition and the conditions of it are
verified in terms of the large deviation theorem for the determinants. In this way one can ``cancel all
common factors''. The first formula~\eqref{eq:APrep} is sufficient for this section, but the second one,
\eqref{eq:APrepII}, will be needed later on. The point is that one has definite knowledge of the signs in the sums there.

\begin{prop}
\label{prop:APrep}
For any positive integer $t$ and $n$ large (depending on~$t$), there is a set $\cB\subset\tor$ with
$\mes(\cB) < n^{-100}$ and so that for all $x\in\tor\setminus\cB$,
\be
&& \log\|M_{n}^{(k_1,\ldots,k_t)}(x,\omega,E)\| = \log\|M_{n}(x,\omega,E)\|\nn \\
&& + \sum_{j=1}^{J} \vep_j\, \Bigl[\log\big\|M_{\ell_j}^{(\nu_{j1},\ldots,\nu_{jt_j})}(x+n_j\omega,\omega,E)\big\|
    - \log\|M_{\ell_j}(x+n_j\omega,\omega,E)\|\Bigr] + O(\frac{1}{n}), \label{eq:APrep}
\ee
with $J\lesssim t$, $\sum_{j=1}^{J_2} t_j \lesssim t$, $\ell_j\lesssim (\log n)^A$, $\vep_j=\pm1$,
$1\le \nu_{j1} < \ldots < \nu_{jt_j}\le \ell_j$, and some integers $n_j$ for $1\le j\le J$. Alternatively, one has for all $x\in\tor\setminus\cB$,
\be
&& \log\|M_{n}^{(k_1,\ldots,k_t)}(x,\omega,E)\| = \log\|M_{n}(x,\omega,E)\|
 + \sum_{j=1}^{J_1} \Bigl[ \Delta_{j}(x+m_j\omega) -  \widetilde{\Delta}_{j}(x+m_j\omega) \Bigr] \nn \\
&&  + \sum_{j=1}^{J_2} \Bigl[\log\big\|M_{\ell_j}^{(\nu_{j1},\ldots,\nu_{jt_j})}(x+n_j\omega,\omega,E)\big\|
    - \log\|M_{\ell_j}(x+n_j\omega,\omega,E)\|\Bigr] + O(\frac{1}{n}), \label{eq:APrepII}
\ee
with $J_1,J_2\lesssim t$, $m_j$ some integers, and with the other parameters satisfying the same restrictions
as before (but without being necessarily identical). Moreover,
\begin{equation}
\label{eq:longmonster}
\Delta_j(x) = \log\|M_{g_j}(x,\omega,E)\| + \log\|M_{h_j}(x+g_j\omega,\omega,E)\| -
\log\|M_{g_j+h_j}(x,\omega,E)\|
\end{equation}
where $g_j, h_j\asymp (\log n)^A$ are integers. $\widetilde{\Delta}_{j}$ is defined in a similar way,
with the same choice of $g_j, h_j$, but with monodromy matrices containing some number of impurities (no more than~$t$).
\end{prop}
\begin{proof}
For simplicity, we suppress both $\omega$ and~$E$ from most of the notation.
We apply the avalanche principle as in Proposition~\ref{prop:AP}  to
$M^{(k_1,\dots, k_t)}_n (x)$ and $M_n (x)$.  To do so,
let $n = \sum\limits_{j=1}^\nu \ell_j $ where
$(\log n)^{C_1} \le \ell_j \le 2 (\log n)^{C_1} $ for every
$j$. Correspondingly, one can write
\begin{eqnarray}
\label{eq3.14} M^{(k_1,\dots,k_t)}_n (x) &=&
\prod\limits_{j=1}^\nu \ A_j (x) {\hbox{\quad and }}\\
\label{eq3.15} M_n (x) &=&
\prod\limits_{j=1}^\nu \ B_j (x)
\end{eqnarray}
where $A_j (x)$ and $B_j (x)$ are matrix products of length
$\ell_j$.  More precisely,
\begin{equation}\nn
B_j (x) = M_{\ell_j} (x+s_j \omega) \ ,
\end{equation}
with $s_j = \sum\limits_{i<j} \ell_i $, and
\begin{equation}\nn
A_j (x) = M^{(k'_1 - s_j ,\dots, k'_{t'} - s_j)}_{\ell_j} \
(x+ s_j \omega )
\end{equation}
where $k'_1,\dots, k'_{t'} $ are precisely
those $k_1,\dots, k_t$ that fall into the interval
$[s_j + 1,\dots, s_{j+1}]$.  We now verify the conditions~\eqref{large} and~\eqref{diff}
for the product~(\ref{eq3.15}).
By the uniform upper bound of~\cite{BG},
\begin{equation}\label{eq3.18}
\sup\limits_{x\in\tor} \| B_j (x) \|
\le \ell_j L_{\ell_j} + C {\ell_j}^{1-\sigma}
\end{equation}
and by the large deviation theorem,
\begin{equation}\label{eq3.19}
\| B_{j+1} (x) B_j (x) \|
\ge \ell_{j+1} L_{\ell_{j+1}} + \ell_j L_{\ell_j} - C\ell^{1-\sigma} \ge
(\ell_j + \ell_{j+1}) L_{\ell_j + \ell_{j + 1}} - C\ell^{1-\sigma}
\end{equation}
up to a set of $x$ of measure not exceeding $\exp (-\ell^\sigma)$.
Here we again used the rate of convergence estimate for the
Lyapunov exponents from~\cite{GS}, i.e.,
\begin{equation}\label{eq:againused}
\ell_{j+1} L_{\ell_{j+1}} + \ell_j L_{\ell_j}
\le (\ell_j + \ell_{j+1}) L_{\ell_j + \ell_{j + 1}} + C \ .
\end{equation}
(\ref{eq3.18}) and~(\ref{eq3.19}) provide the local non-collapsing
condition~\eqref{diff} for the avalanche principle up to to set of~$x\in\tor$ of measure
at most $n \exp (-\ell^\sigma) \le n^{-100}$ by the
choice of $\ell$.  Furthermore, we can ensure that for the same~$x$
\begin{equation}
\nn
\log \| B_j (x) \| \ge \ell_j L_{\ell_j} - C\ell^{1-\sigma}
\ge \gamma (\log n )^{C_1}
\end{equation}
if $n$ is large. In particular, $\min\limits_j \| B_j (x) \| \ge n$ for those~$x$, and~\eqref{large} holds.

\noindent To verify the hypotheses of Proposition~\ref{prop:AP}  for~(\ref{eq3.14}) we use
Proposition~\ref{prop:impureLDT}.  By that lemma, there exists a set $\cB\subset\tor$ with
$|\cB | \le \exp (-\ell^\sigma) \le n^{-100} $ so that for all $x\in \TT \bs \cB$
\begin{equation}\nn
\min_j \| A_j (x) \| \ge n
\end{equation}
provided $n$ is sufficiently large depending on $t$.
By the same lemma we can ensure that for all $x\in \cB$ and
all $1\le j \le \nu -1 $
\begin{equation}\label{eq3.23}
\log \| A_{j+1} (x) A_j (x) \|
\ge  (\ell_{j+1} + \ell_j ) L_{\ell_{j+1} + \ell_j } -
C\ell^{1-\sigma } \ ,
\end{equation}
whereas by Lemma~\ref{lem:impureupper},
\begin{equation}\label{eq3.24}
\max\limits_x \log \| A_j (x) \| \le \ell_j L_{\ell_j }
+ C{\ell}^{1-\sigma} \ .
\end{equation}
The local non-collapsing condition for the avalanche principle
applied to~(\ref{eq3.14}) is given by~(\ref{eq3.23}) and~(\ref{eq3.24}).
We conclude that there is a set $\cB \subset \TT$, $\mes (\cB) \le n^{-100} $, so that
for all $x\in \TT \bs \cB$ the representation~\eqref{eq:AP} holds for both~\eqref{eq3.14} and~\eqref{eq3.15}
with~$\mu=n^2$, say. Subtracting these two representations from each other yields
\begin{eqnarray}
\log \| M^{(k_1,\dots, k_t)}_n (x) \| &=& \log \| M_n (x) \| - \sum\limits_{j\in \cJ_1} \log \| A_j (x) \|
+ \sum\limits_{j\in \cJ_1} \log \| B_j (x) \| \nonumber\\
&&\qquad\qquad + \sum\limits_{j\in \cJ_2} \log \| A_{j+1} (x) A_j (x) \|
- \sum\limits_{j\in \cJ_2} \log \| B_{j+1} (x) B_j (x) \|
+ O(1/n) \ . \label{eq:basicrep}
\end{eqnarray}
Here $\cJ_1$, and $\cJ_2$ are subsets of $\{ 1,\dots, \nu\}$
defined as
\[ \cJ_1 = \{ 1\le j\le \nu\:  |\: A_j \ne B_j \} \text{ \ \  and \ \ }
\cJ_2 = \{ 1\le j \le \nu-1 \:|\: A_{j+1} A_j \ne
B_{j+1} B_j \}.\]
In other words,
\begin{eqnarray}
\nn
\cJ_1 &=& \Big\{ 1\le j\le \nu \ \Big|\ {\hbox{ for some }}
1\le i \le t \ , \quad k_i \in [s_j + 1 , s_{j+1} ] \Big\}\\
\nn
\cJ_2 &=& \Big\{ 1\le j\le \nu-1 \ \Big| \ {\hbox{ for some }}
1\le i \le t-1 \ ,\  k_i \in [s_j + 1 , s_{j+2} ] \Big\},
\end{eqnarray}
and thus
\begin{equation}\nn
\# \cJ_1 \le t {\hbox{\quad and \quad }} \# \cJ_2 \le 2t \ .
\end{equation}
The representation \eqref{eq:basicrep} is exactly the one stated in~\eqref{eq:APrep}.
The obtain~\eqref{eq:APrepII}, one needs to combine (and possibly complete) the sums in~\eqref{eq:basicrep}
so that they form the triples that appear in~\eqref{eq:longmonster}.

\begin{figure}[ht]
\centerline{\hbox{\vbox{ \epsfxsize= 16.0 truecm \epsfysize= 1.0
truecm
\epsfbox{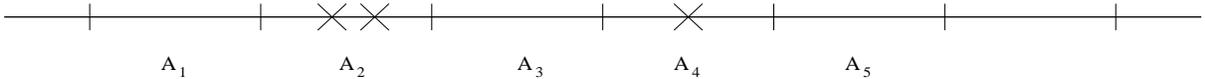}}}}
\caption{An example of impurities}
\label{fig:impure}
\end{figure}

In this process, some terms of the form $\log\|A_{j+1}A_j\|$ might be left over, but they come with positive sign.
The reader should consult Figure~3 
for an example of a distribution of impurities over
some monodromy matrices (the $\times$ marks represent impurities). In the process we just described,
one can let $A_1, A_2, A_1A_2$ and $A_3, A_4, A_3A_4$ form a triple as in~\eqref{eq:longmonster}
(clearly, this is not the only way of grouping the matrices). Notice that in this case the products $A_2A_3$ and $A_4A_5$ are left over.
\end{proof}

\noindent We now turn to a comparison of the averages in~\eqref{eq3.2} to the usual Lyapunov exponents.

\begin{lemma}
\label{lem:sharpavcomp}
For any nonnegative integer $t$ there exists a constant $C(t)$ such that
\begin{equation}\nn
\sup\limits_{1\le k_1 < k_2 < \cdots < k_t \le n} \
\big| L^{(k_1,\dots,k_t)}_n (\omega,E) - L_n (\omega,E) \big|
\le \frac{C}{n}
\end{equation}
for all $n$.
\end{lemma}
\begin{proof}
For simplicity, we suppress both $\omega$ and~$E$ from most of the notation. Let
\begin{equation}
\nn
R_t (n) :=
\sup\limits_{\frac{n}{2} \le m \le n} \
\sup\limits_{1\le k_1 \le k_2 \le \cdots \le k_t \le n} \
\big| L^{(k_1,\dots,k_t)}_m (E) - L_m (E) \big| \ .
\end{equation}
Note that we are allowing some of the $k_j$ to be identical,
which means that the number of the matrices
$\left[ \begin{array}{rr} -1 &0\\ 0&0 \end{array} \right] $
is $\le t$.
Apply Proposition~\ref{prop:APrep}.
Since each of the quantities in~\eqref{eq:APrep} is $O(n)$,
integrating in $x\in \TT$ leads to
\begin{equation}\nn
n \,\big| L^{(k_1,\dots, k_t)}_n (E) - L_n (E) \big|
\le 3t R_t(4(\log n)^{C_1}) + O(1/n)
+ t\, O (n\cdot n^{-100}) \ .
\end{equation}
Therefore, for $n \ge t$,
\begin{equation}
\label{eq3.30}
n \, R_t (n) \le 3t R_t ([4(\log n)^{C_1}]) +
t\, O\Big(\frac{1}{n}\Big) \ .
\end{equation}
Define scales $n_0 = n$, and $n_{j+1} = [4(\log n_j)^{C_1} ] $ for $j\ge0$.
Iterating \eqref{eq3.30} up to some $n_k$ that is determined by~$t$ alone, one obtains
\begin{eqnarray}
\nn
n\, R_t (n)
&\le & \frac{Ct^k}{n_0} +
\frac{Ct^{k-1}}{n_1} + \frac{Ct^{k-2}}{n_2} + \cdots +
\frac{Ct^0}{n_k} + 3tn_k R_t (4n_k) \\
&\le & C(t).\nn
\end{eqnarray}
Thus $R_t (n) \le \frac{C(t)}{n} $, as claimed.
\end{proof}

\noindent We now show how to use the avalanche principle to improve on Lemma~\ref{lem:meanlower}
and Proposition~\ref{prop:thm2.1}.

\begin{corollary}
\label{cor:betterdet}
There exist constants $A$ and $C$
depending on $\omega$ and the potential $V$, so that for every $n\ge1$
\be
\label{eq3.35}
&& \Big| \int_0^1 \log |\det (H_{[1,n]} (x,\omega) - E) |\, dx
- n \, L_n (\omega,E) \Big| \le C \\
\label{eq3.36}
&& \|  \log | \det (H_{[1,n]} (x,\omega) - E) |\, \|_\BMO \le
C(\log n)^A.
\ee
In particular, for every $n\ge1$,
\begin{equation}\label{eq:det_LDT}
\mes \Big[ x\in \TT \:\big|\:
|\log | \det (H_{[1,n]} (x,\omega) - E) | - n\, L_n (\omega,E) |
> H\Big] \le C\exp \left( - \frac{cH}{(\log n )^A} \right)
\end{equation}
for any $H>(\log n)^A$. Moreover, the set on the left-hand side is contained in at most $\les n$ intervals
each of which does not exceed the bound stated in~\eqref{eq:det_LDT} in length.
\end{corollary}
\begin{proof} As usual it suffices to consider the case of $\omega$.
By definition~\eqref{eq3.1} one has
\begin{eqnarray}\label{eq3.39}
M^{(1,n)}_n (x,\omega,E)
&=& \left[ \begin{array}{cc} -1&0\\ 0&0 \end{array} \right]
M_{n-1} (x,\omega,E)
\left[ \begin{array}{cc} -1&0\\ 0&0 \end{array} \right] \\
&=& \left[ \begin{array}{cc} -f_{n-1} (x,\omega,E) & f_{n-2} (x+\omega, E)\\
0&0 \end{array} \right]
\left[ \begin{array}{cc} -1&0\\ 0&0 \end{array} \right] \nonumber \\
&=& \left[ \begin{array}{cc} f_{n-1} (x,\omega,E) &0\\ 0&0 \end{array}\right].
\nonumber
\end{eqnarray}
Hence~(\ref{eq3.35}) follows from Lemma~\ref{lem:sharpavcomp}.
%
To obtain~(\ref{eq3.36}) one applies the avalanche principle to~\eqref{eq3.39}.
More precisely, let $n=\ell_1+(m-2)\ell+\ell_m$ where $\ell \asymp (\log n )^{C_0} $,
$\ell_1  \asymp \ell_n \asymp \ell $, and set $s_1 = 0$, $s_j = \ell_1 + (j-2) \ell $ for $2\le j \le m$.
Then
\[ M^{(1,n)}_n (x,\omega,E) = \prod_{j=m}^1 A_j(x)\]
where $A_j (x) = M_\ell (x+s_j \omega) $, for $2\le j \le m-1$, and $A_1(x)=M_{\ell_1} (x)$,
$A_m(x)= M_{\ell_m} (x+s_m\omega)$. As before, one shows via Proposition~\ref{prop:impureLDT} and
Lemma~\ref{lem:impureupper} that the conditions~\eqref{large} and~\eqref{diff}
hold up to a set of $x$ of measure $< n^{-100}$, say. Hence, by Proposition~\ref{prop:AP},
\begin{equation}
\label{eq3.40}
\log \| M^{(1,n)}_n (x,\omega,E) \| = - \sum\limits_{j=2}^{m-1} \log \| A_j (x) \|
+ \sum\limits_{j=1}^{m-1} \log\| (A_{j+1} A_j) (x) \| + O\left( \frac1{n}\right).
\end{equation}
We now recall the following large deviation theorem for sums of shifts of subharmonic functions,
see~Theorem~3.8 in \cite{GS}: For any subharmonic function $u$ on~$\cA_\rho$ with bounded Riesz mass and harmonic part
\begin{equation}
\label{eq:sumsofshifts}
 \mes\Bigl[x\in\tor \:|\: \bigl|\sum_{k=1}^n u(x-k\omega)- n\langle u\rangle\bigr|>\delta n\Bigr]<\exp(-c\delta n+r_n)
\end{equation}
where $r_n \lesssim (\log n)^A$ (Theorem~3.8 in~\cite{GS} is formulated for bounded subharmonic functions, but
all that is needed are a bound on the Riesz mass and the harmonic part).
The sums in~\eqref{eq3.40} involve shifts by $\ell\omega$  rather than~$\omega$. In order to overcome this,
note that we can take $\ell_n > 2\ell$, say. Repeating the argument that lead to~\eqref{eq3.40} $\ell-1$ times
with the length of $A_1$ increasing by one and that of~$A_m$ decreasing by one, respectively,  at each step leads to
\be
\nn
\log \| M^{(1,n)}_n (x,\omega,E) \| &=& - \frac{1}{\ell}\sum_{k=0}^{\ell-1}\sum\limits_{j=2}^{m-1}
\log \| A_j (x+k\omega) \| + \frac{1}{\ell} \sum_{k=0}^{\ell-1} \sum\limits_{j=2}^{m-2} \log\| (A_{j+1} A_j) (x+k\omega) \| \\
&& \qquad\qquad + \frac{1}{\ell}\sum_{k=0}^{\ell-1} u_k(x) + O\left( \frac1{n}\right) \nn \\
&=& - \frac{1}{\ell}\sum\limits_{j=\ell}^{(m-1)\ell-1} \log \| M_\ell (x+j\omega) \| +
 \frac{1}{\ell}\sum\limits_{j=\ell}^{(m-1)\ell-1} \log\| M_{2\ell} (x+j\omega) \| \nn \\
&& \qquad\qquad + \frac{1}{\ell}\sum_{k=0}^{\ell-1} u_k(x) +     O\left( \frac1{n}\right). \label{eq:nuisance}
\ee
The functions $u_k$ compensate for omitting the terms $j=1$ and $j=m-1$ when
summing~$\log \| A_{j+1} A_j \|$.  They are subharmonic, with Riesz mass and harmonic part bounded by~$(\log n )^{C_0}$.
Estimating the sums involving~$M_\ell$ and~$M_{2\ell}$ by means of~\eqref{eq:sumsofshifts},
and the sums involving~$u_k$ directly by means of Lemma~\ref{lem:riesz} shows that
there exists $\cB\subset\tor$ of measure $\le \exp (-(\log N)^{C_0})$, so that for all $x\in\tor\setminus\cB$,
\begin{equation} \nn
\Big| \log \| M^{(1,n)}_n (x,\omega,E) \| -
\langle \log \| M_n^{(1,n)} (x,\omega,E)\| \rangle \Big|
\le (\log n )^{2C_0}.
\end{equation}
Thus,
\begin{equation}\nn
\log \| M^{(1,n)}_n  (x,\omega,E) \| = u_0 (x) + u_1 (x) \ ,
\end{equation}
where
\[ \| u_0 - \langle \log \| M^{(1,n)}_n(\cdot , E) \| \rangle \|_{L^\infty(\tor)}  \le (\log n)^{2C_0},\]
and
\begin{eqnarray}\nn
\| u_1 -\langle \log \| M^{(1,n)}_n (\cdot , E) \|
\rangle \|_{L^1 (\tor) }
&\lesssim & \| \log \| M^{(1,n)}_n(\cdot , E) \|\;  \|_{L^2 (\tor)} \sqrt{\mes (\cB)} \\
&\lesssim &  n \cdot \sqrt{\mes (\cB)}
\lesssim  \exp \left( - \frac14 (\log n)^{C_0} \right).\nn
\end{eqnarray}
Now apply the splitting lemma, Lemma~2.3 from \cite{BGS}
(see the proof of Proposition~\ref{prop:thm2.1} above for some details concerning the hypotheses of that lemma) one obtains that
\begin{eqnarray}\nn
\Big\| \log \| M^{(1,n)}_n (x,\omega,E) \| \Big\|_{\BMO (\tor)}
&\le& C \left( (\log n)^{2C_0 + 1} +
\sqrt{n \cdot \exp \left( -\frac14 (\log n)^{C_0}\right)} \right) \\
&\le& C\,(\log n)^{2C_0 + 1} \ , \nonumber
\end{eqnarray}
as claimed.
\end{proof}

\begin{remark}
\label{rem:impure}
The same method of proof shows that for any nonnegative integer $t$,
\[ \bigl\| \log\|M^{(k_1,\ldots,k_t)}_n(\cdot,\omega,E)\| _{BMO}\;\bigr\| \le C(t)\, (\log n)^A,\]
for large $n\ge n_0(t)$.
\end{remark}

The following large deviation theorem in the $E$ variable will be applied later
in this paper.

\begin{corollary}
\label{cor:E_LDT}
Given 
$N\gg 1$, there exists $\cB_{N,\omega}
\subset \tor$ with $\mes \cB_{N,\omega} < \exp\left(-(\log N)^A\right)$,
such that for each $x
\in \tor \setminus \cB_{N,\omega}$ there exists $\cE_{N,\omega, x} \subset
\IC$, with $\mes \cE_{N,\omega, x} \le \exp\left(-(\log N)^A\right)$
such that
\begin{equation}
\label{eq:ELDT}
\log \bigm | f_N(x, \omega,E)\bigm | > NL(\omega,E) - (\log N)^C
\end{equation}
for any $E \in \IC \setminus \cE_{N,\omega, x}$.
\end{corollary}
\begin{proof} It follows the from the large deviation theorem for $f_N (z,\omega,E)$ that \eqref{eq:ELDT} holds for any $(x, E) \in (\tor \times \IC)\setminus
\cF$ where $\mes \cF < \exp\left(-2(\log N)^A\right)$.  By Fubini's theorem
there exists $\cB_{N,\omega}\subset \tor$ with $\mes\cB_{N,\omega} <
\exp \left(-(\log N)^A\right)$ such that for any $x \in \tor \setminus
\cB_{N,\omega}$ there exists $\cE_{N,\omega, x} \subset \IC$ with
$\mes\cE_{N,\omega, x}< \exp\left(-(\log N)^A\right)$ such that~\eqref{eq:ELDT}
holds for any $E \in \IC\setminus \cE_{N,\omega, x}$, as claimed.
\end{proof}

\begin{remark}
\label{rem:LDTE}
Note that the large deviation theorem in the $E$-variable from Corollary~\ref{cor:E_LDT}
requires the removal of a small bad set in $x$. This is in contrast to the LDT in $x$,
which holds uniformly in $E$. In addition, in the LDT with respect to the $E$ variable,
we do not have any control over the complexity.
\end{remark}

One immediate application is the following avalanche principle expansion for fixed $x \in \tor$.

\begin{corollary}
\label{cor:AP_exp}
Given 
$N \gg 1$, let  $\cB_{N, \omega}
\subset \tor$ and $\cE_{N,\omega, x} \subset
\IC$ be as in Corollary~\ref{cor:E_LDT}. 
Then for any $(x,E)\subset \tor\setminus\cB_{N,\omega}\times \IC\setminus \cE_{N,\omega, x}$
\begin{equation*}
\begin{split}
\log \bigl | f_N(x, \omega,E)\bigr| & = \sum^{n-1}_{j=1} \log
\bigl\|A_{j+1}(E) A_j(E)\bigr\| -\\
&\quad \sum^{n-1}_{j=2} \log \bigl\|A_j(E)\bigr\| +
O\left(\exp\bigl(-(\log N)^A\bigr)\right)
\end{split}
\end{equation*}
where $A_1(E) = M_{\ell_1}\bigl(e(x), \omega,E\bigr)
\textstyle{\begin{bmatrix} 1 & 0\\ 0 & 0\end{bmatrix}}, A_j(E) =
M_{\ell}\left(e\Bigl(x + \bigl((j -2) \ell +
\ell_1\bigr)\omega\Bigr), \omega,E\right)$, $j = 2, \dots, n-1$,
$A_n(E) = \textstyle{\begin{bmatrix} 1 & 0\\ 0 & 0\end{bmatrix}}
M_\ell \left(e\Bigl(x + \bigl((n-2)\ell + \ell_1\bigr) \omega\Bigr),
\omega,E\right)$, $\ell_1 + (n-1)\ell = N,\, \ell_1 \ge \bigl(\log
N\bigr)^A$.
\end{corollary}
\begin{proof} By Corollary~\ref{cor:E_LDT}, the
conditions of the avalanche principle hold for all $(x,E)\subset \tor\setminus\cB_{N,\omega}\times \IC\setminus \cE_{N,\omega, x}$.
\end{proof}

We conclude this section with some simple observations regarding
continuity in the variables $\omega,E$.

\begin{lemma} For any positive integer $\ell$,
\label{lem:Eomdiff}
\[ \bigl\|M_\ell(z, \omega_1, E_1) - M_\ell(z, \omega_2,
E_2)\bigr\| \le C^\ell\left(1 + \sup |V| + |E_1| + |E_2|\right)^\ell
\left(|\omega_1 - \omega_2| + |E_1 - E_2|\right).\]
\end{lemma}

\begin{lemma} Let 
$E_0 \in \IC$.  Then
\begin{equation}\label{eq:3.35}
\left | \log {\bigl\|M_n(z, \omega,E)\bigr\|\over \bigl\|M_n(z,
\omega_0,E_0)\bigr\|}\right| < \exp \left(-(\log n)^A\right)
\end{equation}
for any $z \notin \cB_{E_0, \omega_0}$, $\mes \cB_{E_0, \omega_0} <
\exp\left(-(\log n)^{C}\right)$, $|\omega - \omega_0| + |E - E_0| <
\exp\left(-(\log n)^{2A}\right)$ where $A\gg1$. 
The same statement also holds for monodromies with impurities.
\end{lemma}
\begin{proof} By the avalanche principle there exists $\cB_{E_0, \omega_0}
\subset \tor$, $\mes\left(\cB_{E_0, \omega_0}\right) \le \exp(-\sqrt
\ell)$ so that for all $z \in \cA_0 \setminus \cB_{E_0, \omega_0}$ one has
\[ \log \bigl\|M_n(z, E_0, \omega_0)\bigr\| = \sum\limits^{m-1}_{j=1}
\log \bigl\|B_{j+1} B_j (z, E_0, \omega_0)\bigr\| -
\sum\limits^{m-1}_{j=2} \log \bigl\| B_j(z, E_0, \omega_0)\bigr\| +
O(e^{-c\ell})
.\]
Here $B_j$ are monodromies of length $\asymp \ell$ where $\ell=(\log n)^C$.
It follows from Lemma~\ref{lem:Eomdiff} that $B_j(z, E,
\omega)$ satisfy the conditions of the avalanche principle for $z \in
\cA_0 \setminus \cB_{E_0, \omega_0}$ and $\omega,E$ so that
\[ |\omega - \omega_0| + |E - E_0| <
\exp\left(-\ell^2\right). \]
Subtracting the avalanche principle representations of $\log
\bigl\|M_\ell(z, E_0, \omega_0)\bigr\|$ and $\log \bigl\|M_\ell(z, E,
\omega)\bigr\|$, and using  Lemma~\ref{lem:Eomdiff}, yields \eqref{eq:3.35}.  The same argument
also applies to monodromies with impurities, and we are done.
\end{proof}

\noindent Set
\begin{align*}
L_N(y, \omega, E) & = \la N^{-1} \log \bigl\|M_N\bigl(e (\cdot +
iy),
\omega, E\bigr)\bigr\|\ra\ ,\\
L(y, \omega, E) & = \lim_{N\to \infty} L_N(y, \omega, E)\ ,\\
L_N(\omega, E) & = L_N(0, \omega, E),\quad L(\omega, E) = L(0,
\omega, E).
\end{align*}

\begin{corollary}
\label{cor:L_comp}
For any positive integer $n$,
$$
\bigl| L_n(y, \omega,E) - L_n(y, \omega_0, E_0)\bigr | < \exp
\left(-(\log n)^A\right)
$$
provided $|\omega - \omega_0| + |E - E_0| < \exp \left(-(\log n)^{2A}\right)$.
\end{corollary}
\begin{proof}
Integrate out \eqref{eq:3.35}.
\end{proof}

\section{Uniform upper estimates on the norms of monodromy matrices}
\label{sec:upper}

\noindent As in the previous section, we only consider the shift model on~$\tor$. We will need the
following version of the large deviation estimates for the monodromy matrices.
Suppose the potential function is analytic on the annulus $\cA_\rho$. Then
\begin{equation}
\label{eq:LDEiy}
\sup_{|y|<\frac{\rho}{2}} \mes\big[x\in\tor\:|\: |\log\|M_N(x+iy,\omega,E)\| -
\langle \log\|M_N(\cdot+iy,\omega,E)\| \rangle | > \delta N\big ] \le C\exp(-c\delta N)
\end{equation}
provided $\delta>N^{-1}(\log N)^C$.
The same proof in~\cite{GS} that leads to \eqref{eq:LDEom} also
establishes~\eqref{eq:LDEiy}. Indeed, if $F$ is analytic on~$\cA_\rho$ and one sets
$V_y(x):=F(e(x+iy))=F(e(x)e^{-2\pi y})$, then the monodromy $M_n$, the Lyapunov exponent~$L_n$, and
the determinants~$f_n$ can all be defined as in Section~\ref{sec:dets} with $V_y$ instead of~$V$.
Generally speaking, $V_y$ is not real if $y\not=0$ (consider, for example, $F(z)=z^{-1}+z$) and
the equation~\eqref{eq:schr} is therefore no longer given in terms of a self-adjoint operator.
However, since the proofs of~\eqref{eq:LDEiy} do not depend on~$V$ being real, but only
rely on subharmonicity and almost invariance of~$\log\|M_N(x,\omega,E)\|$,  they equally well
apply to the case~$y\not=0$.
This will make it necessary to
study the Lyapunov exponent as a function of~$y$.
We  start by showing that these Lyapunov exponents are Lipschitz in~$y$. As shown in the following lemma, this is a
rather general property of averages of subharmonic functions.

\begin{lemma}
\label{lem:liprad}
Let $1>\rho>0$ and suppose $u$ is subharmonic on~$\cA_\rho$ such that
$\sup_{z\in \cA_\rho} u(z)\le 1$ and $\int_{\tor} u(e(x))\,dx\ge0$. Then
for any $r_1,r_2$ so that $1-\frac{\rho}{2} < r_1,r_2 < 1+\frac{\rho}{2}$ one has
\[ |\langle u(r_1 e(\cdot)) \rangle - \langle u(r_2e(\cdot)) \rangle| \le C_\rho\,|r_1-r_2|,\]
here $\la v(\cdot)\ra = \int^1_0 v(\xi) d\xi$.
\end{lemma}
\begin{proof}
By Lemma~\ref{lem:riesz},
\[ u(z) = \int \log|z-\zeta|\, d\mu(\zeta) + h(z),\]
where $\mu$ is a positive measure supported on $\cA_{\rho/2}$, and $h$ is harmonic on~$\cA_{\rho/2}$.
Moreover, by our assumptions on~$u$,
\[ \mu(\cA_{\rho/2}) + \|h\|_{L^\infty(\cA_{\rho/4})} < C_\rho.\]
It is a standard property of harmonic functions on the annulus that
\[ |\langle h(r_1 e(\cdot)) \rangle - \langle h(r_2e(\cdot)) \rangle| \le C_\rho\,|r_1-r_2|.\]
On the other hand, for $v(z)= \int\log|z-\zeta|\,d\mu(\zeta)$ one has
\[
   \int_{|z|=r_j} v(z)\,d\sigma(z) = \int_{|\zeta|>r_j} \log|\zeta|\,d\mu(\zeta) +
   \int_{|\zeta|<r_j} \log r_j \, d\mu(\zeta)
\]
for j=1,2. Suppose $r_1<r_2$. Then subtracting these identities from each other yields
\[ \int_{|z|=r_2} v(z)\,d\sigma(z)-\int_{|z|=r_1} v(z)\,d\sigma(z) =
   -\int_{r_1 < |\zeta| < r_2} \log\frac{|\zeta|}{r_2}\, d\mu(\zeta) + \int_{|\zeta|< r_1} \log \frac{r_2}{r_1}\,d\mu(\zeta),
\]
and the lemma follows.
\end{proof}

Then we have the following corollary regarding the continuity of
$L_N$ in~$y$.

\begin{corollary}
\label{cor:liplap}
Let $L_N(y,\omega,E)$ and $L(y,\omega,E)$ be defined as above. Then
with some constant $\rho>0$ that is determined by the potential,
\[
|L_N(y_1,\omega,E) - L_N(y_2,\omega,E)| \le C|y_1-y_2|
 \text{\ \ \ for all\ \ \ }|y_1|,|y_2| < \rho
\]
uniformly in $N$. In particular, the same bound holds for $L$ instead of~$L_N$ so that
\[ \inf_{E} L(\omega,E)>\gamma>0 \]
implies that
\[ \inf_{E,|y|\ll \gamma} L(y,\omega,E) > \frac{\gamma}{2}.\]
\end{corollary}
\begin{proof} This follows immediately from the definitions and Lemma~\ref{lem:liprad}.
\end{proof}

The following result improves on the uniform upper bound on the monodromy matrices
from~\cite{BG} and~\cite{GS}. The $(\log N)^A$ error here (rather than $N^\sigma$, say, as in~\cite{BG} and~\cite{GS})
will be crucial for the study of the fine properties of the integrated density of states as
well as the distribution of the zeros of the determinants.

\begin{prop}
\label{prop:logupper}
Let $\omega$ be as in \eqref{eq:diophant}. Assume $L(\omega, E) > 0$. Then for all large integers $N$,
\begin{equation}\nn
\sup\limits_{x\in\tor} \log \| M_N (x,\omega,E) \|
\le NL_N(\omega,E)+ C(\log N)^A \ ,
\end{equation}
for some constants $C$ and $A$.
\end{prop}
\begin{proof} As usual, we only consider $\omega$ and suppress $\omega$ and $E$ from most of the notation.
Take $\ell \asymp (\log N)^A$.  Write $N=(n-1) \ell + r$,
$\ell \le r < 2\ell$ and correspondingly
\[ M_N(x) = M_r(x+(n-1)\ell\omega) \prod_{j=n-2}^0 M_\ell(x+j\ell\omega).\]
The avalanche principle and the LDT~\eqref{eq:LDEiy} imply that  for every small $y$
there exists $\cB_y\subset\tor$ so that $\mes (\cB_y) < N^{-100} $ and such that for $x\in [0,1]\bs \cB_y$,
\begin{eqnarray}
\log \| M_N(x + iy)\|
&=& \sum\limits_{j=0}^{n-3} \log \| M_{2\ell} \big( x + j \ell \omega + iy \big) \|  -
\sum\limits_{j=1}^{n-2} \log \| M_\ell(x + j\ell \omega + iy ) \| \nn \\
&& \qquad\qquad + \log\|M_r(x+(n-1)\ell \omega)M_\ell(x+(n-2)\ell\omega)\| + O(1) \nn \\
&=& \sum\limits_{j=0}^{n-3} \log \| M_{2\ell} \big( x + j \ell \omega + iy \big) \|
 - \sum\limits_{j=1}^{n-2} \log \| M_\ell(x + j\ell \omega + iy ) \| + O(\ell). \label{eq4.2}
\end{eqnarray}
Combining the elementary almost invariance property
\begin{equation}\nn
\log \| M_N (x+iy) \| = \ell^{-1}
\sum\limits_{0\le j \le \ell-1} \log \| M_N (x+j\omega + iy) \| + O(\ell)
\end{equation}
with \eqref{eq4.2} yields
\begin{eqnarray}\label{eq4.4}
\log \| M_N (x+iy) \|
&=& \ell^{-1} \sum\limits_{0\le j < N}
\log \| M_{2\ell} (x+j\omega + iy) \| \\
&& \qquad\qquad  - \ell^{-1} \sum\limits_{0\le j < N} \log \| M_\ell (x+j\omega + iy) \| + O(\ell) \ , \nn
\end{eqnarray}
for any $x\in [0,1] \bs B'_y$, where $\mes B'_y < N^{-9}$.
Integrating~\eqref{eq4.4} over~$x$ shows that
\begin{eqnarray}\label{eq4.5}
L_N(y,E) = 2 L_{2\ell}(y,E) -  L_\ell(y,E) + O(\ell/N).
\end{eqnarray}
This identity is basically (5.3) in \cite{GS} (with $y=0$). Since
the Lyapunov exponents are Lipschitz in~$y$, the sub-mean value
property of subharmonic functions on the disk $\cD(x,0;\delta)$
with $\delta=N^{-1}$ in conjunction with~\eqref{eq4.4}
and~\eqref{eq4.5}  implies that, for every~$x\in\tor$,
\begin{eqnarray}
&&\log \| M_N (x) \| - \int_0^1  \log \|M_N (\xi) \|\, d\xi \nonumber \\
&&\le  \strichint\limits_{\cD(x,0;\delta)}\
 \left[ \sum_{0\le j < N} \
u(\xi + j\omega + i\eta ) - N\langle u (\cdot + i\eta) \rangle \right]\,d\xi d\eta
\nonumber \\
&&- \strichint\limits_{\cD(x,0;\delta)}\
 \left[ \sum_{0\le j < N} \
v(\xi + j\omega + i\eta ) - N \langle v (\cdot + i\eta) \rangle \right]\,d\xi d\eta + O(\ell) \label{eq4.6}\ ,
\end{eqnarray}
where $\strichint\limits_{\cD(x,0;\delta)}$ denotes the average over the disk,
\begin{equation}
u(\xi + i\eta) := \ell^{-1}\,\log \| M_{2\ell} (\xi+ i\eta) \| \text{\ \ and\ \ } v(\xi + i\eta) := \ell^{-1}\,\log \| M_{\ell} (\xi+ i\eta) \|,
\end{equation}
and $\langle \cdot \rangle$ denotes averages over the real line. Note that we have included the error
that arises from the bad sets $\cB_y$ into the~$O(\ell)$-term. Passing to the slightly larger
squares $Q(x,0;2\delta)$ of side-length $2\delta$ that contain the disk~$\cD(x,0;\delta)$,
one concludes from~\eqref{eq4.6} that for every $x\in\tor$,
\begin{eqnarray}
&&\log \| M_N (x) \| - \int_0^1  \log \|M_N (\xi) \|\, d\xi \nonumber \\
&&\lesssim  \strichint\limits_{Q(x,0;2\delta)}\
 \left| \sum_{0\le j < N} \
u(\xi + j\omega + i\eta ) - N\langle u (\cdot + i\eta) \rangle \right|\,d\xi d\eta
\nonumber \\
&& + \strichint\limits_{Q(x,0;2\delta)}\
 \left| \sum_{0\le j < N} \
v(\xi + j\omega + i\eta ) - N \langle v (\cdot + i\eta) \rangle \right|\,d\xi d\eta + O(\ell). \label{eq:squares}
\end{eqnarray}
By the large deviation estimate \eqref{eq:sumsofshifts}, the absolute values in these integrals do not
exceed $(\log N)^A$ up to a set of measure~$N^{-100}$, say. Since they cannot exceed $CN$ on this bad set,
we are done.
\end{proof}

We now list some simple consequences of this upper bound. We start
by observing that Proposition~\ref{prop:logupper} applies
uniformly in a small neighborhood of $\omega,E$. The size of this
neighborhood turns out to be much larger than the trivial one,
which would be $e^{-CN}$. This has to do with the fact that we do
not compare matrices of length $N$, but rather those of length
$(\log N)^C$, as given by the avalanche principle.

\begin{corollary}
\label{cor:Eomunif}
Fix $\omega_1$ as in \eqref{eq:diophant} and $E_1 \in \IC$, $|y| < \rho_0$. Assume that $L(y,\omega_1,E_1) > 0$.  Then
\begin{align*}
 & \sup \left\{ \big \| M_N \left(e(x+iy), \omega,E\right) \big \| : |E - E_1| + |\omega - \omega_1|
< \exp\left(-(\log N)^C\right), x \in \tor\right\}\\
&\qquad \lesssim \exp \left(NL_N (y,\omega_1,E_1) + (\log N)^A\right)
\end{align*}
for all $|y| < \rho_0$.
\end{corollary}
\begin{proof} Due to the elementary estimate of Lemma~\ref{lem:Eomdiff} the arguments of the previous proof
apply uniformly to all $\omega,E$ with
\[ |E - E_1| + |\omega - \omega_1|
< \exp\left(-(\log N)^C\right). \]
Therefore, the statement of the Proposition~\ref{prop:logupper} also holds uniformly in this neighborhood.
Finally, by Corollary~\ref{cor:L_comp} we can bound everything in terms of the Lyapunov exponents at the points $\omega_1,E_1$.
\end{proof}

\begin{corollary}
\label{cor:2}
Fix $\omega_1$ as in \eqref{eq:diophant} and $E_1 \in \IC$, $|y| < \rho_0$. Assume that $L(y,\omega_1,E_1) > 0$.
Let $\partial$ denote any of the partial derivatives $\partial_x, \partial_y,
\partial_E$ or $\partial_\omega$.  Then
\begin{align*}
 & \sup \left\{\big \| \partial M_N\left(e(x+ iy), \omega,E\right) \big \|: |E - E_1| + |\omega -
\omega_1| < e^{-(\log N)^C}, x \in \tor\right\}\\
&\qquad \lesssim \exp \left(N L_N(y,\omega_1,E_1) + (\log N)^A\right)
\end{align*}
for all $|y| < \rho_0$.
\end{corollary}
\begin{proof} Clearly, for all $x, y, \omega,E$,
\begin{align*}
\partial M_N \left(e(x+ iy), \omega,E\right) & = \sum^N_{n=1} M_{N-n} \left(e(x + n\omega + iy), \omega,E\right)
\partial \begin{bmatrix}
\lambda V - E  & -1\\ 1 & 0\end{bmatrix} M_{n-1} \left(e(x + iy), \omega,E\right)\ .
\end{align*}
Since $|E - E_1| + |\omega - \omega_1| < e^{-(\log N)^C}$, the statement now follows from
Corollary~\ref{cor:Eomunif} and the estimate~\eqref{eq:againused}
on the Lyapunov exponents.
\end{proof}

\begin{corollary}
\label{cor:4.6}
Under the assumptions of the previous corollary,
\begin{align*}
& \left\| M_N \left(e(x + iy), \omega,E\right) - M_N \left(e(x_1 + iy_1),\omega_1,E_1\right) \right\|\\
\lesssim & \left(|E - E_1| + |\omega - \omega_1| + |x - x_1| + |y - y_1|\right) \cdot
 \exp \left(NL_N \left(y_1, \omega_1, E_1\right) + (\log N)^A\right)
\end{align*}
provided $|E - E_1| + |\omega - \omega_1| + |x - x_1| < e^{-(\log N)^A}$, $|y_1| < \rho_0/2$, $|y - y_1| < N^{-1}$.
In particular
\begin{equation}
\label{eq:fN_comp}
\begin{split}
\left| \log {\bigl | f_N\bigl(e(x+iy), \omega, E\bigr)\bigr|\over \bigl|
f_N\bigl(e(x_1 + iy_1), \omega_1, E_1\bigr)\bigr|} \right| & \lesssim
\left(|E -E_1| + |\omega - \omega_1| + |x - x_1|+ |y - y_1|\right)\\
&\qquad {\exp\left(NL(y_1,\omega_1, E_1\right) + (\log N)^A\bigr)\over \bigl| f_N\bigl(e(x_1 + iy_1), \omega_1,
E_1\bigr)\bigr|}\ ,
\end{split}
\end{equation}
provided the right-hand side of~\eqref{eq:fN_comp} is less than $1/2$.
\end{corollary}
\begin{proof} This follows from Corollaries~\ref{cor:2} and \ref{cor:liplap}.
\end{proof}

\begin{corollary} Using the notation of the previous corollary one has
\begin{equation}
\label{eq:MN_vergl}
\left | \log {\big \|M_N\left(e(x + iy), \omega, E\right) \big \|\over \big \|M_N \left(e(x_1 + iy_1), \omega_1, E_1
\right)\big \|} \right|  < \exp \left(-(\log N)^A\right)
\end{equation}
\begin{equation}
\left| \log {\big | f_N\left(e(x+ iy), \omega, E\right) \big |\over \big | f_N\left(e(x_1 + iy_1),\omega_1, E_1
\right)\big |} \right |  < \exp\left(-(\log N)^A\right)
\label{eq:fN_vergl}
\end{equation}
for any $|E - E_1| + |\omega - \omega_1| + |x - x_1| + |y - y_1| < \exp \left(-(\log N)^{2A}\right)$, $e(x_1 + iy_1)
\in \cA_{\rho_0/2} \setminus \cB_{\omega_1,E_1}$, where $\mes \cB_{\omega_1,E_1} < \exp \left(-(\log N)^{A/2}\right)$,
$\compl(\cB_{\omega_1,E_1}) \lesssim N$.
\end{corollary}
\begin{proof} Due to the large deviation theorem there exists $\cB_{\omega_1,E_1}$ as above such that
$$
\log \big \| M_N\left(e(x_1 + iy_1), \omega_1, E_1\right) \big \| \ge NL_N(y_1, \omega_1, E) - (\log N)^A
$$
for any $e(x_1 + y_1) \in \cA_{\rho_0/2} \setminus \cB_{\omega_1,E_1}$.  Therefore, \eqref{eq:MN_vergl}
follows from Corollary~\ref{cor:4.6}.  The proof
of~\eqref{eq:fN_vergl} is similar.
\end{proof}

\begin{corollary}
\label{cor:4.8}
Under the assumptions of Corollary~\ref{cor:4.6},
\begin{equation}
\label{eq:int_comp}
\iint_{\cA_{\rho_0}} \Big |\log \big | f_N\bigl(z, \omega, E\bigr)\big | - \log \big
|f_N\bigl(z, \omega_1, E_1\bigr)\big |\Big | dz\wedge d\bar{z} \le \exp\bigl(-(\log
N)^{A/2}\bigr)
\end{equation}
for any $|E - E_1| + |\omega - \omega_1| < \exp\bigl(-(\log N)^A\bigr)$.
\end{corollary}
\begin{proof}
Due to Lemma~\ref{lem:basic2}
\begin{equation}
\Big \| \log \big |f_N(\cdot, \tilde\omega, \widetilde E)\big | \Big
\|_{L^2(\cA_{\rho_0/2})} \le N^b
\end{equation}
for any $\tilde\omega, \widetilde E$ and some constant $b>0$.  Therefore, \eqref{eq:int_comp} follows from~\eqref{eq:fN_vergl}.
\end{proof}

\noindent The following proposition presents a typical application of the methods developed so far in this paper.
This result will be considerably refined later in this paper.

\begin{prop}
\label{prop:zero_count}
Let $\omega$ satisfy~\eqref{eq:diophant}. Then
for any $x_0 \in \tor$, $E_0 \in \IR$ one has
\begin{align}
\label{eq:null1}
\# \left\{E \in \IR\::\: f_N\bigl(e(x_0), \omega,E\bigr) = 0, |E - E_0| < \exp \left(-(\log N)^A\right)\right\} &\le (\log N)^{A_1} \\
\# \left\{z \in \IC\::\: f_N(z, \omega, E_0) = 0, |z - e(x_0)| <
N^{-1}\right\} &\le (\log N)^{A_1} \label{eq:null2}
\end{align}
for all sufficiently large $N$.
\end{prop}
\begin{proof} Due to Corollary~\ref{cor:Eomunif}
\[
 \sup \left\{ \log \big | f_N(e(x), \omega,E)\big | \::\: x \in \tor,\, E \in \IC, \,
 |E - E_1| < \exp\left(-(\log N)^A\right) \right\} \le NL_N(\omega,E_1) + (\log N)^B
\]
for any $E_1$.
Due to the large deviation theorem in the $E$ variable, see Corollary~\ref{cor:E_LDT},  there exist $x_1, E_1$ such that $|x_0 - x_1| <
\exp\left(-(\log N)^{2A}\right)$, $|E_0 - E_1| < \exp \left(-(\log N)^{2A}\right)$ so that
$$
\log \big | f_N(e(x_1),  \omega, E_1) \big | > NL_N(\omega,E_1) - (\log N)^{4A}.
$$
By Jensen's formula \eqref{eq:jensen},
$$
\# \left\{E\::\: f_N(e(x_1), \omega,E) = 0, |E - E_1| < \exp \left(-(\log N)^A\right) \right\} \le (\log
N)^C.
$$
Since $\big \| H_N^{(D)} (x_0, \omega) - H_N^{(D)} (x_1, \omega) \big \| \lesssim \exp \left(- (\log N)^{2A}\right)$
and since $H_N^{(D)} (x_0, \omega)$ is self adjoint one has
\begin{equation*}
\begin{split}
& \# \left\{E: f_N(e(x_0), \omega,E) = 0, |E - E_0| < \exp \left(-(\log N)^{2A}\right) \right\}\\
& \le \#\left\{E: f_N(e(x_1), \omega,E) = 0, |E- E_1| < \exp\left(-(\log N)^A\right) \right\} \le (\log N)^C\ .
\end{split}
\end{equation*}
That proves~\eqref{eq:null1}.  The proof of \eqref{eq:null2} is similar. Indeed, due to Corollary~\ref{cor:Eomunif}
\[
 \sup \left\{ \log \big | f_N(e(x+iy), \omega, E_0)\big | \::\: x \in \tor,\, |y|<2N^{-1}
  \right\} \le NL_N(\omega, E_0) + (\log N)^A.
\]
By the large deviation theorem, there is $x_1$ with $|x_0-x_1|< \exp(-(\log N)^C)$ such that
\[ \log \big | f_N(e(x_1), \omega, E_0)\big | > NL_N(\omega, E_0)-(\log N)^A.\]
Hence, by Jensen's formula \eqref{eq:jensen},
\[ \# \left\{z\::\: f_N(z, \omega,E) = 0, |z-e(x_1)| < 2N^{-1} \right\} \le 2(\log N)^A,\]
and \eqref{eq:null2} follows.
\end{proof}

Another application of the uniform upper estimates of this section is the following analogue of
Wegner's estimate from the random case. It will be important that there is only a loss of
$(\log N)^A$ in~\eqref{eq:wegner}.

\begin{lemma}
\label{lem:wegner}
Suppose $\omega$ satisfies \eqref{eq:diophant}.  Then for any $N\gg 1$, $E \in \IR$, $H \ge
(\log N)^A$ one has
\begin{equation}
\label{eq:wegner}
\mes \left\{x \in \tor\::\: \dist\left(\rsp H_N(x,\omega), E\right) <
\exp(-H)\right\} \le \exp \left(-H/(\log N)^A\right)\ .
\end{equation}
Moreover, the set on the left-hand side is contained in the  union of $\les N$ intervals
each of which does not exceed the bound stated in~\eqref{eq:wegner} in length.
\end{lemma}
\begin{proof} By Cramer's rule
$$
\left|\Bigl(H_N(x, \omega) - E\Bigr)^{-1} (k, m)\right| = {\big |f_{[1,
k]}\bigl(e(x), \omega, E\bigr)\big |\, \bigl| f_{[m+1, N]} \bigl(e(x),
\omega, E \bigr)| \over \big | f_N\bigl(e(x), \omega, E\bigr)\big |}\ .
$$
By Proposition~\ref{prop:logupper} and \eqref{eq:rate}
\begin{equation}
\nn
\log \big | f_{[1, k]}\bigl(e(x), \omega, E\bigr)\big | + \log \big
|f_{[m+1, N]}\bigl(e(x), \omega, E\bigr)\big | \le NL(\omega,E) + (\log N)^{A_1}
\end{equation}
for any $x \in \tor$.  Therefore,
$$
\big\| \left(H_N(x, \omega) - E\right)^{-1} \big\| \le N^2\
{\exp\left(NL(\omega,E) + (\log N)^A\right)\over \big |f_N\bigl(e(x), \omega,
E\bigr)\big|}
$$
for any $x \in \tor$.  Since
$$\dist\left(\rsp\bigl(H_N(x,\omega),
E\bigr)\right)^{-1} = \big\| \left(H_N(x,\omega) - E\right)^{-1}\big\|\ ,
$$
the lemma follows from Corollary~\ref{cor:betterdet}.
\end{proof}

We conclude this section with an application of Lemma~\ref{lem:cart_zero} to the
determinants $f_N$.

\begin{corollary}
\label{cor:fN_zeros}
Suppose $\omega$ satisfies~\eqref{eq:diophant}. Given $E_0\in\IC$ and $H > (\log N)^{A}$,
there exists
\[ \cB_{N,E_0, \omega}(H) \subset \IC, \qquad\cB_{N,E_0,\omega}(H)\in \car_1(\sqrt{H},HN^2)\]
such that for any $x \in \tor \setminus \cB_{N,E_0, \omega}(H)$, and large $N$
the following holds: If
\[ \log \big | f_N \left(e(x), \omega, E_1\right) \big | < NL(\omega,E_1) - H(\log N)^A,\quad |E_0-E_1|<\exp(-(\log N)^C), \]
then $f_N\left(e(x), \omega,E\right) = 0$ for some $|E - E_1| \lesssim   \exp(-\sqrt{H})$.
Similarly, given $x_0\in\tor$ and $|y_0|<N^{-1}$, let $z_0=e(x_0+iy_0)$. Then for any $H > (\log N)^{A}$, there exists
\beeq
\label{eq:ENom_est}
 \cE_{N,z_0, \omega}(H) \subset \IC, \qquad\cE_{N,z_0,\omega}(H)\in \car_1\Big(\sqrt{H},H\exp((\log N)^A)\Big)
\eneq
 such that for any  $E \in \cD(0,A) \setminus \cE_{N,z_0,\omega}(H)$,  the following assertion holds: If
\[ \log \big |f_N\bigl(z_1,\omega,E\bigr)\big | < NL(\omega,E) - H(\log N)^A,\quad |z_0-z_1|<\exp(-(\log N)^C), \]
then $f_N\bigl(z, \omega,E\bigr) = 0$ for some $|z-z_1|\lesssim
\exp(-\sqrt{H})$.
\end{corollary}
\begin{proof} Set $r_0=\exp(-(\log N)^C)$ with some large constant $C$. Fix any $z_0$ with $|z_0|=1$
and consider the analytic function
\[ f(z,E)=f_N(z_0+(z-z_0)N^{-1},E_0+(E-E_0)r_0,\omega)\]
on the polydisk $\cP=\cD(z_0,1)\times\cD(E_0,1).$
Then, by Proposition~\ref{prop:logupper},
\[ \sup_{\cP}\log|f(z,E)| \le NL(E_0,\omega) + (\log N)^{2C} =M\]
and by the large deviation theorem, see Corollary~\ref{cor:betterdet},
\[ \log|f(z_1,E_0)|> NL(E_0,\omega) - (\log N)^{2C}=m\]
for some $|z_0-z_1|<1/100$, say. By Lemma~\ref{lem:cart_zero} there exists
\[ \cB_{z_0,E_0,\omega}(H)\subset\IC, \qquad \cB_{z_0,E_0,\omega}(H)\in \car_1(\sqrt{H}, H(\log N)^{3C}) \]
so that for any $z\in \cD(z_0,1/2)\setminus\cB_{z_0,E_0,\omega}(H)$ the following holds: If
\[ \log|f(z,E_1)|< NL(E_0,\omega)-H(\log N)^{3C} \]
for some $|E_1-E_0|<1/2$, then there is $E$ with $|E_1-E|\lesssim \exp(-\sqrt{H})$ such that $f(z,E)=0$.
Now let $z_0$ run over a $N^{-\frac32}$-net on $|z|=1$ and define $\cB_{N,E_0,\omega}(H)$ to be the union of the
sets $z_0+N^{-1}\cB_{z_0,E_0,\omega}(H)$. The first half of the lemma now follows by taking $A$ sufficiently
large and by absorbing some powers of $\log N$ into~$H$ if needed.
The second half of the lemma dealing with zeros in the $z$ variable can be shown analogously.
\end{proof}

\begin{remark}
\label{rem:Enet}
We can draw the following conclusion from the preceding corollary:
Let $\omega\in\tor_{c,a}$ be fixed, and define
\[ \cE_{N,\omega}(H)=\bigcup_{x_0} \cE_{N,e(x_0),\omega}(H) \]
where the union runs over a $N^{-1}$-net of points $x_0\in\tor$. Then, for any $x\in\tor$, if
\[ \log \big |f_N\bigl(x,\omega,E\bigr)\big | < NL(\omega,E) - H(\log N)^A,\quad E\in \cD(0,A)\setminus \cE_{N,\omega}(H)\]
then $f_N\bigl(z, \omega,E\bigr) = 0$ for some $|z-e(x)|\lesssim
\exp(-\sqrt{H})$. Moreover, \eqref{eq:ENom_est} holds for $\cE_{N,\omega}(H)$.
\end{remark}

The following simple observation will be used in the proof of
Theorem~\ref{thm:4}.

Let $V(z)$ be analytic in $\cA_{\rho_0}$ and let $M_N^{(V)}(z,
\omega, E)$ stand for the monodromies with potential $V$. Define
\begin{align*}
L_N^{(V)}(\omega, E) & = \frac{1}{N}\int_\tor \log \big \|M_N^{(V)}\bigl(e(x), \omega, E\bigr) \big \|\,dx\\
L^{(V)}(\omega, E) & = \lim_{N\to \infty}\ L_N(V, \omega, E)\ .
\end{align*}
Assume that $\gamma_0 = L^{(V_0)}(\omega_0, E_0) > 0$ for some $V_0$, $\omega_0 \in \tor_{c,a}$, $E_0 \in \IC$.

\begin{lemma} \label{lem:4.13}
There exist $\tau_0 = \tau_0\bigl(\lambda, V_0, \omega_0, \rho_0, \gamma_0\bigr) > 0$,
$N_0 = N_0\bigl(\lambda, V_0, \omega_0, \rho_0, \gamma_0\bigr)> 0$
as well as $\alpha>0$ such that
\begin{equation}
\label{eq:lnv} \left| L_N^{(V)} (\omega_0, E_0) -
L_N^{(V_0)}(\omega_0, E_0) \right| \lesssim \tau_0^{\alpha}
\end{equation}
for any $N > N_0$ and any $V(z)$ analytic in $\cA_{\rho_0}$ which
satisfies
$$
\sup_{z\in\cA_{\rho_0}} \bigl |V(z) - V_0(z)\big | < \tau_0\ .
$$
\end{lemma}

\begin{proof}
In this proof, we will suppress $E,\omega$ from our notations.
Clearly,
\[ |L_n^{(V)} - L_n^{(V_0)}| \lesssim \|V-V_0\|_{L^\infty(\cA_{\rho_0})}
\exp(C_1\,n)
\]
There exists some large $n_0$ such that $L_{n_0}^{(V_0)}>\gamma_0/2$
and, moreover, so that the conditions of the avalanche principle
hold with $n_0$ and $\mu=\exp(c\,n_0)$ with some sufficiently small
constant~$c>0$. Set
\[ \tau_0 = \exp(-4C_1\, n_0)
\]
Then $|L_n^{(V)}- L_n^{(V_0)}|\lesssim \tau_0^{\frac34} $ for all
$n_0\le n\le 2n_0$ provided~\eqref{eq:lnv} holds. By~\cite{GS} we
have $L^{(V)}>0$ and
\[ |L_N^{(V)} - 2 L_{2n_0}^{(V)} + L_{n_0}^{(V)} | \lesssim
\exp(-c\,n_0/2)
\]
for $N\ge N_0=\exp(c\,n_0)$. Since
\[ |L_{2n_0}^{(V)} - L_{2n_0}^{(V_0)} |, |L_{n_0}^{(V)} - L_{n_0}^{(V_0)}
| < \tau_0^{\frac12} \] we are done.
\end{proof}

\section{A corollary of Jensen's formula}
\label{sec:jensen}

The Jensen formula states that for any function $f$ analytic on a neighborhood of $\cD(z_0,R)$, see~\cite{levin},
\beeq
\label{eq:jensen}
 \int_0^1 \log |f(z_0+Re(\theta))|\, d\theta - \log|f(z_0)| = \sum_{\zeta:f(\zeta)=0} \log\frac{R}{|\zeta-z_0|}
\eneq provided $f(z_0)\ne0$. In the previous section, we showed
how to combine this fact with the large deviation theorem and the
uniform upper bounds to bound the number of zeros of $f_N$ which
fall into small disks, in both the $z$ and $E$ variables. In what
follows, we will refine this approach further. For this purpose,
it will be convenient to average over $z_0$ in~\eqref{eq:jensen}.
Henceforth, we shall use the notation
\begin{align}
\nu_f(z_0, r) &= \# \{z \in \cD(z_0, r): f(z)=0\} \label{eq:nudef}\\
J(u, z_0, r_1, r_2) &=
\mathop{\nint}\limits_{\cD(z_0, r_1)} dx\, dy
\mathop{\nint}\limits_{\cD(z, r_2)} d \xi d
\eta\, [u(\zeta)-u(z)]. \label{eq:Jdef}
\end{align}

\begin{lemma}
\label{lem:Jdef}
Let $f(z)$ be analytic in $\cD(z_0,
R_0)$. Then for any $0<r_2<r_1<R_0-r_2$
\begin{equation*}
\nu_f(z_0, r_1 - r_2) \leq 4
\frac{r_1^2}{r_2^2} J(\log |f|, z_0, r_1,
r_2) \leq \nu_f(z_0, r_1+r_2)
\end{equation*}
\end{lemma}
\begin{proof}
 Jensen's formula yields
\begin{align*}
J(f, z_0, r_1, r_2) &=
\mathop{\nint}\limits_{\cD(z_0, r_1)} dx\, dy \Bigg
[ \frac{2}{r^2_2} \mathop\int\limits_0^{r_2} dr
\Bigg (r \mathop\sum\limits_{f(\zeta)=0, \zeta
\in \cD(z, r)} \log (\frac {r}{|\zeta -
z|})\Bigg ) \Bigg ]\\
&\leq \mathop\sum\limits_{f(\zeta)=0, \zeta \in
\cD(z_0, r_1+r_2)} (\frac{1}{\pi r_1^2}) \Bigg
[ \frac{2}{r^2_2} \mathop\int\limits_0^{r_2} dr
\Bigg ( r \mathop\int\limits_{\cD(\zeta, r)}
\log (\frac{r}{|z - \zeta |}) dx\, dy \Bigg)
\Bigg]\\
&= \frac{1}{4} (\frac{r^2_2}{r^2_1}) \nu_f (z_0,
r_1+r_2),
\end{align*}
which proves the upper estimate for $J(f, z_0,
r_1, r_2)$.  The proof of the lower estimates
is similar.
\end{proof}

\begin{corollary}
\label{cor:5.2}
Let $f$ be analytic in $\cD(z_0, R_0)$, $0 < r_2 < r_1 < R_0 - r_2$.
Assume that $f$ has no zeros in the annulus $\cA = \left\{r_1 - r_2 \le |z
- z_0| \le r_1 + r_2\right\}$.  Then
$$
\nu_f(z_0, r_1) = 4\ {r_1^2\over r_2^2}\ J\left(\log |f|, z_0, r_1,
r_2\right)\ .
$$
\end{corollary}

\begin{corollary}
\label{cor:5.3}
Let $f(z), g(z)$ be analytic in $\cD(z_0,
R_0)$.  Assume that for some
$0<r_2<r_1<R_0-r_2$
\begin{equation*}
|J(f, z_0, r_1, r_2)-J(g, z_0, r_1, r_2)|<
\frac {r^2_2}{4r_1^2}
\end{equation*}
Then
\begin{equation*}
\nu_f(z_0, r_1-r_2) \leq \nu_g(z_0, r_1+r_2), \qquad
\nu_g(z_0, r_1-r_2) \leq \nu_f(z_0, r_1+r_2).
\end{equation*}
\end{corollary}

We shall also need  a simple generalization of these estimates to
averages over general domains. More precisely, set
\begin{align}
\nu_f(\cD) &= \# \{z \in \cD: f(z)=0\} \label{eq:nu2def}\\
J(u, \cD, r_2) &= \mathop{\nint}\limits_{\cD} dx\, dy
\mathop{\nint}\limits_{\cD(z, r_2)} d \xi d \eta\, [u(\zeta)-u(z)].
\nn
\end{align}
Given a domain $\cD$ and $r>0$ , set $ \cD (r)= \{z: \dist(z,\cD) <r
\}$. Let $f(z)$ be analytic in $\cD(R)$. Then for any
$0<r_2<r_1<R-r_2$
\begin{equation}
\label{eq:mes12} \nu_f( \cD(r_1 - r_2)) \leq 4
\frac{\mes(\cD(r_1))}{r_2^2} J(\log |f|, \cD( r_1), r_2) \leq \nu_f(
\cD(r_1+r_2))
\end{equation}
These results generalize in a straightforward way to subharmonic functions. More precisely,
suppose that
\beeq
\label{eq:usubh}
 u(z) = \int \log|z-\zeta|\, \mu(d\zeta) + h(z) \text{\ \ for all\ \ }z\in\Omega
\eneq
where $h$ is harmonic and $\mu$ is a non-negative measure on some domain $\Omega$. In what follows, it will be understood
that all averages are taken inside this domain.
Then the analogue of Lemma~\ref{lem:Jdef} is as follows:
\begin{lemma}
\label{lem:Rr}
Let $u$ be as in~\eqref{eq:usubh}. Then

\[ \mu(\cD(z_0, r_1 - r_2)) \leq 4
\frac{r_1^2}{r_2^2} J(u, z_0, r_1,
r_2) \leq \mu(\cD(z_0, r_1+r_2)).
\]
\end{lemma}

The following lemma is a consequence of some estimates from Section~\ref{sec:upper}.

\begin{lemma}
\label{lem:6.4}
Suppose $z_0 \in \cA_{\rho_0/2}$, $E_0 \in \IC$, and that $\omega_0$ is as in~\eqref{eq:diophant}.
Let $N$ be large and  choose radii $\rho_1,\rho_2$ so that
\[ \exp \left(-(\log N)^{4A}\right) < \rho_2 \ll \rho_1 < \exp \left(-(\log
N)^{A}\right), \qquad \rho_2<\rho_1\exp(-(\log N)^A)\] Then, with
constants $B\gg A\gg1$,
\begin{equation}
 \big | J\left(\log |f_N(\cdot, \omega,E) |, z_0, \rho_1, \rho_2\right) -
 J\left(\log |f_N(\cdot, \omega_0, E_0) |, z_0, \rho_1, \rho_2\right) \big | < e^{-(\log
N)^B}
\label{eq:JN_dif}
\end{equation}
for any
$ |E - E_0| + |\omega - \omega_0| < \exp\bigl(-(\log N)^{2B}\bigr)$.
In particular,
\begin{align*}
\nu_{f_N(\cdot, \omega,E)}(z_0, \rho_1)  &\le \nu_{f_N(\cdot,
\omega_0, E_0)} (z_0, \rho_1 +
\rho_2),\\
\nu_{f_N(\cdot,  \omega_0, E_0)}(z_0, \rho_1) & \le \nu_{f_N(\cdot,
\omega,E)}(z_0, \rho_1 + \rho_2).
\end{align*}
\end{lemma}
\begin{proof}
Corollary~\ref{cor:4.8} implies~\eqref{eq:JN_dif}, which in turn leads to the bounds on $\nu$ via
Corollary~\ref{cor:5.3}.
\end{proof}

Let $\cA_{\rho_1,\rho_2} := \{z\in \IC\::\: \rho_1<|z|<\rho_2 \}$.

\begin{lemma}\label{lem:5.6}
\begin{equation}
\begin{aligned}
& J\Bigl(N^{-1}\log \big |f_N(\cdot, \omega, E)\big |, \cA_{\rho_1,\rho_2}, r_2\Bigr)=\\
& (\rho_2^2-\rho_1^2)^{-1} r_2^{-2}\int^{\rho_1}_{\rho_2}\, \rho\,
d\rho \int_0^{r_2}\, r\,dr
 \int_0^1\, dy \Bigl[L_N(\xi(\rho,r,y), \omega, E ) -
 L_N(\xi(\rho),\omega,E)\Bigr]
\end{aligned}
\end{equation}
where $\xi(\rho,r,y)=\log|\rho+re(y)|$, $\xi(\rho)=\log\rho$.
\end{lemma}
\begin{proof}
Due to the definition of $J(u,{\mathcal D}, r_2)$ one has
\begin{align*}
&J(\log|f_N(\cdot,\omega,E)|, \cA_{\rho_1,\rho_2},r_2) \nn\\
&= \frac{4\pi^2}{ |\cA_{\rho_1,\rho_2}| r_2^{2}}
\int_{\rho_1}^{\rho_2}\, \rho\, d\rho \int_0^{r_2} r\, dr \Biggl \{
\int_0^1\,dx\int_0^1\, dy \Bigl[ \log|f_N(\rho e(x)+r
e(y),\omega,E)|-\log|f_N(\rho e(x),\omega,E)| \Bigr]\Biggr\}\nn \\
&= \frac{4\pi^2}{ |\cA_{\rho_1,\rho_2}| r_2^{2}}
\int_{\rho_1}^{\rho_2}\, \rho\, d\rho \int_0^{r_2} r\, dr \Biggl \{
\int_0^1\,dx\int_0^1\, dy \Bigl[ \log|f_N(|\rho +r
e(y)|e(x),\omega,E)|-\log|f_N(\rho e(x),\omega,E)|
\Bigr]\Biggr\}\nn\\
&= N(\rho_2^2-\rho_1^2)^{-1} r_2^{-2}\int^{\rho_1}_{\rho_2}\, \rho\,
d\rho \int_0^{r_2}\, r\,dr
 \int_0^1\, dy \Bigl[L_N(\xi(\rho,r,y), \omega, E ) -
 L_N(\xi(\rho),\omega,E)\Bigr]
\end{align*}
as claimed.
\end{proof}

\begin{corollary} Let $\rho_-(y) = e(iy)$, $\rho_+(y) = e(-iy)$, $0 < y < \log(1+\rho_0)/4$.  Then
\begin{equation}
\begin{aligned}
& \# \left\{z \in \cA_{\rho_-(y), \rho_+(y)}: f_N(z, \omega, E) = 0 \right\} \le C_1 y^{-1} \cdot\\
& \max \left\{\left| L_N\bigl(e(\cdot + iy_1), \omega, E\bigr) - L_N \bigl(e(\cdot + iy_2), \omega, E\bigr) \right|:
 |y_1|, |y_2| \le 2y\right\} + C_2\, N^{-1}\ .
\end{aligned}
\label{eq:rho-+}
\end{equation}
\end{corollary}
\begin{proof}
By \eqref{eq:mes12}
\begin{align*}
& \#\{ z\in\cA_{\rho_{-}(y),\rho_+(y)}\::\: f_N(z,\omega,E)=0\} \\
& \le 4 \frac{|\cA_{\rho_{-}(y),\rho_+(y)}|}{\rho_+(y)^2} J (\log
|f_N(\cdot,\omega,E)|,\cA_{\rho_{-}(2y),\rho_+(2y)},\rho_+(y))
\end{align*}
Combining this with the representation of Lemma~\ref{lem:5.6}
yields~\eqref{eq:rho-+}.
\end{proof}

\begin{corollary}\label{cor:5.8}
 Using the  notations of Lemma~\ref{lem:4.13} one has
\begin{equation*}
\begin{aligned}
 & \# \left\{z \in \cA_{\rho_0/4} : f_N^{(V)} (z, \omega, E) = 0 \right\} \le\\
& \# \left\{z \in \cA_{\rho_0/2}: f_N^{(V_0)} (z, \omega, E) =
0\right\} + C \tau_0^\alpha\, N
\end{aligned}
\end{equation*}
\end{corollary}

\section{The Weierstrass preparation theorem for Dirichlet determinants}
\label{sec:weier}

We start with a discussion of Weierstrass' preparation theorem for an analytic function
$f(z, w_1, \dots, w_d)$ defined in a polydisk
\begin{equation}
\cP = \cD(z_0, R_0) \times \prod^d_{j=1} \cD(w_{j,0}, R_0),\quad z_0,\ w_{j, 0} \in \IC\qquad
\frac12\ge R_0 > 0\ .
\end{equation}

\begin{lemma}
\label{lem:weier}
Assume that $f(\cdot, w_1, \dots, w_d)$ has no zeros on some circle $\left\{z:
|z-z_0| = \rho_0 \right\}$, $0 < \rho_0 < R_0/2$, for any $\uw = (w_1, \dots, w_d) \in
\cP_1 = \prod\limits^d_{j=1} \cD(w_{j, 0}, r_1)$ where $0<r_1<R_0$.  Then there exist a polynomial $P(z, \uw)
= z^k +a_{k-1} (\uw) z^{k-1} + \cdots + a_0 (\uw)$ with $a_j(\uw)$ analytic in
$\cP_1$ and an analytic function $g(z, \uw), (z, \uw) \in \cD(z_0, \rho_0) \times \cP_1$
so that the following properties hold:
\begin{enumerate}
\item[(a)] $f(z, \uw) = P(z, \uw) g(z, \uw)$ for any $(z, \uw) \in \cD(z_0, \rho_0) \times
\cP_1$.

\item[(b)] $g(z, \uw) \ne 0$ for any $(z, \uw) \in \cD(z_0, \rho) \times \cP_1$

\item[(c)] For any $\uw \in \cP_1$, $P(\cdot, \uw)$ has no zeros in $\IC \setminus \cD(z_0,
\rho_0)$.
\end{enumerate}
\end{lemma}
\begin{proof} By the usual Weierstrass argument, one notes that
$$
b_p(\uw) := \sum^k_{j=1} \zeta^p_j (\uw) = {1\over 2\pi i} \oint\limits_{|z-z_0|= \rho_0} z^p \
{\partial _z f(z, \uw) \over f(z, \uw)}\, dz
$$
are analytic in $\uw \in \cP_1$.  Here $\zeta_j (\uw)$ are the zeros of $f(\cdot, \uw)$
in $\cD(z_0, \rho_0)$.  Since the coefficients $a_j(\uw)$ are linear combinations of the $b_p$, they are analytic in $\uw$.  Analyticity
of $g$ follows by standard arguments.
\end{proof}

\noindent We will repeatedly use the affine maps introduced in the following definition.

\begin{defi}
\label{def:scale} Given $\uw_0 = (w_{1,0}, \dots, w_{d,0}) \in
\IC^d$, $\ur = (r_1, \dots, r_d)$, $r_i > 0$, $i = 1, 2, \dots, d$,
set \[S_{\uw_0, \ur} (w_1, \dots, w_d) = \bigl(r_1^{-1}(w_1 -
w_{1,0}), \dots, r_d^{-1}(w_d - w_{d,0})\bigr).\]
\end{defi}

\noindent Lemma~\ref{lem:6.3} describes a typical situation in which Lemma~\ref{lem:weier} can be applied.

\begin{lemma}
\label{lem:6.3}
Let $f(z, \uw)$ be analytic in a polydisk $\cP = \cD(z_0, R_0) \times
\prod^d_{j=1} \cD(w_{j,0}, R_0)$, $z_0, w_{j,0} \in \IC$, $1 \gg R_0 > 0$.  Let $M \ge
\sup\limits_{z, \uw} \log |f(z, \uw)|$, $m \le \log |f(z_1, \uw_0)|$, where $z_1 \in \cD(z_0,
R_0/2)$, $\uw_0 = (w_{1,0}, \dots, w_{d,0})$.  Then
 there exists a circle $\Gamma_{\rho_0} = \left\{|z - z_0| = \rho_0\right\}$,
$R_0/8 < \rho_0 < R_0/4$ such that for any $\uw \in \prod\limits^d_{j=1} \cD(w_{j,0}, r_1)$,
$r_1\asymp R_0 \exp \left(-C(M-m) \right)$ the function $f(\cdot, \uw)$ has no
zeros on $\Gamma_{\rho_0}$.  In particular, Lemma~\ref{lem:weier} holds for $f(z, \uw)$ with this choice of $\rho_0$ and~$r_1$,
as well as with $k\lesssim M-m$.
\end{lemma}
\begin{proof}
Let $g:= f\circ S_{(z_0,\uw_0),\ur}^{-1}$, where $\ur=(R_0,\ldots, R_0)$. Then $g$ is analytic on the polydisk
\[ \widetilde{\cP}= \cD(z_0,1) \times \prod^d_{j=1} \cD(w_{j,0},1). \]
By Jensen's formula,
\begin{equation}\nn
k = \#\left\{z \in \cD(z_0, R_0/4): f(z, \uw_0) = 0 \right\} \lesssim M - m
\end{equation}
Due to Cartan's estimate one has
$$
\log|g(z, 0)| \ge M - C(M-m)
$$
for any $z \in \cD(0, 1/4) \setminus \cB$, where $\cB \in \car_1\left(C_1, C_1\, k\right)$.
Find $1/8 < \rho_0 < 1/4$ such that $\cB \cap \Gamma_{\rho_0} =
\emptyset$.  Then
\beeq
\label{eq:fzw0}
|g(z, 0)| \ge \exp \left(M - C(M-m) \right)
\eneq
for any $z \in \Gamma_\rho$.  Note that
$$
|g(z, w) - g(z,0)| \lesssim e^M |\uw|
$$
for any $z \in \cD(0, 1)$, $\uw \in \prod\limits^d_{j=1} \cD(0, 1/2)$.  Taking into
account~\eqref{eq:fzw0},  one obtains
$$
|g(z, w)| > {1\over 2} \exp \left(M - C(M - m) \right)
$$
for any $z \in \Gamma_{\rho_0}$, provided $\uw \in \prod\limits^d_{j=1} \cD(0, r_1)$, $r_1 =
C^{-1}\exp \left(-C(M-m)\right)$. The proof is completed by undoing the affine transformation $S_{(z_0,\uw_0),\ur}$
and returning to $f$.
\end{proof}

We are now going to apply Lemma~\ref{lem:6.3} to the Dirichlet determinants $f_N(z, \omega,E)$.
It will be important
to do this in both the $z$ and the $E$ variables.
 We start with the $z$-statement.

\begin{prop}
\label{prop:fN_prep_z}
 Given $z_0 \in \cA_{\rho_0/2}$, $E_0 \in \IC$, and $\omega_0$ as in~\eqref{eq:diophant}, there
exist a polynomial
\[  P_N(z, \omega,E) = z^k + a_{k-1} (\omega,E) z^{k-1} + \cdots + a_0(E,
\omega)\]
 with $a_j(\omega,E)$ analytic in $\cD(E_0, r_1)\times \cD(\omega_0, r_1)$, $r_1 \asymp \exp
\left(-(\log N)^{A_1}\right)$ and an analytic function \[g_N(z,
\omega,E),\quad (z, \omega,E) \in \cP = \cD(z_0, r_0) \times
\cD(E_0, r_1) \times \cD(\omega_0,r_1)\] with $r_0 \asymp N^{-1}$
such that:
\begin{enumerate}
\item[(a)] $f_N(z, \omega,E) = P_N(z, \omega,E) g_N(z, \omega,E)$

\item[(b)] $g_N(z, \omega,E) \ne 0$ for any $(z, \omega,E) \in \cP$

\item[(c)] For any $(\omega,E) \in  \cD(\omega_0, r_1)\times \cD(E_0, r_1) $, the
polynomial
 $P_N(\cdot, \omega,E)$ has no zeros in $\IC \setminus \cD(z_0, r_0)$

\item[(d)] $k = \degg P_N(\cdot, \omega,E) \le (\log N)^A$.
\end{enumerate}
\end{prop}
\begin{proof} Define $\uw=(z_0,\omega_0,E_0)$ and $\ur=(r,r',r')$, where $r=N^{-1}$ and $r'=\exp(-(\log N)^C)$.
Set $f= f_N\circ S^{-1}_{\uw_0,\ur}$ on the polydisk $\cP=\cD(z_0,1/2)\times\cD(\omega_0,1/2)\times \cD(E_0,1/2)$.
Then we can apply Lemma~\ref{lem:6.3} to $f$ with $R_0=\frac12$, and
\[ M=NL_N(\omega_0,E_0)+(\log N)^{3C}, \quad m = NL_N(\omega_0,E_0)-(\log N)^{3C}.\]
The choices of $M,m$ follow by the uniform upper bound from Proposition~\ref{prop:logupper},
and the LDT from Corollary~\ref{cor:betterdet}, respectively.
Hence $f(z,\omega,E)$ has a Weierstrass representation as given by Lemma~\ref{lem:6.3}.
Undoing the change of variables allows one to derive the desired representation for $f_N$, and we are done.
\end{proof}


And now for the preparation theorem relative to $E$. In this case, we will use Lemma~\ref{lem:weier}
directly. This is due to the fact that the LDT in the $E$-variable as stated in Corollary~\ref{cor:E_LDT}, holds
not for all $x_0$, but for $x_0$ outside of some small exceptional set. On the other hand, Lemma~\ref{lem:weier}
requires a lower bound at $\uw=\uw_0$ which might be in this exceptional set.

\begin{prop}
\label{prop:fN_prep_E}
 Given $x_0 \in \tor$, $E_0 \in \IC$, and $\omega_0$ as in~\eqref{eq:diophant}, there exist a
polynomial
\[ P_N(z, \omega,E) = E^k + a_{k-1} (z, \omega)E^{k-1} + \cdots + a_0(z,
\omega) \]
with $a_j(z, \omega)$ analytic in $\cD(z_0, r_1) \times \cD(\omega_0, r_1)$, $z_0=e(x_0)$,  $r_1
\asymp \exp \left(- (\log N)^{A_1}\right)$ and an analytic function $g_N(z, \omega,E)$, $(z,
\omega,E) \in \cP = \cD(z_0, r_1) \times  \cD(\omega_0, r_1)\times \cD(E_0, r_1)$ such that
\begin{enumerate}
\item[(a)] $f_N(z, \omega,E) = P_N(z, \omega,E) g_N(z, \omega,E)$

\item[(b)] $g_N(z, \omega,E) \ne 0$ for any $(z, \omega,E) \in \cP$

\item[(c)] For any $(z, \omega) \in \cD(z_0, r_1) \times \cD(\omega_0, r_1)$, $P_N(z, \cdot, \omega)$
has no zeros in $\IC \setminus \cD(E_0, r_0)$, $r_0 \asymp \exp \left(-(\log N)^{A_0}\right)$

\item[(d)] $k = \degg P_N(z, \cdot, \omega) \le (\log N)^{A_2}$
\end{enumerate}
\end{prop}
\begin{proof} Recall that due to Proposition~\ref{prop:zero_count} one has
$$
\#\left\{E\in \IC: f_N(z_0, \omega_0,E) = 0,\quad |E - E_0| < \exp\left(-(\log N)^A\right) \right\} \le
(\log N)^{A_2}
$$
Find $r_0 \asymp \exp\left(-(\log N)^{A_0}\right)$ such that $f_N(z_0, \omega_0, \cdot)$ has no
zeros in the annulus
\[ \left\{r_0(1-2N^{-2}) < |E - E_0| < r_0(1+2N^{-2})\right\}. \]
  Since $H_N^{(D)}(z_0, \omega_0)$ is self adjoint, $f_N(z,\omega,\cdot)$ has no zeros in the annulus
\[ \left\{r_0 (1-N^{-2}) < |E - E_0| < r_0 (1+N^{-2})\right\},\]
provided $|z - z_0| \ll r_1= r_0 N^{-4}$, $|\omega - \omega_0| \ll r_1 $.  The proposition now follows from Lemma~\ref{lem:weier}.
\end{proof}

We conclude this section with global versions of the previous two propositions. This is done by patching up
the local versions.

\begin{defi}
\label{def:mult}
Let $\cS$ be a system of polydisks \[\cP_m = \prod\limits^d_{j=1} D\left(w_{m,
j, 0}, r_{m,j}\right) \subset \IC^d,\quad
\uw_{m, 0} = \left(w_{m,1,0}, \dots, w_{m, d, 0}\right)
\in \IC^d,\]
$m \in \cJ$, $\#\cJ < +\infty$.  Let $K > 0$ be an integer.  We say that $\cS$ has
multiplicity $\le K$ if no point $\uw \in \IC^d$ belongs to more than $K$ polydisks $\cP_m$.
\end{defi}

\begin{corollary}
\label{cor:patch} There exists a constant $A \gg1$ so that for any constant $B$, and large $N$
there exists a cover of $\tor \times [-B, B]\times \tor_{c,a}$ with a system $\{\cP_m\}_m$ of
polydisks \[\cP_m =
\cD(x_{m, 0}, r) \times \cD(\omega_{m, 0}, r) \times \cD(E_{m, 0},
r)\subset
\IC^3,\; x_{m, 0} \in \tor,\; E_{m, 0} \in [-B, B],\]
and $ \omega_{m, 0} $ as in \eqref{eq:diophant}, and such that
\beeq
\label{eq:rad_size}
  r \asymp \exp\left(-(\log N)^{A}\right)
\eneq
with the following properties: it has multiplicity
$\lesssim 1 $ and the property that for each $(x_{m,0}, \omega_{m, 0}, E_{m, 0})$
Proposition~\ref{prop:fN_prep_E} applies with $r$ in the role of $r_1$.
As similar statement holds for Proposition~\ref{prop:fN_prep_z} with the radius in the $x$-variable
comparable to any fixed radius $\le N^{-1}$. In particular, we can choose the same polydisks as before.
\end{corollary}
\begin{proof}
For any $x_0\in\tor$, $E\in [-B,B]$, and $\omega\in\tor_{c,a}$,
there is a polydisk centered at these points as well as a radius
of size~\eqref{eq:rad_size}, so that
Proposition~\ref{prop:fN_prep_E} holds. Hence, there is a finite
cover of $\tor \times \tor\times [-B, B]$ by such polydisks. By
Wiener's covering lemma (which applies equally well to these
polydisks), we can pass to a sub-collection of disjoint polydisks
so that their dilations by a factor of three remains a cover. This
dilated subcollection has the desired multiplicity by a volume
count. Alternatively, one can just take the union over a net of
points. One proceeds analogously for
Propositions~\ref{prop:fN_prep_z}.
\end{proof}

\section{Eliminating close zeros using resultants}
\label{sec:resultant}

Let $f(z) = z^k + a_{k-1} z^{k-1} + \cdots + a_0$, $g(z) = z^m + b_{m-1} z^{m-1} +
\cdots + b_0$ be polynomials, $a_i, b_j \in \IC$.  Let $\zeta_i$, $1 \le i \le k$
and $\eta_j$, $1 \le j \le m$ be the zeros of $f(z)$ and $g(z)$ respectively.  The
resultant of $f$ and $g$ is defined as follows:
\begin{equation}
\Res(f, g) = \prod_{i,j} (\zeta_i - \eta_j)
\end{equation}
The resultant $\Res(f, g)$ can be found explicitly in terms of the coefficients, see~\cite{lang}, page~200:
\begin{equation}
\Res(f, g) = \left|\begin{array}{ll}
{\overbrace{\begin{array}{lll}
1 & 0 & \cdots\\
a_{k-1}  & 1 & \cdots\\
a_{k-2} & a_{k-1} & \cdots\\
\cdots & \cdots & \cdots\\
\cdots & \cdots & \cdots\\
\cdots & \cdots & \cdots\\
a_0 & a_1 &  \\
0 & a_0 &  \end{array}}^m} &
{\overbrace{\begin{array}{llll}
1 & 0 & \cdots & 0\\
b_{m-1} & 1 & \cdots & \cdots\\
b_{m-2} & b_{m-1} &\cdots & \cdots\\
\cdots & \cdots & \cdots & \cdots\\
\cdots & \cdots & \cdots & \cdots\\
\cdots & \cdots & \cdots& \cdots\\
&&&\\
&&&\\
\end{array}}^k}
\end{array}\right|
\end{equation}
In particular, one has the following property:

\begin{lemma}
Let $f(z; w) = z^k + a_{k-1} (w) z^{k-1} + \cdots + a_0(w)$, $g(z; w) = z^m +
b_{m-1}(w) z^{m-1} + \cdots + b_0(w)$ be polynomials whose coefficients $a_i(w)$,
$b_j(w)$ are analytic functions defined in a domain $G\subset \IC^d$.  Then
$\Res(f(\cdot, w), g(\cdot, w))$ is analytic in $G$.
\end{lemma}

The goal of this section is to separate the zeros of two analytic
functions by means of shifts. This will be reduced to the same
question for polynomials by means of the previous section. We
start with a simple observation regarding the resultant of a
polynomial and a shifted version of another polynomial.

\begin{lemma}
\label{lem:wshift}
 Let $f(z) = z^k + a_{k-1} z^{k-1}+ \cdots + a_0$, $g(z) = z^m +
b_{m-1} z^{m-1} + \cdots + b_0$ be polynomials.  Then
\begin{equation}
\label{eq:wexp}
\Res\left(f(\cdot + w), g(\cdot)\right) = (-w)^n + c_{n-1} w^{n-1} + \cdots + c_0
\end{equation}
where $n = km$, and $c_0, c_1 \cdots$ are some coefficients.
\end{lemma}
\begin{proof} Let $\zeta_j$, $1 \le j \le k$ (resp. $\eta_i, 1 \le i \le m$) be
the zeros of $f(\cdot)$ (resp. $g(\cdot)$).  The zeros of $f(\cdot + w)$ are
$\zeta_j - w$, $1 \le j \le k$.  Hence
\begin{equation}
\label{eq:res_exp}
\Res\left(f(\cdot + w), g(\cdot)\right) = \prod_{i, j} (\zeta_j - w - \eta_i)
\end{equation}
and \eqref{eq:wexp} follows.
\end{proof}

The following lemma gives some information on the coefficients in~\eqref{eq:wexp}.

\begin{lemma}
\label{lem:bjuw}
Let $P_s(z, \uw) = z^{k_s} + a_{s, k_s -1} (\uw)
z^{k_s -1} + \cdots + a_{s,0}(\uw)$, $z \in \IC$, where $a_{s,
j}(\uw)$ are analytic functions defined in some polydisk $\cP =
\prod\limits_i D(w_{i,0}, r)$, $\uw = (w_1, \dots, w_d) \in \IC^d$, $\uw_0
= (w_{1,0}, \dots, w_{d, 0}) \in \IC^d$, $s = 1, 2$.  Set $\chi(\eta,
\uw) = \Res\left(P_1(\cdot, \uw), P_2(\cdot + \eta, \uw)\right)$, $\eta
\in \IC$, $\uw \in \cP$.  Then
\begin{equation}
\label{eq:etauw}
\chi(\eta, \uw) = (-\eta)^k + b_{k-1}(\uw) \eta^{k-1} + \cdots +
b_0(\uw)
\end{equation}
where $k = k_1 k_2$, $b_j(\uw)$ are analytic in $\cP$,  $j = 0, 1, \dots,
k-1$.  Moreover, if the zeros of $P_i(\cdot, \uw)$ belong to the same
disk $D(z_0, r_0)$, $i = 1, 2$ then for all $0 \le j \le k-1$,
\begin{equation}
\label{eq:bjuw}
\big | b_j(\uw)\big | \le \binom{k}{k-j} (2r_0)^{k-j} \le(2r_0 k)^{k-j}
\end{equation}
\end{lemma}
\begin{proof} The relation~\eqref{eq:etauw}  with some coefficients $b_j(\uw)$ follows
from Lemma~\ref{lem:wshift}.  \eqref{eq:bjuw} follows from the expansion~\eqref{eq:res_exp}.
By Lemma~\ref{lem:wshift},
$\chi(\eta, \uw)$ is analytic in $\IC \times \cP$.  Therefore $b_j(\uw) =
(j!)^{-1}(\partial_\eta)^j \chi(\eta,\uw) \Bigm |_{\eta=0}$ are analytic $j = 0,
1, \dots, k-1$.
\end{proof}

The following lemma allows for the separation of the zeros of one polynomial
from those of a shifted version of another polynomial. This will be the main
mechanism for eliminating certain ``bad'' rotation numbers $\omega$.

\begin{lemma}
\label{lem:H2O}
Let $P_s(z, \uw)$ be polynomials in $z$ as in Lemma~\ref{lem:bjuw}, $s = 1, 2$. In particular, $\uw\in\cP$ where $\cP$ is
a polydisk of some given radius $r>0$.  Assume
that $k_s > 0$, $s = 1, 2$ and set $k=k_1k_2$. Suppose that for any $\uw \in \cP$ the zeros
of $P_s(\cdot, \uw)$ belong to the same disk $D(z_0, r_0)$, $r_0 \ll
1$, $s = 1, 2$.  Let $t > 16k\, r_0\, r^{-1}$.  Given
$H \gg 1$ there exists a set $\cB_H\subset \cP$
such that $S_{\uw_0, (16kr_0t^{-1}, r,\dots, r)} (\cB_H) \in \car_d(H^{1/d},
K)$, $K = CHk$ and for any $\uw \in \cD(w_{1,0}, 8kr_0/t) \times \prod\limits^d_{j=2}
\cD(w_{j,0}, r/2) \setminus \cB_H$ one has
\begin{equation}
\label{eq:zero_dist}
\dist\left(\bigl\{\mbox{zeros of
$P_1(\cdot, \uw)$}\bigr\},
\bigl\{\mbox{zeros of $P_2\left(\cdot + t(w_1 - w_{1,0}),
\uw\right)\bigr\}$}\right) \ge e^{-CHk}\ .
\end{equation}
\end{lemma}
\begin{proof} Define $\chi(\eta, \uw)$ as in Lemma~\ref{lem:wshift}.  Note that for any
$\uw \in \cP$ one has
$$
|\chi(\eta, \uw)| \ge |\eta|^k \left[ 1 - \sum^\infty_{j=1} \left({2r_0
k\over |\eta|}\right)^j\right] \ge {1\over 2} |\eta|^k
$$
provided $|\eta| \ge 8r_0\, k$.  Furthermore, for any $\uw\in\cP$,
$$
|\chi(\eta, \uw)| \le |\eta|^k \left[ 1 + \sum^\infty_{j=1} \left({2r_0
k\over |\eta|}\right)^j\right] \le 2 |\eta|^k.
$$
provided $|\eta| \ge 8r_0\, k$.
Hence, by the maximum principle,
$$
\sup \left\{\big | \chi(\eta, \uw) \big |: |\eta| \le 16r_0k\right\} \le 2(16kr_0)^k\ .
$$
Set \[f(\uw) =
\chi\bigl(t(w_1 - w_{1,0}), (w_1,w_2, \dots, w_d)\bigr),\quad w_1 \in \cD(w_{1,0},
16kr_0/t),\quad
(w_2, \dots, w_d) \in \prod\limits^d_{j=2} \cD(w_{j,0}, r). \]
This function is well-defined because $16kr_0/t<r$ by our lower bound on $t$.
By the preceding,
\[ \sup\limits_{\uw} |f(\uw)| \le 2(16kr_0)^k, \quad \big |f\bigl(w_{1,0}
+ 8kr_0/t, w_{2,0}, \dots, w_{d,0}\bigr)\big | > \frac12(8kr_0)^k.\]
We can therefore apply Lemma~\ref{lem:high_cart}  to
\[ \phi=f\circ S_{\uw_0,(16kr_0/t,r,\ldots,r)}^{-1} \text{\ \ with\ \ } M = \log2 + k \log(12kr_0), \quad m=-\log 2 + k\log(8kr_0).\]
Thus, in view of Remark~\ref{rem:diff_rad}, given $H\gg1$ there exists $\cB^{(1)}_H \subset \cP$ such that
$$
S_{\uw_0,(16kr_0t^{-1}, r, \dots, r)} \left(\cB^{(1)}_H\right) \in
\car\nolimits_d
\left(H^{1/d}, K\right), \quad K=CkH,
$$
and such that for any \[(w_1, \dots, w_d) \in D(w_{1,0}, 8kr_0/t) \times
\prod\limits^d_{j=2} D(w_{j, 0}, r/2) \setminus \cB^{(1)}_H\] one has
$|f(\uw)| > e^{-CHk}$.  Recall that due to basic properties of the
resultant
$$
|f(w)| = \prod_{i, j} |\zeta_{i, 1} (\uw) - \zeta_{j, 2} (\uw) |
$$
where $\zeta_{i, 1} (\uw)$, $\zeta_{j, 2} (\uw)$ are the zeros of $P_1(\cdot, \uw)$,  and $P_2(\cdot + t(w_1 -
w_{1,0}), \uw)$, respectively. Since $r_0\ll1$, this implies \eqref{eq:zero_dist}, and we are done.
\end{proof}

Lemma~\ref{lem:H2O} of course applies to polynomials $P_s(z)$ that do not depend on $\uw$ at all.
This example is important, and explains why quantities like $K$ have the stated form.
Next, we combine Lemma~\ref{lem:H2O} with the Weierstrass preparation  theorem from the previous section,
more precisely, Lemma~\ref{lem:6.3}. This allows us to separate the zeros of two analytic functions, one of
which is shifted.

\begin{corollary}
\label{cor:H2O+weier}
 Let $f_s(z, \uw)$ be analytic functions, $s = 1,2$,
defined in a polydisk
\[ \cP = \cD(z_0, R_0) \times \prod\limits^d_{j=1} \cD(w_{j,0}, R_0), \quad z_0, w_{j,0} \in \IC, \quad 1 \gg R_0 > 0.\]
Let, for $s=1,2$,
\begin{align*}
M_s &\ge \sup\limits_{z, \uw} \log |f_s(z, \uw)|, \\
m_s &\le \log |f_s(z_{1s}, \uw_0)| \text{\ \ where\ \  }z_{1s} \in \cD(z_0, R_0/2),\; \uw_0 = (w_{1,0}, \dots, w_{d,0}).
\end{align*}
Now define $M = \max\limits_{s=1,2} M_s$ and $m = \min\limits_{s=1,2} m_s$ and assume that $M-m\ge1$.
Let
\beeq
\label{eq:Rtdef}
 r = R_0\exp \left(-C(M-m)\right), \quad t > \exp \left(2C(M-m)\right)
\eneq
with some large constant $C$. Given $H \gg 1$, there exists a set
\[ \cB_H
\subset \prod\limits^d_{j=1} \cD(w_{j,0}, R_0),\quad S_{\uw_0, (R_0t^{-1},
r, \dots, r)} (\cB_H) \in \car_d \left(H^{1/d}, K\right),\quad K = CH(M-m)^2\]
 such that for any
\[ \uw = (w_1, \dots, w_d) \in \cD(w_{1,0}, R_0/2t) \times \prod\limits^d_{j=2}
\cD(w_{j,0}, r/2) \setminus \cB_H, \]
one has
\begin{align*}
& \dist\left(\bigl\{\mbox{zeros of $f_1(\cdot, \uw)$}\bigr\}\cap \cD(z_0,R_0/2),
\bigl\{\mbox{zeros of $f_2\left(\cdot +
t(w_1 - w_{1,0}), \uw)\right.$}\bigr\}\cap \cD(z_0,R_0/2)\right) \\
&\ge \exp\big(-CH(M-m)^2\big)
\end{align*}
\end{corollary}
\begin{proof} By Lemma~\ref{lem:6.3} there is the representation
$$
f_s(z, w) = P_s (z, w) g_s(z, w),
$$
where $P_s(z, w)$, $g_s(z, w)$ satisfy properties (a)--(c) in Lemma~\ref{lem:weier} with $\rho_0\asymp R_0$ and
$r_1=r$ as in~\eqref{eq:Rtdef}.
Let $k_s$ be the degree of the polynomial $P_s(\cdot, \uw)$, $s = 1,2$.
Then $k_s \lesssim M-m$, $s = 1, 2$. We may assume that $k_1k_2>0$, otherwise the corollary is void.
Now apply Lemma~\ref{lem:H2O} with $r_0=\rho_0$ and the above choice of $r$.
\end{proof}

\section{Eliminating  double resonances for most energies}
\label{sec:double}

We now obtain a  version of Corollary~\ref{cor:H2O+weier} for the special
case of Dirichlet determinants. It will be formulated locally, in the sense that we
cover the parameter space by polydisks, so that locally on each one of them we
obtain the desired separation of the zeros in the $z$-variable.
In what follows, we use the notation
\beeq
\label{eq:Zdef} \cZ(f,z_0,r_0) = \{z\in\IC\::\: f(z)=0,\; |z-z_0|<r_0\}.
\eneq

\begin{lemma}
\label{lem:elim_local}
Let $C_1>1$ be an arbitrary constant.
Given $\ell_1 \ge \ell_2\gg1$, $t > \exp\bigl((\log
\ell_1)^A\bigr)$, $H \gg 1$, there exists a cover of $\tor
\times\tor_{c,a}\times [-C_1, C_1]$ by a system $\cS$ of
polydisks
\beeq
\label{eq:poly_scheib}
\cD(x_m, r)\times \cD(\omega_m, r t^{-1}) \times \cD(E_m, r)  ,\quad x_m \in \tor,\; E_m \in [-C_1, C_1],
\eneq
with $\omega_m\in\tor_{c,a} $, 
and  $r=\exp\bigl(-(\log \ell_1)^{A_2}\bigr)$, and which satisfies
the following properties: $\cS$ has multiplicity $\lesssim 1$, cardinality $\#(\cS)\lesssim
t \exp\bigl((\log\ell_1)^{A_1}\bigr)$ and
for each $m$, there exists a subset $\Omega_{\ell_1, \ell_2, t, H, m} \subset \cD(\omega_m, r t^{-1}/2)$ with
\[
S_{\omega_m, r t^{-1}/2}(\Omega_{\ell_1, \ell_2, t, H,
m}) \in \car_1(H^{1/2}, K),\quad K = (\log \ell_1)^B \]
 such that for any $\omega \in \cD\left(\omega_m, r t^{-1}/2\right) \setminus
\Omega_{\ell_1, \ell_2, t, H, m}$ there exists a subset
\[\cE_{\ell_1, \ell_2, t, H, \omega, m} \subset \cD(E_m, r),\quad S_{E_m, r}\left(\cE_{\ell_1, \ell_2, t, H, \omega, m}\right) \in
\car_1\left(H^{1/2}, K\right)\] such that for any $E \in \cD(E_m,
r)\setminus \cE_{\ell_1, \ell_2, t, H, \omega, m}$ one has
\begin{equation}\label{eq:fN_sep}
\dist\Bigl(\cZ \big( f_{\ell_1}(\cdot, \omega, E), e(x_m),r \big),
\cZ\big( f_{\ell_2}(\cdot e(t\omega), \omega,
E),e(x_m), r \big)\Bigr) > e^{-H(\log \ell_1)^C}\ .
\end{equation}
\end{lemma}
\begin{proof} Take $x_0 \in \tor$, $z_0 = e(x_0)$, $\omega_0 \in \tor_{c,a}$,
$E_0 \in \IR$.  Due to the Weierstrass preparation theorem for Dirichlet
determinants, see Proposition~\ref{prop:fN_prep_z}, one has the representation
\begin{align*}
f_{\ell_1}\bigl(z,  \omega, E\bigr)  &= P_1(z, \omega,E) g_1(z,\omega, E)\\
f_{\ell_2} \bigl(z + t\omega_0, \omega, E\bigr) & = P_2(z, \omega,E)
g_2(z, \omega,E)
\end{align*}
with $z \in \cD(z_0, r)$, $(\omega, E) \in \cD(\omega_0, r) \times \cD(E_0,
r)$, 
$r\asymp \exp\bigl(-(\log \ell_1)^A\bigr)$, where $P_i$, $g_i$ satisfy the properties (a)--(d)
stated in that proposition.  Let $k_i$, $i = 1, 2$ be the degrees of $P_i(\cdot, E,
\omega)$.  We may assume that $k_i > 0$, $i = 1, 2$.  Recall that $k_i < (\log
\ell_1)^A$, $i = 1, 2$ due to property (d).  Due to Lemma~\ref{lem:H2O}, given $t >
\exp \bigl((\log \ell_1)^B\bigr)$, $H > (\log \ell_1)^B$ there exists a
set $\cB_{\ell_1, \ell_2, t, H, z_0, \omega_0, E_0} \subset \cD(\omega_0, r)
\times \cD(E_0, r/2t)$,
$$
S_{(\omega_0, E_0), (r/2t, r)} \left(\cB_{\ell_1, \ell_2, t, H, z_0,
\omega_0, E_0}\right) \in \car_2\left(H^{1/2}, K\right)
$$
such that~\eqref{eq:fN_sep} is valid for any
$$
(\omega, E) \in \cD(\omega_0, r/2t) \times \cD(E_0, r/2) \setminus
\cB_{\ell_1, \ell_2, t, H, z_0, \omega_0, E_0}\ .
$$
By the definition of $\car_2$ sets there exists $\Omega_{\ell_1, \ell_2,
t, H, z_0, E_0} \subset \cD(\omega_0, r/2t)$,
$$
S_{\omega_0, r/2t}\left(\Omega_{\ell_1, \ell_2, t, H, z_0, \omega_0,
E_0}\right) \in \car_1\left(H^{1/2}, K\right)
$$
such that for any
$\omega\in \cD(\omega_0, r/2t)\setminus
\Omega_{\ell_1, \ell_2, t, H, z_0, \omega_0, E_0}
$
the set
\[
\cE_{\ell_1,
\ell_2, t, H, z_0, \omega_0, \omega, E_0} = \left\{E \in \cD(E_0, r_0):
(\omega, E) \in \cB_{\ell_1, \ell_2, t, H, z_0, \omega_0, E_0}\right\}
\]
satisfies
$$
\cS_{E_0, r} \left(\cE_{\ell_1, \ell_2, t, H, z_0, \omega_0, E_0,
\omega}\right) \in \car_1\left(H^{1/2}, K\right)\ .
$$
Letting $(x_0, \omega_0, E_0)$ run over an appropriate net in $\tor
\times \tor_{c,a}\times \left[-C_1, C_1\right]$ completes the
proof of the lemma. See the proof of Corollary~\ref{cor:patch} for the details of this.
\end{proof}

We now transfer the separation of the zeros to the separation of the eigenvalues.

\begin{corollary}
\label{cor:ev_sep}
Let $A\gg B\gg1$.
Given $\ell_1 \ge \ell_2\gg1$, $t > \exp\bigl((\log
\ell_1)^A\bigr)$, $H \gg 1$,
there exist a set $\Omega_{\ell_1,\ell_2,t,H}\subset\tor$ with
\[ \mes(\Omega_{\ell_1,\ell_2,t,H})\lesssim e^{-\sqrt{H}}, \quad \compl(\Omega_{\ell_1,\ell_2,t,H}) \lesssim t H\exp((\log \ell_1)^B)\]
and for each $\omega\in\tor_{c,a}\setminus\Omega_{\ell_1,\ell_2,t,H}$ there is  a set $\cE_{\ell_1, \ell_2, t, H,\omega}\subset \IC$ with
\[ \mes(\cE_{\ell_1, \ell_2, t, H,\omega})
\les t e^{-\sqrt{H}}, \quad \compl(\cE_{\ell_1, \ell_2, t, H,\omega})\les t H^2\exp((\log \ell_1)^B)\]
 with the following property: For any $\omega\in\tor_{c,a}\setminus\Omega_{\ell_1,\ell_2,t,H}$
and any $x_0 \in \tor$,
$$\dist\Bigl(\rsp\bigl(H_{\ell_1}^{(D)} (x_0, \omega)\bigr)
\setminus\cE_{\ell_1, \ell_2, t, H, \omega}\;,\;
\rsp\bigl(H_{\ell_2}^{(D)} (x_0 + t\omega,
\omega)\bigr)\Bigr) \ge e^{-H^2(\log \ell_1)^A}.
$$
Similarly, if $z_1=e(x_1+iy_1)$ with $|y_1|\le r$ and $E\in [-C,C]\setminus \cE_{\ell_1, \ell_2, t, H,\omega}$,
then
\[ f_{\ell_1}(z_1,\omega,E)=0,\quad f_{\ell_2}(z_2e(t\omega),\omega,E)=0\]
implies that $|z_1-z_2|> e^{-H(\log \ell_1)^A}$.
\end{corollary}
\begin{proof} Set $\tilde{H}=H(\log \ell_1)^{\frac{B}{2}}$. Let $\cE_{\ell_s,\omega}(\tilde H^2)$ be the sets defined in Remark~\ref{rem:Enet}, $s=1,2$.
With $\Omega_{\ell_1, \ell_2, t, \tilde H,m}$ and $\cE_{\ell_1, \ell_2, t, \tilde H, \omega, m}$ as in Lemma~\ref{lem:elim_local},
set
\[ \Omega_{\ell_1,\ell_2,t,H}  = \bigcup_m \Omega_{\ell_1, \ell_2, t, \tilde H,m}, \quad
\cE_{\ell_1, \ell_2, t, H, \omega} = \cE_{\ell_1,\omega}(\tilde{H}^2)\cup
\cE_{\ell_2,\omega}(\tilde H^2)\cup \bigcup_m \cE_{\ell_1, \ell_2, t, \tilde H, \omega, m} \]
where the union runs over all $m$ in \eqref{eq:poly_scheib}.
Then, with $r$ as in \eqref{eq:poly_scheib},
\[ \mes (\Omega_{\ell_1,\ell_2,t,H})\les \sum_m r\,t^{-1}e^{-\sqrt{\tilde H}}\les t\,r^{-4}r\,t^{-1}e^{-\sqrt{\tilde H}} \les e^{-\sqrt{H}},\]
and, with $K$ as in Lemma~\ref{lem:elim_local},
\[  \compl (\Omega_{\ell_1,\ell_2,t,H})\les tr^{-4}K\les tr^{-4}\tilde{H}\, (\log \ell_1)^B \les tH\exp((\log \ell_1)^B). \]
Similarly, invoking \eqref{eq:ENom_est} as well as Lemma~\ref{lem:elim_local}, it follows that
\begin{align*}
  \compl (\cE_{\ell_1,\ell_2,t,H,\omega}) &\les tr^{-4}K + \tilde{H}^2\exp((\log\ell_1)^B)\les t\,H^2\exp((\log \ell_1)^B) \\
  \mes (\cE_{\ell_1,\ell_2,t,H,\omega}) &\les (1+t\,r^{-4}\,r^{2})e^{-\sqrt{\tilde H}}\les te^{-\sqrt{H}},
\end{align*}
as claimed.
Now assume that $f_{\ell_1}(z_1, \omega, E_1) = 0$, $f_{\ell_2}\bigl(z_1
e(t\omega), \omega, E_2 \bigr) = 0$ for arbitrary $z_1=e(x_0)$, and
\beeq
\label{eq:E1E2} |E_1 - E_2| < e^{-H^2(\log \ell_1)^A}, \qquad E_1\in [-C,C]\setminus
\cE_{\ell_1,\ell_2,t,H,\omega},\;\omega\in\tor_{c,a}\setminus\Omega_{\ell_1,\ell_2,t,H}\,.
\eneq
Fix some $m$ from \eqref{eq:poly_scheib} such that $(x_0,\omega, E_1)\in \cD(x_m,r)\times\cD(E_m,r)\times \cD(\omega_m,r)$.
Then
\begin{align*}
 \big | f_{\ell_2}(z_1,  \omega, E_1)\big | &\lesssim
|E_1 - E_2| \exp\left(\ell_2 L(\omega, E_1) + (\log \ell_2)^B\right) <
\exp\left(\ell_2 L(\omega,E_1) - H^2(\log \ell_1)^{A}/2\right)\\
&\les \exp\left(\ell_2 L(\omega, E_1) - \tilde H^2(\log \ell_1)^{B}\right)
\end{align*}
By our choice of $E_1$,  there exists
$z_2$ so that $|z_2 - z_1| < \exp(-H(\log \ell_1)^B)$, for which
\[ f_{\ell_2}\bigl(z_2 e(t\omega), \omega, E_1\bigr) = 0,\]
see Lemma~\ref{lem:cart_zero} and Remark~\ref{rem:Enet}. But this would mean that
\[
\dist\Bigl(\bigl\{\mbox{zeros of}\ f_{\ell_1}(\cdot,  \omega, E_1)\bigr\}\cap
\cD(e(x_m), 2r), \bigl\{\mbox{zeros of}\ f_{\ell_2}(\cdot e(t\omega),\omega, E_1)\bigr\}\cap \cD(e(x_m),2 r)\Bigr) < e^{-H(\log \ell_1)^B}
\]
which contradicts \eqref{eq:fN_sep} by our choice of $\omega$. The final statement of the corollary
is an immediate consequence of~\eqref{eq:fN_sep}.
\end{proof}

By means of the Wegner-type estimate from Lemma~\ref{lem:wegner} we can formulate a version
of Corollary~\ref{cor:ev_sep} which is based on the removal of a set of $x\in\tor$ rather than
a set of energies.

\begin{corollary}
\label{cor:8.3}
Let $A\gg B\gg1$.
Given $\ell_1 \ge \ell_2\gg1$, $t > \exp\bigl((\log \ell_1)^A\bigr)$, $H \gg 1$,
there exists a set $\Omega_{\ell_1,\ell_2,t,H}\subset\tor$ with
\[ \mes(\Omega_{\ell_1,\ell_2,t,H})\lesssim e^{-\sqrt{H}}, \quad \compl(\Omega_{\ell_1,\ell_2,t,H}) \lesssim t H\exp((\log \ell_1)^B)\]
and for each $\omega\in\tor_{c,a}\setminus\Omega_{\ell_1,\ell_2,t,H}$ there is  a set $\cB_{\ell_1, \ell_2, t, H,\omega}\subset \tor$ with
\[ \mes(\cB_{\ell_1, \ell_2, t, H,\omega})
\les t e^{-\sqrt{H}}, \quad \compl(\cB_{\ell_1, \ell_2, t, H,\omega})\les t H^2\exp((\log \ell_1)^B)\]
with the following property: For any $\omega\in\tor_{c,a}\setminus\Omega_{\ell_1,\ell_2,t,H}$
and any $x_0 \in \tor\setminus\cB_{\ell_1, \ell_2, t, H,\omega} $,
$$
\dist\Bigl(\rsp\bigl(H_{\ell_1}^{(D)}(x_0, \omega)\bigr),
\rsp\bigr(H_{\ell_2}^{(D)} (x_0 + t\omega, \omega)\bigr)\Bigr) \ge \exp
\bigr(-H (\log \ell_1)^{A_1}\bigr)\ .
$$
\end{corollary}
\begin{proof}
We need to remove the set
\[ \cB_{\ell_1, \ell_2, t, H,\omega} = \{ x\in\tor\::\: \rsp\bigl(H_{\ell_1}^{(D)} (x, \omega)\bigr)
\cap\cE_{\ell_1, \ell_2, t, H, \omega} \ne\emptyset \},
\]
where the set $\cE_{\ell_1, \ell_2, t, H, \omega}$ is as in \eqref{cor:ev_sep}. The corollary now follows
from Lemma~\ref{lem:wegner}.
\end{proof}

\section{Localized Dirichlet eigenfunctions on a finite interval}
\label{sec:localize}

In this section we apply the results of the previous sections to the study of the
eigenfunctions of the Hamiltonian restricted to intervals on the integer lattice.

\begin{lemma}
\label{lem:Green}
 Let $\omega \in \tor_{c,a}$.  Suppose $L(\omega,E_0) > 0$,
\begin{equation}
\label{eq:fN_unter}
\log \big | f_N(z_0, \omega,E_0) \big | > NL(\omega,E_0) - K/2
\end{equation}
for some $z_0 = e(x_0)$, $x_0 \in \tor$, $E_0 \in \IR$, $N \gg 1$, $K >
(\log N)^A$.  Then
\begin{align}
\label{eq:Gjk}
\big | \cG_{[1, N]} (z_0,\omega,E) (j, k)\big |  & \le \exp\left(-
{\gamma\over 2}(k - j) + K\right)\\
\big \| \cG_{[1, N]} (z_0, \omega,E) \big \| & \le \exp(K)
\label{eq:normG}
\end{align}
where $\cG_{[1, N]}(z_0, \omega,E_0) = \left(H(z_0, \omega) -
E_0\right)^{-1}$ is the Green's function, $\gamma = L(\omega,E_0)$, $1
\le j \le k \le N$.
\end{lemma}
\begin{proof}
By Cramer's rule and the uniform upper bound of Proposition~\ref{prop:logupper} as well
as the rate of convergence estimate~\eqref{eq:rate},
\begin{equation}
\begin{split}
\big | \cG_{[1, N]}(z_0, \omega,E) (j, k) \big | & = \big | f_{j-1}
(z_0, \omega, E_0)\big | \cdot \big | f_{N - k} \bigl(z_0e(k\omega),
\omega, E_0\bigr) \big | \cdot\\
&\quad \big | f_N(z_0, \omega, E_0)\big |^{-1} \le \big | f_N(z_0,
\omega, E_0)\big |^{-1} \\
&\quad \exp \left(NL(\omega, E_0) - (k-j) L(\omega, E_0) + (\log N)^C\right)
\end{split}
\end{equation}
Therefore, \eqref{eq:Gjk} follows from condition \eqref{eq:fN_unter}.  Estimate \eqref{eq:normG}
follows from \eqref{eq:Gjk}.
\end{proof}

Any solution of the equation
\begin{equation}
\label{eq:hamilton}
-\psi(n+1) - \psi(n-1) + v(n)\psi(n) = E\psi(n)\ ,\quad n \in \IZ\ ,
\end{equation}
obeys the relation
\beeq
\label{eq:poisson}
\psi(m) = \cG_{[a, b]} (E)(m, a-1)\psi(a-1) + \cG_{[a, b]} (E)(m,
b+1)\psi(b+1),\quad m \in [a, b].
\eneq
where $\cG_{[a,b]} (E) = \left(H_{[a,b]} -E\right)^{-1}$ is the Green's
function, $H_{[a,b]}$ is the linear operator defined by \eqref{eq:hamilton}
for $n \in [a, b]$ with zero boundary conditions.
In particular, if $\psi$ is a solution of equation \eqref{eq:hamilton},  which satisfies a
zero boundary condition at the left (right) edge, i.e.,
\begin{equation}
\nn
\psi(a-1) = 0 \quad \mbox{(resp. $\psi(b+1) = 0$)}\ ,
\end{equation}
then
\begin{align}
\psi(m) & =\cG_{[a,b]}(m, b+1)\psi(b+1) \nn \\
\bigl(\mbox{resp. $\psi(m)$} & = \cG_{[a,b]}(m,a-1) \psi(a-1)\; \bigr)
\nonumber
\end{align}
If, for instance, in addition   $\big | \cG(m, b+1)\big | <1$,
then $|\psi(m)| < |\psi(b+1)|$.

The following lemma states that after removal of certain rotation numbers $\omega$ and energies $E$,
but uniformly in $x\in\tor$, only  one choice
of $n\in[1,N]$ can lead to a determinant $f_\ell(x+n\omega,\omega,E)$ with $\ell \asymp (\log n)^C$ which is
not large. This relies on the elimination results of Section~\ref{sec:double} and is of crucial importance
for all our subsequent work.

\begin{lemma}
\label{lem:waschno}
Given $N$, there exists $\Omega_N \subset \tor$ with $\mes(\Omega_N) <
\exp\left(-(\log N)^{C_2}\right)$, $\compl(\Omega_N) < \exp\left((\log N)^{C_1}\right)$,
$C_1 \ll C_2$ such that for all $\omega \in \tor_{c,a} \setminus \Omega_N$ there is $\cE_{N, \omega}
\subset \IR$, $\mes(\cE_{N,\omega}) < e^{-(\log N)^{C_2}}$, $\compl(\cE_{N,\omega}) < e^{(\log
N)^{C_1}}$, with the following property: For any $x \in \tor$ and any $\omega \in
\tor_{c,a}\setminus \Omega_N$, $E \in \IR \setminus\cE_{N,\omega}$ either
\begin{equation}\label{eq:star}
\log \big | f_\ell\bigl(e(x+n\omega), \omega, E\bigr)\big | > \ell L(\omega,E) - \sqrt\ell
\end{equation}
for all $\ell \asymp (\log N)^C$ and all $1  \le n \le N$, or there exists $n_1 = n_1(x, \omega,
E) \in [1, N]$ such that (\ref{eq:star}) holds for all $n \in [1, N] \setminus [n_1 - L, n_1 +
L]$, $L \asymp \exp \left((\log\log N)^A\right)$, but not for $n = n_1$.  However, in this case
\begin{equation}
\label{eq:f1n}
\big | f_{[1,n]} \bigl(e(x), \omega, E\bigr) \big | > \exp \left(nL(\omega, E) - (\log N)^C\right)
\end{equation}
for each $1 \le n \le n_1 - L$ and
\begin{equation}
\label{eq:fnN}
\big | f_{[n, N]} \bigl(e(x), \omega, E\bigr)\big | > \exp \left((N-n)L(\omega, E) - (\log N)^C\right)
\end{equation}
for each $n_1 + L \le n \le N$.
\end{lemma}
\begin{proof} Define $\Omega_N = \bigcup\, \Omega_{\ell_1, \ell_2, t, H}$ where the union runs over
$\ell_1, \ell_2 \asymp (\log N)^C$, $N > t > \exp \left((\log\log N)^A\right)$ with fixed
$H\asymp (\log N)^{C/100}$.  Here $\Omega_{\ell_1, \ell_2, t, H}$ is as in Corollary~\ref{cor:ev_sep}.  Similarly,
for any $\omega \in \tor_{c,a} \setminus \Omega_N$ set
\[ \cE_{N,\omega} = \bigcup_{(\log N)^C\le k\le N}\cE_{k,\omega}(H)\cup\bigcup\, \cE_{\ell_1, \ell_2, t, H, \omega}\]
where the second union is the same as before, and where $\cE_{k,\omega}(H)$ are as in Remark~\ref{rem:Enet}.
The measure and complexity estimates follow from Corollary~\ref{cor:ev_sep}.
Now suppose \eqref{eq:star} does not hold.  Then
$$
\log \big | f_{\ell_1}\bigl(e(x + n_1 \omega), \omega, E\bigr)\big | < \ell_1 L(\omega,E) - \sqrt{\ell_1}
$$
for some $1 \le n_1 \le N$ and $\ell_1 \asymp (\log N)^C$.  By Lemma~\ref{lem:cart_zero} and Remark~\ref{rem:Enet} there exists $z_1$ with
$|z_1 - e(x + n_1 \omega) | < e^{-\ell_1^{{1\over 4}}}$ and
$$
f_{\ell_1} (z_1, \omega, E) = 0\ .
$$
If
$$
\log \big | f_{\ell_2}\bigl(e(x + n_2 \omega), \omega, E\bigr)\big | < \ell_2 L(\omega,E) -
\sqrt{\ell_2}
$$
for some $\ell_2 \asymp (\log N)^C$ and $|n_2 - n_1| > \exp \left((\log\log
N)^A\right)$, then for some $z_2$, and $t = n_1 - n_2$
$$
f_{\ell_2} \left(z_2 e(t\omega), \omega, E\right) = 0
$$
with $|z_1 - z_2| < e^{-(\log N)^C}$, which contradicts our choice
of $(\omega, E)$, see Corollary~\ref{cor:ev_sep}.
Thus~\eqref{eq:star} holds for all $\ell \asymp (\log N)^C$ and $1
\le n \le N$, $|n - n_1| > \exp \left((\log\log N)^A\right)$, as
claimed.  This allows one to apply the avalanche principle at
scale $\ell \asymp (\log N)^C$ to $f_{[1, n]} \bigl(e(x), \omega,
E\bigr)$ with $(\log N)^C \ll n \le n_1 - L$.  It yields that
$$
\log \big | f_{[1, n]} \bigl(e(x), \omega, E\bigr)\big | \ge nL(\omega, E) - C{n\over (\log N)^C} > 0\ .
$$
Note that by Lemma~\ref{lem:Eomdiff}, if \eqref{eq:star} holds at $x$, then also for all $z \in \cD\bigl(e(x), e^{-\ell}\bigr)$. Thus,
\begin{equation}
\label{eq:fN_notzer}
f_{[1, n]} (z, \omega, E) \ne 0
\end{equation}
for those $z$ by the avalanche principle.  Now suppose
$$
\log \big | f_{[1, n]} \bigl(e(x), \omega, E\bigr) \big | \le nL(\omega, E) - (\log N)^B
$$
for some large constant B.  By our choice of $E$,
$$
f_{[1, n]} (z, \omega, E) = 0
$$
for some $|z - e(x)| < \exp \left(-(\log N)^{B/2}\right)$.  This contradicts \eqref{eq:fN_notzer} provided $B$ is
sufficiently large.  Hence, \eqref{eq:f1n} holds and \eqref{eq:fnN} follows from a similar argument.
\end{proof}

\begin{remark}
\label{rem:expstab}
It follows from Lemma~\ref{lem:Eomdiff} that \eqref{eq:star} is stable under perturbations of $E$ by an
amount $< e^{-C\ell}$.  More precisely, if \eqref{eq:star} holds for $E$, then
\begin{equation}
\nn
\log \big | f_\ell \bigl(e(x + n\omega), \omega, E'\bigr) \big | > \ell L(E', \omega) - 2 \sqrt{\ell}
\end{equation}
for any $E'$ with $|E' - E| <  e^{-C\ell}$.  Inspection of the previous proof now shows that
\eqref{eq:f1n} and \eqref{eq:fnN} are also stable under such perturbations.
\end{remark}

The previous lemma yields the following finite volume version of Anderson localization.

\begin{prop}
\label{prop:9.4}
For any $x , \omega \in \tor$, let $\left\{E_j^{(N)} (x, \omega)\right\}_{j=1}^N$ and
$\left\{\psi_j^{(N)} (x, \omega, \cdot)\right\}_{j=1}^N$ denote the eigenvalues and normalized
eigenvectors of $H_{[1, N]}(x, \omega)$, respectively.  Let $\Omega_N$ and $\cE_{N, \omega}$ be as in
the previous lemma.
If $\omega \in \tor_{c,a} \setminus \Omega_N$ and for some $j$, $E_j^{(N)}(x, \omega) \notin \cE_{N, \omega}$,
then there exists a point $\nu_j^{(N)} (x, \omega) \in [1, N]$ (which we call the center of
localization) so that for any $\exp \left((\log\log N)^A\right) \le Q \le N$ and with $\Lambda_Q
:= [1, N] \cap \left[\nu_j^{(N)}(x, \omega) - Q, \nu_j^{(N)} (x, \omega) + Q\right)$ one has
\begin{enumerate}
\item[{\rm(i)}] $\dist\left(E_j^{(N)}(x, \omega), \spec\left(H_{\Lambda_Q}(x, \omega)\right)\right) <
e^{-(\log N)^C}$

\item[{\rm(ii)}] $\sum\limits_{k \in [1, N] \setminus \Lambda_Q} \big | \psi_j^{(N)} (x, \omega; k)
\big |^2 < e^{-Q\gamma/4}$, where $\gamma > 0$ is a lower bound for the Lyapunov exponents.
\end{enumerate}
\end{prop}
\begin{proof} Fix $N$, $\omega \in \tor_{c,a} \setminus \Omega_N$ and $E_j^{(N)} (x, \omega) \notin
\cE_{N, \omega}$.  Let $n_1 = \nu_j^{(N)} (x, \omega)$ be such that
$$
\big | \psi_j^{(N)} (x, \omega; n_1)\big | = \max_{1 \le n \le N} \big | \psi_j^{(N)} (x, \omega;
n)\big |\ .
$$
Fix some $\ell \asymp (\log N)^C$ and suppose that, with $E = E_j^{(N)} (x, \omega)$, and
$\Lambda_0 := [1, N] \cap [n_1 -\ell, n_1 +
\ell]$,
\begin{equation}
\label{eq:9.14}
\log \big | f_{\Lambda_0} (x, \omega, E) \big | > |\Lambda_0| L(\omega,E) - \sqrt{\ell}
\end{equation}
By Lemma~\ref{lem:waschno}
$$
\big | G_{\Lambda_0} (x, \omega, E) (k, j) \big | < \exp \left(- {\gamma\over 2} \big |k - j\big
| + C\sqrt{\ell}\right)
$$
for all $k, j \in \Lambda_0$.  But this contradicts the maximality of $\big | \psi_j^{(N)} (x,
\omega; n_1) \big |$ due to \eqref{eq:poisson}.  Hence \eqref{eq:9.14} above fails, and we conclude from Lemma~\ref{lem:waschno} that
$$
\log \big | f_{\Lambda_1} (x, \omega, E) \big | > |\Lambda_1| L(\omega,E) - \sqrt{\ell}
$$
for every $\Lambda_1 = [k - \ell, k + \ell] \cap [1, N]$ provided $|k - n_1 | > \exp
\left((\log\log N)^A\right)$.  Since \eqref{eq:9.14} fails, we conclude that $f_{\Lambda_0} (z_0,
\omega, E) = 0$ for some $z_0$ with $|z_0 - e(x)| < e^{-\ell^{1/4}}$.  By self-adjointness of
$H_{\Lambda_0} (x, \omega, E)$ we obtain
$$
\dist\left(E, \spec\left(H_{\Lambda_0} (x, \omega)\right)\right) < e^{-\ell^{{1/4}}}\ ,
$$
as claimed (the same arguments applies to the larger intervals $\Lambda_Q$ around $n_0$).

From \eqref{eq:f1n} of the previous lemma with $n = n_1 - Q/2$ (if $n_1 - Q/2 < 1$, then proceed to
the next case) one concludes that
\begin{equation}
\label{eq:9.15}
\big | G_{[1, n_1 - {1\over 2} Q]} (x, \omega, E) (k, m) \big | < \exp \left(-\gamma |k-m| +
(\log N)^C\right)
\end{equation}
for all $1 \le k, m \le n_1 - {1\over 2} Q$.  In particular,
$$
\big | \psi_j^{(N)} (x, \omega; k) \big | < e^{-{\gamma\over 2} | n_1 -
{1\over 2} Q - k|}
$$
for all $1 \le k \le n_1 - Q$.
Finally, the same reasoning applies to
$$
G_{[n_1 + {1\over 2} Q, N]} (x, \omega, E)
$$
via \eqref{eq:fnN} of the previous lemma, and (ii) follows.
\end{proof}

The following corollary deals with the stability of the localization statement of
Proposition~\ref{prop:9.4} with respect to the energy. As in previous stability results
of this type in this paper,  the most important issue is the relatively large size of the perturbation, i.e.,
$\exp(-(\log N)^B)$ instead of $e^{-N}$, say.

\begin{corollary}
\label{cor:Estab}
Let $\Omega_N$, $\cE_{N,\omega}$, $\left\{E^{(N)}_j(x, \omega) \right\}^N_{j=1}$,
and $\left\{\psi_j^{(N)}(x, \omega; \cdot)\right\}^N_{j=1}$, be as in the previous proposition.
Then for any $\omega \in \tor_{c,a} \setminus \Omega_N$, any $x \in \tor$, $E_j^{(N)}(x, \omega)
\notin \cE_{N,\omega}$  let $\nu_j^{(N)}(x,\omega)$ be as in the previous proposition.  For such
$\omega, E_j^{(N)}(x, \omega)$, if $|E - E_j^{(N)}(x, \omega)| < e^{-(\log N)^B}$ with $B$
sufficiently large, then
\begin{equation}
\label{eq:9.16}
\sum^{\nu_j^{(N)}(x,\omega) - Q}_{n=1} \big | f_{[1, n]} \bigl(e(x), \omega, E\bigr) \big |^2 < e^{-c\gamma Q}
\sum_{n \in \Lambda_Q} \big | f_{[1, n]}\bigl(e(x), \omega, E\bigr)\big |^2
\end{equation}
where $\Lambda_Q = \left[\nu_j^{(N)} (x, \omega) - Q, \nu_j^{(N)}(x, \omega) + Q\right] \cap [1,
N]$.  Similarly,
\begin{equation}
\label{eq:9.17}
\sum^N_{n = \nu_j^{(N)}(x, \omega)+ Q} \big | f_{[n, N]}(x, \omega, E)\big |^2 < e^{-c\gamma Q}
\sum_{n \in \Lambda_{Q}} \big | f_{[n, N]}(x, \omega, E) \big |^2
\end{equation}
Finally, under the same assumptions one has
\begin{align}
\label{eq:9.18} &\big | f_{[1, n]} \left(e(x), \omega, E\right) -
f_{[1, n]}\bigl(e(x), \omega, E_j^{(N)} (x, \omega)\bigr) \big |
\\
&\le \exp\left((\log N)^C\right) \big | E - E_j^{(N)}\bigl(x,
\omega)\big |\, \big |f_{[1, n]}(e(x), \omega, E_j^{(N)} (x,
\omega)\bigr)\big |\nn
\end{align}
provided $1 \le n \le \nu_j^{(N)} (x, \omega) - Q$, and similarly for $f_{[n, N]}$.
\end{corollary}
\begin{proof} For each $j$ there exists a constant $\mu_j(x, \omega)$ so that
$$
\psi_j^{(N)}(x, \omega; n) = \mu_j(x, \omega) f_{[1, n-1]}\left(x, \omega; E_j^{(N)} (x,
\omega)\right)
$$
for all $1 \le n \le N$ (with the convention that $f_{[1, 0]} = 1$).  A
similar formula holds for
\[ f_{[n+1, N]}\left(e(x), \omega, E_j^{(N)} (x,
\omega)\right).\]
As in the previous proof, one obtains
estimate \eqref{eq:9.15} with $E = E_j^{(N)} (x, \omega)$.  Thus, for $1 \le n \le
\nu_j^{(N)}(x, \omega) - Q$
$$
\Big | f_{[1, n]}\left(e(x), \omega, E_j^{(N)} (x, \omega)\right)\Big | <
e^{-c\gamma|\nu_j^{(N)} (x, \omega) - n|}\ \Big | f_{[1, \nu_j^{(N)}(x,
\omega)]} \left(e(x), \omega, E_j^{(N)} (x, \omega)\right)\Big |\ ,
$$
which implies \eqref{eq:9.16} for $E = E_j^{(N)}(x, \omega)$, and \eqref{eq:9.17} follows by a similar argument for this $E$.
Corollary~\ref{cor:4.6} implies that
\begin{align*}
& \Big | f_{[1, n]}\bigl(e(x), \omega, E\bigr) - f_{[1, n]}\left(e(x), \omega,
E_j^{(N)}(x, \omega)\right)\Big | \\
& \le \exp \left((\log N)^C\right) \Big | E - E_j^{(N)} (x, \omega)\Big
|\ \Big | f_{[1, n]} \left(e(x), \omega, E_j^{(N)}(x, \omega)\right)\Big |
\end{align*}
for all $1 \le n \le \nu_j^{(N)}(x, \omega) - Q$,
and \eqref{eq:9.18} follows for all $\big | E - E_j^{(N)}(x, \omega)\big | <
\exp \bigl(-(\log N)^B\bigr)$.
\end{proof}

\section{Minimal distance between the Dirichlet eigenvalues
on a finite interval}
\label{sec:minimal}

In this section it will be convenient for us to work with the
 operators $H_{[-N, N]}(x, \omega)$ instead of
$H_{[1, N]}(x, \omega)$ as we did in Section~\ref{sec:localize}.
Abusing our notation somewhat, we use the symbols $E_j^{(N)}, \psi_j^{(N)}$ to denote the
eigenvalues and normalized eigenfunctions of $H_{[-N,N]}(x, \omega)$, rather than
the eigenvalues and normalized eigenfunctions of $H_{[1, N]}(x, \omega)$, as in the previous section.
A similar comment applies to
$\Omega_N, \cE_{N, \omega}.$

The following proposition states that the eigenvalues $\{
E_j^{(N)}(x,\omega)\}_{j=1}^{2N+1}$  are separated from each other
by at least $e^{-N^\delta}$ provided $\omega\not\in\Omega_N$ and
provided we delete those eigenvalues that fall into a bad set
$\cE_{N,\omega}$ of energies. We remind the reader that
\[ \mes(\cE_{N,\omega})\les \exp(-(\log N)^{A_2}),\quad \compl(\cE_{N,\omega})\les \exp((\log N)^{A_1}), \]
where $A_2\gg A_1$, and the same for $\Omega_N$, see Lemma~\ref{lem:waschno}.

\begin{prop}
\label{prop:Ej_sep}
For any $\omega \in\tor_{c,a}\setminus \Omega_N$ and all $x$ one has for all $j, k$ and any small
$\delta > 0$
\begin{equation}
\label{eq:Ej_sep}
\big | E_j^{(N)} (x, \omega) - E_k^{(N)} (x, \omega) \big | >
e^{-N^\delta}
\end{equation}
provided $E_j^{(N)}(x, \omega) \notin \cE_{N,\omega}$ and $N \ge
N_0(\delta)$.
\end{prop}
\begin{proof} Fix $x \in \tor, E_j^{(N)} (x, \omega) \notin
\cE_{N,\omega}$.  Let $Q\asymp \exp \left((\log\log N)^C\right)$.  By
Proposition~\ref{prop:9.4} there exists
$$
\Lambda_Q := \left[\nu_j^{(N)} (x, \omega) - Q, \nu_j^{(N)} (x, \omega)
+ Q\right] \cap [-N, N]
$$
so that
\begin{equation}
\label{eq:10.2}
\begin{split}
& \sum_{n \in [-N, N]\setminus \Lambda_Q} \big | f_{[-N, n]} \bigl(e(x),
\omega; E_j^{(N)}(x, \omega)\bigr)\big |^2 \\
& < e^{-2Q \gamma} \sum^N_{n=-N} \big |f_{[-N, n]} \bigl(e(x), \omega;
E_j^{(N)}(x, \omega)\bigr)\big|^2\ .
\end{split}
\end{equation}
Here we used that with some $\mu = \const$
$$
\psi_j^{(N)} (x, \omega; n) = \mu \cdot f_{[-N, n-1]}\left(e(x), \omega;
E_j^{(N)}(x, \omega)\right)
$$
for $ -N \le n \le N$.
Here we use the convention that
\[ f_{[-N,-N-1]}=0, \quad f_{[-N,-N]}=1.\]
One can assume $\nu_j^{(N)}(x, \omega)
\ge 0$ by symmetry.
Using Corollary~\ref{cor:4.6} and \eqref{eq:9.18}, we conclude that
\begin{equation}
\label{eq:10.3}
\begin{split}
& \sum_{n= -N}^{\nu_j^{(N)}(x, \omega) - Q} \big | f_{[-N, n]} (e(x),
\omega, E) - f_{[-N, n]}\bigl(e(x), \omega, E_j^{(N)} (x, \omega)\bigr)\big
|^2 \\
& \le e^{-2\gamma Q} \big |E - E_j^{(N)}(x, \omega) \big |^2 e^{(\log
N)^C} \sum_{n \in \Lambda_Q} \big |f_{[-N, n]} \bigl(e(x), \omega,
E_j^{(N)} (x, \omega)\bigr)\big |^2
\end{split}
\end{equation}
Let $n_1 = \nu_j^{(N)} (x, \omega) - Q -1$.  Furthermore,
\begin{equation}
\label{eq:10.4}
\begin{split}
 &\Bigg \| \binom{f_{[-N, n+1]}(e(x), \omega, E)}{f_{[-N, n]}(e(x), \omega, E)}
- \binom{f_{[-N, n+1]} \bigl(e(x), \omega, E_j^{(N)} (x,
  \omega)\bigr)}{f_{[-N, n]}\bigl(e(x), \omega, E_j^{(N)}(x, \omega)\bigr)}
  \Bigg \|\\
  &= \Bigg \| M_{[n_1+1, n]} (e(x), \omega, E) \binom{f_{[-N, n_1
  +1]}(e(x),
  \omega, E)}{f_{[-N, n_1]}(e(x), \omega, E)} \\
  & \qquad\qquad- M_{[n_1 +1,
  n]}\bigl(e(x),
  \omega, E_j^{(N)}(x, \omega)\bigr) \binom{f_{[-N, n_1+1]}\bigl(e(x),
  \omega, E_j^{(N)}\bigr)}{f_{[-N, n_1]}\bigl(e(x), \omega,
  E_j^{(N)}\bigr)}\Bigg \|\\
& \le e^{C(n - n_1)} e^{-\gamma Q} \big |E - E_j^{(N)} (x, \omega) \big
| e^{(\log N)^C} \biggl(\sum_{n \in \Lambda_Q} \big | f_{[-N, n]}
\bigl(e(x), \omega, E_j^{(N)}(x, \omega)\bigr) \big |^2\biggr)^{{1\over
2}}\ .\\
  \end{split}
  \end{equation}
Now suppose there is $E_k^{(N)}(x, \omega)$ with $\big |E_k^{(N)}(e(x),
\omega) - E_j^{(N)} (x, \omega) \big | < e^{-N^\delta}$ for some small
$\delta > 0$.  Then (\ref{eq:10.3}), (\ref{eq:10.4}) imply that
\begin{equation}
\label{eq:10.5}
\begin{split}
& \sum^{\nu_j^{(N)}(x, \omega) + Q}_{n= -N} \big | f_{[-N, n]}
\bigl(e(x),
\omega, E_j^{(N)} (x, \omega)\bigr) - f_{[-N, n]}\bigl(e(x), \omega,
E_k^{(N)} (x, \omega)\bigr)\big |^2\\
& < e^{-{1\over 2} N^\delta} \sum_{n \in \Lambda_Q} \big |f_{[-N, n]}
\bigl(e(x), \omega, E_j^{(N)}(x, \omega)\bigr)\big |^2\ ,
\end{split}
\end{equation}
provided $N^\delta > \exp\bigl((\log\log N)^A\bigr)$.
Let us estimate the contributions of $\left[\nu_j^{(N)} (x,
\omega) + Q, N\right]$ to the sum terms in the left-hand side of (\ref{eq:10.5}).

For both $E = E_j^{(N)}$ and $E_k^{(N)}$ one has
$$
f_{[-N, n]}(e(x), \omega, E) = G_{[\nu_j^{(N)} (x, \omega) + {Q\over 2},
N]} (e(x), \omega, E) \bigl(n ,\nu_j^{(N)}(x, \omega) + {Q\over 2}\bigr)
f_{[-N, \nu_j^{(N)}(x, \omega) + {Q\over 2} -1]}(e(x), \omega, E)
$$
due to the zero boundary condition at $N +1$, i.e.,
$$
f_{[-N, N]} \bigl(e(x), \omega, E_j^{(N)} (x, \omega)\bigr) = f_{[-N, N]}
\bigl(e(x), \omega, E_k^{(N)} (x, \omega) \bigr) = 0\ .
$$
Therefore, 
\begin{equation}
\label{eq:10.6}
\sum^N_{n = \nu_j^{(N)} + Q} \big | f_{[-N, n]}(e(x), \omega, E)\big |^2
\le e^{-{\gamma Q\over 4}} \sum_{k \in \Lambda_Q}\big |f_{[-N, k]}(e(x),
\omega, E)\big |^2
\end{equation}
again for both $E = E_j^{(N)}(x, \omega)$ and $E = E_k^{(N)}(x,
\omega)$.  Finally, in view of (\ref{eq:10.5}) and (\ref{eq:10.6}),
\begin{equation}
\label{eq:10.7}
\begin{split}
 &\sum^N_{n=-N} \big | f_{[-N, n]}\bigl( e(x), \omega, E_j^{(N)} (x,
\omega)\bigr) - f_{[-N, n]}\bigl(e(x), \omega, E_k^{(N)} (x,
\omega)\bigr)\big |^2\\
& < e^{-{\delta Q\over 4}} \biggl[\sum_{n \in \Lambda_Q} \big |
f_{[-N, n]} \bigl(e(x), \omega, E_j^{(N)} (x, \omega)\bigr)\big |^2\\
&\quad +
\sum_{n \in \Lambda_Q} \big | f_{[-N, n]}\bigl(e(x), \omega, E_k^{(N)} (x,
\omega)\bigr) \big |^2 \biggr ]
\end{split}
\end{equation}
By orthogonality of $\left\{f_{[-N, n]}\bigl(e(x), \omega, E_j^{(N)} (x,
\omega)\bigr)\right\}^N_{n=-N}$ and $\left\{ f_{[-N, n]} \bigl(e(x),
\omega, E_k^{(N)} (x, \omega)\bigr)\right\}^N_{n=-N}$, we obtain a
contradiction from (\ref{eq:10.7}).
\end{proof}

By the well-known Rellich theorem, the eigenvalues $E_j^{(N)}(x,\omega)$ of the
Dirichlet problem on $[-N,N]$ are analytic functions of $x$ and can therefore be extended analytically to
a complex neighborhood of $\tor$. Moreover, by simplicity of the eigenvalues of the Dirichlet
problem, the graphs of these functions of $x$ do not cross. Proposition~\ref{prop:Ej_sep}
makes this non-crossing quantitative, up to certain sections of the graphs where we loose
control. These are the portions of the graph that intersect horizontal strips corresponding
to those energies in~$\cE_{N,\omega}$. The quantitative control provided by~\eqref{eq:Ej_sep}
allows us to give lower bounds on the radii of the disks to which the functions $E_j^{(N)}(x,\omega)$
extend analytically.

\begin{corollary}
\label{cor:10.2}
Let $\Omega_N$, $\cE_{N,\omega}$ be as above. Take
arbitrary $x_0 \in \tor$.  Assume $f_N(x_0,\omega_0, E_0) = 0$ for some $\omega_0\in\tor_{c,a}\setminus\Omega_N$ and
$E_0\notin \cE_{N, \omega_0}$.  Then there exist $r_0, r_1$, $r_1 =
e^{-N^\delta}$, $r_0 =c r_1$, such
that (with $\omega_0$ fixed)
\begin{equation}
\label{eq:b0z}
f_N(z,\omega_0, E) = \bigl(E - b_0(z)\bigr) \chi(z, E)
\end{equation}
for all $z \in \cD(x_0, r_0)$, $E \in \cD(E_0, r_1)$. Moreover, $b_0(z)$ is
analytic on $\cD(x_0, r_0)$, $\chi(z, E)$ is analytic and nonzero
on $\cD(x_0, r_0) \times \cD(E_0, r_1)$, $b_0(x_0) = E_0$.
\end{corollary}
\begin{proof} By Proposition~\ref{prop:Ej_sep}, $f_N(x_0,\omega_0, E) \ne 0$ if $E \in \cD(E_0,
r_1)$, $E \ne E_0$.  Since $H_N(x_0,\omega_0)$ is self adjoint and
\[ \big\| H_N(z,\omega_0)
- H_N(x_0,\omega_0)\bigr\| \lesssim |z - x_0|,
\] it follows that $f_N(z,\omega_0, E) \ne 0$
for any $|z - x_0| \ll r_1$, $r_1/2 < |E - E_0| < {3\over 4}r_1$.  The representation~\eqref{eq:b0z}
is now obtained by the same arguments that lead to the  Weierstrass
preparation theorem, see Lemma~\ref{lem:weier}.
\end{proof}

As an application of Proposition~\ref{prop:9.4} combined with
Proposition~\ref{prop:Ej_sep} we now illustrate how to relate the
localized eigenfunctions of consecutive scales. Indeed, by
Proposition~\ref{prop:9.4} any eigenfunction
$\psi_j^{(N)}(x,\omega,\cdot)$ is exponentially localized around
some interval $\Lambda$ of size $N'\asymp \exp((\log\log N)^A)$
provided $E_j^{(N)}(x,\omega)$ is outside of some
set~$E_{N,\omega}$. Due to this fact and the separation of
eigenvalues, the restricition of $\psi_j^{(N)}(x,\omega,\cdot)$ to
$\Lambda$ closely resembles some eigenfunction
$\psi_{j'}^{(N')}(x',\omega,\cdot)$. In particular, it is
exponentially localized around some interval $\Lambda'$ of size
$N''=\exp((\log\log N')^A)$.

\begin{lemma} Using the notations of Proposition~\ref{prop:Ej_sep}
 assume that $\omega \in \tor_{c,a} \setminus \left(\Omega_N \cup \Omega_{N'}\right)$,
 where $N' \asymp \exp\left((\log\log N)^{C_1}\right)$, $C_1 \gg C$, and with
  $Q = \exp\left((\log\log N)^C\right)$.
If \[E_j^{(N)}(x,\omega) \notin \cE_{N,\omega},\qquad
\dist\left(E_j^{(N)} (x,\omega), \cE_{N',\omega}\right) >
\exp\left(-(N')^{1/2}\right),\] then there exists $\nu \in \IZ$,
$|\nu - \nu_j^{(N)} (x,\omega)| \le Q$ and
\begin{equation}
\nn E_{j'}^{(N')}(x + \nu\omega, \omega) \in
\left(E_j^{(N)}(x,\omega) - \exp(-\gamma_1 N'), E_j^{(N)}(x,\omega)
+ \exp(-\gamma_1 N')\right)\ ,
\end{equation}
where $\gamma_1 = c\gamma$, $\gamma = \inf L(E, \omega)$.  Moreover,
the corresponding normalized eigenfunctions \[\psi_j^{(N)}(x,\omega,
k),\qquad \psi_{j'}^{(N')}(x + \nu\omega, \omega, k - N' + \nu)\]
satisfy
\begin{equation}
\label{eq:1010}
 \sum_{k \in [\nu - N', \nu + N']} \left|\psi_j^{(N)} (x, \omega, k) - \psi_{j'}^{(N')}(x + \nu\omega, \omega, k - N' + \nu)\right|^2
 \le \exp(-\gamma_1 N')\ .
\end{equation}
\end{lemma}
\begin{proof} Assume first $-N + N' < \nu_j^{(N)}(x, \omega) < N - N'$.  Then with $\nu = \nu_j^{(N)} (x, \omega)$ one has:
\begin{align}
& \label{eq:1011}\left\|\left(H_{[\nu - N', \nu + N']} (x, \omega) - E_j^{(N)} (x, \omega)\right) \psi_j^{(N)}(x, \omega, \cdot)
\right\| \le \exp(-\gamma N'/4)\ ,\\
& \label{eq:1012} 1 - \sum_{k \in [\nu - N', \nu + N']}
\left|\psi_j^{(N)}(x, \omega, k)\right|^2 < \exp(-\gamma N'/4)
\end{align}
due to Proposition~\ref{prop:Ej_sep}.  Hence, there exists
\[E_{j'}^{(N')} (x + \nu\omega, \omega) \in \left(E_j^{(N)} (x,
\omega) - \exp(-\gamma_1 N'), E_j^{(N)} (x,\omega) + \exp(-\gamma_1
N')\right).\]  Moreover, due to assumptions on $E_j^{(N)}(x,
\omega)$, one has $E_{j'}^{(N')}(x + \nu\omega, \omega) \notin
\cE_{N',\omega}$. Hence,
\begin{equation}\label{eq:1013}
\left|E_{j'}^{(N')}(x + \nu\omega, \omega) - E_k^{(N')} (x + \nu\omega, \omega) \right| > \exp\left(-(N')^\delta\right)
\end{equation}
for any $k \ne j'$.  Then (\ref{eq:1011})--(\ref{eq:1013}) combined
imply (\ref{eq:1010}) (expand in the orthonormal basis
$\{\psi^{(N')}_k\}_k$). If $\nu_j^{(N)}(x, \omega) \le - N + N'$
(resp $\nu_j^{(N)}(x, \omega) \ge N - N'$), then
(\ref{eq:1011})--(\ref{eq:1013}) are valid with $\nu_j^{(N)}(x,
\omega) = - N + N'$ (respectively with $\nu_j^{(N)}(x, \omega) = N -
N'$).
\end{proof}

Next, we iterate the construction of the previous lemma to obtain
the following.

\begin{corollary}\label{cor:inducm}
Given integers $m^{(1)}, m^{(2)}, \dots, m^{(t)}$ such that
\begin{equation}
\log m^{(s+1)} \asymp \exp\left(\bigl(\log m^{(s)}\bigr)^\delta\right)\ ,\quad s = 1, 2,\dots, t-1
\end{equation}
there exist subsets $\Omega^{(s)} \subset \tor$, $s = 1, 2,\dots$,
\[\mes\left(\Omega^{(s)}\right) < \exp \left(- \left(\log
m^{(s)}\right)^{A_2}\right),\qquad \compl (\Omega^{(s)}) < \exp
\left(\left(\log m^{(s)}\right)^{A_1}\right),\] $1 \ll A_1 \ll A_2$,
such that for any $\omega \in \tor_{c,a} \setminus \bigcup\limits_s
\Omega^{(s)}$ there exist subsets $\cE^{(s)}_\omega\subset \IR$ with
\[ \mes \cE^{(s)}_\omega < \exp\left(- \left(\log
m^{(s)}\right)^{A_3}\right),\qquad  \compl
\left(\cE^{(s)}_\omega\right) < \exp\left(\left(\log
m^{(s)}\right)^{A_4}\right)\] with $A_4\ll A_3$ such that for any $x
\in \tor$ and any $E \in \rsp\left(H_{m^{(t)}} (x, \omega)\right)
\setminus \bigcup\limits_s \cE^{(s)}_\omega$, the corresponding
eigenfunction $\psi(n)$, $1 \le n \le m^{(t)}$ of
$H_{m^{(t)}}(x,\omega)$ has the following property: there exists an
integer $\nu^{(t)} (x, \omega) \in \left[1, m^{(t)}\right]$ such
that
$$
\sum_{|n - \nu^{(t)}(x, \omega)| > Q}  |\psi(n)|^2  \lesssim
\exp(-\gamma'\, Q)
$$
where $ Q  = \exp\left(\left(\log\log m^{(1)}\right)^A\right)$ and
$\gamma'=c\gamma>0$.
\end{corollary}

\begin{proof} The proof goes by induction over $t = 1, 2,\dots$.
For $t = 1$, the assertion is valid due to
Proposition~\ref{prop:Ej_sep}. So, assume that it is valid for
$H_{m^{(t-1)}}(\tilde x, \omega)$ for any $\widetilde E \notin
\bigcup\limits^{t-1}_{s=1} \widetilde\cE^{(s)}_\omega$, $\tilde x
\in \tor$.  Let $E, \psi$ be as in the statement.  By the previous
lemma  there exist an interval $\Lambda = [a, a+\widetilde N]$,
$\widetilde N = m^{(t-1)}$, $1 \le a \le N - \widetilde N$ and a
normalized eigenfunction $\widetilde\psi$ such that $H_{[a,
a+\widetilde N]}(x, \omega) \widetilde\Psi = \widetilde E\,
\widetilde\Psi$, $|\widetilde E - E|< \exp\left(-\gamma \widetilde
N\right)$, $\big |\widetilde\psi(n) - \psi(n)\big | <
\exp\bigl(-\gamma \widetilde N\bigr)$, $n \in \Lambda$.  Applying
now the inductive assumption to $H_{m^{(t-1)}}(x+ a\omega, \omega)$
one obtains the assertion.
\end{proof}

\section{Mobility of Dirichlet eigenvalues and the separation of zeros of $f_N$ in~$z$}
\label{sec:mobil}

In this section, we will use the separation of the eigenvalues from Section~\ref{sec:minimal}
to obtain lower bounds on the derivatives of the Rellich functions off some small bad set
of phases. In particular, this will use Corollary~\ref{cor:10.2}.

\begin{lemma}
\label{lem:sard}
Let $\varphi(z)$ be analytic in some disk $\cD(0, r)$, $r>0$.  Then
\begin{equation}
\label{eq:11.1}
\mes \left\{ w: w = \vz,\ z \in \cD(0, r),\ |\varphi'(z)| < \eta \right\}
\le \pi r^2\eta^2
\end{equation}
\end{lemma}
\begin{proof} Set $A = \left\{z = x+iy \in \cD(0, r): |\varphi'(z)| < \eta
\right\}$.  By the general change of variables formula, see Theorem~3.2.3 in~\cite{fed},
$$
\int_A \Big | {\partial(u, v)\over \partial(x,y)}\Big | dx\, dy =
\int_{\IR^2} \#\left\{(x, y) \in A: \varphi(x+ iy) = u+ iv\right\}du\,
dv \ge \mes\varphi(A)
$$
where $\varphi(x + iy) = u(x, y) + iv(x, y)$.  On the other hand,
$$
\Big | {\partial(u, v)\over \partial(x, y)} \Big | = \left|\varphi'(x +
iy)\right|^2
$$
since $\varphi$ is analytic.
\end{proof}

The following lemma will allow us to transform the separation of the eigenvalues
into a lower bound on the derivative of the Rellich functions.
The logic behind Lemma~\ref{lem:2weier} is as follows: Let $b_0$ be as in (i).
By Lemma~\ref{lem:sard}, the measure of those $w$ which satisfy $w=b_0(z)$ with $b_0'(z)$ small,
is small. However, we also require a bound on the complexity of this set of $w$ which is only
logarithmic in $r_0$ and $r_1$. This is where property (ii) comes into play, and the complexity
will be proportional to a power of the degree $k$ as well as to $\log[(r_0r_1)^{-1}]$.

\begin{lemma}
\label{lem:2weier}
Let $f(z, w)$ be an analytic function defined in $\cD(0,1)\times \cD(0,1)$.
Assume that one has the following representations:
\begin{enumerate}
\item[(i)] $f(z, w) = (w - b_0(z))\chi(z, w)$,
for any $z \in \cD(0, r_0)$, $w \in \cD(0, r_1)$, where $b_0(z)$ is analytic
in $\cD(0, r_0)$, $\sup|b_0(z)| \le 1$,  $\chi(z, w)$ is analytic and non-vanishing on $\cD(0, r_0)
\times \cD(0, r_1)$, where $0<r_0,r_1<\frac12$

\item[(ii)] $f(z, w) = P(z, w)\theta(z, w)$, for any $z \in \cD(0, r_0)$,
$w \in \cD(0, r_1)$ where
$$
P(z, w) = z^k + c_{k-1}(w) z^{k-1} + \cdots +
c_0(w)\ ,
$$
$c_j(w)$ are analytic in $\cD(0, r_0)$, and $\theta(z, w)$ is
analytic and non-vanishing on $\cD(0, r_0) \times \cD(0, r_1)$, and all the
zeros of $P(z, w)$ belong to $\cD(0, 1/2)$.
\end{enumerate}
Then given $H \gg k^2 \log [(r_0r_1)^{-1}]$ one can find
a set $\cS_H \subset \cD(w_0, r_1)$ with the property that
\[
\mes(\cS_H) \les r_1^2 \exp\left(-cH/k^2 \log [(r_0r_1)^{-1}]\right), \text{\ \ and \ \ } \compl(S_H)\lesssim k^2 \log [(r_0r_1)^{-1}]
\]
such that for any $w \in \cD(0, r_1/2) \setminus \cS_H$ and $z \in \cD(0, r_0)$
for which $w = b_0(z)$ one has
$$
\big | b'_0 (z) \big | > e^{-kH} 2^{-k} r_1\ .
$$
Moreover, for those $w$ the distance between any two zeros of $P(\cdot,w)$ exceeds $e^{-H}$.
\end{lemma}
\begin{proof} Assume that $k\ge2$ and set $\psi(w) = \disc P(\cdot, w)$. If $k=1$, then skip to~\eqref{eq:11.6}.
Then $\Psi(w)$ is analytic in $\cD(0, r_1)$.  Assume that $|\psi(w)| < \tau$ for some
$\tau> 0$, $w\in \cD(0, r_1)$.  Recall that for any $w$
\begin{equation}
\label{eq:11.2}
\psi(w) = \prod_{i\ne j} \left(\zeta_i(w) - \zeta_j(w)\right)\ ,
\end{equation}
where $\zeta_i(w)$, $i = 1, 2, \dots, k$ are the zeros of $P(\cdot, w)$.
Then $|\zeta_i(w) - \zeta_j(w)| < \tau^{2/k(k-1)}$ for some $i \ne j$.  Set
$\zeta_i = \zeta_i(w)$, $\zeta_j = \zeta_j(w)$.  Assume first $\zeta_i
\ne \zeta_j$.  Then
$$
f(\zeta_i, w) = 0\,\ f(\zeta_j, w) = 0,\quad 0 < |\zeta_i - \zeta_j| <
\tau^{2/k(k-1)}\ .
$$
Due to (i) one has $w = b_0(\zeta_i) = b_0(\zeta_j)$.  Hence,
\begin{equation}\label{eq:11.3}
|b'_0(\zeta_i)| \le {1\over 2} |\zeta_i - \zeta_j| \max |b''_0(z) |
\lesssim |\zeta_i - \zeta_j| r_0^{-2} < r_0^{-2} \tau ^{2/k(k-1)}\ .
\end{equation}
If $\zeta_i = \zeta_j$ then $P(\zeta_i, w) = 0$, $\partial_z P(\zeta_i,
w) = 0$.  Then $f(\zeta_i, w) = 0$, $\partial_z f(\zeta_i, w) = 0$ due
to the representation~(ii).  Then $w - b_0(\zeta_i) = 0$, $b'(\zeta_i) = 0$
due to the representation~(i).  Thus (\ref{eq:11.3}) holds at any event.
Combining that bound with the  estimate (\ref{eq:11.1}) one obtains
\begin{equation}
\label{eq:question}
\mes \left\{w \in \cD(0, r_1): |\psi(w)| < \tau\right\} \lesssim r_0^{-2} \tau^{2/k(k-1)}.
\end{equation}
On the other hand, due to \eqref{eq:11.2} one obtains
$$
\sup\left\{|\psi(w)|: w \in \cD(0, r_1)\right\} \le 1.
$$
Take $\tau \ll (r_0r_1)^{k(k-1)/2}$. Then one obtains from (\ref{eq:question}) that
$$
|\psi(w) | \ge \tau
$$
for some $|w| < \frac{r_1}{2}$.  By Cartan's estimate there exists a set
$\cT_H \subset \cD\left(0, \frac{r_1}{2}\right)$ with
$$\mes \cT_H \lesssim r_1^2\exp\left(-cH/k^2 \log[
(r_0r_1)^{-1}]\right)
$$
and of complexity $\lesssim k^2 \log [(r_0r_1)^{-1}]$ such that
\begin{equation}
\label{eq:11.5}
\log |\psi(w)| > -H
\end{equation}
for any $w \in \cD(0, {r_1\over 2}) \setminus \cT_H$.

In particular, (\ref{eq:11.5}) implies that
\beeq
\label{eq:zetaij}
|\zeta_i(w) - \zeta_j(w) | > e^{-H}
\eneq
for any $w \in \cD(0, {r_1\over 2}) \setminus \cT_H$, $i \ne j$.  Take arbitrary
$w_0$ such that $\dist(w_0, \cT_H) > 2e^{-H}$, $w_0 = b_0(z_0)$ for some
$z_0 \in \cD(0, r_0)$. Then
\[ |P(z,w_0)|\ge (2e^{H})^{-k} \text{\ \ for all\ \ }|z-z_0|= e^{-H}/2\]
by the separation of the zeros~\eqref{eq:zetaij}.   By our assumption on the zeros of $P(z,w)$,
\[ \sup_{z\in\cD(0,r_0)}\sup_{w\in\cD(0,r_1)} |\partial_wP(z,w)| \lesssim r_1^{-1}.\]
Thus,
\[ |P(z,w)|>\frac12 2^{-k}e^{-kH} \text{\ \ if\ \ }|z-z_0|= e^{-H}/2,\quad |w-w_0|\ll 2^{-k}e^{-kH}r_1.\]
Then due to the Weierstrass preparation theorem, see Lemma~\ref{lem:weier},
\begin{equation}
\label{eq:11.6}
P(z, w) = \bigl(z - \zeta(w)\bigr) \lambda (z, w)
\end{equation}
for any $z \in \cD(z_0, r'_0)$, $w \in \cD(w_0, r'_1)$, where $r'_0 = e^{-H}/2$,
$r'_1 \ll e^{-kH} 2^{-k} r_1$, and  $\zeta(w)$ is an analytic function in
$\cD(w_0, r'_1)$, $\lambda(z, w)$ is analytic and non-vanishing on $\cD(z_0, r'_0)\times
\cD(w_0,r'_0)$.  Comparing the representation (i) and (\ref{eq:11.6}) one obtains
\begin{equation}
\label{eq:11.7}
\begin{cases}
w - b_0(z) = 0 & \text{iff}\\
z - \zeta(w) = 0
\end{cases}
\end{equation}
for any $z \in \cD(z_0, r'_0)$, $w \in \cD(w_0, r'_1)$.  It follows from \eqref{eq:11.7} that
$$
\big | b'_0 \bigl(\zeta(w)\bigr)\big | \ge \big |\zeta'(w)\big |^{-1}
\gtrsim r'_1 \gtrsim e^{-kH} 2^{-k} r_1,
$$
as claimed.
\end{proof}

Now choose arbitrary $\omega_0 \in \tor_{c, a} \setminus \Omega_N$, $E_0
\in\left(-C(V), C(V)\right) \setminus \widetilde{\cE}_{N, \omega_0}$,
where
$$
\widetilde{\cE}_{N, \omega_0} = \left\{E: \dist(E, \cE_{N,\omega_0}) < \exp(-N^{\delta/2})\right\}\ ,
$$
$\Omega_N$, $\cE_{N,\omega_0}$ are the same as in Proposition~\ref{prop:Ej_sep}.
Then for any $x \in \tor$ one has
\begin{equation}
\begin{split}
 & \min \Bigl\{ \big | E_{j}^{(N)}(x, \omega_0) - E_{i}^{(N)}(x, \omega_0)\big |:
E_{j}^{(N)}(x, \omega_0), E_{i}^{(N)}(x, \omega_0) \in\\
&  \left(E_0 - \exp(-N^{\delta/2}), E_0 + \exp(-N^{\delta/2})\right), i
\ne j\Bigr\} \ge \exp(-N^{\delta})
\end{split}
\end{equation}
Here $E_{j}^{(N)}(x,\omega)$ stand for the eigenvalues of $H_N^{(D)}(x, \omega)$
as usual.

Now assume that there is $x_0 \in \tor$ such that $E_{j_0}^{(N)}(x_0, \omega_0) \in \left(E_0 - \exp(-N^{\delta/2}), E_0 +
\exp(-N^{\delta/2})\right)$ for some $j_0$.  Then, as in Corollary~\ref{cor:10.2},
$$
f_N(z,  \omega_0, E) = \left(E - b_0(z)\right) \chi(z, E)
$$
where $(z, E) \in \cP = \cD(x_0, r_0) \times \cD(E_{j_0}^{(N)}(x_0, \omega_0), r_0)$, $r_0 = \exp(-N^{\delta_1})$
with $\delta_1\gg\delta$, and the analytic functions $b_0(z)$,
$\chi(z,E)$ satisfy the properties stated in Corollary~\ref{cor:10.2}.  On the
other hand,  due to the Weierstrass preparation theorem in the $z$-variable, see Proposition~\ref{prop:fN_prep_z},
\beeq
\label{eq:11.40}
f_N(z,  \omega, E)= P_N(z, \omega,E) g_N(z, \omega,E)
\eneq
$(z,  \omega, E) \in \cP_1 = \cD(x_0, r_1)  \times
\cD(\omega_0, r_1)\times \cD(E_0, r_1)$, $r_1 \asymp \exp\bigl(-(\log N)^C\bigr)$,
where $P_N, g_N$ satisfy conditions (a)--(d) of Proposition~\ref{prop:fN_prep_z}.  Thus, all
conditions needed to apply Lemma~\ref{lem:2weier} are valid for $f_N(z, \omega_0,E)$.
So, using the notations of the previous two paragraphs we obtain the following

\begin{corollary}
\label{cor:11.3}
There exist constants $\delta_1 \ll \delta_2 \ll 1$ with
the following properties:
Set $E_1=E_{j_0}^{(N)}(x_0, \omega_0)$ where $\omega_0\in\tor_{c,a}\setminus\Omega_N$ and $x_0\in\tor$.
There exists a subset $\cE'_{N, \omega_0,x_0,j_0} \subset \IC$, with
\[ \mes(\cE'_{N,\omega_0,x_0,j_0})\le \exp(-N^{\delta_2}), \quad \compl(\cE'_{N, \omega_0,x_0,j_0})\le N\]
such that for any $E \in \cD(E_1,r_1)\setminus\cE'_{N,\omega_0,x_0,j_0}$ and $z \in D(x_0, r_1)$, $r_1 = \exp(-N^{\delta_1})$,
for which $E=b_0(z, \omega_0)$ one has
$$
|\partial_z b_0(z)| > \exp(-N^{2\delta_2})\ .
$$
Moreover, for any $E \in \cD(E_1,r_1)\setminus\cE'_{N,\omega_0,x_0,j_0}$
the distance between any two
zeros of the polynomial $P_N(\cdot,  \omega_0, E)$ which fall into the disk $\cD(x_0,r_1)$ exceeds
$\exp(-N^{2\delta_2})$.
\end{corollary}

As usual, we can go from an exceptional set in the energies to one in the phases $x$ by
means of the Wegner-type bound of Lemma~\ref{lem:wegner}.

\begin{corollary}
\label{cor:11.4}  Let us use the notations of the previous corollary.
Let $\omega_0\in\tor_{c,a}\setminus\Omega_N$ and $x_0\in\tor$.  Then
there exists a subset $\cB'_{N, \omega} \subset (x_0 - r_1, x_0 + r_1)$
\[ \mes(\cB'_{N,\omega_0})\le \exp(-N^{\delta_2}),\quad \compl(\cB'_{N, \omega_0})\le N^2\]
such that for any $x \in (x_0 - r_1, x_0 + r_1)\setminus\cB'_{N,\omega_0}$
one has
\begin{equation}
\big | \partial_x b_0(x, \omega)\big | \gtrsim e^{-N^{2\delta_2}}
\end{equation}
\end{corollary}

\begin{proof} Let $E\in \cD(E_1,r_1)$. Suppose $x\in (x_0-r_1,x_0+r_1)$.
Then $E = b_0(x)$ iff $E_1 \in {\rm sp}\left(H_N(x,
\omega_0)\right)$.  Due to Lemma~\ref{lem:wegner} there exists $\cB'_{N, \omega_0} \subset
(x_0 - r_1, x_0 + r_1)$ with the stated measure and complexity bounds
such that for any $x \in (x_0 - r_1, x_0 + r_1)\setminus \cB'_{N, \omega_0}$ one has
\begin{equation}
\nn
{\rm sp}\left(H_N(x, \omega_0)\right) \cap \cE'_{N, \omega_0,x_0,j_0} = \emptyset
\end{equation}
Here $\cE'_{N, \omega_0,x_0,j_0}$ is the same as in Corollary~\ref{cor:11.3}.
\end{proof}

With $\omega_0 \in \tor_{c,a} \setminus \Omega_N$ fixed as above, we take the union of the sets $\cE_{N,\omega_0,x_0,j_0}$
in $x_0,j_0$ with $x_0
\in \tor$ running over an appropriate net, to conclude the following

\begin{corollary}
\label{cor:11.6}
There exists a set
$\cE''_{N,\omega_0} \subset \IR$
with
\[ \mes(\cE''_{N,\omega_0}) \le \exp\bigl(-N^{2\delta_1}\bigr), \qquad \compl(\cE''_{N,\omega_0})\le \exp\bigl(N^{\delta_1}\bigr), \]
such that for any $E \in \bigl(-C(V), C(V)\bigr)\setminus\cE''_{N,\omega_0} $ and any $|\eta|\le \exp(-N^{2\delta_1})$ one has
\begin{quote}
the distance between any two zeros of $f_N(\cdot, \omega_0, E+i\eta)$  exceeds $\exp(-N^{\delta_2})$
\end{quote}
where $\delta_1\ll\delta_2\ll 1$.
\end{corollary}

\section{Harnack's inequality for norms of monodromies}
\label{sec:harnack}

Let $M(z) = \bigl(a_{ij}(z)\bigr)_{1\le i, j \le m}$ be an analytic
matrix function defined in some disk $\cD(z_0, r_0) \subset \IC$, $r_0
\ll 1$.  Let $ K:= \sup \left\{\|M(z)\|: z \in \cD(z_0, r_0) \right\} <
\infty$, and assume that for any $H \ge (\log\log K)^A$ one has
\begin{equation}
\log \bigl |a_{ij}(z)\big | > \log K - H
\tag{\rm{I}}
\end{equation}
for any entry $a_{ij}$ which is not identically zero and all
\[ z \in \cD(z_0, r_0\exp(-(\log\log K)^A)) \setminus \cB, \quad \cB =
\bigcup\limits^J_{j=1} \cD(\zeta_j, r),\quad r= r_0 \exp \left({-H\over
(\log\log K)^A}\right),\]
and $J \lesssim (\log\log K)^A$. Moreover, we assume that $\log r_0^{-1}<(\log K)^{\frac12}$ and $K\gg1$.

\begin{lemma}
\label{lem:14.1}
Suppose some entry of $M(z)$ has no zeros in $\cD(z_0,
r_1)$,
\[ r_0\exp(-\sqrt{\log K})\le r_1 \le r_0 \exp \bigl(-(\log\log K)^C\bigr)\] where $C \gg A$ is
some constant.  Then
\begin{equation}
\label{eq:14.1}
\Big | \log {\|M(z)\|\over \|M(z_0)\|}\Big | \lesssim |z - z_0|
r_2^{-1}
\end{equation}
for any $|z - z_0| \ll r_2$, where $r_2 = r_1 \exp\bigl(-(\log\log
K)^{2C}\bigr)$ provided $K$ is large.
\end{lemma}

\begin{proof} By assumption, some  entry $a_{i_0, j_0}(z)$ has no zeros in
$\cD(z_0, r_1)$. Define $\tilde r_1<r_1$ and $(\log\log K)^C\le H\le \log K$ via
\[ \tilde r_1 = r_1\exp(-(\log\log K)^C)=r_0\exp \left({-H\over
(\log\log K)^{2A}}\right). \]
By (I) there exists $z_{i_0,\, j_0} \in \cD(z_0, \tilde r_1)$ so that
$$
\log \big | a_{i_0, j_0}(z_{i_0\, j_0})\big | > \log K - H
$$
Apply Harnack's inequality to the non-negative harmonic function $-\log
\left(K^{-1} \big | a_{i_0, j_0}(z)\big | \right)$.  Then, for any $z
\in \cD(z_0, \tilde r_1)$,
\begin{equation}
\nn
\begin{split}
\big |a_{i_0, j_0}(z)\big | & \ge K \exp \left[\left(1 + {\tilde
r_1\over r_1}\right) \log \left(K^{-1} \big|a_{i_0, j_0}(z_{i_0\,
j_0})\big | \right)\right]\\
& \ge \big |a_{i_0, j_0}(z_{i_0\, j_0})\big | \exp\left(-\exp(-(\log\log K)^C) H \right) \ge {1\over 2} \big |a_{i_0, j_0}(z_{i_0\, j_0})\big |\ ,
\end{split}
\end{equation}
since $H\le \log K$, and similarly $\big | a_{i_0, j_0}(z)\big | \le 2 \big |a_{i_0,
j_0}(z_{i_0\, j_0})\big |$.  In particular,
\begin{equation}
\nn
\max_{|z - z_0| \le \tilde r_1} \big | a_{i_0, j_0}(z) \big | \le 4
\big | a_{i_0, j_0}(z_0)\big |
\end{equation}
and also
\begin{equation}
\label{eq:14.4}
\big |a_{i_0, j_0}(z_0)\big | \ge {1\over 2} K e^{-H}\ .
\end{equation}
Define
\begin{equation}
\nn
\begin{split}
\Gamma_1  & = \left\{(i, j) \big |1 \le i,j \le m,\ (i,j) \ne (i_0, j_0),\
a_{ij}(z)\ \mbox{has no zeros in $\cD(z_0, \tilde r_1)$}\right\}\\
\Gamma_2 & = \left\{(i, j) \big | 1 \le i, j \le m, \ (i, j) \ne (i_0,
j_0),\ a_{ij}(z) \ \mbox{has a zero in $\cD(z_0, \tilde r_1)$}\right\}\ .
\end{split}
\end{equation}
For each $(i, j)$ there is $z_{ij} \in \cD(z_0,\tilde r_1)$ with
$$
\log \big | a_{ij}(z_{ij}) \big | > \log K - H\ .
$$
Let $(i, j) \in \Gamma_1$.  Applying Harnack's inequality as before
yields
\begin{equation}
\nn
\begin{split}
\big | a_{ij}(z) \big | & \ge K \exp\left[\left(1 + {2r_2\over
\tilde r_1}\right) \log \left(K^{-1} \big | a_{ij} (z_{ij})\big |
\right)\right]\\
& \ge \big |a_{ij}(z_{ij})\big | \exp \left(-\exp\left(-\log\log K)^{C}\right)H\right)\\
& \ge {1\over 2} \big |a_{ij} (z_{ij})\big |
\end{split}
\end{equation}
and also  $\big |a_{ij}(z) \big | \le 2 \big | a_{ij}(z_{ij})\big |$
for all $z \in \cD(z_0, 2r_2)$.  In particular,
\begin{equation}
\label{eq:14.7}
\max_{|z- z_0| \le 2r_2} \big | a_{ij}(z)\big | \le 4 \big
|a_{ij}(z_0)\big |\ .
\end{equation}

Now let $(i,j) \in \Gamma_2$.  Then
$
a_{ij} (\zeta_{ij}) = 0$ for some $\zeta_{ij} \in \cD(z_0, \tilde r_1)$.  By
the maximum principle,
$$
\big |a_{ij}(z) \big | \le {K\over r_0/2} \big |z - \zeta_{ij}\big |
$$
for all $z \in \cD(z_0, r_0)$.  In particular,
\begin{equation}
\label{eq:14.8}
\begin{split}
\max_{|z - z_0| \le 2r_2} \big |a_{ij} (z) \big | & \le {4Kr_2\over
r_0} = 4K \exp\bigl(-(\log\log K)^{2C}\bigr)\\
& < 8 \big |a_{i_0\, j_0} (z_0)\big |
\end{split}
\end{equation}
where the last inequality uses (\ref{eq:14.4}).  In view of  \eqref{eq:14.7} and
(\ref{eq:14.8}), one obtains
\begin{equation}
\nn
\begin{split}
\big \| M(z) - M(z_0)\big \|  &\le \max_{|w - z_0| \le r_2} \| M'(w)\|\,
|z - z_0|\\
& \lesssim {1\over r_2} |z - z_0| \max_{|w - z_0| \le 2r_2} \|M(w)\|
 \lesssim \|M(z_0)\| {|z - z_0|\over r_2}\ .
\end{split}
\end{equation}
Thus,
\begin{equation}
\nn
1 - C {|z - z_0| \over r_2} \le {\|M(z)\|\over \|M(z_0)\|} \le 1 + C
{|z - z_0| \over r_2}\ ,
\end{equation}
and the lemma follows.
\end{proof}

We assume now that $M(z)$ be $ 2\times 2$
unimodular, i.e.
\begin{equation}
\label{eq:14.11}
M(z) = \begin{bmatrix}
a_{11}(z) & a_{12}(z)\\
a_{21}(z) & a_{22}(z)
\end{bmatrix}\ ,
\end{equation}
$\det M(z) = 1$.  By polar decomposition,
\begin{equation}
\nn
M(z_0) = U_0 \begin{bmatrix}
\mu_1^{(0)} & 0\\
0 & \mu_2^{(0)}\end{bmatrix}
\end{equation}
where $U_0$ is a unitary matrix, and where $\mu_i^{(0)}$ are the singular values
of $M(z_0)$, i.e., the eigenvalues of $\left(M(z_0)^*
M(z_0)\right)^{1/2}$, $\mu_1^{(0)} = \|M(z_0)\|$, $\mu_2^{(0)} =
\bigl(\mu_1^{(0)}\bigr)^{-1}$.  Recall that for any unitary matrix $U$
and arbitrary matrix $A$ one has
\begin{equation}
\label{eq:14.13}
\|UA \| = \|A\|\ .
\end{equation}
We can draw the following  conclusions.

\begin{lemma}
\label{lem:14.2}
Consider the  matrix (\ref{eq:14.11}). Let $\widehat M(z) = \bigl(\mu_1^{(0)}\bigr)^{-1}
U_0^{-1} M(z)$.  Then $\widehat M(z)$ is analytic in $\cD(z_0, r_0)$,
$$
\| \widehat M(z) \| = \|M(z_0)\|^{-1} \|M(z)\|\ .
$$
\end{lemma}

\begin{lemma}
\label{lem:14.3}
 Assume that the conditions of Lemma~\ref{lem:14.1} are valid for the matrix~\eqref{eq:14.11}.
Then for $z\in \cD(z_0, r_2)$ one has
$$
\widehat M(z) = \begin{bmatrix}
1 & 0\\ 0 & \hat \mu^{(0)}\end{bmatrix} + (z - z_0) B_0 + \widehat R(z)
$$
where $\hat \mu^{(0)} = \bigl(\mu_1^{(0)}\bigr)^{-2}$,
\begin{equation}
\label{eq:14.14}
\begin{split}
\|B_0\| & \le r_1^{-1} \exp \left((\log\log K)^B\right)\\
\|\widehat R_0(z) \| & \le r_1^{-2} |z - z_0|^2 \exp\left((\log\log
K)^B\right)\ .
\end{split}
\end{equation}
\end{lemma}

\begin{proof} By Lemma~\ref{lem:14.1}
$$
\| \widehat M(z)\| = \|M(z)\|\, \|M(z_0)\|^{-1} \le 1 + r_1^{-1} \exp \bigl((\log\log K)^C\bigr) |z -
z_0| \le 2
$$
for any $z \in \cD\big(z_0, r_1\exp(-(\log\log K)^{2C})\big)$.
Therefore, (\ref{eq:14.14}) follows from the Cauchy estimates for analytic
functions.
\end{proof}

Now define
\begin{equation}
\nn
\check M(z) = \begin{bmatrix}
1 & 0\\ 0 & 0\end{bmatrix} + (z - z_0) B_0\ .
\end{equation}
Then (\ref{eq:14.13}) combined with (\ref{eq:14.14}) of Lemma~\ref{lem:14.3} implies

\begin{lemma}
\label{lem:14.4}
Under the conditions of Lemma~\ref{lem:14.3}
\begin{equation}
\nn
\log {\|\check M(z)\|\over \| \widehat M(z)\|} \lesssim r_1^{-2} |z -
z_0|^2
\exp\left(\bigl(\log\log K\bigr)^B\right) + K^{-2}\ .
\end{equation}
\end{lemma}

Now observe that
\begin{equation}
\nn
\begin{split}
\log \|\check M(z)\| & = {1\over 2} \log \| \check M(z)^*
\check M(z)\|\\
& = {1\over 2} \log \Big \| \begin{bmatrix}
1 & 0\\ 0 & 0\end{bmatrix} + (z - z_0) \begin{bmatrix}
1 & 0 \\ 0 & 0\end{bmatrix} B_0 + \overline{(z - z_0)} B_0
\begin{bmatrix}
1 & 0\\ 0 & 0\end{bmatrix} + |z - z_0|^2 B_0^* B_0\Big \|\
\end{split}
\end{equation}
Consequently, we have the following lemma.

\begin{lemma}
\label{lem:14.5}
\begin{equation}
\log \|\check M(z)\| = {1\over 2} \log \Big \| \begin{bmatrix}
1 + 2\Ree\bigl(a_0(z - z_0)\bigr) & b_0(z - z_0)\\
\bar b(z - z_0) & 0 \end{bmatrix} \Big \| + \check R(z)\ ,
\end{equation}
where $a_0, b_0 \in \IC$
\begin{align}
|a_0|, |b_0| & \le r_1^{-1} \exp \left((\log\log K)^{B_2} \right)\ ,\\
| \check R(z)| & \le r_1^{-2} \exp\left((\log\log K)^{B_2}\right) |z
- z_0|^2\ .
\end{align}
\end{lemma}

Note that
\begin{equation}
\nn
\begin{bmatrix}
1 + 2 \Ree\bigl(a_0 (z - z_0)\bigr) & b_0(z - z_0)\\[5pt]
\bar b_0(\overline{z - z_0}) & 0
\end{bmatrix}
\end{equation}
is self-adjoint and its eigenvalues $\tilde \mu_i(z)$, $i = 1, 2$ are
as follows
\begin{equation}
\label{eq:14.22}
\tilde \mu_1(z) = \big | 1 + a_0(z - z_0)\big |^2 + \tilde\nu_1, \quad
|\tilde \nu_1 | \lesssim \left(|a_0|^2 + |b_0|^2\right) |z - z_0|^2
\end{equation}
and
\begin{equation}
\label{eq:14.23}
|\tilde \mu_2(z)| \lesssim \left(|a_0|^2 + |b_0|^2\right) |z - z_0|^2\ .
\end{equation}
Summarizing Lemmas~\ref{lem:14.2}--\ref{lem:14.5} and estimates (\ref{eq:14.22}), (\ref{eq:14.23}) one obtains the
following proposition. The main feature here is the fact that the logarithm on the right-hand
side of~\eqref{eq:14.40} is harmonic.

\begin{prop}
\label{prop:14.6}
Let $M(z) = \bigl(a_{ij}(z)\bigr)$ be $2\times 2$
unimodular matrix-function $z \in \cD(z_0, r_0)$.  Assume that
the conditions of Lemma~\ref{lem:14.1} are valid.  Then for any $z \in \cD(z_0,
r_2)$, $r_2 = r_1 \exp \bigl(-(\log\log K)^A\bigr)$ one has
\begin{equation}
\label{eq:14.40}
\log {\|M(z)\|\over \|M(z_0)\|} = \log \big | 1 + a_0(z - z_0)\big | +
R(z)\ ,
\end{equation}
where
\begin{equation}
\nn
\begin{split}
|a_0| &\les r_1^{-1}\exp\big((\log\log K)^B\big) \\
|R(z)| &\le r_1^{-2} \exp \bigl((\log\log K)^B\bigr) |z - z_0|^2+K^{-2}\ .
\end{split}
\end{equation}
\end{prop}

We now consider the case when all entries $a_{ij}(z)$ have zeros in the disk
$\cD(z_0, r_0)$.  Let $M(z) = \bigl(a_{ij}(z)\bigr)_{1 \le i, j \le m}$ be an analytic matrix
valued function for all $z \in \cD(z_0, r_0)$.  The following lemmas
lead up to our main result, Proposition~\ref{prop:14.10}.

\begin{lemma}
\label{lem:14.7}
Assume condition (I) is valid for some
$H\asymp \bigl(\log\log K\bigr)^B$.  Then
\begin{equation}
\label{eq:14.26}
\left\{ z \in \cD\left(z_0, {r_0\over 4}\right) : a_{i,j}(z) =
0\right\} \le (\log\log K)^{B_1}
\end{equation}
for any entry $a_{i,j} (z)$.
\end{lemma}

\begin{proof} It follows from condition (I) that
$$
\nint\nolimits_{\cD(x_1+iy_1, {r_0\over 2})} \left[\log \big |a_{i,j} (\xi +
i\eta)\big | - \log \big |a_{i,j}(x_1 + iy_1)\big |\right] d\xi\, d\eta
\le H
$$
for any $z_1 = x_1 + iy_1 \in \cD(z_0, r_0) \setminus \cB$, where $\cB$
is the same as in condition (I).  Since
\[ \mes \cB \lesssim J \, r_0^2 \, \exp \left(-H/(\log\log K)^A\right) \ll r_0^2,\]
such $z_1 = (x_1+ iy) \in \cD(z_0, r_0/8)$ exists.  Then (\ref{eq:14.26}) follows from Jensen's
formula.
\end{proof}

\begin{lemma}
\label{lem:14.8}
Assume that condition (I) is valid.  Assume further that
$a_{ij}(z) = b_{ij}(z) \prod\limits^{k_{ij}}_{k=1} (z - \zeta_{i,j,k})$
for some $\zeta_{i, j, k} \in \cD(z_0,r_0/8)$.  Set $\widehat M(z) =
\left(b_{ij}(z)\right)_{1 \le i, j \le m}$.  Then condition (I) is
valid for $\widehat M(z)$, $z \in \cD(z_0, r_0/4)$, with
\[
\log K + k_{ij}\log (2r_0)^{-1}  \le
\log \widehat K\\
 = \sup \left\{\log \|\widehat M(z)\| : z \in \cD(z_0,
cr_0)\right\} \le \log K + \bigl(\log r_0^{-1}\bigr) \bigl(\log\log
K\bigr)^B\ .
\]
\end{lemma}

\begin{proof} By Lemma \ref{lem:14.7}, $k_{ij} \le (\log\log K)^B$.
Then
$$
\min_{i, j, k} \big | z - \zeta_{i, j, k}\big | > r_0/8 \ ,
$$
for any $|z - z_0| = r_1=r_0/4$.  Then
\begin{equation}
\label{eq:14.27}
\log \big |b_{ij}(z) \big | \le \log \big |a_{ij}(z)\big | + Ck_{ij}
\log r_0^{-1} \le \log K + C\bigl(\log r_0^{-1}\bigr)\bigl(\log\log
K\bigr)^B\
\end{equation}
for any $|z - z_0| = r_1$.  By the maximum principle, (\ref{eq:14.27}) is valid for any
$|z - z_0| \le r_1$.  On the other hand, $|b_{ij}(z) | >(2r_0)^{-k_{ij}} |a_{ij}(z)|$
for any $z \in \cD(z_0, r_0)$.
\end{proof}

In the following lemma we use the notation~$\cZ$ from \eqref{eq:Zdef}.

\begin{lemma}
\label{lem:14.9}
Assume that $M(z) = \bigl(a_{ij}(z)\bigr)_{1 \le j, j\le
m}$ satisfies condition (I) in $\cD(z_0, r_0)$.
Assume that there
exists $\zeta_0 \in \cD(z_0, r_0/2)$ such that the following conditions
are valid:
\begin{enumerate}
\item[{\rm{(a)}}] each entry $a_{ij}(z)$ has at least one zero in
$\cD(\zeta_0, \rho_0)$,
\[ r_0\exp(-\sqrt{\log K})\le \rho_0 \le r_0\exp\left(-(\log\log
K)^{B_0}\right),\quad B_0 \gg 1
\]

\item[{\rm{(b)}}] no entry $a_{ij}(z)$ has zeros in $\cD(\zeta_0,
\rho_1) \setminus \cD(\zeta_0, \rho_0)$,
\[ \rho_0 \exp\left((\log\log
K)^{B_1}\right)\le \rho_1 \ll r_0,\]
$B_0 \gg B_1 \gg 1$.  Let $k_0 =
\min\limits_{ij} \#\cZ (a_{ij},\zeta_0, \rho_0)$.  Then for any
\[ z \in \cD(\zeta_0, \rho_1')
\setminus \cD(\zeta_0, \rho_2),\; \rho_1'=\exp\left(-(\log\log
K)^{B_2}\right)\rho_1,\; \rho_2 = \exp\left((\log\log
K)^{B_2}\right)\rho_0,
\]
$B_1 \gg B_2 \gg 1$, one has
\end{enumerate}
\begin{equation}
\Big | \log {\|M(\zeta)\|\over \|M(z)\|} - k_0 \log {|\zeta -
\zeta_0|\over |z - \zeta_0|} \Big | <
\exp\bigl(-(\log\log K)^{C_1}\bigr)\ .
\end{equation}
\end{lemma}

\begin{proof} We can write
\begin{equation}\nn
a_{ij}(z) = b_{ij}(z) P_{ij}(z)\ ,
\end{equation}
where $b_{ij}(z)$ is analytic and does not vanish in
$\cD(\zeta_0, \rho_1)$, $P_{ij}(z) = \prod\limits^{k_{ij}}_{k=1}\bigl(z -
\zeta_{i, j, k}\bigr)$, $\zeta_{i,j,k} \in \cD(\zeta_0, \rho_0)$.
By Lemma~\ref{lem:14.7}, $k_{ij} \le (\log\log K)^B$.
Since $k_{ij} \ge k_0$, one can split $P_{ij}$ as follows:
\[ P_{ij}(z) =
\widetilde P_{ij}(z)Q_{ij}(z), \qquad \degg \widetilde P_{ij} = k_0,\;
\degg Q_{ij} \ge 0,\; i, j = 1, 2, \dots, m.
\]
Set
\[\widehat P_{ij}(z)
= (z - \zeta_0)^{k_0} Q_{ij} (z),\quad \hat a_{ij}(z) = b_{ij}(z) \widehat
P_{ij}(z),\; i, j = 1, 2, \dots, m,\quad \widehat M(z) = \bigl(\hat
a_{ij}(z)\bigr)_{1 \le i, j \le m}.
\]  Then for any $z \in \cD(\zeta_0,\rho_1)$ one has
\begin{equation}
\nn
\begin{split}
 \big \| M(z) - \widehat M(z) \big \| &\les m\biggl(\max_{i, j} \big |
b_{ij}(z) \big |\, \big | Q_{ij}(z) \big |\biggr)\,k_0\rho_0 \cdot \big | z - \zeta_0\big |^{k_0 -1}\\
& \les m k_0 \rho_0 \big |z - \zeta_0\big |^{-1} \big\|\widehat M(z)
\big\|\ .
\end{split}
\end{equation}
Hence, for all $z \in \cD(\zeta_0,\rho_1) \setminus \cD(\zeta_0, \rho_2)$,
\begin{equation}
\label{eq:14.31}
\log {\|M(z)\|\over \|\widehat M(z)\|} \les m  k_0 \rho_0 |z -
\zeta_0|^{-1} <  \exp\bigl(-(\log\log K)^{C_1}\bigr)
\end{equation}
Note
\begin{equation}
\label{eq:14.30}
\log \big \| \widehat M(z) \big \| = k_0 \log |z - \zeta_0| + \log \big
\| \widehat B(z)\big \|\ ,
\end{equation}
where $\widehat B(z) = \bigl(\hat b_{ij}(z)
\bigr)$, $\hat b_{ij}(z) = b_{ij}(z)Q_{ij}(z)$.
Since $M(z) = \bigl(a_{ij}(z)\bigr)$ satisfies
conditions (I), $\widehat B(z) = \bigl(\hat b_{ij}(z)\bigr)$ also
satisfies this condition in $\cD(z_0, r_0)$ with
\[ \log \widehat K = \sup
\left\{\log \big \| \widehat B(z) \big \|\ z \in \cD(z_0, cr_0)\right\}
\asymp \log K.\]
This follows from Lemma~\ref{lem:14.8} and the condition $\log r_0^{-1}<(\log K)^{\frac12}$. There is some entry
$(i_0, j_0)$ with $Q_{i_0\, j_0}(z) = 1$, i.e., $\hat b_{i_0\, j_0}(z) =
b_{i_0\, j_0}(z)$.
Therefore, the lemma
follows from Lemma~\ref{lem:14.1}, (\ref{eq:14.30}) and (\ref{eq:14.31}).
\end{proof}

Let $M_N(z, \omega, E)$ be the usual monodromy matrix. Then
for any $z_0 \in \cA_{\rho_0/2}$, $M_N(\cdot, \omega, E)$
satisfies condition (I) in $\cD(z_0, r_0)$ with $r_0 \asymp
\exp\bigl(-(\log N)^A\bigr)$,
\begin{equation}
\label{eq:14.50}
\log K = \sup\left\{\log \big \| M_N(z, \omega, E)\big \|: z \in \cD(z_0,
r_0)\right\} = NL(\omega, E) + O((\log N)^A)
\end{equation}
provided $\omega \in \tor_{c,a}$.  Therefore, Proposition~\ref{prop:14.6} and Lemma~\ref{lem:14.9}
apply to $M_N(\cdot, \omega, E)$.  Let us summarize
our conclusions in the following proposition.

\begin{prop}
\label{prop:14.10}
\begin{enumerate}
\item[{\rm{(i)}}] Suppose that one of the Dirichlet determinants
\[ f_{[1,
N]}(\cdot, \omega, E),\; f_{[1, N-1]}(\cdot, \omega, E),\; f_{[2,
N]}(\cdot, \omega, E),\; f_{[2, N-1]}(\cdot, \omega, E)
\]
has no zeros in
$\cD(z_0, r_1)$, $\exp(-\sqrt{N})\le r_1 \le \exp\bigl(-(\log N)^C\bigr)$.  Then
\end{enumerate}
\begin{equation}
\nn
\Big | \log {\big \|M_N(z, \omega, E)\big \|\over \big \| M_N(z_0,
\omega, E)\big \|} - \log \big |1 + a_0(z - z_0)\big | \Big | \le |z -
z_0|^2 r_2^{-2}
\end{equation}
\begin{enumerate}
\item[]
for any $z \in \cD(z_0, r_2)$,  $r_2 = r_1\exp\bigl(-(\log
N)^{2C}\bigr)$, and with $|a_0| \lesssim r_2^{-1}$.

\item[{\rm{(ii)}}] Assume that the following conditions are valid
\begin{enumerate}
\item[{\rm{(a)}}] each of the determinants $f_{[a, N-b]}(\cdot, \omega,
E)$, $a = 1, 2$; $b = 0, 1$ has at least one zero in $\cD(\zeta_0,
\rho_0)$, where $e^{-\sqrt{N}}\le \rho_0 \le \exp\bigl(-(\log N)^{B_0}\bigr)$

\item[{\rm{(b)}}] no determinant $f_{[a, N-b]} (\cdot, \omega, E)$ has a
zero in $\cD(\zeta_0, \rho_1)\setminus \cD(\zeta_0, \rho_0)$, $\rho_1 \ge
\exp\bigl((\log N)^{B_1}\bigr)\rho_0$, $B_0 \gg B_1 +A$.
\end{enumerate}
Let $k_0 = \min\limits_{a, b} \cZ(f_{[a, N-b]} (\cdot, \omega, E),\zeta_0,\rho_0)$.
Then for any
\[ z, \zeta \in \cD(\zeta_0, \rho_1')\setminus \cD(\zeta_0,
\rho_2),\;\rho_1'=\exp\bigl(-(\log N)^{B_2}\bigr)\rho_1,\;\rho_2 = \exp\bigl((\log N)^{B_2}\bigr)\rho_0,\quad B_1 \gg
B_2\gg
1\]
 one has
$$
\Big | \log {\|M(\zeta)\|\over \|M(z)\|} - k_0 \log {|\zeta -
\zeta_0|\over |z - \zeta_0|} \Big | \le \exp\bigl(-(\log N)^{C}\bigr)
$$
\end{enumerate}
\end{prop}

\section{Jensen's averages of norms of monodromies}
\label{sec:Jav}

Let us start with some corollaries to the results of the
previous section.  Let $M(z) = \bigl(a_{ij}(z)\bigr)_{1 \le i, j \le
m}$ be an analytic matrix function defined in disk $\cD(z_0, r_0)$.
Assume that condition (I) of Section~\ref{sec:harnack} is valid.

\begin{lemma}
\label{lem:15.1}
Impose the conditions of Lemma~\ref{lem:14.1}.  Then
\begin{equation}
\label{eq:15.1}
4 {\rho_1^2\over \rho_2^2} J\left(\log \|M(z)\|, z_0, \rho_1,
\rho_2\right) < \rho_1^2 \rho_2^{-1} r_0^{-1} \exp\bigl((\log\log
K)^{C_1}\bigr)
\end{equation}
for any $0 < \rho_2 < \rho_1 < r_0\exp \bigl(-(\log\log
K)^{C_2}\bigr)$. Here $C_1, C_2 \gg 1$ and $r_2$ is as in Lemma~\ref{lem:14.1}.
In particular, estimate (\ref{eq:15.1}) is valid for $\log\|M_N(z, \omega, E)\|$ and $\log| f_N(z,\omega,E)|$,
$\omega \in \tor_{c,a}$ with $\log K = NL(\omega, E)+(\log N)^C$ and $r_0=\exp(-(\log N)^A)$.
\end{lemma}

\begin{proof} Recall that
\begin{equation}
\label{eq:15.2}
J\bigl(u, z_0, \rho_1, \rho_2\bigr) = \nint\nolimits_{\cD(z_0,
\rho_1)}dx\,dy\ \nint\nolimits_{\cD(x + iy, \rho_2)} \left[u(\xi + i\eta) - u(x
+ iy)\right] d\xi\, d\eta\ .
\end{equation}
By Lemma 14.1, $u(z) = \log \|M(z)\|$ satisfies
\begin{equation}
\label{eq:15.3}
\big | u(\xi + i\eta) - u(x + iy)\big | < \big | (\xi + i\eta) - (x +
iy) \big | r_0^{-1} \exp\bigl((\log\log K)^{C_1}\bigr)
\end{equation}
for any $(x +iy) \in \cD(z_0, \rho_1)$, $(\xi + i\eta) \in \cD(z_0,
\rho_1 + \rho_2)$.  Evaluating the averages in (\ref{eq:15.2}) with use of
(\ref{eq:15.3}) one obtains (\ref{eq:15.1}).
\end{proof}

Note that condition (I) implies the following assertion

\begin{lemma}
\label{lem:15.2}
Assume that $M(z)$ satisfies condition (I).  Then
\begin{equation}
\nn
\big \| \log \| M(\cdot)\| \big \|_{L^2\bigl(\cD(z_0, r_0/2)\bigr)}
\lesssim r_0 \log K
\end{equation}
\end{lemma}

\begin{lemma}
\label{lem:15.3}
 Assume that $M(z)$ satisfies condition (I).  Then
\begin{equation}
\label{eq:15.5}
 J \left(\log \|M(\cdot)\|,\ z_0, \rho_1,
\rho_2 \right) \le \bigl(\log\log K\bigr)^{C_4}
\end{equation}
for any $r_0\,\exp\bigl(-(\log\log K)^{C_3}\bigr) < \rho_2 < \rho_1 <
r_0\,\exp\bigl(-(\log\log K)^{C_2}\bigr)$, and provided $\rho_2^{-1}\rho_1< (\log\log K)^{C_3}$. Here $1 \ll C_2 \ll C_3 \ll C_4$.
\end{lemma}

\begin{proof} Take $H = \bigl(\log\log K)^{C_4/2}$ in (I).  Let $\cB$
be the set provided by condition (I).  For any $z \in \cD(z_0,
r_0)\setminus \cB$ and any $\zeta \in \cD(z_0, r_0)$ one has
\begin{equation}
\nn
\log \|M(\zeta)\| \le \log K \le \log \| M(z) \| + H\ .
\end{equation}
Thus,
\begin{equation}
\nn
\nint\nolimits_{\cD(x+iy, \rho_2)} \left[\log \|M(\xi + i\eta)\| - \log
\|M(x+iy)\| d\xi\, d\eta\right] \le H
\end{equation}
for any $(x + iy) \in \cD(z_0, {r_0\over 2}\bigr) \setminus \cB$.

Recall that
$\mes\cB < r_0^2\exp \bigl(-H(\log\log K)^{-A}\bigr) < r_0^2\exp
\bigl(-(\log\log K)^{C_4\over 4}\bigr)$.
Hence, due to Lemma~\ref{lem:15.2},
\begin{equation}
\nn
\begin{split}
 &
\bigl(\pi\rho_1^2\bigr)^{-1} \cdot \int_{\cD(z_0, \rho_1)\cap \cB}
dx\, dy \left[\nint_{\cD(x+iy, \rho_2)} \left |\Bigl[\log \|M(\xi + i\eta)\| -
\log \|M(x + iy)\|\Bigr]\right| d\xi\, d\eta\right]\\
&\les \bigl(\pi\rho^2_2\bigr)^{-1} \bigl(r_0 \log K (\mes \cB)^{1/2}\bigr)
\lesssim \exp\bigl((\log\log K)^{2C_3} + (\log\log K) - {1\over 2}(\log\log
K)^{C_4\over 4}\bigr)
\le 1
\end{split}
\end{equation}
Therefore, the left-hand side  of (\ref{eq:15.5}) is $\le H+1$.
\end{proof}

\begin{prop}
\label{prop:15.4}
\begin{enumerate}
\item[{\rm{(i)}}] Assume that one of the Dirichlet determinants $f_{[a,
N-b]}(\cdot, \omega, E)$, $a = 1, 2$, $b = 0, 1$ has no zeros in
$\cD(z_0, r_1)$, $\exp(-\sqrt{N}) \le r_1 \le \exp\bigl(-(\log N)^{C_1}\bigr)$.  Then
\begin{equation}
\label{eq:15.9}
4{\rho_1^2\over \rho_2^2} J \left(\log \|M_N(\cdot, \omega, E)\|, z_0,
\rho_1, \rho_2\right) \le \rho_1^2 r_1^{-2} \exp\bigl((\log N)^B\bigr)
\end{equation}
for any $r_1\exp(-\sqrt{N}) \le \rho_1 \le r_1\exp\bigl(-(\log N)^A\bigr)$, $\rho_2 =c \rho_1$
\item[{\rm{(ii)}}] Assume that for some $\zeta_0$ the following
conditions are valid
\begin{enumerate}
\item[{\rm{(a)}}] each of the determinants $f_{[a, N-b]} (\cdot,
\omega, E)$, $a = 1, 2;$ $b = 0, 1$ has at least one zero in
$\cD(\zeta_0, \rho_0)$, $\exp(-\sqrt{N})<\rho_0 \le \exp\bigl(-(\log
N)^{B_0}\bigr)$.

\item[{\rm{(b)}}] no determinant $f_{[a, N-b]} (\cdot, \omega, E)$ has
a zero in $\cD(\zeta_0, \rho_1) \setminus \cD(\zeta_0, \rho_0)$, $\rho_1
\ge \exp\bigl((\log N)^{B_1}\bigr)\rho_0$, $B_0 > B_1$.
\end{enumerate}
\end{enumerate}
Let $k_0 = \min\limits_{a, b} \# \cZ(f_{[a, N-b]}(\cdot, \omega, E),\zeta_0,\rho_0)$.  Then for any
\[z_1 \in \cD(\zeta_0, \rho_1') \setminus \cD(\zeta_0, \rho_2),\;\rho_1'=\exp\bigl(-(\log N)^{B_2}\bigr)\rho_1,\;\rho_2
\asymp \exp\bigl((\log N)^{B_2}\bigr) \rho_0,\]
$B_1>B_2$, one has
$$
\Big | 4{r^2_1\over r_2^2} J\left(\log \|M_N(\cdot, \omega, E)\|, z_1,
r_1, r_2\right) - k_0 \Big | \le \exp \bigl(-(\log N)^C\bigr)
$$
where $|z_1-\zeta_0|(1+2c)<r_1 < \rho_1'$, $r_2 = c r_1$, and $0<c\ll 1$ is some constant.
\end{prop}

\begin{proof}
Due to part (i) of Proposition~\ref{prop:14.10} one has
\begin{equation}
\label{eq:15.10}
\Big | \log {\|M_N(\zeta, \omega, E)\|\over \|M_N(z, \omega, E)\|} -
\log |1 + a_0(\zeta - z)| \Big |
\lesssim |z - \zeta|^2\,r_1^{-2}\ ,
\end{equation}
for any $z, \zeta \in \cD(z_0, r_2)$, $r_2 \asymp \exp \bigl(-(\log N)^C\bigr)r_1$.
Evaluating the averages on the right-hand side
of (\ref{eq:15.10}) one obtains (\ref{eq:15.9}).

To prove (ii), recall that by Proposition~\ref{prop:14.10}
the functions $u(z) = \log \|M(z)\|$, $v(z) = \log \big
|(z - \zeta_0)^k\big |$ satisfy
\begin{equation}
\nn
\Big | \left[u(\zeta) - u(z) \right] - \left[v(\zeta) - v(z)\right]\Big
| \le \exp\bigl(-(\log N)^{C_1}\bigr)
\end{equation}
for any $z, \zeta \in \cD(\zeta_0, \rho_1) \setminus \cD(\zeta_0,
\rho_2)$, $\rho_2 = \exp\bigl((\log N)^{B_2}\bigr) \rho_0$.  Hence,
\begin{equation}
\label{eq:15.12}
\begin{split}
 & \bigg | \bigl(\pi r_1^2\bigr)^{-1} \bigl(\pi r_2^2\bigr)^{-1}
\int_{\cD(z_1, r_1) \setminus \cD(\zeta_0, \rho_2)} dx\, dy\  \int_{\cD(x+iy,
r_2)\setminus \cD(\zeta_0, \rho_2)} d\xi\, d\eta\\
& \Bigl\{\bigl[u(\xi + i\eta) - u(x + iy)\bigr] - \bigl[v(\xi + i\eta)
- v(x + iy)\bigr]\Bigr\} \bigg |
 \le \exp\bigl(-(\log N)^{C_1} \bigr)\ .
\end{split}
\end{equation}
Note that
\begin{equation}
\nn
\big |u(\xi + i\eta) - u(x + iy)\big |, \big |v(\xi + i\eta) - v(x +
iy)\big | \lesssim H
\end{equation}
for any  $(x + iy), (\xi + i\eta) \in \cD(z_0, r_0/2) \setminus \cB'_H$,
$\mes\cB'_H < \exp \bigl(-H/(\log N)^C\bigr)$, for any $H >
\bigl(\log{\rho_2}^{-1}\bigr) \bigl(\log N\bigr)^B$.  Hence,
\begin{equation}
\label{eq:15.14}
\int_{\cD(z_1, r_1)} \int_{\cD(\zeta_0, \rho_2)} \bigl(|u(x+ iy) - u(\xi
+ i\eta)| + |v(x+iy) - v(\xi + i\eta)|\bigr) dx\, dy\, d\xi\, d\eta
\lesssim r_1^2 \cdot \rho^2_2 \bigl(\log \rho^{-1}_2\bigr)\bigl(\log
N\bigr)^{B_1}\ .
\end{equation}
Combining (\ref{eq:15.12}), (\ref{eq:15.14}) implies
\begin{equation}
\nn
 \bigg | 4 {r_1^2\over r_2^2} J\bigl(u(\cdot) - v(\cdot), z_1, r_1,
r_2\bigr) \bigg |
\lesssim 4 {r_1^2\over r_2^2}\cdot \exp\bigl(-(\log
N)^{C'}\bigr)
 \le \exp \bigl(-(\log N)^{C_0}\bigr)\ .
\end{equation}
By Corollary~\ref{cor:5.2} one has
$$
4 {r_1^2\over r_2^2} J\bigl(v(\cdot), z_1, r_1, r_2 \bigr) = k_0
$$
and we are done.
\end{proof}

We turn now to the evaluation of the Jensen averages of the norms of
monodromies with the use of the avalanche principle expansion.  The first
issue we examine here is the ``positivity of contributions'' of
the terms in this expansion.  Recall that
the terms under consideration are of the following form
\begin{equation}
\nn
\log \big \| M_{[a, b]} (z, \omega, E) \big \| - \log \big \|
M_{[a, c]} (z, \omega, E)\big \|
\end{equation}
$a < c < b$, or
\begin{equation}
\nn
\log \big\|M_{[a,b]} (z, \omega, E)\big \| - \log \big \| M_{[c,
b]} (z, \omega, E\big \|
\end{equation}
It is important to show that the Jensen averages of such terms are
almost non-negative.  Instead of that we show that one can always add
up such terms in the avalanche principle expansion in such a way that
the Jensen averages of these sums are almost non-negative.

\begin{lemma}
\label{lem:15.5}
 Consider
\beeq
\label{eq:vsum}
v(z) = \sum^{m_1}_{m=1} \left\{\log \big \|A_{m+1}(z) A_m(z)\big \| -
\log \big \|A_m(z)\big\|\right\}
\eneq
where $A_m(z) = M_{\ell_m} \bigl(ze(s_m\omega), \omega, E)\bigr)$, $s_m
= \sum\limits_{j<m} \ell_j$, $\ell_j \asymp \ell_1$, $1 \ll m_1 \le
\exp\bigl((\log \ell_1)^A\bigr)$.  Take some $z_0 \in \cA_{\rho/2}$.
Assume that
for each $m \in [1, m_0]\cup [m_1 - m_0, m_1]$
the monodromies $A_m(z)$ and $A_{m+1}(z) A_m(z)$ satisfy the conditions of part
(i) of Proposition~\ref{prop:15.4} in some disk $\cD(z_0, r^{(1)})$ with
\[\exp\big(-\sqrt{\ell_1}\big)< r^{(1)} <
 \exp\bigl(-(\log\ell_1)^C\bigr).\]
Then
\begin{equation}
\label{eq:15.18}
4{r_1^2\over r_2^2} \bigg | J\bigl(v(\cdot), z_0, r_1, r_2 \bigr)
- J\Bigl(\log \Big \| \prod^{m_0 +1}_{m = m_1 - m_0} A_m (\cdot)\Big
  \|, z_0, r_1, r_2\Bigr) \bigg | \le \exp \bigl((\log \ell_1)^C\bigr) r_1^2 (r^{(1)})^{-2}
\end{equation}
for any $r^{(1)}\exp(-\sqrt{\ell_1}) < r_1 <r^{(1)}\exp \bigl(- (\log \ell_1)^{C_1}\bigr)$, $r_2 \ll r_1$.  In particular,
\begin{equation}
\label{eq:15.19}
  4 {r_1^2\over r_2^2} J \bigl(v(\cdot), z_0, r_1, r_2\bigr) \ge - \exp
  \bigl((\log \ell_1)^C\bigr)r_1^2 (r^{(1)})^{-2}\ .
\end{equation}
\end{lemma}
\begin{proof} By the avalanche principle expansion
\begin{equation}
\label{eq:15.20}
  \begin{split}
  \log \Big \| \prod^{m_0 +1}_{m=m_1 - m_0} A_m(\cdot) \Big \|  & =
  \sum^{m_1 - m_0 -1}_{m=m_0 +1} \log \big \| A_{m+1}(z) A_m(z) \big \|\\
&\quad - \sum^{m_1 - m_0 -1}_{m = m_0 +2} \log \big \| A_m(z)\big \| +
O\left(\exp(-\sqrt{\ell_1})\right)
\end{split}
\end{equation}
for any $z \notin \cB$, with $\mes \cB \le
\exp\bigl(-\sqrt{\ell_1}\bigr)$.  It follows from (\ref{eq:15.20}) that
\begin{equation*}
\begin{split}
  J\left(\log \Big \| \prod^{m_0 +1}_{m = m_1 - m_0 +1} A_m(\cdot)\Big\|,
z_0, r_1, r_2\right)
& = \sum^{m_1 - m_0 -1}_{m = m_0 + 1} J\left(\log \big \|
A_{m+1}(\cdot) A_m(\cdot)\big \|, z_0, r_1, r_2\right) \\
&\quad - \sum^{m_1 - m_0 - 1}_{m = m_0 + 2} J\left(\log \big \|
A_m(\cdot) \big \|, z_0, r_1, r_2\right) +
O\left(\exp(-\sqrt{\ell})\right)\ .
\end{split}
\end{equation*}
Due to (i) in Proposition~\ref{prop:15.4} one has
\begin{equation}
\nn
\begin{split}
& 4 {r_1^2\over r_2^2} \biggl\{\sum_{m \in [1, m_0] \cup [m_1 - m_0,
m_1 -1]} J\Bigl(\log \big \| A_{m+1}(\cdot) A_m(\cdot) \big \|, z_0,
r_1, r_2\Bigr) +\\
&\qquad + \sum_{m \in [2, m_0]\cup [m_1 - m_0, m_1 -1]} J\Bigl(\log \big \|
A_m(\cdot) \big \|, z_0, r_1, r_2\Bigr) \biggr\} \le
r_1^2\bigl(r^{(1)}\bigr)^{-2} \exp\bigl((\log \ell)^C\bigr)\ .
\end{split}
\end{equation}
Note that we do not need absolute values here, since the Jensen averages
of subharmonic functions are non-negative.
That proves (\ref{eq:15.18}).  Since $\log \Big\| \prod\limits^{m_0 +1}_{m = m_1
- m_0} A_m(z) \Big \|$ is subharmonic its Jensen's averages are
  non-negative and \eqref{eq:15.19} follows.
\end{proof}

%

\begin{remark}
\label{rem:15.6}
The same statement and proof applies to slightly modified functions $v$
in \eqref{eq:vsum}. Indeed, in the definition
\[
v(z) = \sum^{m_1}_{m=1} \log \big \|A_{m+1}(z) A_m(z)\big \| - \sum_{m=1}^{m_1}
\log \big \|A_m(z)\big\|
\]
we can omit a finite number of terms in both sums from the edges $m=1$ and $m=m_1$, respectively.
\end{remark}

We now introduce the notion of ``adjusted''.

\begin{defi}
\label{def:adj}
 Let $\ell\gg 1$ be some integer, and $s\in\ZZ$. We say that $s$ is {\em adjusted }
to a disk $\cD(z_0,r_0)$ at scale $\ell$ if for all $k\asymp\ell$
\[ \cZ(f_{k}(\cdot e((s+m)\omega),\omega,E),z_0,r_0)=\emptyset \qquad \forall\;|m|\le C\ell. \]
\end{defi}

\noindent  We will now prepare the way for our main assertion concerning the Jensen
  averages of norms of monodromies. This will be done by means of their avalanche
  principle expansions.  Consider the avalanche
  principle expansion of $\log \big | f_N(z, \omega, E)\bigr |$:
  \begin{equation}
\label{eq:15.25}
  \log \big | f_N(z, \omega, E+i\eta)\big | = \sum^{n-1}_{m=1} \log
  \big \|A_{m+1}(z) A_m(z) \big \| - \sum^{n-1}_{m=2} \log \big \| A_m(z)\big \| +
  O\left(\exp\bigl(-\ell^{1/2}\bigr)\right)\ ,
  \end{equation}
  for any $z \in \cA_{\rho_0/2}\setminus \cB_{E,\eta,\omega}$, $\mes
  \cB_{E,\eta, \omega} \le \exp\bigl(-\ell^{1/2}\bigr)$, where $A_m(z)
  = M_\ell\bigl(ze(s_m\omega), \omega, E+i\eta\bigr)$, $m = 2,\dots,
  n -1$, $A_1(z) = M_{\ell_1}(z, \omega, E) \begin{bmatrix} 1 & 0\\ 0
  & 0\end{bmatrix}$, $A_n(z) = \begin{bmatrix} 1 & 0\\ 0 &
  0\end{bmatrix}M_{\ell_n}\bigl(ze(s_n\omega), \omega, E\bigr)$,
  $\ell_m = \ell$, $m = 1, 2,\dots, n-1$, $\ell_n = \tilde\ell$,
  $(n-1)\ell + \tilde\ell = N$, $\ell, \tilde \ell \asymp (\log N)^A$,
  $s_m = \sum\limits_{j< m} \ell_j$.

\begin{lemma}
\label{lem:15.8}
Assume that
$\{s_{m_j}\}_{j=1}^{j_0}$ is adjusted to
$\cD(z_0, r_0)$ at scale $\ell$. Set $m_0=0$, $m_{j_0+1}=n$, and
\[
w_j(z)   = \log \Big \| \prod^{m_j+1}_{m = m_{j+1}} A_m(z)\Big\| \text{\ \ for any\ \ }0\le j\le j_0
\]
Then
\begin{equation}
 4{r_1^2\over r_2^2} \Big | J\left(\log \big |f_N(\cdot, \omega, E)\big
|, z_0, r_1, r_2\right)  - \sum_{j=0}^{j_0} J(w_j(\cdot), z_0, r_1, r_2)
\Big |
\le N\exp\bigl((\log \ell)^C\bigr)\, r_1^2 r_0^{-2}
\label{eq:15.29}
\end{equation}
for any $e^{-\sqrt{\ell}}<r_1\les\exp(-(\log \ell)^A)r_0$, and $r_2=cr_1$.
In particular,
\begin{equation}
 4 {r_1^2\over r_2^2}\ J\left(\log \big | f_N(\cdot, \omega, E)\big |,
z_0, r_1, r_2\right)  \ge \sum_{j\in\cJ}\ J\bigl(w_j(\cdot), z_0, r_1,
r_2\bigr) -N\exp\bigl((\log
\ell)^C\bigr)\,r_1^2 r_0^{-2}
\end{equation}
for any $\cJ\subset [0,j_0]$.
\end{lemma}

\begin{proof}
This follows immediately from Lemma~\ref{lem:15.5} and Remark~\ref{rem:15.6}.
\end{proof}

\section{Proof of Theorem~\ref{thm:4}}
\label{sec:Th3}

The proof of Theorem~\ref{thm:4} is based on Section~\ref{sec:Jav} on Jensen
averages of norms of monodromies.  To make use of Proposition~\ref{prop:15.4} we
consider again the avalanche principle expansion (\ref{eq:15.25}). We first address
the issue of defining sequences $\{s_{m_j}\}_{j=1}^{j_0}$ which are adjusted
to a given disk $\cD(z_0, r_1)$, $r_1\asymp\exp(-(\log \ell)^A)$, $0<\vep \ll 1$, where $\ell \asymp
(\log N)^C$, see Definition~\ref{def:adj}.

\begin{lemma}
\label{lem:181} Given $\ell$ and $r_1 \asymp \exp(-(\log \ell)^C)$,
 $\omega \in
\tor_{c,a} $,  $x_0 \in \tor$, $E \in \IR$, and
 $s_0\in\IZ$
there exists (with $B\gg1$)
\beeq
\label{eq:m1m0}
s_0'\in [s_0-B {\ell}^2 ,s_0+B {\ell}^2]
\eneq
such that with $z_0 = e(x_0)$,
\beeq\label{eq:goodell}
f_\ell \bigl(\cdot e(s\omega), \omega, E\bigr)\ \mbox{has no zero in
$\cD(z_0, r_1)$}
\eneq
for any $\big |s - s_0'\big | \les\ell$.
\end{lemma}
\begin{proof} Recall that the total number of zeros of $f_\ell \bigl(\cdot, \omega, E\bigr)$ does not exceed $C \ell$.
Since the zeros of $f_\ell \bigl(\cdot e(s\omega), \omega, E\bigr)$ are the shifts of the zeros of
$f_\ell \bigl(\cdot, \omega, E\bigr)$ by $e(-s\omega)$ , the assertion follows from the Diophantine condition on $\omega$.
\end{proof}

This lemma gives  us a lot of room to define sequences $\{s_{m_j}\}_{j=1}^{j_0}$ which are adjusted
to a given disk.

\begin{corollary}
\label{cor:18.2}
Consider the avalanche principle expansion (\ref{eq:15.25}).  Assume that
$\omega \in \tor_{c,a}$.
Given a disk $\cD(z_0, r_1)$, $r_1 \asymp
\exp(-(\log \ell)^A)$ and an increasing sequence $\{\tilde m_j\}_{j=1}^{j_0}$
such that
$\tilde m_{j+1}-\tilde m_j>\exp \left((\log \ell)^{2B}\right)$ for $1\le j<j_0$,
there exists an increasing sequence $\{s_{m_j}\}_{j=1}^{j_0}$
which is adjusted to $\cD(z_0, r_1)$ at scale $\ell$ and such that
\begin{equation}
\label{eq:184} \big | m_j - \tilde m_j\big | < \exp
\left((\log\ell)^{B}\right),\quad 1\le j\le j_0.
\end{equation}
\end{corollary}
\begin{proof}
Apply Lemma~\ref{lem:181} to each $s_{\tilde m_j}$, $1\le j<j_0$,
and let $m_j$ be such that~\eqref{eq:goodell} holds for each
$|s-s_{m_j}|\les \ell$. This implies that at least one of the
entries of the monodromies $A_m$ where $|m-m_j|\le C$ has no zeros
in $\cD(z_0, r_1)$. But this is precisely the requirement of
Definition~\ref{def:adj} for the sequence $\{s_{m_j}\}$ to be
adjusted. Finally, \eqref{eq:184} follows from~\eqref{eq:m1m0}.
\end{proof}

We can now draw the following conclusion from Corollary~\ref{cor:18.2}.

\begin{lemma}
\label{lem:184} Assume $\omega \in \tor_{c,a}$.  Given a disk
$\cD(z_0, r_1)$, $r_1 \asymp \exp(-(\log \ell)^A)$ there exists a
sequence $\{s_{m_j}\}_{j=1}^{j_0}$ with $0\le j_0< n$ so that
\begin{enumerate}
\item[\rm{(a)}] it is adjusted to $\cD(z_0, r_1)$ at scale $\ell$

\item[\rm{(b)}] $m_{j+1} - m_j \le \exp \left((\log
\ell)^{2B}\right)$ for $0\le j\le j_0$ with $m_0=0,\;m_{j_0+1}=n$
\end{enumerate}
\end{lemma}
\begin{proof}
Take a net of points in $[1,n]$ of step-size $\exp((\log N)^{2B})$
and denote it by $\{\tilde m_j\}_j$. Then apply the corollary to
this sequence. Property (b) follows from~\eqref{eq:184}.
\end{proof}

Set
\[ \widehat A_j(z) = \prod^{m_j +1}_{m = m_{j+1}} A_m(z), \quad
w_j(z) = \log \big \|\widehat A_j(z) \big \|
\]
for $0\le j\le j_0$.

\begin{lemma}
\label{lem:18.5}
Let $0<c\ll1$. Then
\begin{equation}
\label{eq:18.11}
\begin{split}
& \# \Bigl\{z \in \cD\left(z_0, \rho_1 (1 - c)\right): f_N(z, \omega, E) = 0\Bigr\}\\
&\qquad \le 4 {\rho_1^2\over
\rho_2^2} \sum\limits^{j_0}_{j=0}\ J\bigl(w_j(\cdot), z_0, \rho_1,\rho_2\bigr) + N\rho_1^{5/4} \le \\
& \#\Bigl\{z \in \cD\left(z_0, \rho_1 (1 + c)\right): f_N(z, \omega, E) = 0\Bigr\} + 2N\rho_1^{5/4}\ .
\end{split}
\end{equation}
provided  $e^{-\sqrt{\ell}}<\rho_1\les\exp(-(\log \ell)^A)r_1$, $\rho_2=c\rho_1$, and $\rho_1 \le r_1^4$.
\end{lemma}
\begin{proof}
By Lemma~\ref{lem:15.8},
\begin{equation}
\label{eq:18.10}
 4 {\rho_1^2\over \rho_2^2} \Big | J\left(\log \big |f_N(\cdot, \omega,
E)\big |, z_0, \rho_1, \rho_2\right) - \sum_{j=0}^{j_0}\ J\bigl(w_j(\cdot), z_0,
\rho_1, \rho_2\bigr) \Big |
\le N \exp\bigl((\log \ell)^{C_1}\bigr) \rho_1^2
r_1^{-2}
\end{equation}
The lemma follows by
combining (\ref{eq:18.10}) with Lemma~\ref{lem:Jdef} on Jensen averages.
\end{proof}

We will now use the following notation for matrix-valued  functions $M(z)=\{a_{p,q}\}_{p,q=1}^m\,$:
\beeq
\label{eq:k_def}
 k(M,z, r)=\min_{p,q}\cZ(a_{pq},z, r)
\eneq
for any $z\in\IC$ and $r>0$.

\begin{defi}
\label{def:contrib}
Let $w_j(z)$ and $\cD(z_0,r_1)$ be as above.  Assume that the following
conditions are valid:
\begin{enumerate}
\item[\rm{(1)}] no entry  of the monodromy $\widehat A_j(z)$ has a zero in
\[ \cD\bigl(\zeta_j, r^{(1)}\exp((\log\ell)^C)\bigr) \setminus \cD\bigl(\zeta_j,
r^{(2)}\bigr),\quad \exp(-\sqrt{\ell})\le r^{(1)} \le r_1^4,\; r^{(2)} = r^{(1)}\exp(-(\log \ell)^C),
\]
where $\zeta_j =e(\xi_j+i\eta_j) \in \cD\bigl(z_0, (1-c_0)r^{(1)}\bigr)$, with $c_0>0$ being some small constant,
and with $|\eta_j|<c_0r^{(1)}$.

\item[\rm{(2)}] $k(\widehat A_j, \zeta_j, r^{(2)})\ge1$
\end{enumerate}
Under these conditions we say that $w_j(\cdot)$ is a {\em contributing term} for $\cD(z_0,r^{(1)})$.
\end{defi}

We now relate the integer $k(j, z_0, r^{(1)})$ from Definition~\ref{def:contrib} to the Jensen averages.

\begin{lemma}
\label{lem:18.7}
Assume that $w_j(\cdot)$ is a contributing
term for the disk $\cD(z_0, r^{(1)})$.  Then
\begin{equation}
\nn
\Big | 4 {\rho_1^2\over \rho_2^2}\ J\bigl(w_j(\cdot), z_0, \rho_1,
\rho_2\bigr) - k\bigl(\widehat A_j(\cdot), \zeta_j, r^{(2)}\bigr) \Big | \le  \exp\bigl(-(\log \ell)^A\bigr)
\end{equation}
where $\rho_1=r^{(1)}$ and $\rho_2=c_1\rho_1$ and $c_1\ll c_0$.
\end{lemma}
\begin{proof}
Apply part (ii) of Proposition~\ref{prop:15.4} to the Jensen averages
$
J\bigl(\log \big \| \widehat A_j(\cdot) \big\|, z_0, \rho_1,\rho_2\bigr)\ .
$
\end{proof}

Assume now that there is at least one contributing term
$w_{j_0}(\cdot)$ for the disk $\cD(z_0,r^{(1)})$ where $z_0=e(x_0)$
and $x_0\in\tor$. We will now show that in this case one can modify
the subsequence $\{m_j\}$ from Lemma~\ref{lem:184} in such a way
that it gives rise to a large collection of contributing terms
without changing $w_{j_0}$.  Choose an arbitrary $s \in\cS_{j_0}^+$
where \beeq \label{eq:sxij} \cS_{j_0}^+=\Big\{s\in  (s_{m_{j_0}},
N]\::\:-r^{(1)}(1-2c_0) < x_0 -\xi_j -\{(s-s_{m_{j_0}})\omega\} <
r^{(1)}(1-2c_0)\Big\}. \eneq We have $\cS_{j_0}^+\ne\emptyset$ since
$r^{(1)}\asymp \exp\left(-(\log \ell)^A\right)$. Note that due to
the Diophantine condition $\omega\in\tor_{c,a}$
\begin{equation}
\label{eq:ssmj_sep}
s-s_{m_{j_0}} > \exp(\ell^\eps) \gg \exp\bigl((\log \ell)^A\bigr)\ .
\end{equation}
Assume now that $s < N - \exp\bigl((\log \ell)^B\bigr)$.  Note that
\begin{equation}
\nn
M_{[s, s_{m_{j_0+1}}-s_{m_{j_0}} +s]} (z, \omega, E)  =
M_{[s_{m_{j_0}}, s_{m_{j_0+1}}]}\bigl(ze((s-s_{m_{j_0}})\omega), \omega,
E\bigr)
 = \widehat A_{j_0} \bigl(ze((s-s_{m_{j_0}})\omega)\bigr)\ .
\end{equation}

The following lemma is a shifted form of Definition~\ref{def:contrib}.

\begin{lemma}
\label{lem:18.8}
Let $t=s-s_{m_j^{(0)}}$. Then
\begin{enumerate}
\item[\rm{(1)}] no entry of $M_{[s_{m_{j_0}}, s_{m_{j_0+1}}]}\bigl(\cdot e(t\omega), \omega,E\bigr)$ has
a zero in $\cD\left(\zeta_{j_0} e(-t\omega),
r_1/2\right) \setminus \cD\left(\zeta_{j_0} e(-t\omega),
r^{(2)}\right)$

\item[\rm{(2)}] $k \left(M_{[s_{m_{j_0}}, s_{m_{j_0+1}}]}\bigl(\cdot e(t\omega), \omega,E\bigr), \zeta_{j_0} e(-t\omega), r^{(2)}\right) =
k\bigl(\widehat A_{j_0}(\cdot), \zeta_{j_0}, r^{(2)}\bigr)$
\end{enumerate}
Moreover, $s_{m_{j_0}}+t$ and $s_{m_{j_0+1}}+t$ are adjusted to $\cD(z_0,r_1/2)$ at scale $\ell$.
\end{lemma}
\begin{proof}
These are basically just (1) and (2) of Definition~\ref{def:contrib} shifted by $e(-t\omega)$.
The only difference is that in property (1) the outer radius needs to be replaced by $r_1-r^{(1)}$,
see~\eqref{eq:sxij}. Since $r^{(1)}\ll r_1$, this is larger than $r_1/2$, as claimed.
The claim about the adjustedness is also a consequence of the small size of the shift by $t\omega$.
\end{proof}

Recall that we assumed that $s_{m_{j_0}}<s<N$. Analogously to~\eqref{eq:sxij}, we can now consider
\beeq
\label{eq:sxij'}
 \cS_{j_0}^{-}=\Big\{s\in[1,s_{m_{j_0}})\::\:-r^{(1)}(1-2c_0) <
 x_0 -\xi_j + \{(s_{m_{j_0}}-s)\omega\} < r^{(1)}(1-2c_0)\Big\}.
\eneq
Clearly, there will be a version of Lemma~\ref{lem:18.8} in this case.
We arrive at the following conclusion.

\begin{lemma}
\label{lem:18.9}
The set
\beeq
\label{eq:small_corr}
\Big\{s\in\cS_{j_0}^+\cup\cS_{j_0}^-\::\: \exp\bigl((\log \ell)^B\bigr) <  s <
N - \exp\bigl((\log \ell)^B\bigr) \Big\}
\eneq
is adjusted to the disk $\cD(z_0,r_1/2)$ at scale $\ell$ and the distance between any two
distinct elements of this sequence
exceeds $\exp\left((\log \ell)^{A_1}\right)$.
\end{lemma}

Now consider Lemma~\ref{lem:18.5} with this choice of adjusted sequence.
Then we obtain the following lower bound for the number of zeros.

\begin{corollary}
\label{cor:null_unten}
Assume that there is at least one
 term $w_{j_0}(\cdot)$ on the right-hand side of
(\ref{eq:18.11}) which is contributing for $\cD(z_0,r^{(1)})$.
Then
\begin{equation}
\nn
 \#\left\{z \in \cD(z_0, \rho_1(1 + c_0)): f_N(z, \omega, E) = 0 \right\}
 \ge 2N k\bigl(\widehat A_{j_0}, \zeta_{j_0}, r^{(2)}\bigr) \rho_1
(1-3c_0) - N\rho_1^{5/4}.
\end{equation}
where the constants are the same as in Definition~\ref{def:contrib}.
\end{corollary}

\begin{proof} Since $\rho_1 > \exp(-(\log N)^\delta)$ and $\omega$ satisfies the Diophantine
condition~\eqref{eq:diophant}, the total number of integers which satisfy (\ref{eq:sxij}) or
\eqref{eq:sxij'} as well as~\eqref{eq:small_corr}
is equal to
\begin{equation}
\tilde N=2\rho_1(1-2c_0) N\bigl(1 + O(\rho_1^{c_2})\bigr)\ .
\end{equation}
Now apply the upper bound
in~\eqref{eq:18.11} by omitting those terms in the sum that do not arrive as shifts of $\widehat A_{j_0}$.
This yields
\beeq
\nn
4 {\rho_1^2\over\rho_2^2}{ \sum_j}'\ J\bigl(w_j(\cdot), z_0, \rho_1,\rho_2\bigr) - N\rho_1^{\frac54} \le
 \#\Bigl\{z \in \cD\left(z_0, \rho_1(1 + c_0)\right): f_N(z, \omega, E) = 0\Bigr\}
\eneq
where $\sum_j'$ denotes the sum over those terms that arise as described in Lemma~\ref{lem:18.8}.
Lemma~\ref{lem:18.8} and Proposition~\ref{prop:15.4} imply that
\[
|4 {\rho_1^2\over\rho_2^2}J\bigl(w_j(\cdot), z_0, \rho_1,\rho_2\bigr) - k\bigl(\widehat A_{j_0}(\cdot), \zeta_{j_0}, r^{(2)}\bigr)| < \exp(-(\log \ell)^A).
\]
Note carefully that this requires our assumption that $|\eta_j|< c_0 r^{(1)}$,
see Definition~\ref{def:contrib}.
Hence,
\[
\Big|4 {\rho_1^2\over\rho_2^2}{ \sum_j}'\ J\bigl(w_j(\cdot), z_0, \rho_1,\rho_2\bigr)  - \tilde N k\bigl(\widehat A_{j_0}(\cdot), \zeta_{j_0}, r^{(2)}\bigr) \Big| < \tilde N \exp(-(\log \ell)^A).
\]
In view of the preceding,
\begin{align*}
& \#\Bigl\{z \in \cD\left(z_0, \rho_1(1 + c_0)\right): f_N(z, \omega, E) = 0\Bigr\} \\
& \ge \tilde N [k\bigl(\widehat A_{j_0}(\cdot), \zeta_{j_0}, r^{(2)}\bigr) - \exp(-(\log \ell)^A)] - N \rho_1^{\frac54}  \\
& \ge 2\rho_1(1-3c_0)N - N \rho_1^{\frac54},
\end{align*}
as claimed.
\end{proof}

Next, we want to show that under the assumption of Corollary~\ref{cor:null_unten}
we can produce many centers $z_1=e(x_1)$ which give rise to a contributing term in~\ref{eq:18.11}.
Note that this will require changing the underlying adjusted sequence as before. This, however,
is not important since we will only be interested in a lower bound on the number of zeros as
in Corollary~\ref{cor:null_unten}.

Thus, choose an arbitrary $x_1 \in \tor_0$, $z_1 = e(x_1)$.  There exists
$t_1 \in [1, N]$ such that
$$
\| x_1 + t_1\omega - x_0\| < c_1\rho_1
$$
where $c_1\ll c_0$. Recall that $\rho_1=r^{(1)}$.
Then in analogy with Lemma~\ref{lem:18.8} we obtain the following properties:
\[ \zeta_{j_0}e(-t_1\omega) \in \cD\bigl(z_1, (1-c_0)r^{(1)} + c_1r^{(1)}\bigr),\]
as well as
\begin{enumerate}
\item[(1)] no entry of the monodromy $M_{[s_{m_{j_0}}+t,
s_{m_{j_0+1}}+t]} (z, \omega, E)$ has a zero in
\[ \cD\left(\zeta_{j_0} e\bigl(-t_1\omega\bigr), r^{(1)}\right)\setminus
\cD\left(\zeta_{j_0} e\bigl(-t_1\omega\bigr), r^{(2)}\right)\]

\item[(2)] $k\left(M_{[s_{m_{j_0}}+t, s_{m_{j_0+1}}+t]} (\cdot, \omega,
E), \zeta_{j_0} e\bigl(-t_1\omega), r^{(2)}\right) =
k\left(\widehat A_{j_0}(\cdot), \zeta_{j_0}, r^{(2)}\right)$
\end{enumerate}
These properties allow us to conclude the following result, which is
analogous to Corollary~\ref{cor:null_unten}.

\begin{lemma}
\label{lem:null_unten2}
Assume that for some $x_0 \in \tor$ there is at least one
term $w_{j_0}(\cdot)$ on the right-hand side of relation
(\ref{eq:18.11}) which is contributing for the disk $\cD(z_0,r^{(1)})$.
Then for any $x_1 \in \tor$, $z_1 = e(x_1)$ one has
\begin{equation}
\label{eq:unt2}
 \#\left\{z \in \cD\left(z_1, \rho_1(1+c_0)\right): f_N(z, \omega, E) = 0 \right\}
 \ge 2Nk\left(\widehat A_{j_0}(\cdot), \zeta_{j_0}, r^{(2)}\right) \rho_1 \left(1 -3c_0\right) - N\rho_1^{\frac54}
\end{equation}
where $\rho_1$ is as above.  In particular,
\begin{equation}
\label{eq:all0}
\#\left\{z: 1-2\rho_1 < |z| < 1 + 2 \rho_1: f_N(z, \omega, E) = 0
\right\}
\ge Nk\left(\widehat A_{j_0}(\cdot), \zeta_{j_0}, r^{(2)}\right)\left(1 - 4c_0\right)\ .
\end{equation}
\end{lemma}
\begin{proof}
The estimate \eqref{eq:unt2} is proved as in Corollary~\ref{cor:null_unten}.
This then leads to~\eqref{eq:all0} by means of a covering argument.
\end{proof}

Let $V_0(e(x))$ be a trigonometric polynomial, i.e.,
$$
V_0\bigl(e(x)\bigr) = \sum^{k_0}_{k = - k_0} \hat v_0(k) e(kx)\ ,\quad \hat v_0(-k) = \overline{v_0(k)}\ .
$$
We refer to $k_0$ as the degree of $V_0$ and denote it by $\degg
V_0$. Assume that the Lyapunov exponent $L(\omega_0, E)$ relative to
the potential $V_0$ and some $\omega_0 \in \tor_{c,a}$ is positive
for all $E$, $\gamma_0 = \inf L(\omega_0, E)$.  Due to
Corollary~\ref{cor:5.8}, given $\rho_0$, there exists $\tau_0 =
\tau_0(\lambda, V_0, \omega_0, \gamma_0, \rho_0) > 0$ such that
$$
\#\left\{z \in \cA_{\rho_0/2}\ ,\ f_N(z, \omega_0, E) = 0\right\} \le N(2 \degg V_0 + c)\ ,\ c < 1
$$
(with $f_N$ defined in terms of $V$ rather than $V_0$) provided
\begin{equation}
\sup_{\cA_{\rho_0}} \bigl| V(z) - V_0(z)\big | \le \tau_0
\end{equation}
with sufficiently small $\tau_0$.  Assume now that $V(z)$ satisfies
this condition. In what follows, all determinants, monodromy
matrices etc.~are defined using $V$ rather than $V_0$.

\begin{corollary}
\label{cor:deg}
Assume that for some $x_0 \in \tor$ there is at least
one term $w_{j_0}(\cdot)$ in~\eqref{eq:18.11} which is contributing for $\cD(z_0,r^{(1)})$.
Then
\begin{equation}
\label{eq:deg_est}
k\left(\widehat A_{j_0}(\cdot), \zeta_{j_0}, r^{(2)}\right)\le 2\degg V_0
\end{equation}
\end{corollary}

Next, we produce a zero-free annulus for the determinants.

\begin{lemma}
\label{lem:zerofree_ann}
Given $x_0 \in \tor$ and large $n$, there exists
\beeq
\label{eq:rNeps}
\exp(-(\log n)^{C_1})<r < \exp(-(\log n)^{C_2}),
\eneq
 such that all $f_{[a,n-b]}(\cdot, \omega, E)$, $a,b=0,\pm1$,  have no zeros in $\cD\bigl(\zeta_0, r\bigr)\setminus
\cD\bigl(\zeta_0, r\exp(-(\log n)^C)\bigr)$, $\zeta_0=e(x_0)$.
\end{lemma}
\begin{proof}
Set $\rho_0=\exp(-(\log n)^{C_1})$
and $\rho^{(m)} = \rho_0\exp(-m(\log n)^C)$ for all $m \ge0$.
By Proposition~\ref{prop:zero_count},
\[
\# \left\{ z \in \cD\left(\zeta_0, n^{-1} \right): f_{[a,n-b]}(z,\omega, E) = 0 \text{\ \ for some\ \ }a,b=0,\pm1\right\} \le \bigl(\log n\bigr)^{A}\ .
\]
Therefore, there exists $0\le m \le \bigl(\log n\bigr)^{A}$ such that each $f_{[a,n-b]}(\cdot, \omega, E)$, $a,b=0,\pm1$,  has no
zeros in \[ \cD\bigl(\zeta_0, \rho^{(m+1)}\bigr) \setminus
\cD\bigl(\zeta_0, \rho^{(m)}\bigr).\]  Set $r = \rho^{(m+1)}$,
and we are done.
\end{proof}

Now we show how to obtain a contributing term.

\begin{lemma}
\label{lem:contrib}
Let $\hat{A}(z)=M_{[s',s'']}(z,\omega,E)$ where $s'<s''$ are adjusted to $\cD(z_0,r_1)$ at
scale $\ell$, and with $r_1=\exp(-(\log n)^{C_3})$ . Assume that $\ell\ll s''-s'<\exp((\log\ell)^C)$ and that
\[ k(\hat{A},z_0,\rho_0)\ge 1, \quad \rho_0=\exp(-(\log \ell)^{C_1}).\]
Then $\hat{A}$ is contributing for $\cD(z_0,r^{(1)})$ for some $r^{(1)}\le r_1^4$ and
$\rho_0\ll r^{(2)}$. In particular,
\[ k(\hat{A},z_0,\rho_0)\le 2\deg V_0.\]
\end{lemma}
\begin{proof}
Due to Lemma~\ref{lem:zerofree_ann}, there exists
\[ r^{(1)}\le r_1^4, \qquad 2\rho_0<r^{(2)}<r^{(1)}\exp(-(\log\ell)^C),\]
such that $\hat{A}$ is contributing to $\cD(z_0,r^{(1)})$.
In particular, Corollary~\ref{cor:deg} implies that
\[ k(\hat{A},z_0,\rho_0)\le 2\degg V_0\]
as claimed.
\end{proof}

\begin{proof}[Proof of Theorem~\ref{thm:4}.]
Let $\ell \asymp \exp((\log s)^\delta)$, with $\delta>0$ small.
 Assume $\omega\in\tor_{c,a}$
 There exist $-s^{-}<0<s^{+}$ adjusted to $\cD(z_0,r_1)$
at scale $\ell$
with $r_1=\exp(-(\log \ell^C))$, such that $|s^{\pm}-s|\le \exp((\log\ell)^B)$,
$B\ll C$. Assume that
\[ k_1=k(M_{[-s^{-},s^{+}]}(\cdot,\omega,E),z_0,\rho)>0\]
for $\rho_0=\exp(-(\log \ell^{C_1}))$. Then by Lemma~\ref{lem:contrib}, $k_1\le2\deg V$.
\end{proof}

Theorem~\ref{thm:4} has the technical disadvantage that its estimate
is guaranteed not for the Dirichlet determinant $f_\ell(z, \omega,
E)$ of arbitrary size $\ell$ but rather for a determinant $f_s(z,
\omega, E)$ of a specially chosen size $|s - \ell|<(\log \ell)^A$.
The $(\log N)^A$ upper bound on the number of zeros of $f_N(\cdot,
\omega, E)$ with arbitrary $N$, guaranteed by
Proposition~\ref{prop:zero_count}, is not sufficient for our goals.
However, the following simple corollary of Theorem~\ref{thm:4}
resolves this technical problem.

\begin{prop}\label{prop:1414}
 For any $z_0 = e(x_0 + iy_0)$, $|y_0| < \rho_0/2$, $E \in \IC$ and $N > C$ one has
\begin{equation}
\nu_{f_N(\cdot, \omega, E)} \left(z_0, \exp\bigl(-(\log N)^{
A}\bigr)\right) \le \bigl(\log\log N\bigr)^B
\end{equation}
\end{prop}

\begin{proof} Due to Theorem~\ref{thm:4}, with $s$ replaced by $N$, there exist $s^\pm$, such that
$|s^\pm -N|<\exp((\log N)^\delta)$, $\nu_{f_{[-s^-, s^+]}(\cdot,
\omega, E)} \left(z_0, \exp\bigl(-(\log N)^{ A}\bigr)\right) \le
k_0(V)$. Due to the avalanche principle expansion with $N_1=[N/2]$
\begin{align*}
& \log|f_{[-N+N_1,N-N_1-1]}(z,\omega,E)|-
\log|f_{[-s_-,s_+]}(z,\omega,E)|  \\
& =\Big[ \log\Big\| \Mat M_{[s_+',s_+''']}(z,\omega,E)\Big\| +\log
\Big \| \Mat M_{[s_+',s_+'']}(z,\omega,E) \Big\| \Big] \\
& \quad + \Big[ \log \Big\| M_{[-s_-',-s_-''']}(z,\omega,E)\Mat
\Big\| + \log \Big\|M_{[-s_-',-s_-'']}(z,\omega,E)\Mat \Big\| \Big]
+ O(N^{-B})
\end{align*}
for any $z\not\in \cB_{N,E}$, $\mes(\cB_{N,E})<\exp(-(\log N)^C)$,
where $s_-'=\max\{N-N_1,s_-\}$, $s_+'=\max\{N-N_1-1,s_+\}$,
\[ s_+'''-(\log N)^{C_1} < s_+' < s_+''< s_+''', \qquad  s_-'-(\log
N)^{C_1} < s_-'''< s_-'' < s_-'\] Recall that
\[ 4\frac{r_1^2}{r_2^2} J\Big(\log\Big\| \Mat
M_\ell(\cdot,\omega,E)\Big\|,\zeta_0,r_1,r_2\Big) \lesssim
(\log\ell)^{A_1}
\]
for any $\ell\gg 1$, $\zeta_0\in \cA_{\rho_0/2}$,
$\exp(-\ell^\delta)\le r_1\le \exp(-(\log \ell)^A)$, $r_2=cr_1$.
Hence,
\begin{equation}
4\, {\rho_1^2\over \rho_2^2} \left|J\left(\left(\log \big
|f_{[-N_1,N_1]}(\cdot, \omega, E)\big | - \log \big |f_{[-s^-,
s^+]}(\cdot, \omega, E)\big |\right), z_0e(-N_1\omega), \rho_1,
\rho_2\right)\right | < (\log\log N)^C\ ,
\end{equation}
where $\rho_1 \asymp \exp \left(-(\log N)^{A}\right)$, $\rho_2 =
c\rho_1$.  By Corollary 5.2 relation (14.16) implies
$$
\nu_{f_N(\cdot, \omega, E)} (z_0, \rho_1 - \rho_2) \le \nu_{f_{[-s^-, s^+]}(\cdot, \omega, E)} (z_0, \rho_1 + \rho_2)\ .
$$
\end{proof}

\section{Concatenation terms and the number of eigenvalues falling into an interval}
\label{sec:monster}

 Consider the following concatenation terms
\begin{equation}\label{eq:151}
\cW_{N, k}\bigl(e(x), E + i\eta\bigr) =
{\big \|M_{[1, k]}\bigl(e(x), \omega, E+i\eta\bigr)\big \|\, \big \|M_{[k+1, N]}\bigl(e(x), \omega, E + i\eta\bigr)\big \|\over
\big \|M_{[1, N]}\bigl(e(x), \omega, E+i\eta\bigr)\big \|}
\end{equation}
$1 \le k \le N$, where $\omega$ is fixed.

\begin{lemma}
\label{lem:151}
 Let $x \in \tor$, $E \in \IR$, $\eta > 0$, and let
\begin{equation}\label{eq:152}
\left | f_{[a, N-b+1]} \bigl(e(x), \omega, E+i\eta\bigr) \right | = \max_{1 \le a', b' \lesssim 2}
\left |f_{[a', N- b' +1]} \bigl(e(x), \omega, E+ i\eta\bigr)\right |
\end{equation}
for some $1 \le a$, $b \lesssim 2$.  Then
\begin{equation}
\label{eq:153} \#\left(\rsp H_{[a, N-b +1]} \bigl(e(x), \omega\bigr)
\cap \bigl(E - \eta, E+\eta\bigr) \right) \le 4 \eta \sum_{1 \le k
\le N}\, \cW_{N, k} \bigl(e(x), E + i\eta\bigr)
\end{equation}
\end{lemma}

\begin{proof} Recall that
\begin{align}
 & \left(H_{[a, N']} \bigl(e(x),\omega\bigr) - E - i\eta\right)^{-1}(k, k) =
{f_{[a, k]} \mape f_{[k+2, N']}\mape\over f_{[a, N']} \mape} \label{eq:154}\\[6pt]
& M_{[a, N']} \mape = \begin{bmatrix}
f_{[a, N']} \mape & - f_{[a+1, N']} \mape\\
f_{[a, N'-1]} \mape & - f_{[a+1, N'-1]}\mape
\end{bmatrix} \label{eq:155}
\end{align}
Due to (\ref{eq:152})
\begin{equation}
\label{eq:156} \big \|M_N \mape \big \| \le 2 \Big | f_{[a, N-b +1]}
\mape \Big |\ .
\end{equation}
Combining (\ref{eq:154}), (\ref{eq:155}), (\ref{eq:156}) one obtains
\begin{equation}
\label{eq:157}
\begin{aligned}
 & \left| \tr \left(\Bigl(H_{[a, N-b +1]} (x,\omega) -E - i\eta\Bigr)^{-1}\right)\right |\\
 & \le \sum_{a \le k \le N -b +1}\ {\Big |f_{[a, k]} \mape\Big |\, \Big |f_{[k+2, N-b +1]}\mape \Big |\over
 \Big |f_{[a, N-b +1]} \mape \Big |}\\
 & \le \sum_{a \le k \le N-b+1}\ 2 \cW_{N, k} (e(x),E+i\eta)
 \end{aligned}
 \end{equation}
 On the other hand,
 \begin{equation}\nn
 \begin{aligned}
 & \left| \tr \left(H_{[a, N-b +1]} (x, \omega) -E - i\eta\right)^{-1} \right|\\
 & \ge (2\eta)^{-1} \#\left(\rsp H_{[a, N-b+1]} \bigl(e(x), \omega\bigr) \cap (E - \eta, E+\eta)\right)
 \end{aligned}
 \end{equation}
 and we are done.
 \end{proof}

\begin{lemma}\label{lem:152} Using the notations of Lemma \ref{lem:151} one has for any $K$
\begin{equation}\nn
\begin{aligned}
& \#\left(\rsp\left(H_{[a, N-b+1]} \bigl(e(x),\omega\bigr)\right)\cap
\bigl(E - \eta, E+\eta\bigr)\right)\\
& \le 4 \eta \sum_{k \notin K}\ \cW_{N,k} \bigl(e(x), E + i\eta\bigr) + \#(K)
\end{aligned}
\end{equation}
\end{lemma}

\begin{proof} Just as in (\ref{eq:157}) one has
\begin{equation}\label{eq:1510}
\begin{aligned}
& \left | \tr \left(M_{[a, N-b +1]} \mapen\right)^{-1}\right |\\
& \le \sum_{k \notin K }\ 2\cW_{N,k} \bigl(e(x), E + i\eta\bigr) + \sum_{k \in K}
\left| \left(H\bigl(e(x),\omega\bigr) - E - i\eta\right)^{-1}(k, k) \right |
\end{aligned}
\end{equation}
Recall that
\begin{equation}\label{eq:1511}
\begin{aligned}
\left | \left(H\mapen\right)^{-1} (k, k)\right| & \le
\left \| \left(H\mapen\right)^{-1}\right\|\\
& \le \eta^{-1}
\end{aligned}
\end{equation}
The assertion follows from (\ref{eq:1510}), (\ref{eq:1511}) and
(\ref{eq:153}).
\end{proof}

\begin{corollary}\label{cor:153}
 Using the notations of Lemma \ref{lem:151} one has
\begin{equation*}
\begin{aligned}
& \# \left(\rsp \left(H_{[1, N]}\bigl(e(x),\omega\bigr)\right)\cap \bigl(E - \eta, E+\eta\bigr)\right)\\
& \le 4 \eta \sum_{ k \in K}\ \cW_{N,k}\bigl(e(x), E + i\eta\bigr) + \# (K) + 2
\end{aligned}
\end{equation*}
\end{corollary}

\begin{proof} Due to Weyl's Comparison Lemma, see~\cite{Bhat},
\begin{equation*}
\begin{aligned}
&\#\left(\rsp\left(H_{[1, N]}\bigl(e(x),\omega\bigr)\right) \cap \bigl(E - \eta, E+ \eta\bigr) \right) \le\\
& \#\left(\rsp\left(H_{[a, N-b+1]} \bigl(e(x), \omega \bigr) \right)\cap \bigl(E - \eta, E + \eta\bigr) \right) + 2
\end{aligned}
\end{equation*}
as claimed.
\end{proof}

\begin{lemma}\label{lem:154} Let $A$ be $n\times n$ hermitian matrix.  Let $\Psi^{(1)}, \Psi^{(2)},\dots, \Psi^{(n)} \in \IC^n$
be an orthonormal basis of eigenvectors of $A$ and $E^{(1)},
E^{(2)}, \dots, E^{(n)}$ be the corresponding eigenvalues.  Then for
any $E + i\eta$, $E \in \IR$, $\eta > 0$ one has
\begin{equation*}
\sum_{1 \le k \le n} \left| \left(\bigl(A - E - i\eta\bigr)^{-1} e_k, e_k \right)\right |^2  \ge \sum_{1 \le j \le n} \Biggl(\sum_{1 \le k \le n} \left |\bigl(e_k, \Psi^{(j)}\bigr) \right |^4 \Biggr) \cdot
\left(\imm \left(E^{(j)} - E - i\eta\right)^{-1}\right)^2
\end{equation*}
where $e_1, e_2, \dots, e_n$ is arbitrary orthonormal basis in $\IC^n$.
\end{lemma}

\begin{proof} One has
\begin{equation*}
\begin{aligned}
\left(\left(A - E - i\eta\right)^{-1} e_k, e_k \right) & = \sum_{1 \le j \le n} \left| \bigl(e_k, \Psi^{(j)}\bigr)\right |^2 \left(E^{(j)} - E - i \eta\right)^{-1}\\
\left|\left(\bigl(A - E - i\eta\bigr)^{-1} e_k, e_k \right) \right | & \ge \imm \left(\bigl(A - E - i\eta\bigr)^{-1} e_k, e_k \right)\\
& = \sum_{1 \le j \le n} \left|\bigl(e_k, \Psi^{(j)}\bigr) \right|^2 \imm \bigl(E^{(j)} - E - i\eta\bigr)^{-1}\ .
\end{aligned}
\end{equation*}
Since $\imm\left(E^{(j)} - E - i \eta\right)^{-1} > 0$, $j = 1,
2,\dots, n$, the assertion follows (use $(\sum_j a_j)^2\ge \sum_j
a_j^2$ if $a_j\ge0$).
\end{proof}

\begin{corollary} Using the notations of the previous lemma assume that the following condition is valid for some $E, \eta$:

\bigskip\noindent
{\bf (L)\ } for each eigenvector $\Psi^{(j)}$ with $\left| E^{(j)} -
E\right| < \eta$ there exists a set $\cS(j) \subset
 \left\{1, 2, \dots, n\right\}$, $\#\cS(j) \le \ell$ such that $\sum\limits_{k \notin \cS(j)}
 \left|\bigl(e_k, \Psi^{(j)}\right)|^2 \le 1/2$.

\bigskip\noindent
 Then
$$
\#\left\{j: \left|E^{(j)} - E\right| < \eta\right\} \le 8\ell \eta^2
\sum_{1 \le k \le n} \left|\left(\bigl(A - E - i\eta\bigr)^{-1} e_k,
e_k \right) \right |^2
$$
\end{corollary}

\begin{proof} Recall that for any positive $\alpha_1, \dots, \alpha_\ell$ with $\sum\limits_j \alpha_j = 1$
\begin{equation}\nn
\sum_j \alpha^2_j \ge \sum_j\ {1\over \ell^2} = {1\over \ell}
\end{equation}
by Cauchy-Schwarz. Due to the assumptions of the corollary
$$
1 = \left(\Psi^{(j)}, \Psi^{(j)}\right) = \sum_{1 \le k \le n}
\left|\bigl(e_k, \Psi^{(j)})\right|^2 \le \sum_{k \in \cS(j)}
\left|\bigl(e_k, \Psi^{(j)}\bigr) \right|^2 + 1/2
$$
for any $\left|E^{(j)} - E\right | < \eta$.  Hence, for such
$E^{(j)}$ we have
\begin{equation}
\sum_{k \in \cS(j)} \left|(e_k, \Psi^{(j)})\right|^4 \ge 1/4\ell
\end{equation}
and the assertion follows from the previous lemma.
\end{proof}

\section{The Riesz measure of concatenation terms}
\label{sec:16}

Consider the concatenation term
\begin{equation}\nn
w_m(z) = \log\cW_m(z)= \log {\big \|M_{2m}(z,\omega, E)\big \|\over
\big \|M_m \bigl(ze(m\omega), \omega, E\bigr) \big \|\, \big
\|M_m(z,\omega, E)\big \|}
\end{equation}
Assume that the following condition holds:

\bigskip
{\bf (I)} no determinant $f_{[a, m-b]}\bigl(\cdot e(nm\omega),
\omega, E\bigr)$, $f_{[a, 2m-b]}(\cdot, \omega, E)$, $a = 1, 2$; $b
= 0, 1$; $n = 0, 1$, has a zero in some annulus $\cD(\zeta_0,
\rho_1)\setminus \cD(\zeta_0, \rho_0)$, where $\rho_0 \asymp
\exp\bigl(-m^\delta\bigr)$, $\rho_0 < \rho_1 < \exp\left(-(\log
m)^A\right)$, $0 < \delta \ll 1$.
\bigskip

 Set
\begin{equation}\nn
\bar k_n = \min_{a, b}\, \nu_{f_{[a, m-b]}\bigl(\cdot
e(nm\omega),\omega,E\bigr)} (\zeta_0, \rho_0)\ ,
\end{equation}
$n = 0, 1$,
\begin{equation}\nn
\bar k = \min_{a, b}\, \nu_{f_{[a, 2m-b]}(\cdot, \omega,
E)}(\zeta_0, \rho_0)\ .
\end{equation}

\begin{lemma}\label{lem:16.1} Assume that $\rho_1 \ge \rho_0^{(\log m)^{-B_0}}$.  Then
\begin{equation}\nn
\bar k_0 + \bar k_1 \le \bar k \le \bar k_0 + \bar k_1 +
k_1(\lambda, V)
\end{equation}
provided $B_0 \gg 1$. Here $k_1(\lambda,V)$ is some integer
constant.
\end{lemma}

\begin{proof} Due to Proposition \ref{prop:15.4} one has
\begin{equation}\label{eq:16.5}
\left | \log {\big \|M_{[1, m]}\bigl(\zeta e(nm\omega), \omega,
E\bigr) \big \|\over \big \|M_{[1, m]}\bigl(ze(nm\omega), \omega,
E\bigr) \big \|} - \bar k_n \log {|\zeta - \zeta_0|\over |z -
\zeta_0|} \right | < 1\ ,\quad n = 0, 1,
\end{equation}

\begin{equation}\label{eq:16.6}
\left |\log {\big \|M_{[1, 2m]} (\zeta, \omega, E)\big \|\over
\big \| M_{[1, 2m]}(z, \omega, E) \big \|} - \bar k_0 \log {|\zeta - \zeta_0|\over |z - \zeta_0|}\right | < 1\ ,
\end{equation}
for any $z, \zeta \in \cD(\zeta_0, \rho'_1) \setminus \cD(\zeta_0, \rho_2)$,
where $\rho'_1 = \exp\left(-(\log m)^{B_1}\right)\rho_1$, $\rho_2 = \exp\left((\log m)^{B_1}\right)\rho_0$.
 There exists $z = e(x + iy) \in \cD(\zeta_0, \rho'_1) \setminus \cD(\zeta_0, \rho'_1/2)$ such that
\begin{equation}\nn
\left| \log \big \|M_{[1, m]}\bigl(ze(nm\omega),\omega, E\bigr) \big
\| - m L(y, E) \right| \le m^\delta\bigl(\log m\bigr)^{-B_2}\ ,\ n =
0, 1,
\end{equation}
\begin{equation}\nn
\left | \log \big \|M_{[1, 2m]}(z, \omega, E) \big \| - 2m L(y, E)
\right| \le m^\delta \bigl(\log m\bigr)^{-B_2}
\end{equation}
with $1\ll B_2 < B_0$. Combining these relations one obtains
\begin{equation}\nn
\begin{aligned}
& \biggl | \log {\big \|M_{[1, m]} \bigl(\zeta e(m\omega), \omega, E\bigr) \big\|\, \big \|M_m(\zeta, \omega, E)\big \|\over
\big \|M_{[1,2m]}(\zeta, \omega, E)\big \|} -\\
&\quad \left(\bar k_0 + \bar k_1 - \bar k\right) \log {|\zeta - \zeta_0|\over |z - \zeta_0|} \bigg| \le
Cm^\delta\bigl(\log m\bigr)^{B_1}
\end{aligned}
\end{equation}
for any $\zeta \in \cD(\zeta_0, \rho'_1) \setminus \cD(\zeta_0,
\rho_2)$. Since $|z - \zeta_0| \asymp \rho'_1$, $\rho'_1 \asymp
\exp\left(-m^\delta (\log m)^{-B_0} - (\log m)^{B_1}\right)$, one
can pick $\zeta \in \cD(\zeta_0, \rho'_1) \setminus \cD(\zeta_0,
\rho_2)$ such that $|\zeta - \zeta_0| = |z - \zeta_0|^{(\log
m)^{B_0/2}}$.  Then
\begin{equation}\label{eq:16.10}
\begin{aligned}
& \bigg | \log {\big \|M_{[1, m]} \bigl(\zeta e(m\omega), \omega, E\bigr) \big \|\, \big \| M_m(\zeta, \omega, E)\big \|\over
\big \|M_{2m} (\zeta, \omega, E)\big \|} +\\
& m^\delta (\log m)^{B_0/2} \left(\bar k_0 + \bar k_1 - \bar k\right) \bigg | \le
Cm^\delta (\log m)^{B_1}\ .
\end{aligned}
\end{equation}
Recall that
\begin{equation}\label{eq:16.11}
\big \|M_{2m} (\zeta, \omega, E) \big \| \le \big \|M_{m}\bigl(\zeta e(m\omega), \omega, E\bigr) \big \|\, \big \|M_m (\zeta, \omega, E)\big \|
\end{equation}

Relations (\ref{eq:16.10}), (\ref{eq:16.11}) imply $\bar k_0 + \bar
k_1 - \bar k \le 0$.  Removing the absolute values
in~\eqref{eq:16.5} and~\eqref{eq:16.6} and taking Jensen's averages
one obtains the following:
\begin{equation}\nn
\nn \bigg | 4\, {r_2^2\over r_1^2}\ J\biggl (\log {\big
\|M_{2m}(\cdot, \omega, E)\big \|\over \big \|M_{[1, m]}\bigl(\cdot
e(m\omega), \omega, E\bigr)\big \|\, \big \|M_m(\cdot, \omega,
E)\big \|}\ ,\ \zeta_0, r_1, r_2\biggr ) - \left(\bar k - \bar k_1 -
\bar k_2\right)\bigg | < 2
\end{equation}
where $r_1=\rho_1'/2$, $r_2=cr_1$.  Hence, $\bar k - \bar k_1 - \bar
k_2 \le k_1(\lambda, V)$.
\end{proof}

\begin{defi} Let $m, E$ be as above.  Fix $0 < \delta \ll 1$, $B_0 \gg 1$.  Set
\begin{equation}\nn
\rho^{(0)}_m = \exp\bigl(-m^\delta\bigr)\ ,\quad \rho_m^{(t)} = \bigl(\rho^{(0)}_m\bigr)^{(\log m)^{-tB_0}}\ ,\quad t = 1, 2,\dots\ .
\end{equation}
Given $\zeta_0\in \cA_{\rho_0/2}$, define $t_m(\zeta_0)$ as the minimal integer $t = 0, 1,\dots$ such that condition (I) is valid in the annulus
$$
\cD\left(\zeta_0, \rho_m^{(t+1)}\right)\setminus \cD\left(\zeta_0, \rho_m^{(t)}\right)\ .
$$
\end{defi}

\noindent Recall that by Proposition~\ref{prop:1414},
$\nu_{f_N(\cdot, \omega, E)} (\zeta_0$, $\exp\left(-(\log
N)^A\right))\le \bigl(\log\log N\bigr)^{A_1}$ for any $N$, $\zeta_0
\in \cA_{\rho_0/2}$.

\begin{lemma} $t_m (\zeta_0) \le \bigl(\log\log m\bigr)^{A_1}$ for any $m$, $\zeta_0 \in \cA_{\rho_0/2}$.
In particular, $\rho_m^{\bigl(t_m(\zeta_0) +1\bigr)} \le
\exp\bigl(-m^{\delta/2}\bigr)$, provided $m > m_0(\delta, B_0)$.
\end{lemma}

\begin{proof} Note that
$$
\log\log \left(\rho_m^{(t)}\right)^{-1} = \delta\log m - t B_0\bigl(\log\log m\bigr) > \delta/2\log m
$$
provided $t \ll \bigl(\log\log m\bigr)^{C_0}$, $m \ge m_0(\delta,
B_0, C_0)$.  Therefore, both assertions of the lemma follow.
\end{proof}

Set
\begin{equation}\nn
\begin{aligned}
& \bar k_{m, n}(\zeta) = \min_{a, b}\ \nu_{f_{[a, m-b]\bigl(\cdot e(nm\omega), \omega, E\bigr)}}
\left(\zeta, \rho^{(t_m(\zeta))}\right)\ ,\quad n = 0, 1,\\
& \bar k_m(\zeta) = \min \nu_{f_{[a, 2m-b](\cdot, \omega, E)}} \bigl(\zeta, \rho^{(t_m(\zeta))}\bigr)
\end{aligned}
\end{equation}

\begin{lemma} There exist $\zeta_{1, m}, \zeta_{2, m}, \dots, \zeta_{j_m, m}\in \cA_{\rho_0/2}$
such that the following conditions are valid:
\begin{enumerate}
\item[(a)] $\cA_{\rho_0/2} \subset \bigcup\limits_j \cD\left(\zeta_{j, m}, \bar \rho_{j, m}\right)$,
$\bar\rho_{j, m} = \left(\rho^{(t_m(\zeta_{j, m}) +1)}\right)^{(\log m)^{B_0/4}}$

\item[(b)] $\dist\left(\cD\left(\zeta_{j_1, m}, \underline\rho_{j_1, m}\right), \cD\left(\zeta_{j_2, m},
\underline\rho_{j_2, m}\right)\right) \ge \underline\rho_{j_1, m} +
\underline\rho_{j_2, m}$, where $j_1 \ne j_2$, $\underline\rho_{j,
m} = (\bar \rho_{j, m})^{(\log m)^{B_0/4}}$, $j_1 = 1, 2,\dots, j_m$

\item[(c)] no determinant $f_{[a, m-b]}\bigl(\cdot e(nm), \omega, E\bigr), f_{[a, 2m-b]}
\bigl(\cdot, \omega, E\bigr)$, $a = 1, 2$; $b = 0, 1$; $n = 0$, has
a zero in $\cD\left(\zeta_{j, m}, \bar{\bar\rho}_{j,
m}\right)\setminus \cD\left(\zeta_{j, m},
\underline{\underline\rho}_{j, m}\right)$, $\bar{\bar\rho}_{j, m} =
\left(\bar \rho_{j, m}\right)^{(\log m)^{-B_0/4}}$,
$\underline{\underline\rho}_{j, m} = \left(\underline\rho_{j,
m}\right)^{(\log m)^{B_0/4}}$
\end{enumerate}
\end{lemma}

\begin{proof} Let $t_1 = \max \left\{t_m(\zeta): \zeta \in \cA_{\rho_0/2}\right\}$ and let
$\zeta_{1, m}$ be any point such that $t_m\bigl(\zeta_{1, m}\bigr) = t_1$.  Let
$$
t_2 = \max \left\{t_m(\zeta): \zeta \in \cA_{\rho_0/2} \setminus \cD\left(\zeta_{1, m}, \bar\rho_{1, m}\right)\right\}
$$
Let $\zeta_{2, m} \in \cA_{\rho_0/2} \setminus \cD\left(\zeta_{1,
m}, \bar\rho_{1, m}\right)$ be an arbitrary point such that
$t_m(\zeta_2) = \bar t_2$, etc. One obtains $\zeta_{1, m}, \zeta_{2,
m}, \dots$ such that $t_m \left(\zeta_{j, m}\right) \le
t\left(\zeta_{j+1, m}\right)$, $j = 1, 2,\dots, j_m -1$,
$\cA_{\rho_0/2} \subset \bigcup\limits_{1 \le j \le j_m}
\cD\left(\zeta_{j, m}, \bar\rho_{j, m}\right)$.  If $j_{1,m} < j_{2,
m}$, then $\left| \zeta_{j_1, m} - \zeta_{j_2, m}\right| \ge
\bar\rho_{j_1, m} > 4 \underline\rho_{j_1, m} \ge
2\underline\rho_{j_2, m}$.  Hence, \[\dist\left(\cD\left(\zeta_{j_1,
m}, \underline\rho_{j_1, m}\right)\right), \cD\left(\zeta_{j_2, m},
\rho_{j_2, m}\right) \ge\underline\rho_{j_1, m} +
\underline\rho_{j_2, m}.\] Property (c) follows from the definition
of the integers $t_m(\zeta)$.
\end{proof}

Note that one can apply Lemma~\ref{lem:16.1}  to each annulus
$\cD\left(\zeta_{j, m}, \bar{\bar\rho}_{j, m}\right)\setminus
\cD\left(\zeta_{j, m}, \underline{\underline\rho}_{j, m}\right)$.
That allows one to establish  important properties of the disks
$\cD\left(\zeta_{j, m}, \bar\rho_{j, m}\right)$ in addition to
(a)--(c) stated in the last lemma.  We summarize all these
properties in the following statement.

\begin{prop} \label{prop:165}
Given $E \in \IC$ and integer $m \gg 1$, there exists a cover of $\cA_{\rho_0/2}$ by disks $\cD\left(\zeta_{j, m}, \bar\rho_{j, m}\right)$, $\zeta_{j, m} \in \cA_{\rho_0/2}$, $j = 1, 2,\dots, j_m$ such that the following conditions are valid:
\begin{enumerate}
\item[(1)] $\exp(-m^\delta) \le \bar \rho_{j, m} \le \exp(-m^{\delta/2})$, $j = 1, 2,\dots, j_m$, $0 < \delta \ll 1$,

\item[(2)] $\dist\left(\cD\left(\zeta_{j_1, m}, \underline \rho_{j_1, m}\right), \cD\left(\zeta_{j_2, m},\underline\rho_{j_2, m}\right)\right) \ge \underline\rho_{j_1, m} + \underline\rho_{j_2, m}$ where $\underline\rho_{j, m} = \bar\rho_{j, m}^{(\log m)^{B_1}}$, $B_1 \gg 1$, $j = 1, 2,\dots, j_m$, provided $j_1 \ne j_2$,

\item[(3)] no determinant $f_{[a, m-b]}\left(\cdot e(nm\omega), \omega, E\right)$ or $f_{[a, 2m-b]}(\cdot, \omega, E)$, $a = 1, 2$; $b = 0, 1$; $n = 0$, has a zero in $\cD\left(\zeta_{j, m}, \bar{\bar\rho}_{j, m}\right) \setminus \cD\left(\zeta_{j, m}, \underline{\underline\rho}_{j, m}\right)$, where $\underline{\underline\rho}_{j, m} = \underline\rho_{j, m}^{(\log m)^{B_1}}$, $\bar{\bar\rho}_{j, m} = \bar\rho_{j, m}^{(\log m)^{-B_1}}$, $j = 1, 2,\dots, j_m$,

\item[(4)] for each $\zeta_{j, m}$ there is an integer $k(j, m)$,
\begin{equation}\nn
0 \le k(j, m) \le \nu_{f_{2m}(\cdot, \omega, E)}\left(\zeta_{j, m}, \underline\rho_{j, m}\right)
\end{equation}
such that for any $z, \zeta \in \cD\left(\zeta_{j, m}, 2\bar\rho_{j, m}\right)\setminus \cD\left(\zeta_{j, m}, \underline\rho_{j, m}/2\right)$ holds
\begin{equation}\nn
\left| \left(w_m(\zeta) - w_m(z)\right) - k(j, m) \log {|\zeta -
\zeta_{j, m}|\over |z - \zeta_{j, m}|} \right| \le |\zeta - z|^2
\cdot \left(\bar\rho_{j, m}\right)^{-2}\ .
\end{equation}
\end{enumerate}
\end{prop}

\begin{proof} All properties except (4) follow from the previous lemma.  Property (4) follows from Proposition~\ref{prop:14.10}
and Lemma~\ref{lem:Rr}.
\end{proof}

We turn now to the Riesz representation of the concatenation term
$w_m(z)$.  Let $d\mu^{(0)}_m (z), d\mu^{(1)}_m (z)$ and $d\mu(z)$ be
the Riesz measures of $u^{(0)}_m(z) = \log \big \|M_m(z, \omega,
E)\big\|$, $u^{(1)}_m(z) = \log \big \|M_m\bigl(ze(m\omega), \omega,
E\bigr) \big \|$ and $u_m(z) = \log \big \|M_{2m}(z, \omega, E)\big
\|$, respectively, in the annulus $\cA_{\rho_0/2}$.  Since
\begin{equation}\nn
w_m(z) = u_m(z) - u^{(0)}_m(z) - u^{(1)}_m(z)
\end{equation}
we call $d\nu_m = d\mu_m - d\mu_m^{(0)} - d\mu_m^{(1)}$ the Riesz
measure of $w_m(z)$ in the annulus.  Due to Jensen's formula  one
can relate the measure $d\nu_m(z)$ to the integers $k(j, m)$ defined
in Proposition~\ref{prop:165}, using the results on Jensen averages
of the monodromies from Section~\ref{sec:Jav}.

\begin{lemma} Using the notations of Proposition~\ref{prop:165} one has for any $\zeta_{j, m}$:
\begin{equation}\label{eq:1618}
\begin{aligned}
\mu_m \left(\cD\left(\zeta_{j, m}, C\bar\rho_{j, m}\right)\setminus \cD\left(\zeta_{j, m},
C\underline\rho_{j, m}\right)\right) & \lesssim \bar\rho^2_{j, m}\big / \bar{\bar\rho}^2_{j,m}\ ,\\
\left| \mu_m \left(\cD\left(\zeta_{j, m}, \underline\rho_{j, m}/2\right)\right) -
\bar k_{j, m}\bigl(\zeta_{j, m}\bigr)\right | & \lesssim \bar\rho^2_{j, m}\big / \bar{\bar\rho}^2_{j, m}\ .
\end{aligned}
\end{equation}
Similar estimates are valid for $\mu^{(0)}_m, \mu^{(1)}_m$.  In particular,
\begin{align}
\left | \nu_m \left(\cD\left(\zeta_{j, m}, \underline\rho_{j, m}/4\right)\right) - k(j, m)\right| &
\lesssim \bar\rho^2_{j, m}\big / \bar{\bar\rho}^2_{j, m} \nn\\
\left|\nu_m(\cD)\right| & \lesssim \bar\rho^2_{j, m} \big /
\bar\rho^2_{j, m} \nn
\end{align}
for any $\cD \subset \cD\left(\zeta_{j, m}, C\bar\rho_{j, m}\right) \setminus \cD \left(\zeta_{j, m}, C\underline\rho_{j, m}\right)$.
\end{lemma}

\begin{proof} Due to property (3) of Proposition~\ref{prop:165}
 no determinant $f_{[a, 2m-b]}(\cdot, \omega, E)$, $a = 1, 2$; $b = 0, 1$,
 has a zero in $\cD\left(\zeta_{j, m}, \bar{\bar\rho}_{j, m}\right)\setminus
 \cD\left(\zeta_{j, m}, \underline{\underline\rho}_{j, m}\right)$.  Therefore (\ref{eq:1618}) follows from Proposition~\ref{prop:15.4}.
\end{proof}

\noindent Assume that the following condition is valid:

\bigskip \noindent{\bf (II.m)}  no determinant $f_{[a,
m-b]}\bigl(\cdot e(nm\omega), \omega, E\bigr), f_{[a, 2m-b]}(\cdot,
\omega, E)$ has more than one zero in any disk \[\cD\bigl(z_0,
r_m\bigr),\; r_m = \exp\left(-(\log m)^A\right),\; z_0 \in
\cA_{\rho_0/2},\; n=0,1,\; a=1,2,\; b=0,1 \]

\medskip

\begin{lemma} If condition (II.m) is valid then:
\begin{enumerate}
\item[(a)] $0 \le k(j, m) \le 1$ for any $j$

\item[(b)] the set
\begin{align*}
&J\bigl(z_0, m\bigr) = \biggl\{j: \big |\zeta_{j, m} - z_0 \big | <
\exp \left(-(\log m)^A\right),\\
& \max\Bigl\{\nu_{f_{[a, m-b]}\bigl(\cdot e(nm\omega), \omega,
E\bigr)}\left(\zeta_{j, m}, \bar\rho_{j, m}\right), \nu_{f_{[a,
2m-b]}(\cdot, \omega,E)}\left(\zeta_{j, m}, \bar\rho_{j,
m}\right)\::\: a = 1,2;\ b = 0,1;\ n=0,1\Bigr\} > 0\biggr\}
\end{align*} satisfies
$\#J(z_0, m) \le 1$ for any $z_0\in \cA_{\rho_0/2}$

\item[(c)] if $j \in J\bigl(z_0, m\bigr)$, then
\[ \left|\nu\left(\cD\left(\zeta_{j, m}, r\right)\right) - k(j,
m)\right | \le r^2 \exp\left(2(\log m)^A\right) \qquad \forall\quad
\underline\rho_{j,m}/4 < r <\exp\left(C(\log m)^A\right) \]
\end{enumerate}
\end{lemma}

\begin{proof} Assertion (a), (b) follow from property (4) of Proposition~\ref{prop:165} and condition~(II.m).
 The proof of property  (c) is analogous to the argument of the previous lemma.
\end{proof}

\section{Relation between concatenation terms of consecutive scales}
\label{sec:17}

 In this section we establish a relation between the
concatenation terms $w_m(z)$ and $w_\ulm(z)$ (see
Section~\ref{sec:16}) with $m \asymp \exp
\left(\ulm^{\delta_1}\right)$, $0 < \delta_1 \ll 1$. Let
$\cD\left(\zeta_{j,m}, \bar\rho_{j, m}\right)$, $j = 1, 2,\dots,
j_m$ and $\cD\left(\zeta_{j, \ulm}, \bar\rho_{j, \ulm}\right)$, $j =
1, 2,\dots, j_{\ulm}$ be the disks defined in
Proposition~\ref{prop:165} for $w_m(z)$ and $w_\ulm (z)$,
respectively. To relate the integers $k(j, m)$ and $k(j, \ulm)$
defined in Proposition~\ref{prop:165} note that due to the avalanche
principle expansion the following assertion is valid:

\begin{lemma}\label{lem:171} There exists $\cF_{\ulm, \omega, E} \subset \capo$ with
$\mes \cF_{\ulm, \omega, E} \le \exp\bigl(-\ulm^{1/2}\bigr)$ such that
\begin{equation}\nn
\Bigm |w_m(z) - w_\ulm\left(ze(m - \ulm)\omega\right)\Bigm | < \exp \left(-\ulm^{1/2}\right)
\end{equation}
for any $z \in \cA_{\rho_0/2} \setminus \cF_{\ulm, \omega, E}$.
\end{lemma}

Combining this assertion with Lemma~\ref{lem:Jdef} one obtains the
following corollary (using the notations of
Proposition~\ref{prop:165}).

\begin{corollary}\label{cor:172}
Let $\nu_m, \nu_\ulm$ be the Riesz measures of $w_m$ and $w_\ulm$,
respectively, which were defined in Section~\ref{sec:16}.  Then
\begin{equation}\nn
\Bigm | \nu_m\bigl(\cD(z, r)\bigr) - \nu_\ulm
\left(\cD\bigl(ze(\ulm-m)\omega\bigr), r\right)\Bigm | \le
\exp\left(- \ulm^{1/2}/2\right)
\end{equation}
for any disk $\cD(z, r) \subset \cA_{\rho_0/2}$, $r \ge
\exp\left(-cm^{1/2}\right)$.
\end{corollary}

For convenience we define $\underline\zeta_{j, \ulm} = \zeta_{j,
\ulm} e\bigl((\ulm - m)\omega\bigr)$.

\begin{lemma}\label{lem:17.3} Assume that condition (II.m) is valid.  Then
\begin{enumerate}
\item[(i)] $\#\left\{j: \zeta_{j, m} \in \cD\left(\underline \zeta_{j_1, \ulm}, \bar\rho_{j_1, \ulm}\right),\;
k(j, m) \ne 0 \right\} \le \min \left(1, k(j_1, \ulm)\right)$ for any $j_1 = 1, 2, \dots, j_\ulm$,

\item[(ii)] $\#\left\{j: \zeta_{j, m} \in \cD\left(\underline\zeta_{j_1, \ulm},
\bar{\bar\rho}_{j_1, \ulm}\right)\setminus \cD\left(\zeta_{j_1, \ulm}, \underline\rho_{j_1, \ulm}/2\right):
k(j, m) \ne 0\right\} = 0$.
\end{enumerate}
\end{lemma}

\begin{proof} Recall that $\bar{\bar\rho}_{j_1,\ulm} \le\exp
\left(-\ulm^{\delta/2}\right)$.  Hence, $\bar\rho_{j_1, \ulm} \le
\exp\left(-\bigl(\log m\bigr)^A\right)$. If $k(j_1, \ulm) = 0$, then
\[ \left |\nu_\ulm\left(\cD\left(\zeta, \bar\rho_{j_1,
\ulm}\right)\right)\right | < \exp\left(-\ulm^\delta\right)\qquad
\forall\; \zeta \in \cD\left(\zeta_{j_1, \ulm}, \bar\rho_{j_1,
\ulm}\right).\] So,\[\left |\nu_\ulm\left(\cD\left(\zeta_{j, m}\cdot
\exp\left(\bigl(m - \bar m\bigr)\omega\right), \bar\rho_{j_1,
\ulm}\right)\right) \right | < \exp\left(-\ulm^\delta\right),
\]
provided $\zeta_{j, m} \in \cD\left(\underline\zeta_{j_1, \ulm},
\bar\rho_{j_1, \ulm}\right)$ in this case. By
Corollary~\ref{cor:172}, $\left |\nu_m \left(\cD\left(\zeta_{j, m},
\bar\rho_{j_1, \ulm}\right)\right)\right | <
2\exp\left(-\ulm^\delta\right)$ which implies $k(j, m) = 0$. Thus
(i) is valid regardless of the value of $k(j_1, \ulm)$.  The proof
of (ii) is similar.
\end{proof}

The following statement is the main result of this section.

\begin{prop}\label{prop:17.4}
Assume that conditions (II.$\ulm$), (II.m) are valid.  Then, using
the notations of the previous lemma one has
\begin{enumerate}
\item[(0)] if $k(j_1, \ulm) = 0$, then there exists $\zeta_{j, m} \in \cD\left(\underline\zeta_{j_1, \ulm},
\underline\rho_{j_1, \ulm}\right)$ with $k(j, m) = 0$ such that
$$
\left |w_m(z) - w_\ulm \left(z_1 e\bigl((m - \ulm)\omega\bigr)\right)\right | \lesssim \exp\left(-\ulm^{1/2}\right)
$$
for any $z \in \cD\left(\underline\zeta_{j_1, \ulm}, \bar\rho_{j_1,
\ulm}\right)\setminus \cD\left(\zeta_{j, m}, \underline\rho_{j,
m}\right)$ and any \[ z_1 \in \cD\left(\zeta_{j_1, \ulm},
\bar\rho_{j_1, \ulm}\right)\setminus \cF_{\ulm, \omega, E},\quad
\mes\cF_{\ulm, \omega, E} < \exp\left(-\ulm^{1/2}\right)\]

\item[(1)] if $k(j_1, \ulm) = 1$, then there exists $\zeta_{j, m} \in \cD\left(\underline\zeta_{j_1, \ulm}, \bar\rho_{j_1, \ulm}\right)$ with $k(j, m) = 1$ such that
$$
\left |w_m(z) - w_\ulm \left(\zeta e\bigl((m -\ulm)\omega\bigr)\right) - \log \left(\tfrac{|z - \zeta_{j, m}|}{|\zeta e\bigl((m - \ulm)\omega\bigr) - \zeta_{j, m}|}\right)\right|
< \exp\left(-\ulm^{1/2}\right)\ ,
$$
for any $z \in \cD\left(\underline\zeta_{j_1, \ulm}, \bar\rho_{j_1, \ulm}\right)\setminus \cD\left(\zeta_{j, m},
\underline\rho_{j, m}\right)$, $\zeta \in \cD\left(\underline\zeta_{j_1, \ulm},
\bar\rho_{j_1, \ulm}\right)\setminus \cF_{\ulm, \omega, E}$.
\end{enumerate}
\end{prop}

\begin{proof} (0)~~By previous lemma $k(j, m) = 0$ for any $\zeta_{j, m} \in \cD\left(\underline \zeta_{j_1, \ulm},
\bar\rho_{j_1, \ulm}\right)$.  Assume $j \in
J\left(\underline\zeta_{j_1, \ulm}, m\right)$. Then condition (I.m)
is valid in the annulus $\cD\left(\zeta_{j, m}, \bar{\bar\rho}_{j,
\ulm}\right) \setminus \cD\left(\zeta_{j, m},
\underline{\underline\rho}_{j, m}\right)$.  Due to
Proposition~\ref{prop:14.10}
\begin{equation}
\nn \left |w_m (\zeta) - w_m(z) \right | \le \underline\rho^2_{j,
m}\Big / \bar\rho^2_{j_1, \ulm} \le \exp\left(-m^{\delta/2}\right)
\end{equation}
for any $z, \zeta \in \cD\left(\zeta_{j, m}, 2\bar\rho_{j,
\ulm}\right)\setminus \cD\left(\zeta_{j, m}, \underline\rho_{j,
m}\right)$, since $k(j, m) = 0$.  By Lemma~\ref{lem:171}
$$
\left |w_m(z) - w_\ulm \left(ze\bigl((m-\ulm)\omega\bigr)\right)\right | \le \exp\left(-\ulm^{1/2}\right)
$$
for any $z \in \cA_{\rho_0/2} \setminus \cF_{\ulm, \omega, E}$.  Therefore,
\begin{equation}
\left |w_m(z) - w_\ulm\left(z_1 e\bigl((m - \ulm)\omega\bigr)\right)\right | \lesssim \exp\left(-\ulm^{1/2}\right)
\end{equation}
for any $z \in \cD\left(\zeta_{j_1, \ulm}, \bar\rho_{j_1,
\ulm}\right)\setminus \cD\left(\zeta_{j, m}, \underline\rho_{j,
m}\right)$, $z_1 \in \cD\left(\zeta_{j, m}, \rho_{j_1,
m}\right)\setminus \cF_{\ulm, \omega, E}$.  If
$J\left(\underline\zeta_{j_1, m}, m\right) = \emptyset$, then the
assertion holds for any $z_{j, m} \in \cD\left(\underline\zeta_{j_1,
\ulm}, \bar\rho_{j_1, \ulm}\right)$. That proves~(0). The proof of
part~(1) is similar.
\end{proof}

In order to prove Theorem~\ref{thm:1}  we use the assertion of
Theorem~\ref{thm:4} to produce the needed estimates on concatenation
terms in this case.  Assume that the following condition is valid
for some $m$ and $z_0 = e(x_0)$, $E \in \IC$:
\begin{equation}\label{eq:175}
\nu_{f_{2m}(\cdot, \omega, E)}(z_0, R_0) \le k_0
\end{equation}
where $R_0 \asymp \exp\left(-(\log m)^{A_0}\right)$, $k_0$ is the
constant of Theorem~\ref{thm:4}. To estimate the concatenation term
$w_m(z)$ recall first of all that
\begin{equation}
\nn \log \big \|M_m(z, \omega, E) \big \|\, \big
\|M_m\bigl(ze(m\omega), \omega, E\bigr) \big \| \le 2L(\omega, E) m
+ (\log m)^B
\end{equation}
for any $z \in \cD(z_0, R_0)$ due to
Proposition~\ref{prop:logupper}.  The same upper estimate is valid
for $\log \big |f_{2m}(z, \omega, E)\big |$. On the other hand there
always exists $z_1 \in \cD(z_0, R_0/4)$ such that
$$
\log \big |f_{2m}(z_1, \omega, E) \big | > 2L(\omega, E) m - (\log m)^{B_1}\ .
$$
Hence, due to Cartan's estimate for analytic functions
\begin{gather*}
\log \big |f_{2m}(z, \omega, E)\big | > 2m L(\omega, E) - (\log m)^C\\
 + k_0 \log\left(\min_j |z - \zeta_j|\right)
\end{gather*}
for any $z \in \cD(z_0, R_0/2)$, where $\zeta_j$ are the zeros of $f_{2m}(\cdot, \omega, E)$ belonging to $\cD(z_0, R_0)$.  Thus the following estimate is valid:

\begin{lemma}\label{lem:17.5} If condition (\ref{eq:175}) is valid, then
$$
w_m(z) \ge -(\log m)^C + k_0 \log \left(\min_j |z - \zeta_j|\right)
$$
where $\zeta_j \in \cD(z_0, R_0)$, $j = 1,2,\dots, j_0$, $j_0 \le
k_0$.
\end{lemma}

\section{Proof of Theorem~\ref{thm:1}}\label{sec:18}

The first part of Theorem~\ref{thm:1} will follow from
Corollary~\ref{cor:153} on concatenation terms.  To use this
corollary we consider concatenation terms
$$
\cW_{N, k}\bigl(e(x)\bigr) = {\big \|M_{[1, k]}\bigl(e(x), \omega, E + i\eta\bigr) \big \|\, \big \|M_{[k+1, N]}\bigl(e(x), \omega, E + i\eta\bigr)\big \|\over \big \|M_N\bigl(e(x), \omega, E+i\eta\bigr)\big \|}
$$
with $\eta = N^{-1-\beta}$, $0 < \beta \ll 1$. Given $x_0$, due to
Theorem~\ref{thm:4} there exists $m = m(x_0)$, $m \asymp(\log N)^B$
such that
\begin{equation}\label{eq:181}
\nu_{f_{[-m+1, m]}(\cdot, \omega, E + i\eta]}\bigl(e(x_0),
\exp\bigl(-(\log m)^A\bigr)\bigr) \le k_0\ .
\end{equation}

\begin{lemma}\label{lem:1810}
 There exist $\cB_{N, \omega, E, \eta, m} \subset \tor$, $\mes\cB_{N, \omega, E, \eta, m} < \exp\bigl(-m^{1/2}\bigr)$
such that
\begin{equation}\label{eq:18.2}
\left| \log \cW_{N, k}\bigl(e(x)\bigr) - \log \cW_{m}\left(e\bigl(k - m)\omega + x\bigr)\right) \right| \le 1
\end{equation}
for any $x \in \tor \setminus \cB_{N, \omega, E, \eta, m}$, provided
$Cm < k < N - Cm$, here $\cW_m(z)$ stands for concatenation term
\begin{equation}\nn
{\big \|M_{m}(z, \omega, E + i \eta)\big \|\, \big \|
M_m\bigl(ze(m\omega), \omega, E+ i\eta\bigr) \big \| \over \big
\|M_{2m}(z, \omega, E + i\eta)\big \|}\ .
\end{equation}
\end{lemma}

The proof of this lemma is based on the avalanche principle
expansion. Due to condition (\ref{eq:181}) one can apply
Lemma~\ref{lem:17.5} to evaluate $w_m(z) = - \log \cW_m(z)$,
provided $z \in \cD\left(e(x_0), R_1\right)$, $R_1 =
\exp\left(-(\log m)^A\right)$ :
\begin{equation}\label{eq:18.40}
\cW_m(z) \le \left((\log m)^{A_1}\right) \left(\min_{1 \le j \le j_0} |z - \zeta_j|\right)^{-k_0}
\end{equation}
where $j_0 \le k_0$.  Recall that
$$
\min\limits_{1 \le k \le N} \|k\omega \| = \vep(\omega, N) \lesssim N^{-1} (\log N)^{-2}
$$
since $\omega \in \tor_{c,a}$.  In particular
\begin{equation}\label{eq:18.5}
\#\left\{k: \|e(x + k\omega ) - \zeta\| < \kappa \right\} \le \kappa \cdot \vep(\omega, N)^{-1}
\end{equation}
for any $x \in \tor$, $\zeta \in \IC$, $\kappa > 0$.  Finally, note that
\begin{equation}\label{eq:18.6}
\sum_{|e(x + k\omega) - \zeta| > \kappa} \bigl |e(x + k\omega) - \zeta\big |^{-k_0} \le C\vep(\omega, N)^{-1} \kappa^{1- k_0}\ .
\end{equation}

\begin{proof}[Proof of Theorem 1.1] Using the above notations set $\kappa = \eta^{1/k_0}$,
\[\cK(x_0) = \left\{k: \min\limits_j \bigl|e(x + (k-m)\omega\bigr) - \zeta_j \big| < \kappa\right\}.\]
Combining (\ref{eq:18.2}), (\ref{eq:18.40}), (\ref{eq:18.5}),
(\ref{eq:18.6}) one obtains for $x \notin \cB_{N, \omega, E, \eta,
m}$, $\mes\cB_{N,\omega, E, \eta, m} \le \exp\left(-(\log
N)^A\right)$
\begin{align}
& \sum_{e\bigl(x + (k-m) \omega\bigr) \in \cD\bigl(e(x_0), R_1), k \notin \cK(x_0)} \cW_{N, k}\bigl(e(x)\bigr)\nn \\
&\quad \le \exp\left((\log m)^{A_2}\right) \vep(\omega, N)^{-1} \kappa^{1-k_0}\nonumber\\
&\quad \le \exp\left((\log\log N)^{B_1}\right)\cdot \eta^{\left({1\over k_0} -1\right)}\label{eq:18.7}\\
& \#\cK(x_0) \le N \eta^{1/k_0} \bigl(\log N\bigr)^2\ ,\quad R_1
\asymp \exp\left((\log\log N)^{-B_1}\right)\label{eq:18.8}
\end{align}
Since (\ref{eq:18.7}), (\ref{eq:18.8}) are valid for any $x_0 \in
\tor$,
\begin{equation}
\label{eq:18.9} \sum \cW_{N, k} \bigl(e(x)\bigr) \le
\exp\left((\log\log N)^{B_2}\right) \eta^{\left({1\over k_0}
-1\right)}
\end{equation}
for any $x \in \tor \setminus \cB_{N, \omega, E, \eta}$, $\mes
\cB_{N, \omega, E, \eta} \le \exp\left(-(\log N)^A\right)$.  Due to
Lemma~\ref{lem:152}, (\ref{eq:18.9}) implies
$$
\#\left\{E_j^{(N)} (x) \in (E - \eta, E + \eta)\right\} \le
\eta^{\frac{1}{k_0}-\eps} N
$$
$x \in \tor\setminus \cB_{N, \omega, E, \eta}$. That established the
first part of Theorem~\ref{thm:1}.
\end{proof}

To prove the second part of Theorem~\ref{thm:2} we need the
following simple fact:

\begin{lemma}
\label{lem:12.7}
For any $N, t, N'$, any interval $(E', E'')$, and any $x
\in \tor$ one has
\begin{multline}
\nn
\#\left(\rsp\Bigl(H^{(D)}_{Nt + N'} (x, \omega)\Bigr)\cap(E',
E'')\right)\le \sum^{t-1}_{k=0} \#\left(\rsp\Bigl(H^{(D)}_N (x +
kN\omega, \omega)\Bigr)\cap(E', E'')\right) + 2t + N'
\end{multline}
\end{lemma}

\begin{proof} Clearly
\begin{equation}
\label{eq:12.29}
H_N(x, \omega) = \bigoplus^{t-1}_{k=0}\ H_N^{(D)} (x + kN\omega, \omega)
+ T + S
\end{equation}
where $\rank T  = 2t$, $\rank S = N'$.  Since the operators in (\ref{eq:12.29}) are
Hermitian, the assertion follows from the minimax principle (see~\cite{Bhat}).
\end{proof}

\begin{corollary}
\label{cor:12.8} Let $N, \omega, E, \eta$ be as in Theorem 1.  Then for
any $\widehat N = tN + N'$, $t > N$, $N' \le N$ one has
\begin{equation}
\label{eq:12.30} {1\over \widehat N} \int_\tor \#\left(\bigl(\rsp
H_{\widehat N}(x, \omega)\bigr)\cap \bigl(E - \eta, E +
\eta\bigr)\right) dx \le \eta^{1-\eps}\ .
\end{equation}
\end{corollary}
The second part of Theorem~\ref{thm:1} follows from (\ref{eq:12.30})
by letting $\widehat N\to \infty$.

\section{Proof of Theorems \ref{thm:2}, \ref{thm:3}}\label{sec:19}

Let $m^{(1)}, m^{(2)}, \dots, m^{(t)}$ be integers such that
\begin{equation}
\label{eq:19.1} m^{(s+1)} \asymp
\exp\left(\bigl(m^{(s)}\bigr)^{\delta_1}\right)\ ,\quad s = 1,
2,\dots, t-1\ .
\end{equation}
Assume that
\begin{equation}\label{eq:19.2}
\text{for some $\omega, E$ condition (II. $m^{(s+1)}$) of
Section~\ref{sec:16} is valid for any $s = 1, 2,\dots, t-1$.}
\end{equation}

\begin{lemma}
\label{lem:19.1}
 Let $\cD\left(\zeta_{j, m^{(s)}}, \bar\rho_{j, m^{(s)}}\right)$,  $j = 1, 2,\dots, j_{m^{(s)}}$
be the disks defined in Section~\ref{sec:16} for the concatenation
term $\cW_{m^{(s)}}(z)$. For each $\zeta_{j^{(1)}, m^{(1)}}$ with
$k\left(j^{(1)}, m^{(1)}\right) = 0$, there exist $\zeta_{j_n^{(t)},
m^{(t)}} \in \cD\left(\underline\zeta_{j^{(1)},m^{(1)}}, (t-1)
\underline\rho_{j^{(1)}_1, m^{(1)}}\right)$, $n = 1, 2,\dots$,
$\underline\zeta_{j^{(1)}, m^{(1)}} = \zeta_{j^{(1)}, m^{(1)}} \cdot
e\left((m^{(1)} - m^{(t)})\omega\right)$, such that $k\left(j^{(t)},
m^{(t)}\right) = 0$,
\begin{equation}\label{eq:19.3}
\left|w_{m^{(t)}}(z) - w_{m^{(t)}}\left(z_1 e\bigl((m^{(t)} -
m)\omega\bigr)\right)\right| \lesssim \sum^{t-1}_{s=1}\,
\exp\left(-(m^{(s)})^{1/2}\right)
\end{equation}
for any $z \in \cD\left(\underline\zeta_{j^{(1)}, m^{(1)}}, \bar\rho_{j^{(1)}, m^{(1)}}\right)\setminus \left(\cup \cD\left(\zeta_{j_n^{(t)}, m^{(t)}}, \rho_{j_n^{(t)}, m^{(t)}}\right)\right)$ and $z_1 \in \cD\left(\zeta_{j^{(1)}, m^{(1)}}, \bar\rho_{j^{(1)}, m^{(1)}}\right) \setminus \cF_{m^{(1)}, \omega, E}$, $\mes\left(\cF_{m^{(1)}, \omega, E}\right) \lesssim \exp\left(-(m^{(1)})^{1/2}\right)$.
\end{lemma}

\begin{proof}
The proof goes by induction over the $t = 2, 3,\dots$.  If $t = 2$,
then the assertion is valid due to part (0) of
Proposition~\ref{prop:17.4}. So, there is $\zeta_{j_1^{(2)},
m^{(2)}} \in \cD\left(\underline\zeta'_{j^{(1)}, m^{(1)}},
\underline\rho_{j^{(1)}, m^{(1)}}\right)$,
$\underline\zeta'_{j^{(1)}, m^{(1)}} = \zeta_{j^{(1)}, m^{(1)}}
e\left((m^{(1)} - m^{(2)})\omega\right)$, with $k \left(j_1^{(2)},
m^{(2)}\right) = 0$, such that
\begin{equation}\label{eq:19.4}
\left| w_{m^{(2)}}(z') - w_{m^{(1)}} \left(z'_1e\bigl((m^{(2)} -
m^{(1)})\omega\bigr)\right)\right| \le \exp
\left(-(m^{(1)})^{1/2}\right)
\end{equation}
for any $z' \in \cD\left(\underline\zeta'_{j^{(1)}, m^{(1)}}, \bar\rho_{j^{(1)}, m^{(1)}}\right)
\setminus \cD\left(\zeta_{j_1^{(2)}, m^{(2)}}, \underline\rho_{j_1^{(2)}, m^{(2)}}\right)$
and any $z'_1 \in \cD\left(\zeta'_{j^{(1)}, m^{(1)}}, \bar\rho_{j^{(1)}, m^{(1)}}\right)\setminus
\cF_{m^{(1)}, \omega, E}$, $\mes \cF_{m^{(1)}, \omega, E} < \exp\left(-(m^{(1)})^{1/2}\right)$.
Since $k\left(j^{(2)}, m^{(2)}\right) = 0$, one can use inductive assumption and find
$\zeta_{j_n^{(t)}, m^{(t)}} \in \cD\left(\underline\zeta_{j_1^{(2)}, m^{(2)}},
(t-2)\bar\rho_{j^{(2)}, m^{(2)}}\right)$, $n = 1, 2,\dots$,
$\underline\zeta_{j_1^{(2)}, m^{(2)}} = \zeta_{j_1^{(2)}, m^{(2)}}\cdot e\left((m^{(2)} - m^{(t)})\omega\right)$
with $k \left(j_n^{(t)}, m^{(t)}\right) = 0$ such that
\begin{equation}\label{eq:19.5}
\left| w_{m^{(t)}}(z) - w_{m^{(2)}}\left(z_1''e\bigl((m^{(t)} -
m^{(2)})\omega\bigr) \right)\right | \lesssim \sum^{t-1}_{s=2}
\exp\left(-(m^{(s)})^{1/2}\right)
\end{equation}
for any $z \in \cD\left(\underline\zeta_{j^{(2)}, m^{(2)}}, \bar\rho_{j^{(2)}, m^{(2)}}\right)
\setminus \left(\bigcup\limits_n \cD\left(\zeta_{j_n^{(t)}, m^{(t)}}, \underline\rho_{j_1^{(t)},
 m^{(t)}}\right)\right)$ and any $z''_1 \in \cD\left(\zeta_{j^{(2)}, m^{(2)}}, \bar\rho_{j^{(2)}, m^{(2)}}\right)
 \setminus \cF_{m^{(2)}, \omega, E}$, $\mes\cF_{m^{(2)}, \omega, E} \lesssim \exp\left(-(m^{(2)})^{1/2}\right)$.
 Since $\zeta_{j^{(t)}, m^{(t)}} \in \cD\left(\zeta_{j^{(2)}} e\left((m^{(2)} - m^{(t)})\omega\right),
 (t-2)\underline\rho_{j^{(2)}, m^{(2)}}\right)$, $\zeta_{j^{(2)}, m^{(2)}} \in
 \cD\left(\zeta_{j^{(1)}, m^{(1)}} e\left((m^{(1)} - m^{(2)})\omega\right), \underline\rho_{j^{(1)}, m^{(1)}}\right)$, one has
$$
\zeta_{j^{(t)}, m^{(t)}} \in \cD\left(\zeta_{j^{(1)}, m^{(1)}}
e\left((m^{(1)} - m^{(t)})\omega\right), (t-1)\rho_{j^{(1)},
m^{(1)}}\right)\ .
$$
Let $z''_1 \in \cD\left(\underline\zeta_{j^{(2)}, m^{(2)}},
\bar\rho_{j^{(2)}, m^{(2)}}\right)\setminus
\left(\cD\left(\underline\zeta_{j_1^{(2)}, m^{(2)}},
\underline\rho_{j_1^{(2)}, m^{(2)}}\right)\cup \cF_{m^{(2)}, \omega,
E}\right)$ and $z_1 \in \cD\left(\zeta_{j^{(1)}, m^{(1)}},
\bar\rho_{j^{(1)}, m^{(1)}}\right)\setminus \cF_{m^{(1)},\omega, E}$
be arbitrary. Then (\ref{eq:19.5}) is valid and (\ref{eq:19.4}) is
also valid with $z' = z''_1 e\left((m^{(t)} -
m^{(2)})\omega\right)$, $z_1'=z_1e\left((m^{(t)} -
m^{(2)})\omega\right)$. Hence, (\ref{eq:19.3}) is valid for any $z
\in \cD\left(\underline\zeta_{j_1^{(2)}, m^{(2)}}, \bar
\rho_{j_1^{(2)}, m^{(2)}}\right)\setminus \left(\bigcup\limits_n
\cD\left(\zeta_{j_n^{(t)}, m^{(t)}},\underline\rho_{j_n^{(t)},
m^{(t)}}\right)\right)$ and any $z_1 \in \cD\left(\zeta_{j^{(1)},
m^{(1)}}, \bar\rho_{j^{(1)}, m^{(1)}}\right)\setminus \cF_{m^{(1)},
\omega, E}$.  By Proposition~\ref{prop:165} the disk
$\cD\left(\underline\zeta_{j^{(1)}, m^{(1)}},
\underline\rho_{j^{(1)}, m^{(1)}}\right)$ can be covered by some
$\cD\left(\zeta_{j_q^{(2)}, m^{(2)}}, \bar\rho_{j_q^{(2)},
m^{(2)}}\right)$, $q = 1, \dots$, $\zeta_{j_q^{(2)}, m^{(2)}} \in
\cD\left(\underline\zeta_{j^{(1)}, m^{(1)}},
2\underline\rho_{j^{(1)}, m^{(1)}}\right)$.  By
Lemma~\ref{lem:17.3}, $k\left(j_q^{(2)}, m^{(2)}\right) = 0$, $q =
1,\dots$.  Therefore, one can apply inductive arguments to each disk
$\cD\left(\zeta_{j_q^{(2)}, m^{(2)}}, \bar\rho_{j_q^{(2)},
m^{(2)}}\right)$.
\end{proof}

\begin{lemma} For each $\zeta_{j^{(1)}, m^{(1)}}$ with $k\left(j^{(1)}, m^{(1)}\right) = 1$ there exist
$\zeta_{j_n^{(t)}, m^{(t)}} \in \cD\left(\underline\zeta_{j^{(1)}, m^{(1)}}, \bar\rho_{j^{(1)}, m^{(1)}}\right)$,
$n = 1, 2,\dots$, $k\left(j_1^{(t)}, m^{(t)}\right) = 1$, $k\left(j_n^{(t)}, m^{(t)}\right) = 0$, $n = 2,\dots$ such that
\begin{equation}
\label{eq:19.6}
\begin{aligned}
 & \biggl| w_{m^{(t)}}(z) - w_{m^{(1)}}\left(\zeta e\bigl((m^{(t)} - m^{(1)})\omega\bigr)\right)\\
 &\quad - \log {\big |z - \zeta^{(t)}_{j_1, m^{(t)}}\big |\over
\big |\zeta e\bigl((m^{(t)} - m^{(1)})\omega\bigr) - \zeta^{(t)}_{j_1^{(t)}, m^{(t)}}\big |}\biggr |
\le \exp\left(-(m^{(1)})^{1/2}\right)
\end{aligned}
\end{equation}
for any $z \in \cD\left(\underline\zeta_{j^{(1)}, m^{(t)}}, \bar\rho_{j^{(1)}, m^{(1)}}\right)\setminus
\left(\cup \cD\left(\zeta_{j_n^{(t)}, m^{(t)}}, \rho_{j_n^{(t)}, m^{(t)}}\right)\right)$ and any
$\zeta \in \cD\left(\zeta_{j^{(1)}, m^{(1)}}, \bar\rho_{j^{(1)}, m^{(1)}}\right)\setminus \cF_{m^{(1)}, \omega, E},
\mes\cF_{m^{(1)}, \omega, E} < \exp\left(-(m^{(1)})^{1/2}\right)$.
\end{lemma}
\begin{proof} The proof is similar to the one in the previous lemma. There exists unique
\[ \zeta_{j_1^{(2)}, m^{(2)}} \in \cD\left(\underline\zeta'_{j^{(1)}, m^{(1)}}, \underline\rho_{j^{(1)},
m^{(1)}}\right) \text{\  \ where\ \ } \underline\zeta'_{j^{(1)},
m^{(1)}} = \zeta_{j^{(1)}, m^{(1)}} e\left(\bigl(m^{(1)} -
m^{(2)}\bigr)\omega\right) \]
 such that $k\left(j_1^{(2)}, m^{(2)}\right) = 1$ and
\begin{equation}
\label{eq:19.7}
\begin{aligned}
&\biggl|
w_{m^{(2)}}(z') - w_{m^{(1)}}\left(\zeta'e \bigl((m^{(2)} - m^{(1)})\omega\bigr)\right) \\
&\quad -
\log {\big |z' - \zeta_{j_1^{(2)}, m^{(2)}}\big |\over
\big |\zeta'e \bigl((m^{(2)} - m^{(1)})\omega\bigr) - \zeta_{j^{(2)}, m^{(2)}}\big |}
\biggr | \le \exp\left(-(m^{(1)})^{1/2}\right)
\end{aligned}
\end{equation}
for any $z' \in \cD\left(\underline\zeta_{j^{(1)}, m^{(1)}},
\bar\rho_{j^{(1)},m^{(1)}}\right)\setminus\cD\left(\zeta_{j_1^{(2)},
m^{(2)}}, \underline\rho_{j^{(2)}, m^{(2)}}\right)$ and any $\zeta'
\in \cD\left(\zeta_{j^{(1)},
m^{(1)}},\bar\rho_{j^{(1)},m^{(1)}}\right) \setminus \cF_{m^{(1)},
\omega, E}$, $\mes\cF_{m^{(1)}, \omega, E} <
\exp\left(-(m^{(1)})^{1/2}\right)$.  By the inductive assumption
there exist $\zeta_{j^{(t)}_n, m^{(t)}}$, $n = 1, 2,\dots$ such that
$k\left(j_1^{(t)}, m^{(t)}\right) = 1$, $k \left(j_n^{(t)},
m^{(t)}\right) = 0$, $n = 2,\dots$ and
\begin{equation}
\label{eq:19.8}
\begin{aligned}
& \left|w_{m^{(t)}}(z) - w_{m^{(2)}} \left(\zeta''e\bigl((m^{(t)} -
m^{(2)})\omega\bigr)\right) - \log {\big |z - \zeta_{j_1^{(t)},
m^{(t)}}\big |\over
\big |\zeta''e \bigl((m^{(t)} - m^{(2)})\omega\bigr) - \zeta_{j_1^{(t)}, m^{(t)}}\big |}\right|\\
& \le \sum^{t-1}_{s=2}\, \exp\left(-m^{(s)}\right)
\end{aligned}
\end{equation}
for any $z \in \cD\left(\underline\zeta_{j_1^{(2)}, m^{(2)}}, \bar\rho_{j_1^{(2)}, m^{(2)}}\right)\setminus
\left(\bigcup\limits_n \cD\left(\zeta_{j_n^{(t)}, m^{(t)}}, \underline\rho_{j_n^{(t)}, m^{(t)}}\right)\right)$
and any $\zeta'' \in \cD\left(\zeta_{j_1^{(2)}, m^{(2)}}, \bar\rho_{j_1^{(2)}, m^{(2)}}\right) \setminus \cF_{m^{(2)}, \omega, E}$,
where
$\underline\zeta_{j_1^{(2)}, m^{(2)}} = \zeta_{j_1^{(2)}, m^{(2)}} e\left(\bigl(m^{(t)} - m^{(2)}\bigr)\omega\right),
 \mes \cF_{m^{(2)}, \omega, E} < \exp\left(-(m^{(2)})^{1/2}\right)$.
 Substituting $\zeta' = \zeta e\left(\bigl(m^{(t)} - m^{(2)}\bigr)\omega\right)$
 and $z' = \zeta''e \left(\bigl(m^{(t)} - m^{(2)}\bigr)\omega\right)$ in (\ref{eq:19.7}) and combining it with (\ref{eq:19.8})
 one obtains
\begin{equation*}
\begin{aligned}
& \biggr|w_{m^{(t)}} (z) - w_{m^{(1)}} \left(\zeta e\bigl((m^{(t)} -
m^{(1)})\omega\bigr)\right) -
 \log {\big |z - \zeta_{j_1^{(t)}, m^{(t)}}\big |\over
\big |\zeta e\left(\bigl(m^{(t)} - m^{(1)}\bigr)\omega\right) - \zeta_{j_1^{(t)}, m_1^{(t)}}\big |}\biggl |
\le \sum^{t-1}_{s=1}\, \exp\left(-(m^{(s)})^\sigma\right)\\
&\quad + \left| \log {\big |\zeta'' e\bigl((m^{(t)} - m^{(2)})\omega\bigr) - \zeta_{j_1^{(2)}, m^{(2)}}\big |\over
\big |\zeta'' e\bigl((m^{(t)} - m^{(2)})\omega\bigr) - \zeta_{j_1^{(t)}, m^{(t)}}\big |}\right|
\end{aligned}
\end{equation*}
for any $z \in \cD\left(\underline\zeta_{j_1^{(2)}, m^{(2)}},
\bar\rho_{j_1^{(2)}, m^{(2)}}\right)\setminus \left(\bigcup\limits_n
\cD\left(\zeta_{j_n^{(t)}, m^{(t)}}, \underline\rho_{j_n^{(t)},
m^{(t)}}\right)\right)$,
\[\zeta \in \cD\left(\zeta_{j_1^{(2)},
m^{(2)}}, \bar\rho_{j_1^{(2)}, m^{(2)}}\right)\setminus
\left(\cF_{m^{(1)}, \omega, E} \cup \cD\left(\zeta_{j_1^{(2)},
m^{(2)}}, \underline\rho_{j_1^{(2)}, m^{(2)}}\right)\right), \zeta''
\in \cD\left(\zeta_{j_1^{(2)}, m^{(2)}}, \bar\rho_{j_1^{(2)},
m^{(2)}}\right)\setminus \cF_{m^{(2)}, \omega, E}.\] Since
$\left|\zeta_{j_1^{(t)}, m^{(t)}} - \zeta_{j_1^{(2)},
m^{(2)}}\right| < \underline\rho_{j_1^{(2)}, m^{(2)}}$, the
logarithmic term on the right-hand side here does not exceed
$\exp\left(-(m^{(2)})^\sigma\right)$, provided $\zeta'' \notin
\cD\left(\zeta_{j_1^{(2)}, m^{(2)}},
\sqrt{\underline\rho_{j_1^{(2)}, m^{(2)}}}\right)$ (recall that
$\underline\rho_{j, m} < \bar\rho_{j, m}^A$).  This proves
(\ref{eq:19.6}) for such $z, \zeta$. To finish the proof recall that
$\cD\left(\zeta_{j^{(1)}, m^{(1)}}, \bar\rho_{j^{(1)},
m^{(1)}}\right)$ can be covered by disks $\cD\left(\zeta_{j_q^{(2)},
m^{(2)}}, \bar\rho_{j_q^{(2)}, m^{(2)}}\right)$, $q = 1, 2,\dots$
with $k \left(j_q^{(2)}, m^{(2)}\right) = 0$ for $q = 2, 3,\dots$.
In each disk $\cD\left(\zeta_{j_q^{(2)}, m^{(2)}},
\bar\rho_{j_q^{(2)}, m^{(2)}}\right)$, $q \ge 2$ one can use the
assertion of this lemma applied to $m^{(2)},\dots, m^{(t)}$. Thus
there exist $\zeta_{j_n^{(t)}, m^{(t)}}$, $n \in \cN_q$ with $k
\left(j_n^{(t)}, m^{(t)}\right) = 0$ such that
$$
\left| w_{m^{(t)}}(z) - w_{m^{(2)}}\left(z_1 e\bigl((m^{(t)} -
m^{(2)})\omega\bigr)\right) \right | \le \sum^t_{s=2}\,
\exp\left(-(m^{(s)})^\sigma\right)
$$
for any $z\in \cD\left(\underline\zeta'_{j_q^{(2)},
m^{(2)}},\bar\rho_{j_q^{(2)}, m^{(2)}}\right) \setminus
\left(\bigcup\limits_{n \in \cN_q}\cD\left(\zeta_{j_n^{(t)},
m^{(t)}}, \underline\rho_{j_n^{(t)}, m^{(t)}}\right)\right)$, $z_1
\in \cD\left(\zeta_{j_q^{(2)}, m^{(2)}}, \bar\rho_{j_q^{(2)},
m^{(2)}}\right) \setminus \cF_{m^{(2)}, \omega, E}$, where
$\zeta'_{j_q^{(2)}, m^{(2)}} = \zeta_{j_q^{(2)}, m^{(q)}} \cdot
e\left(\bigl(m^{(2)} - m^{(2)}\bigr)\omega\right)$,
$\mes\cF_{m^{(2)}, \omega, E} < \sum\limits_{2 \le s \le t}\,
\exp\left(-(m^{(s)})^{1/2}\right)$. That implies (\ref{eq:19.6}) for
$z \in \cD\left(\underline\zeta'_{j_q^{(2)}, m^{(2)}},
\bar\rho_{j_q^{(2)}, m^{(2)}}\right)$.
\end{proof}

\begin{corollary}\label{cor:19.3}
 For each disk $\cD\left(\underline\zeta_{j^{(1)}, m^{(1)}}, \bar\rho_{j^{(1)}, m^{(1)}}\right)$,
with $k\left(j^{(1)}, m^{(1)}\right) = 1$, there exist $\zeta_n
\left(j^{(1)}, m^{(1)}\right)$ belonging to this disk such that
$f_{m^{(t)}}\left(\zeta_0(j^{(1)}, m^{(1)}),\omega, E\right) = 0$
and for any \[ z \in \cD\left(\underline\zeta_{j^{(1)}, m^{(1)}},
\bar\rho_{j^{(1)}, m^{(1)}}\right)\setminus
\left(\bigcup\limits_{n=0,\dots}\cD\Bigl(\zeta_n\left(j^{(1)},
m^{(1)}\right), \exp\left(-(m^{(t)})^\sigma\right)\Bigr)\right)\]
one has
\begin{equation}\nn
\left|\cW_{m^{(t)}} (z) \right| \lesssim \left| z - \zeta_0\left(j^{(1)}, m^{(1)}\right)\right|^{-1} \exp\left((\log m^{(1)})^A\right)
\end{equation}
where $\cW_m(z) = \exp\left(-\cW_m(z)\right)$, i.e.
$$
\cW_m(z) = {\big \|M_m\left(ze(m\omega), \omega, E\right) \big \|\cdot \big \| M_m(z, \omega, E)\big \|\over
\big \|M_{2m} \left(ze(m\omega), \omega, E\right)\big \|}\ .
$$
\end{corollary}

\begin{remark}
\label{rem:194} Recall that since $V(e(x))$ assumes only real values
for $x\in\tor$, one has
\[ \dist\Big({\rm
sp}\Big(H_m(e(x_0+iy_0),\omega)\Big),\IR\Big)\lesssim |y_0|
\]
for any $m$, $x_0\in\tor$, $|y_0|<\rho_0$. Therefore, if
$E=E_0+i\eta_0$, $e_0,\eta_0\in\IR$ and if
$f_m(e(x_0+iy_0),\omega,E)=0$, then $|\eta_0|\lesssim |y_0|$.
\end{remark}

\begin{proof}[Proof of Theorem~\ref{thm:2}]
Let $m^{(1)}, m^{(2)}, \dots, m^{(t)}$ be as in (\ref{eq:19.1}).
Recall that due to Proposition~\ref{prop:Ej_sep} given $m$, there
exists $\Omega _m \subset \tor$, $\mes \Omega_m < \exp\left(-(\log
m)^{A_2}\right)$, $\compl(\Omega_m) < \exp\left((\log
m)^{A_1}\right)$, $A_1 \ll A_2$ such that for any $\omega \in
\tor_{c,a} \setminus \Omega_m$ there exists $\cE_{m, \omega} \subset
\IR$, $\mes \cE_{m, \omega} < \exp\left(-(\log m)^{A_2}\right)$,
$\compl \cE_{m, \omega} < \exp\left((\log m^{(t)})^{A_1}\right)$
such that for any $E = E_0 + i\eta$, $E_0 \in \IR \setminus \cE_{m,
\omega}$, $\eta \asymp \exp\left(-(m^{(t)})^\delta\right)$ condition
(II.m) is valid.  Let $\Omega\left(m^{(t)}\right) =
\bigcup\limits^t_{s=1} \Omega_{m^{(s)}}$, $\cE_\omega
\left(m^{(t)}\right) = \bigcup\limits^t_{s=1} \cE_{m^{(s)},\omega}$
for $\omega \in \tor_{c, a} \setminus \Omega\left(m^{(t)}\right)$.
Thus, the hypotheses of Proposition~\ref{prop:17.4} are valid for
such $E$. Let $\cD\left(\zeta_{j^{(1)}, m^{(1)}}, \bar\rho_{j^{(1)},
m^{(1)}}\right)$ be an arbitrary disk defined by
Proposition~\ref{prop:17.4} with $k\left(j^{(1)}, m^{(1)}\right) =
1$. By Corollary~\ref{cor:19.3}
\begin{equation}\nn
\left|\cW_{m^{(t)}}(z) \right|^2 \lesssim \left|z -
\zeta_0(j^{(1)},m^{(1)})\right|^{-2} \exp\left((\log
m^{(1)})^A\right)
\end{equation}
for any $z \in \cD\left(\underline\zeta_{j^{(1)}, m^{(1)}},
\bar\rho_{j^{(1)}, m^{(1)}}\right) \setminus \left(\bigcup\limits_n
\cD\left(\zeta_n(j^{(1)},
m^{(1)})\right)\exp\left(-(m^{(t)})^\sigma\right)\right)$. Hence,
for any $z \in \cA_{\rho_0/2}$ such that $z \notin \bigcup\limits_{0
\le j \le \exp\left((m^{(t)})^{\sigma/2}\right)}
\left(\bigcup\limits_n \cD\left(\zeta_n(j^{(1)},
m^{(1)})e(-j\omega)\right),
\exp\left(-(m^{(t)})^\sigma\right)\right)$
one has
\begin{equation}
\nn
\begin{aligned}
& \sum_{\ ze(j\omega) \in \cD\left(\underline\zeta(j^{(1)}, m^{(1)}), \rho_{j^{(1)}, m^{(1)}}\right)}
\Bigl |\cW_{m^{(t)}}\left(ze(j\omega)\right)\Bigr|^2\\
&\lesssim \left(\min_{j} \left|ze(j\omega) - \zeta_0 \left(j^{(1)},
m^{(1)}\right)\right|\right)^{-1} \cdot \left(\min_{j_1 \ne j_2}
\big \|(j_1 - j_2)\omega \big \|\right)^{-1} \exp\left((\log
m^{(1)})^A\right)\ .
\end{aligned}
\end{equation}
Recall that by Remark~\ref{rem:194},
$|e(x+j\omega)-\zeta_0(j^{(1)},m^{(1)})|\gtrsim \eta^{-1}$  for any
$x\in\tor$ and $j\in\IZ$. Let $0 < \delta \ll \sigma \ll 1$ and $N =
\left[\eta^{-1-\kappa}\right]$, where $0 < \kappa$ is arbitrary, be
such that $N = q_r$ where $q_r^{-1}p_r$ is a convergent for
$\omega$.  Then $\min\limits_{0 < j < q} \|j\omega \| \gtrsim
q_r^{-1}$.  Hence
\begin{equation}\nn
\sum_{0 \le j \le q_r} \left|\cW_{m^{(t)}} \left(e(x +
j\omega)\right) \right |^2 \lesssim \eta^{-1} q_r \exp\left((\log
m^{(1)})^A\right)
\end{equation}
for any $e(x) \in \tor \setminus \bigcup\limits_{n, j^{(1)}, j}\,
\cD\left(\zeta_n \left(j^{(1)}, n^{(1)}\right) e(j\omega)\right),
\exp\left(-(m^{(t)})^\sigma\right)$. Due to
Corollary~\ref{cor:inducm} and Lemma~\ref{lem:154} , there exist
subsets $\Omega_{m^{(s)}} \subset \tor$, $s = 1, 2,\dots$,
$\mes\Omega'_{m^{(s)}} < \exp \left(-(\log m^{(s)})^{A_2}\right)$,
$\compl \Omega '_{m^{(s)}} < \exp\left((\log m^{(s)})^{A_1}\right)$
such that for any $\omega \in \tor_{c, a} \setminus \bigcup\limits_s
\Omega'_{m^{(s)}}$, there exist subsets $\cE'_{m^{(s)},\omega}
\subset \IR$, $\mes \cE'_{m^{(s)},\omega} < \exp\left(-(\log
m^{(s)})^{A_2}\right)$, $\compl\left(\cE'_{m^{(s)},\omega}\right) <
\exp\left((\log m^{(s)})^{A_1}\right)$ such that for any $E \in \IR
\setminus \bigcup\limits_s\cE'_{m^{(s)},\omega}$, $x \in \tor$ one
has
\begin{equation}
\begin{aligned}
& \#\biggl(\Bigl(\rsp H_N(x,\omega)\Bigr)\cap \bigl(E_0 - \eta, E_0 + \eta\bigr)\biggr) \lesssim\\
&\quad \eta^2\cdot \bigl(\log m^{(1)}\bigr)^B \sum_{1 \le k \le N}
\left| \left(\left(H_N(x,\omega) - E_0 - i\eta\right)^{-1} e_k,
e_k\right)\right|^2
\end{aligned}
\end{equation}
 The first part of Theorem~\ref{thm:2} follows now.
 The second part  follows from Lemma~\ref{lem:12.7}.
\end{proof}

\begin{proof}[Proof of Theorem~\ref{thm:3}]  Let $\left(E'_n, E''_n\right)$,
$1 \le n \le \bar n$ be disjoint intervals with $\varepsilon =
\sum\limits_n \left(E''_n - E'_n\right) \ll 1$. Set $\tau =
\min\limits_n \left(E''_n - E'_n\right)$.  Let $\omega_r = p_r
q_r^{-1}$ be a convergent of $\omega$ with $q_r > \tau^{-4}$.  Let
$m^{(s)}$, $s = 1, 2,\dots, t+1$ be integers such that: (1)~ $\log
\bigl( m^{(s+1)}\bigr) \asymp \big(m^{(s)}\big)^\delta$, $s = 1,
2,\dots,t$, (2)~$\varepsilon >
\exp\left(-m^{(1)}\right)>\sqrt{\varepsilon}$, $m^{(t+1)} = q_r=:N$.
Using the notations of the proof of Theorem~\ref{thm:1}, one has
$$
{1\over N} \int \#\Bigl(\rsp H_N(x,\omega)\cap \Bigl({\ell\over
N^{1/2}}\,  {\ell+1\over N^{1/2}}\Bigr)\Bigr)\, dx \lesssim m^{(1)}
N^{-1/2}
$$
for any interval $\left({\ell\over N^{1/2}}\ , {\ell +1\over
N^{1/2}}\right)\subset \IR \setminus \bigcup\limits^{t+1}_{s=1}\,
\cE_\omega^{(s)}$, where  $\ell \in \IZ$, provided $\omega \in
\tor_{c,a} \setminus \bigcup\limits_s \Omega^{(s)}$. Let
$\left\{\cI_\ell: \ell \in \cL\right\}$ be the collection of such
intervals.  Then
\begin{equation}\nn
{1\over N} \int \#\left(\rsp H_N(x,\omega) \cap \Biggl(\bigcup_{\ell \in \cL, \cI_\ell
\subset \bigcup\bigl(E'_n, E''_n\bigr)} \cI_\ell\Biggr)\right) dx \le m^{(1)} \varepsilon\ .
\end{equation}
Let $\cL' = \left\{\ell \in \cL: \cI_\ell\cap \left\{E'_n,
E''_n\::\: n = 1, 2,\dots, \bar n\right\} \ne \emptyset \right\}$.
Then $\#\cL' \le 2\bar n$.  Since $\bar n \lesssim \tau^{-1}$ one
obtains:
\begin{equation}\nn
{1\over N} \int \#\left(\rsp H_N(x,\omega)\cap \Biggl(\bigcup_{\ell
\in \cL'} \cI_\ell\Biggr)\right) dx \lesssim m^{(1)} N^{-1/2}\bar n
< m^{(1)} \tau\ .
\end{equation}
Finally, using the H\"older bound of Theorem~\ref{thm:1} as well as
the measure and complexity bounds on the exceptional sets
$\cE^{(s)}$ yields
$$
{1\over N} \int \#\left(\rsp H_N(x,\omega)\cap
\Biggl(\bigcup^{t+1}_{s=1} \cE^{(s)}_\omega\Biggr)\right) dx <
\exp\left(-(\log m^{(1)})^A\right)
$$
and we are done.
\end{proof}

\end{document}